\documentclass[structabstract]{aa}
\pdfoutput=1
\usepackage{txfonts}
\usepackage{pictex}
\usepackage{latexsym}
\usepackage{graphics}
\usepackage{afterpage}
\usepackage{graphicx}
\usepackage{color}

\begin{document}

\title
{
Broad-band spectral energy distribution of 3000 \AA\ break quasars from the Sloan Digital Sky Survey
}

\author{H. Meusinger      \inst{1,2}
         \and 
        P. Schalldach     \inst{1}
         \and 
        A. Mirhosseini    \inst{2}
         \and
        F. Pertermann     \inst{1} 
         }
         
    \institute{
         \inst{1} Th\"uringer Landessternwarte Tautenburg, Sternwarte 5, D--07778
         Tautenburg, Germany,
               e-mail: meus@tls-tautenburg.de \\               
         \inst{2} University Leipzig, Faculty of Physics and Geosciences, Linn\`estr. 5, 04103 Leipzig, Germany       
        }
\date{Received / Accepted }

\abstract 
  {In past decades, huge surveys have confirmed the existence of populations of exotic and 
   hitherto unknown quasar types. The discovery and investigation of these rare peculiar 
   objects is important because they may represent links to special evolutionary stages and hold 
   clues to the evolution of quasars and galaxies.
   }
   {
   The Sloan Digital Sky Survey (SDSS) discovered the unusual quasars \object{J010540.75-003313.9} and 
   \object{J220445.27+003141.8} and a small number of similar objects.
   Their spectra are characterised by a break in the continuum around 3000 \AA\ that 
   neither shows the typical structure of broad absorption line (BAL) troughs nor is 
   explained by typical intrinsic dust reddening.  The main aim of the present paper was twofold.
   First, a new target-oriented search was performed in the spectra database of the SDSS to construct a sizable
   sample of such 3000 \AA\ break quasars. Second, their broad-band spectral energy distribution (SED) 
   was compared with SEDs of BAL quasars.   
   }
   {
   We used the method of Kohonen self-organising maps for data mining in 
   the SDSS spectra archive to search for more quasars with properties comparable to the prototypes 
   \object{J010540.75-003313.9} and \object{J220445.27+003141.8}. We constructed a sample of 3000 \AA\ break quasars and
   comparison samples of quasars with similar properties, to some extent, but also showing indications
   for typical BAL features. 
   Particular attention was payed to a possible contamination by rare stellar spectral types, in particular DQ white 
   dwarfs. We construct ensemble-averaged broad-band  SEDs based on archival data from SDSS, GALEX, 2MASS, 
   UKIDSS, WISE, and other surveys. The SEDs were corrected for dust absorption at the systemic redshifts 
   of the quasars by the comparison with the average SED of normal quasars.
   }
   {
   We compiled a list of 23 quasars classified as 3000 \AA\ break quasars with
   properties similar to \object{010540.75-003313.9} and \object{J220445.27+003141.8}.
   Their de-reddened arithmetic median composite SED is indistinguishable 
   from that of the unusual BAL quasars. We conclude that 3000 \AA\ break quasars are 
   most likely extreme versions of BAL quasars. Assuming that the intrinsic SED 
   of the continuum source is represented by the quasar composite SED, the 3000 \AA\ break quasars 
   tend to be intrinsically more luminous than ordinary quasars.
   }
  {}      

\titlerunning{3000 \AA\ break quasars}
\authorrunning{H. Meusinger et al.}

\maketitle

%
%
\section{Introduction}\label{sec:intro}
%
%

In the past two decades, huge spectroscopic surveys have produced a tremendous increase in 
the database of known quasars. This exciting evolution also led to the discovery of 
a variety of quasar spectra that dramatically differ from the standard spectral 
energy distribution (SED) of active galactic nuclei (AGN) and confirmed the existence of populations 
of exotic and hitherto unknown quasar types. These rare peculiar objects possibly represent links to 
special evolutionary stages and provide a laboratory for studying the processes 
of potential importance for the cosmic structure evolution, such as
feedback mechanisms between accretion activity and star formation. 
With its efficient quasar selection technique and huge number 
of high-quality spectra, the Sloan Digital Sky Survey (SDSS; York et al. \cite{York2000}), 
is uniquely qualified to significantly increase the database of unusual quasars. 
Based on early SDSS data, Hall et al. (\cite{Hall2002a}) discussed in detail the spectra of about 20 unusual 
quasars that are characterised by unusual broad absorption line (BAL) properties, in some cases 
combined with heavily reddened continua. 

The present paper focuses on unusual SDSS spectra of the type of two objects 
referred to as ``mysterious objects'' by Hall et al. (\cite{Hall2002a}),
\object{J010540.75-003313.9} and \object{J220445.27+003141.8}.
Their characterising spectral features are a lack of both substantial typical emission lines and
obvious absorption troughs in combination with a drop-off around 3000 \AA\footnote{The spectra
appear unusually red at $\lambda \la 3000$ \AA, yet they show the typical blue continuum at longer wavelengths.},
which appears too steep to be caused by dust reddening with
reddening laws taken as typical of quasars. Following Jiang et al. (\cite{Jiang2013}), we call this
feature  ``3000 \AA\ break'' throughout the present paper.
Both two prototypical 3000 \AA\ break quasars show associated \ion{Mg}{ii} $\lambda\lambda 2796.3, 2803.5$  
absorption, are unresolved FIRST radio sources, and are much more luminous than any galaxy. 
Regardless of their physical interpretation, these objects almost certainly represent quasar 
types that are extremely rare in presently available samples.
The most convincing explanation to date is unusually broad structures from line absorption 
with broad smooth outflow velocities, probably in combination with moderate dust reddening 
(Hall et al. \cite{Hall2002a}; Green \cite{Green2006}). 

Quasar broad absorption lines (BALs), usually seen as blue-shifted troughs, are 
naturally explained as footprints of powerful, sub-relativistic (up to $\sim 0.2c$) outflows 
(Foltz et al. \cite{Foltz1983}; 
Weymann et al. \cite{Weymann1991}), 
probably related to radiation-driven winds from the accretion disc 
(Murray \& Chiang \cite{Murray1998}; 
Elvis \cite{Elvis2000}; 
Everett et al. \cite{Everett2002}) 
and are thus the most obvious manifestation of matter accelerated 
by the central engine of AGNs (Fabian \cite{Fabian2012}; Arav et al. \cite{Arav2013}). 
BAL quasars are usually classified into those showing broad absorption lines
from only high-ionisation species, such as \ion{C}{iv} and \ion{N}{v} (HiBALs),
and those showing in addition also absorption from low-ionisation species, such
as \ion{Mg}{ii} and \ion{Al}{iii} (LoBALs). The comparison of samples of LoBAL 
and HiBAL quasars showed that the former have substantially redder ultraviolet (UV) 
continua and stronger UV \ion{Fe}{ii} and \ion{Fe}{iii} emission 
(Weymann et al. \cite{Weymann1991}).

A rare subclass of the LoBAL quasars are the FeLoBAL quasars showing low-ionisation 
BALs from metastable excited states of \ion{Fe}{ii}\, and \ion{Fe}{iii}.
The catalogued FeLoBAL quasars constitute only $\sim$ 2\% of the 
BAL quasars and $\sim$ 0.3\% of the total quasar population, where the 
rate of incidence measured strongly depends on the selection criteria
and the true FeLoBAL quasar fraction may be considerably higher
(e.g., Trump et al. \cite{Trump2006};
Urrutia et al. \cite{Urrutia2009}; 
Dai et al. \cite{Dai2012}; 
Lucy et al. \cite{Lucy2014}).
Unlike HiBAL quasars, LoBALs  may not be explained by orientation alone 
(e.g.,  Ghosh  \& Punsly \cite{Ghosh2007};
Montenegro-Montes et al. \cite{Montenegro2009};
Zhang et al. \cite{Zhang2010}).

BAL variability studies of FeLoBALs suggest that the outflowing optically thick 
gas resides on a range of scales 
(Hall et al. \cite{Hall2011}; 
Filiz Ak et al. \cite{FilizAk2012};
Vivek et al. \cite{Vivek2012};
Capellupo et al. \cite{Capellupo2013};
McGraw et al. \cite{McGraw2015}).
From the analysis of spectral variability in 28 BAL quasars,
Zhang et al. (\cite{Zhang2015}) conclude that short distances from the central 
engine are particularly characteristic of FeLoBALs with overlapping absorption troughs,
perhaps including matter in a rotating disk wind. Rotationally dominated outflows
may also produce redshifted absorption seen in a small fraction of BAL quasars, 
though infall of relatively dense gas at radii as small as several hundred Schwarzschild 
radii provides an alternative explanation (Hall et al. \cite{Hall2013}).

There is positive evidence that FeLoBAL quasars may represent an evolutionary stage of 
the quasar phenomenon when a dusty obscuring cocoon is being expelled by a powerful 
quasar-driven wind
(Voit et al. \cite{Voit1993}; 
Egami et al. \cite{Egami1996};
Becker et al. \cite{Becker1997}; 
Canalizo \& Stockton \cite{Canalizo2001};
L\'ipari et al. \cite{Lipari2009};
Dai et al. \cite{Dai2012};
Farrah et al. \cite{Farrah2007}, \cite{Farrah2012};
Morabito et al. \cite{Morabito2011};
Leighly et al. \cite{Leighly2014}
).
From a sample of 31 FeLoBAL quasars at $0.8 < z < 1$, selected on the basis of their
rest-frame UV spectral properties, Farrah et al. (\cite{Farrah2012}) concluded that
the radiatively driven outflows from FeLoBAL quasars act to curtail obscured star 
formation in the host galaxies to less than $\sim$ 25\% of the total 
infrared (IR) luminosity.
Studies of BAL quasar samples dominated by HiBALs
(Willott et al. \cite{Willott2003}; 
Gallagher et al. \cite{Gallagher2007}; 
Cao Orrjales et al. \cite{Cao2012}; 
Pu \cite{Pu2015}) 
indicate that BAL quasars are not significantly different from non-BAL quasars in terms of their 
IR to UV spectral energy distribution (SED) and their IR luminosities. A similar result
was published for LoBAL quasars by Lazarova et al. (\cite{Lazarova2012}). However,
this finding does not necessarily mean that LoBAL quasars are not related to a rapid transition 
state from a  ultra-luminous infrared galaxy (ULIRG) phase to a more quiescent phase of non-BAL quasars.
Lazarova et al.(\cite{Lazarova2012}) argue that it could be that the majority of the LoBAL quasars
in their sample could have already passed through the short phase of quenching the star formation rate 
to the normal level.

Hall et al. (\cite{Hall2002a}) quoted only two other objects with spectra that are probably similar 
to the two prototypical 3000 \AA\ quasars, namely the low-$z$  quasars \object{FBQS 1503+2330} ($z=0.40$) 
and \object{FBQS 1055+3124} ($z=0.49$) from the FIRST Bright Quasar Survey (White et al. \cite{White2000}). 
Another four quasars with similar spectra were discovered by Plotkin et
al. (\cite{Plotkin2008}, \cite{Plotkin2010}) and independently by Meusinger et al. (\cite{Meusinger2012}; 
hereafter Paper I).
All these objects were targeted for spectroscopy because they were detected as radio sources. The 
first similar object that was selected neither as a radio source nor by the SDSS colour selection was 
\object{J134246.24+284027.5} from the Tautenburg-Calar Alto variability and (zero-) 
proper motion survey (Meusinger et al. \cite{Meusinger2005}). 
The small sample of these sources, enlarged by the similar object J085502.20+280219.6, 
was presented in Paper I. 

The first aim of the present study was to increase the sample of quasars with 
properties characteristic of 3000 \AA\ break quasars. Rather than trying to perform an unbiased search,
our approach is focused on selecting a larger sample from the SDSS spectra archive. 
The tremendous amount of data produced by the SDSS requires efficient methods for browsing the huge archive. 
We developed the software tool ASPECT to produce large Kohonen Self-Organising Maps (SOMs) for up 
to one million SDSS spectra (in der Au et al. 
\cite{inderAu2012}). In previous studies, we computed SOMs for the quasar spectra from the 
SDSS data release 7 (DR7; Abazajian et al. \cite{Abazajian2009}) to select unusual quasars. 
A resulting sample of about 1000 unusual quasars of different types was 
published and discussed in Paper I, a large set of 365 Kohonen-selected weak-line quasars was 
analysed by Meusinger \& Balafkan (\cite{Meusinger2014}). 
Thereafter we applied the SOM technique also to  galaxies, stars, and unknowns in the SDSS DR7, and 
to the quasar spectra in the SDSS DR10 (Ahn et al. \cite{Ahn2014}) with the main aim to search
for further 3000 \AA\ break quasar candidates. The present sample 
combines objects from Paper I and newly discovered objects from SDSS DR7 and DR10.

Section\,\ref{sec:selection} gives an overview of the quasar selection.
Our main goal is to construct a composite broad-band SED of 3000 \AA\ break quasars 
from the extreme UV to the mid IR and to compare it with the SEDs of ordinary
quasars and of unusual BAL quasars. The SEDs are the subject of Sect.\,\ref{sec:SED}.
We combined photometric data from GALEX, SDSS, 2MASS, UKIDSS, {\it WISE}, {\it Spitzer}, 
and {\it Herschel} with the main aim to search for differences between the SEDs of different types.
A particular question is the role of internal reddening at the quasar redshift.
In Sect.\,\ref{sec:disc}, we discuss three scenarios for interpreting the 3000 \AA\ break.
Finally, the results are summarised in Sect.\,\ref{sec:summary}. Individual descriptions and SEDs for the
quasars are given in two Appendixes.
 
Throughout this paper we assume a spatially flat cosmology with
$\Omega =1$, $\Omega_\Lambda = 0.7$, and $H_0 = 70$ km\,s$^{-1}$\,Mpc$^{-1}$.

%
%
\section{Quasar sample}\label{sec:selection}
%
%

\begin{table*}[hbpt]
\caption{Quasars from samples A, B, and C.}
\begin{tabular}{rcccccccccccr}
\hline
Nr& 
Name&
Source$^{\ a}$&
Spectrum&
Sample&
$z_{\rm em}$&
$z_{\rm abs}$&
f$^{\ b}$&
E$_{B-V}^{\rm smc}$&
E$_{B-V}^{\rm gask}$&
log$ L_{\rm IR}^{\,c}$&
log$ L_{3000}^{\,c}$&
$R$\\
\hline
 1 & \object{J$105528.80+312411.3$} & PI  & SDSS & A & 0.500 & 0.492 &   & 0.08 & 0.04 & 12.39 & 12.08 &$  6.2$\\
 2 & \object{J$140025.53-012957.0$} & here& BOSS & A & 0.585 & 0.585 &   & 0.08 & 0.05 & 11.87 & 11.76 &$  4.6$\\
 3 & \object{J$134050.80+152138.7$} & here& SDSS & A & 1.141 & 1.141 &   & 0.15 & 0.10 & 12.55 & 12.36 &$  1.7$\\
 4 & \object{J$091613.59+292106.2$} & PI  & SDSS & A & 1.143 & 1.143 &   & 0.07 & 0.04 & 12.62 & 12.23 &$< 1.4$\\
 5 & \object{J$010540.75-003313.9$} & PI  & BOSS & A & 1.179 & 1.179 &   & 0.15 & 0.10 & 12.98 & 12.79 &$  2.6$\\
 6 & \object{J$160827.08+075811.5$} & PI  & SDSS & A & 1.182 & 1.182 &   & 0.25 & 0.15 & 13.54 & 13.36 &$  2.2$\\
 7 & \object{J$152928.53+133426.6$} & here& SDSS & A & 1.230 & 1.230 &   & 0.15 & 0.10 & 13.02 & 12.77 &$< 0.4$\\
 8 & \object{J$134246.24+284027.5$} & PI  & VPMS & A & 1.255 & 1.255 & u & 0.05 & 0.03 & 12.51 & 12.08 &$  2.4$\\
 9 & \object{J$091940.97+064459.9$} & PI  & SDSS & A & 1.351 & 1.351 &   & 0.12 & 0.08 & 12.80 & 12.53 &$ 24.5$\\
10 & \object{J$220445.27+003141.8$} & PI  & BOSS & A & 1.353 & 1.335 &   & 0.20 & 0.14 & 13.60 & 13.37 &$  0.5$\\
11 & \object{J$111541.01+263328.6$} & here& SDSS & A & 1.355 & 1.327 & u & 0.25 & 0.18 & 12.52 & 12.24 &$  7.3$\\
12 & \object{J$161836.09+153313.5$} & PI  & SDSS & A & 1.359 & 1.158 &   & 0.22 & 0.16 & 12.99 & 12.72 &$  2.4$\\
13 & \object{J$130941.35+112540.1$} & PI  & BOSS & A & 1.362 & 1.362 &   & 0.12 & 0.07 & 12.98 & 12.78 &$  1.1$\\
14 & \object{J$152423.18+115312.0$} & here& SDSS & A & 1.420 & 1.420 & u & 0.07 & 0.04 & 12.70 & 12.46 &$  2.3$\\
15 & \object{J$215950.30+124718.4$} & PI  & BOSS & A &   -   & 1.514 &   & 0.00 & 0.00 & 12.76 & 12.59 &$< 0.9$\\
16 & \object{J$085502.20+280219.6$} & PI  & SDSS & A & 1.515 & 1.515 &   & 0.13 & 0.09 & 12.46 & 12.35 &$< 1.6$\\
17 & \object{J$134408.32+283932.0$} & PI  & BOSS & A &   -   & 1.767 &   & 0.35 & 0.20 & 13.58 & 13.27 &$  3.8$\\
18 & \object{J$145045.56+461504.2$} & PI  & BOSS & A &   -   & 1.877 &   & 0.10 & 0.06 & 13.27 & 12.96 &$  1.4$\\
19 & \object{J$110511.15+530806.5$} & PI  & BOSS & A & 1.929 & 1.925 &   & 0.22 & 0.13 & 13.26 & 13.19 &$< 0.3$\\
20 & \object{J$075437.85+422115.3$} & PI  & BOSS & A & 1.950 & 1.950 &   & 0.23 & 0.17 & 13.13 & 12.89 &$  1.7$\\
21 & \object{J$092602.98+162809.2$} & here& SDSS & A & 2.090 & 2.090 & u & 0.20 & 0.15 & 12.91 & 13.01 &$< 0.6$\\
22 & \object{J$013435.66-093103.0$} & here& SDSS & A & 2.220 & 0.766 &   & 0.42 & 0.27 & 14.11 & 14.15 &$ 64.8$\\
23 & \object{J$100933.22+255901.1$} & here& BOSS & A & 2.235 & 2.218 & u & 0.20 & 0.12 & 13.75 & 13.27 &$  1.9$\\
\hline
24 & \object{J$141428.30+185646.0$} & PI  & SDSS & B & 0.609 & 0.609 &   & 0.42 & 0.25 & 12.20 & 12.18 &$< 0.5$\\
25 & \object{J$143821.39+094623.1$} & PI  & BOSS & B & 0.804 & 0.804 &   & 0.20 & 0.12 & 12.31 & 11.98 &$ 14.8$\\
26 & \object{J$094225.42+565613.0$} & PI  & SDSS & B &   -   & 0.822 & u & 0.10 & 0.07 & 12.33 & 12.06 &$  1.9$\\
27 & \object{J$134951.93+382334.1$} & PI  & BOSS & B & 1.093 & 1.093 &   & 0.17 & 0.11 & 12.57 & 12.29 &$  2.6$\\
28 & \object{J$152438.79+415543.0$} & PI  & BOSS & B & 1.229 & 1.090 &   & 0.18 & 0.13 & 12.79 & 12.65 &$  1.8$\\
29 & \object{J$120337.91+153006.6$} & PI  & SDSS & B & 1.235 & 1.130 & u & 0.02 & 0.00 & 12.81 & 12.54 &$  1.4$\\
30 & \object{J$164941.87+401455.9$} & here& SDSS & B & 1.266 & 1.266 & u & 0.07 & 0.05 & 12.67 & 12.56 &$< 0.7$\\
31 & \object{J$153341.88+150059.5$} & here& SDSS & B &   -   & 1.420 & u & 0.32 & 0.22 & 12.61 & 12.58 &$< 0.9$\\
32 & \object{J$165117.31+240836.3$} & here& SDSS & B &   -   & 1.785 & u & 0.12 & 0.10 & 12.59 & 12.56 &$  3.1$\\
33 & \object{J$101723.04+230322.1$} & PI  & SDSS & B &   -   & 1.794 &   & 0.10 & 0.07 & 13.32 & 13.05 &$  3.2$\\
34 & \object{J$151627.40+305219.7$} & PI  & SDSS & B &   -   & 1.846 &   & 0.05 & 0.03 & 13.41 & 13.10 &$  0.7$\\
35 & \object{J$073816.91+314437.0$} & PI  & BOSS & B &   -   & 2.010 &   & 0.10 & 0.07 & 13.09 & 12.83 &$  1.4$\\
36 & \object{J$103453.55+241332.8$} & here& SDSS & B &   -   & 2.080 & u & 0.00 & 0.00 & 13.04 & 12.77 &$< 1.0$\\
37 & \object{J$021102.33-081007.4$} & PI  & BOSS & B & 2.200 & 2.015 &   & 0.00 & 0.00 & 12.74 & 12.51 &$< 2.0$\\
38 & \object{J$012412.47-010049.7$} & here& BOSS & B & 2.884 & 1.115 & u & 0.00 & 0.00 & 13.79 & 13.75 &$< 0.2$\\
\hline
39 & \object{J$030000.57+004828.0$} & here& SDSS & C &   -   & 0.877 &   & 0.00 & 0.00 & 13.05 & 12.82 &$< 0.2$\\
40 & \object{J$123058.37+275924.2$} & here& SDSS & C & 0.972 & 0.952 &   & 0.03 & 0.02 & 12.25 & 11.87 &$< 2.4$\\
41 & \object{J$115436.60+030006.3$} & here& BOSS & C &   -   & 1.390 & u & 0.16 & 0.10 & 13.18 & 12.98 &$  0.3$\\
42 & \object{J$145729.62+042655.8$} & here& SDSS & C &   -   & 1.423 &   & 0.28 & 0.16 & 12.77 & 12.79 &$  1.6$\\
43 & \object{J$164226.61+205955.2$} & PI  & SDSS & C & 1.892 & 1.800 &   & 0.05 & 0.04 & 13.14 & 12.69 &$< 1.0$\\
44 & \object{J$100237.22+270056.5$} & PI  & BOSS & C & 1.984 & 1.724 &   & 0.00 & 0.00 & 13.28 & 13.29 &$< 0.3$\\
45 & \object{J$163313.26+352050.7$} & here& BOSS & C & 2.000 & 2.000 &   & 0.07 & 0.04 & 13.12 & 12.98 &$< 0.6$\\
46 & \object{J$173049.10+585059.5$} & PI  & SDSS & C &   -   & 2.000 &   & 0.20 & 0.13 & 13.55 & 13.33 &$< 0.3$\\
47 & \object{J$140259.67+145615.5$} & here& BOSS & C &   -   & 2.075 &   & 0.00 & 0.00 & 12.81 & 12.71 &$< 1.1$\\
48 & \object{J$080202.69+140315.1$} & here& SDSS & C &   -   & 2.120 &   & 0.00 & 0.00 & 13.00 & 12.96 &$< 0.7$\\
49 & \object{J$170044.00+175417.5$} & here& SDSS & C &   -   & 2.139 & u & 0.20 & 0.08 & 13.39 & 13.47 &$< 0.2$\\
50 & \object{J$140419.53+342807.5$} & PI  & BOSS & C &   -   & 2.160 &   & 0.20 & 0.20 & 12.97 & 12.82 &$< 0.9$\\
51 & \object{J$094317.59+541705.1$} & PI  & BOSS & C &   -   & 2.230 & u & 0.20 & 0.10 & 13.58 & 13.66 &$  0.4$\\
52 & \object{J$162527.73+093332.8$} & PI  & SDSS & C &   -   & 2.290 &   & 0.15 & 0.10 & 13.26 & 13.20 &$< 0.4$\\
53 & \object{J$093224.48+084008.1$} & PI  & BOSS & C &   -   & 2.300 &   & 0.00 & 0.00 & 12.73 & 12.74 &$< 1.2$\\
54 & \object{J$113212.92+010441.3$} & here& BOSS & C & 2.343 & 2.330 &   & 0.10 & 0.07 & 13.45 & 13.12 &$< 0.5$\\
55 & \object{J$094328.94+140415.6$} & here& BOSS & C &   -   & 2.370 &   & 0.13 & 0.08 & 13.38 & 13.35 &$< 0.3$\\
56 & \object{J$101912.84+410807.4$} & PI  & BOSS & C & 2.415 & 2.145 & u & 0.07 & 0.05 & 13.12 & 13.30 &$< 0.4$\\
57 & \object{J$153702.77+062203.8$} & PI  & BOSS & C &   -   & 2.470 &   & 0.15 & 0.12 & 13.41 & 13.49 &$< 0.2$\\
58 & \object{J$103014.21+262534.9$} & here& BOSS & C &   -   & 2.640 & u & 0.10 & 0.12 & 12.89 & 12.59 &$< 2.1$\\
59 & \object{J$100424.89+122922.2$} & here& BOSS & C & 2.650 & 2.600 &   & 0.30 & 0.18 & 13.73 & 13.87 &$  2.0$\\
\hline
\end{tabular}
\label{tab:sample}
\end{table*}

\addtocounter{table}{-1}
\begin{table*}[hbpt]
\caption{Quasars from samples A, B, and C (continued).}
\begin{tabular}{rcccccccccccr}
\hline
Nr&
Name&
Source\tablefootmark{\,a}&
Spectrum&
Sample&
$z_{\rm em}$&
$z_{\rm abs}$&
f\tablefootmark{\,b}&
E$_{B-V}^{\rm smc}$&
E$_{B-V}^{\rm gask}$&
log$ L_{\rm IR}$\tablefootmark{\,c}&
log$ L_{3000}$\tablefootmark{\,c}&
$R$\\
\hline
60 & \object{J$083257.57+264528.1$} & here& BOSS & C &   -   & 2.700 &   & 0.00 & 0.00 & 13.21 & 12.94 &$< 1.0$\\
61 & \object{J$124140.50+012228.5$} & here& BOSS & C &   -   & 2.760 &   & 0.13 & 0.08 & 13.06 & 13.19 &$< 0.6$\\
62 & \object{J$113734.05+024159.3$} & here& SDSS & C &   -   & 2.767 &   & 0.00 & 0.00 & 13.34 & 13.14 &$< 0.6$\\
63 & \object{J$094719.38+140809.8$} & here& BOSS & C &   -   & 3.010 & u & 0.20 & 0.20 & 13.00 & 13.10 &$< 0.8$\\
64 & \object{J$005041.59+143755.9$} & here& BOSS & C &   -   & 3.520 &   & 0.00 & 0.00 & 13.19 & 12.82 &$ -\  $\\
\hline
\end{tabular}
\tablefoot{
\tablefoottext{a}{PI = paper I, here = present study;}
\tablefoottext{b}{redshift uncertainty flag: u = uncertain;}
\tablefoottext{c}{in $L_\odot$}
}
\label{tab:sample}
\end{table*}

\subsection{Sample selection}\label{sec:sample}

In Paper I, we presented a sample of nine quasars with properties similar to 
\object{J010540.75-003313.9} and \object{J220445.27+003141.8} from Hall et al. (\cite{Hall2002a})
in combination with 11 possibly
related quasars that may represent links between that type and other odd quasar types such as unusual
BAL quasars. With the only exception of \object{J134246.24+284027.5} from Meusinger et al. (\cite{Meusinger2005}), all 
quasars were selected from a systematic search for spectroscopic outliers among the objects from SDSS Data Release 7 
(DR7; Abazajian et al. \cite{Abazajian2009}) classified by the spectroscopic pipeline as quasars. 

While the spectroscopic pipeline of the SDSS works accurately and efficiently for the vast majority of the
spectra, it tends to fail in case of very unusual ones. We developed the software package ASPECT, which is able to project a huge number of spectra into a sorted topological map (in der Au et al. \cite{inderAu2012}).
The approach is based on the Kohonen method (Kohonen \cite{Kohonen2001}) of self-organising maps (SOMs). 
This is an artificial network algorithm that uses unsupervised 
learning to produce a two-dimensional mapping of higher-dimensional input data (spectra), where the 
spectra are sorted (`clustered') by their relative similarity. Because of its clustering properties, the
resulting SOM provides an efficient tool for selecting certain spectral types or spectroscopic outliers.
In Paper I, we computed 37 SOMs for about 10$^5$ SDSS quasars binned in narrow redshift intervals. The final 
selection of unusual quasar spectra was based on a combination of the clustering power of the SOMs with their
quick visual inspection where special attention was paid to the problem of contamination by other, rare spectral 
types.

\begin{figure}[h]
\includegraphics[viewport=120 20 560 780,angle=270,width=9.0cm,clip]{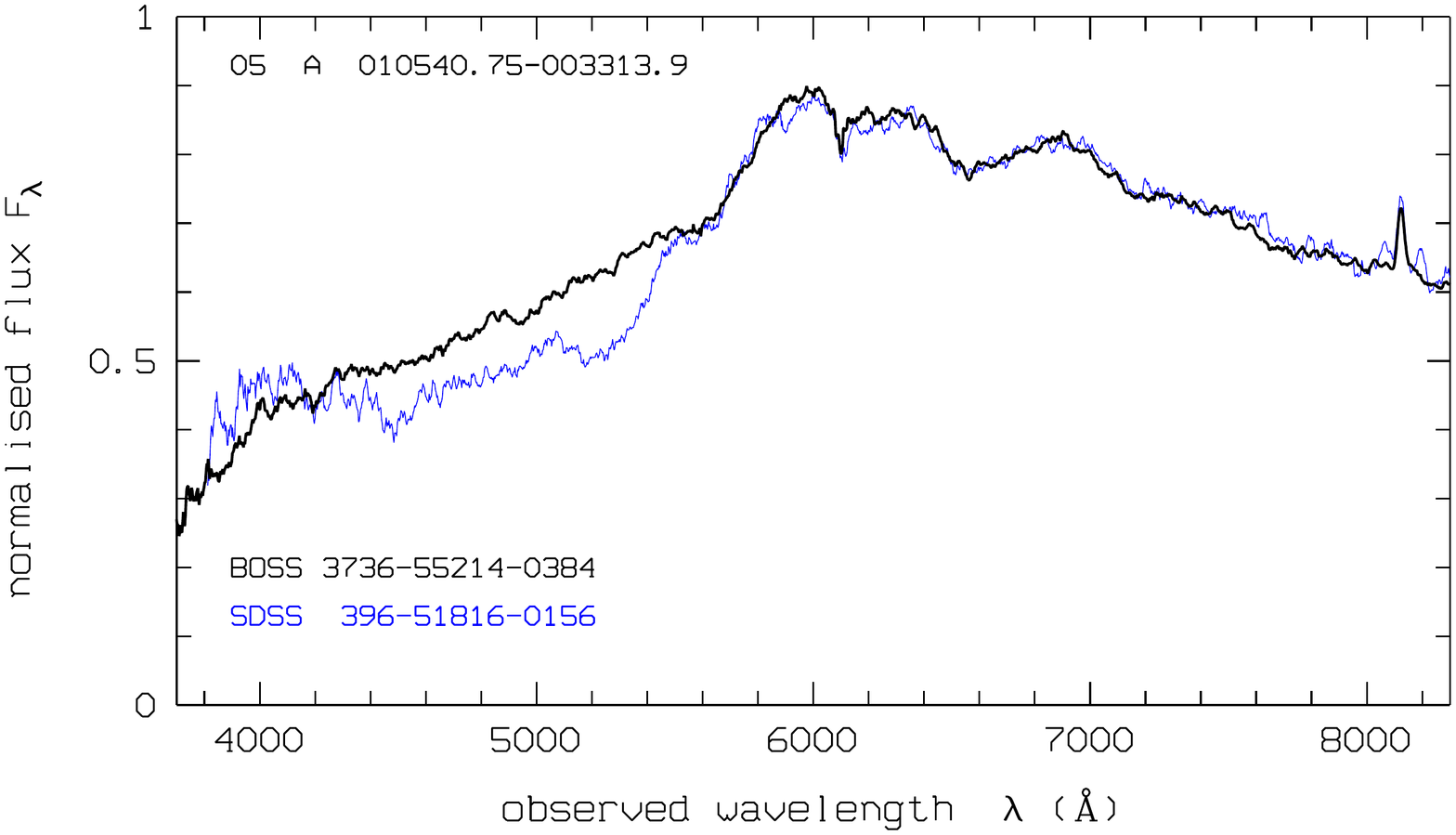}
\includegraphics[viewport=120 20 560 780,angle=270,width=9.0cm,clip]{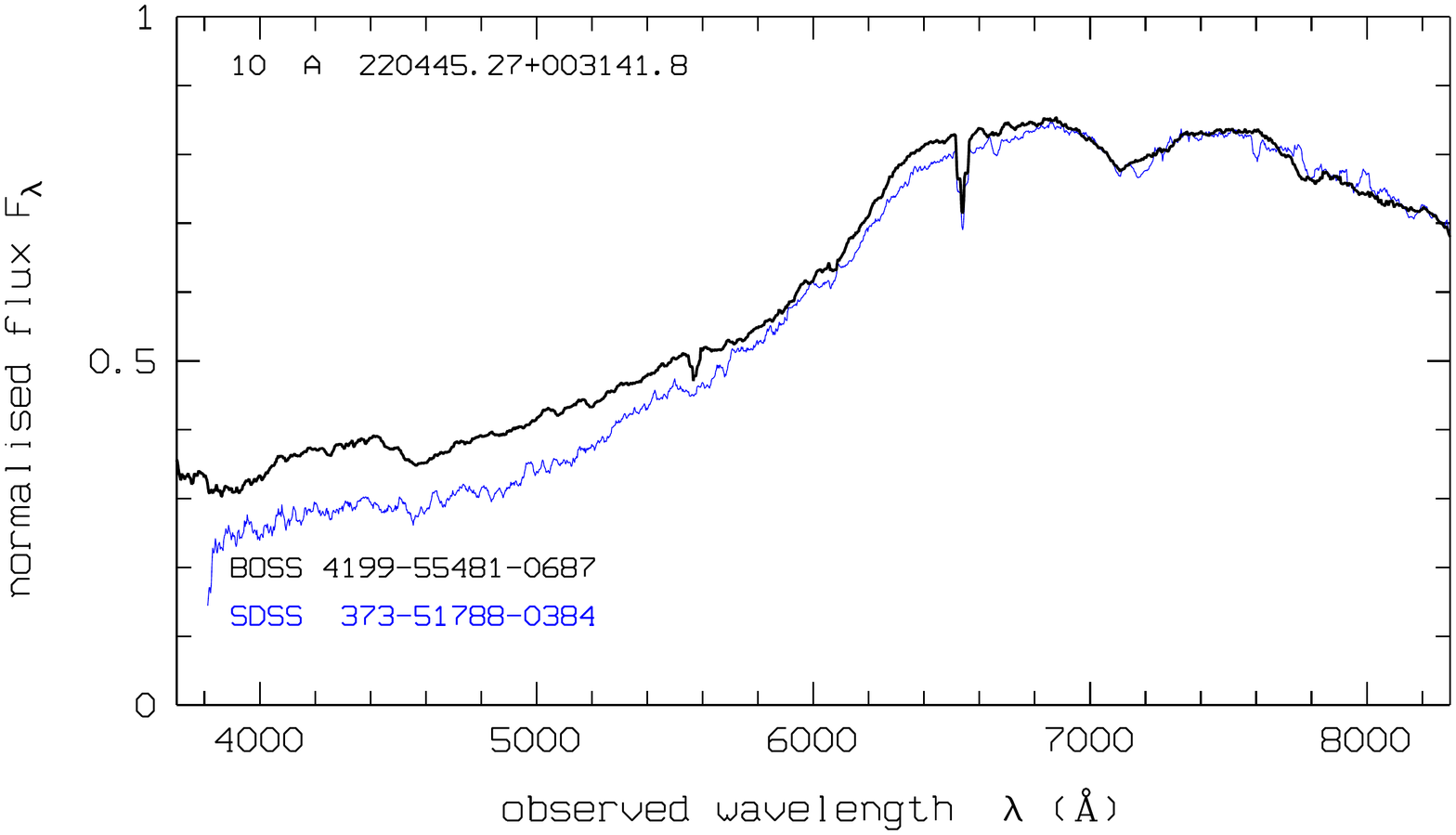}
\caption{BOSS (black) and SDSS (blue) spectra of the two prototypical objects \object{J010540.75-003313.9}
and \object{J220445.27+003141.8}, arbitrarily normalised in the wavelength interval 6500 - 7000 \AA.
The numbers at the bottom of each panel give plate number -- MJD -- fibre number.}
\label{fig:two_spectra}
\end{figure}

In the present study, we aimed to create a larger sample of 3000 \AA\ break quasars by a systematic search 
in the SDSS spectra archive. The selection method is basically the same as in Paper I. We continued computing
SOMs for different sets with large numbers of spectra. First, in addition to the objects classified as
quasars, we computed SOMs for other spectral classes from the SDSS DR7:  $\sim4\cdot10^4$ stars (40 SOMs), 
$\sim8\cdot10^5$ galaxies (160 SOMs), and $\sim10^4$ unclassified objects (spectral class = {\sc unknown}) with 
sufficiently high signal-to-noise ratios (5 SOMs).\footnote{
The SOM for the nearly one million spectra from the SDSS DR7 will be made available to the community at
http://aspect-ui.de/sdssdr7/.
}
In a second step, we clustered $\sim 3\cdot10^5$ quasar spectra 
from SDSS DR10 (Ahn et al. \cite{Ahn2014}). All SOMs were inspected by eye to select candidates for 3000 \AA\ 
break quasars and odd BAL quasars. After examining all selected spectra individually and rejecting contaminants 
(see below), a sample of 30 quasars was selected from this new search. 

The final sample for the present study was compiled from the newly selected quasars in combination 
with 34 quasars from Paper I. Following Paper I, we defined three subsamples:
\vspace{-0.2cm}
\begin{itemize}
\item[(A)]{Objects with optical-to-UV spectra similar to \object{J010540.75-003313.9} and \object{J220445.27+003141.8.}
The characteristic spectral features are:
  \begin{itemize}
  \item[(a)] the lack of both substantial typical quasar emission lines (except broad \ion{Fe}{ii} emission
  and [\ion{O}{ii}] 3730 in some cases) and obvious BAL troughs,
  \item[(b)] a typical blue continuum at restframe wavelengths $\ga 3000$ \AA,
  \item[(c)] a continuum drop-off at around 3000 \AA\ that appears too steep
  to be caused by dust reddening (see Hall et al. \cite{Hall2002a} for a more detailed description).
  \end{itemize} 
The sample contains 23 quasars, including the nine objects from the upper 
part of Table 8 in Paper I. These objects are referred to as 3000 \AA\ break quasars 
throughout this paper.}
\item[(B)]{Fifteen quasars with spectra that resemble those in sample A to some degree but not in all aspects. 
The sample includes the objects from the lower part of Table 8 in Paper I.}
\item[(C)]{A comparison sample of 26 BAL quasars with extreme BALs and probably strong dust reddening, 
including the objects shown in Fig.\,11 of Paper I.}
\end{itemize}
Some quasars from samples B and C show a pronounced peak in the spectrum around 3000 \AA. 
Nevertheless, they were not classified as sample A objects because the break can be 
clearly attributed to LoBALs. The most extreme objects of this type \object{J173049.10+585059.5} (\#46) and 
\object{J094317.59+541705.1} (\#52) have already been discussed in Paper I.

The SDSS and BOSS spectra of the two prototypical sample A objects from Hall et al. (\cite{Hall2002a}), 
\object{J010540.75-003313.9} and \object{J220445.27+003141.8}, are shown in Fig.\,\ref{fig:two_spectra}. 
The time lag between the two spectra corresponds to 4.3 yr in the rest frame for both quasars.
Quasars are variable on time scales of months to years (Kelly et al. \cite{Kelly2009};
MacLeod et al. \cite{MacLeod2010}) and 3000 \AA\ break quasars are also known to be
variable (Meusinger et al. \cite{Meusinger2005}, \cite{Meusinger2011}). It is thus not 
surprising that the two spectra are not fully identical. However, the variability properties 
of the two quasars appear to be different (see Sect.\,\ref{sec:disc_a}).

The quasars from the three samples are listed in Table\,\ref{tab:sample}. 
Remarks on single objects are given in Appendix A. The optical spectra are shown on the 
left-hand side of Fig.\,\ref{fig:spectra_and_seds} in Appendix B (see Sect.\,\ref{sec:individual_SEDs}).
The redshifts $z$ of the new quasars were estimated as described in Paper I. The process requires detailed 
analysis of each spectrum individually. For most quasars in samples A and B, 
emission lines could be identified (80\% and 53\%, respectively). On the other hand, it is in the nature of 
the unusual BAL quasars in sample C that for most objects $z$ could be measured from absorption lines only (73\%).
Nevertheless, we feel that redshifts should be correct to within a few percent for the vast majority. 
Quasars with uncertain redshift estimates were flagged in Table\,\ref{tab:sample}.  

\begin{figure}[h]
\includegraphics[viewport=80 20 560 780,angle=270,width=9.0cm,clip]{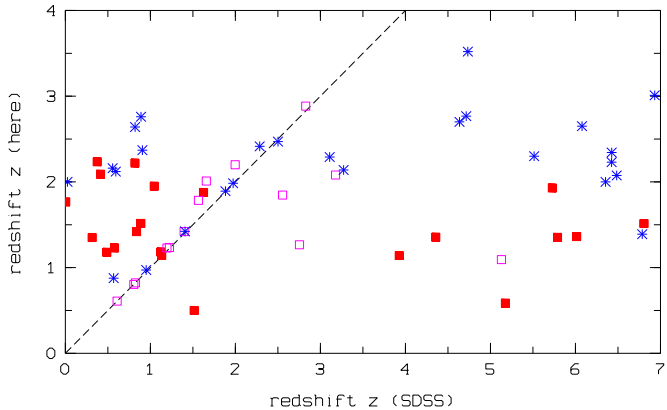}
\caption{Redshifts estimated in the present study compared with the redshifts from the SDSS DR12 
spectroscopic pipeline for the present samples (A: filled red squares; B: open magenta squares; 
C: blue asterisks).}
\label{fig:z_here_z_sdss}
\end{figure}

In Fig. \ref{fig:z_here_z_sdss}, we compare the redshifts from the present study with those from 
the SDSS spectroscopic pipeline as provided by the SDSS DR12 explorer page. To guide the eye, the 1:1 relation 
(dashed line) is over-plotted. Although the pipeline is efficient and accurate for the overwhelming majority of
sources, there is a substantial danger of misinterpretations in the case of unusual spectra like those from the
present study. It has long been known that BAL quasars mimic high-$z$ quasars (Appenzeller et al.
\cite{Appenzeller2005}). In fact, 41\% of the SDSS quasars in sample A were targeted as high-$z$ quasars
(SDSS target flag PrimTarget = {\sc target\_qso\_hiz}) and the 3000 \AA\ break in the spectrum 
was obviously mismatched by the pipeline with the depression of the continuum shortwards of the Lyman $\alpha$ 
line due to either the Lyman forest or the Lyman break for many of the quasars from Table\,\ref{tab:sample}. 
Others were wrongly identified as galaxies at relatively small $z$.

\begin{figure}[h]
\includegraphics[viewport=200 20 560 780,angle=270,width=9.0cm,clip]{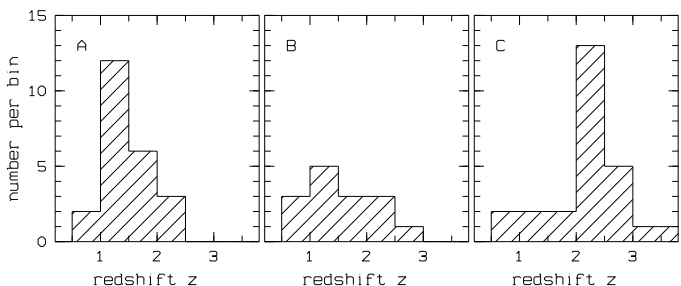}
\caption{Histograms of the redshift distribution in samples A, B, and C. 
}
\label{fig:z_hist}
\end{figure}

The redshift distributions are shown in Fig.\,\ref{fig:z_hist}. The distributions are similar for 
samples A and B with mean values $\overline{z} =$ 1.457 and 1.54, respectively. 
At such redshifts the \ion{Mg}{ii} 2800\AA\ line is in 
the middle of the SDSS and BOSS spectra so that the continuum at either side of the break is 
clearly visible. Sample C is weighted towards higher redshifts ($\overline{z} \approx 2.2$) 
where the optical spectrum is completely 
dominated by the intrinsic UV part where the absorption lines of FeLoBAL quasars are concentrated.

\subsection{Uncertain classifications and possible contamination} 

\begin{figure*}[bhpt]
\begin{tabbing}
\includegraphics[viewport=50 20 490 650,angle=270,width=4.9cm,clip]{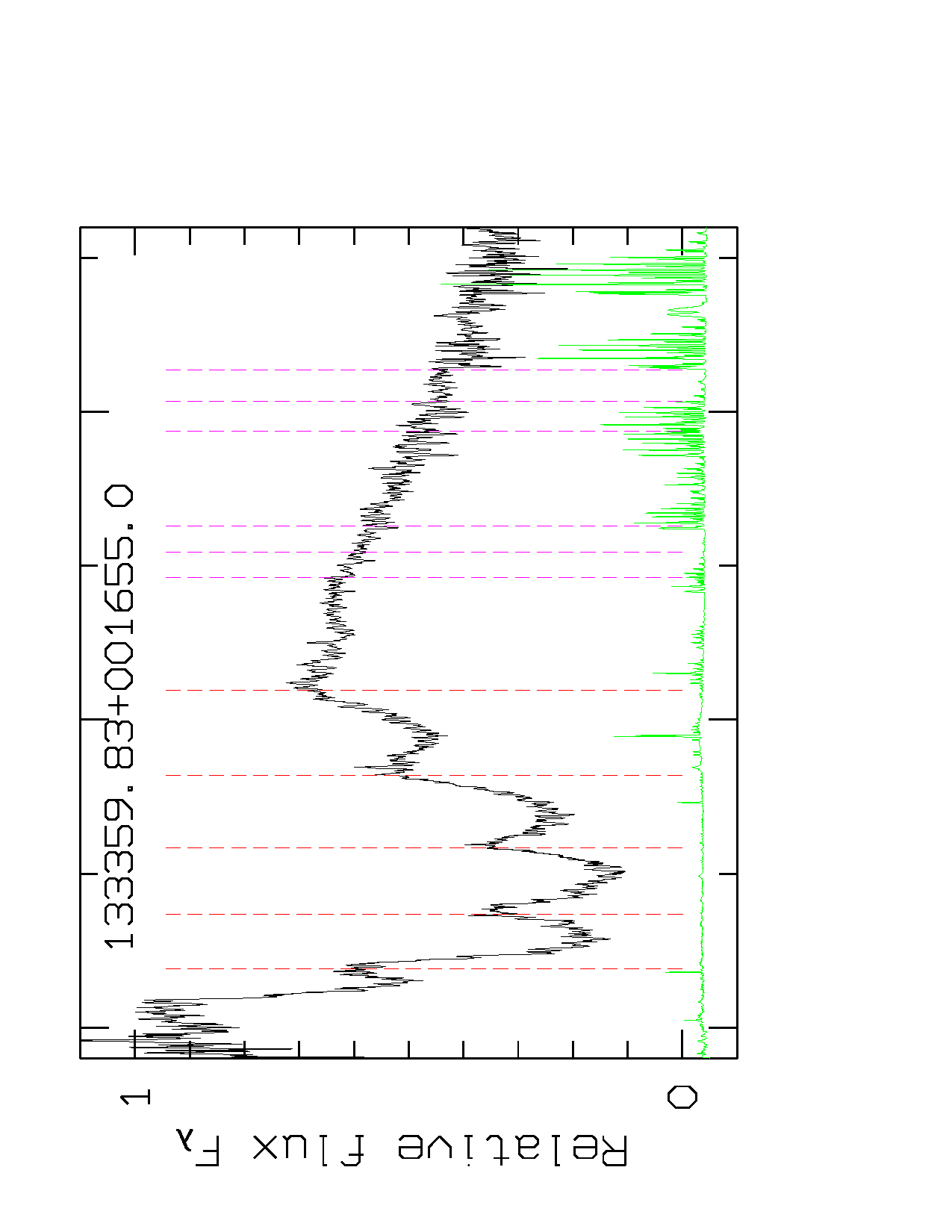}\hfill \=
\includegraphics[viewport=50 70 490 650,angle=270,width=4.5cm,clip]{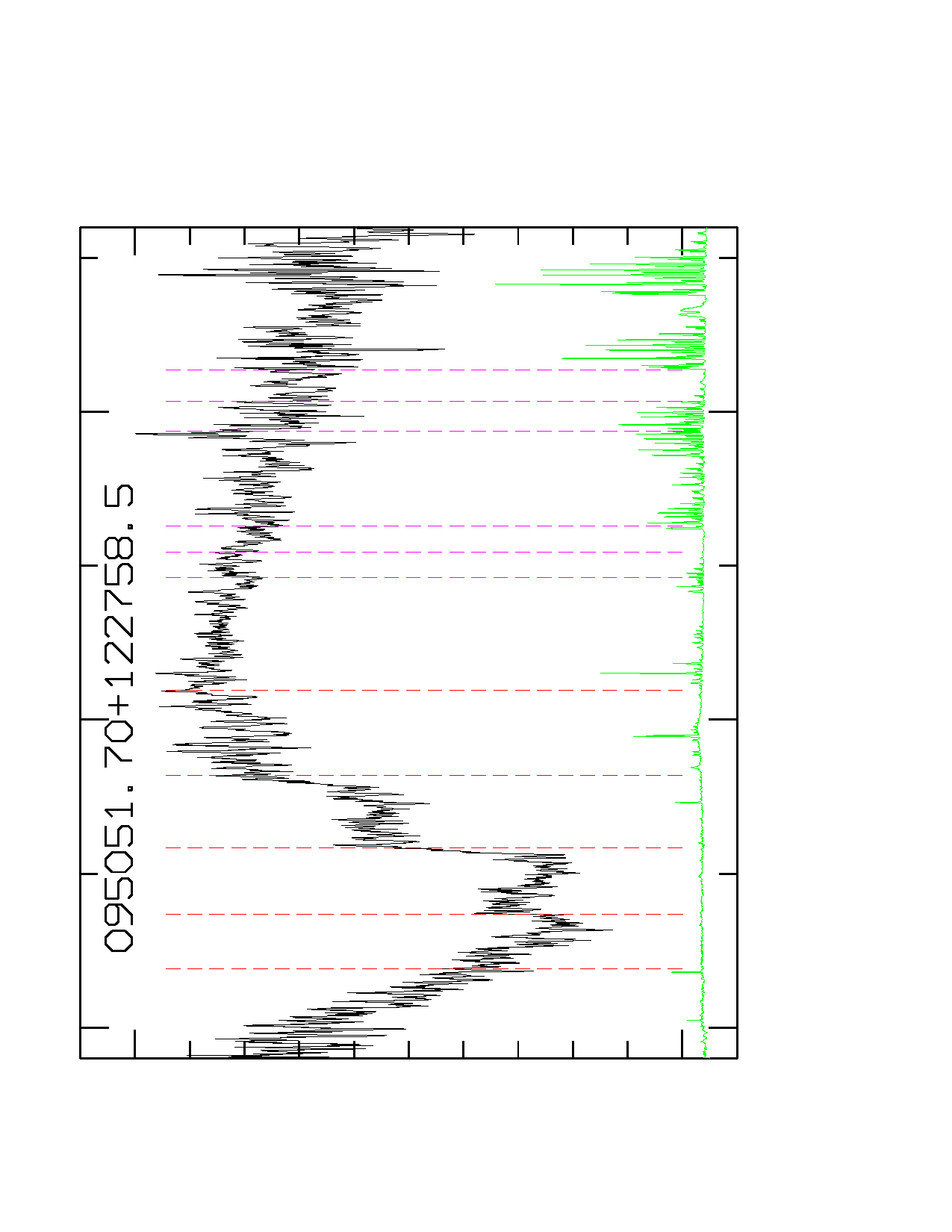}\hfill \=
\includegraphics[viewport=50 70 490 650,angle=270,width=4.5cm,clip]{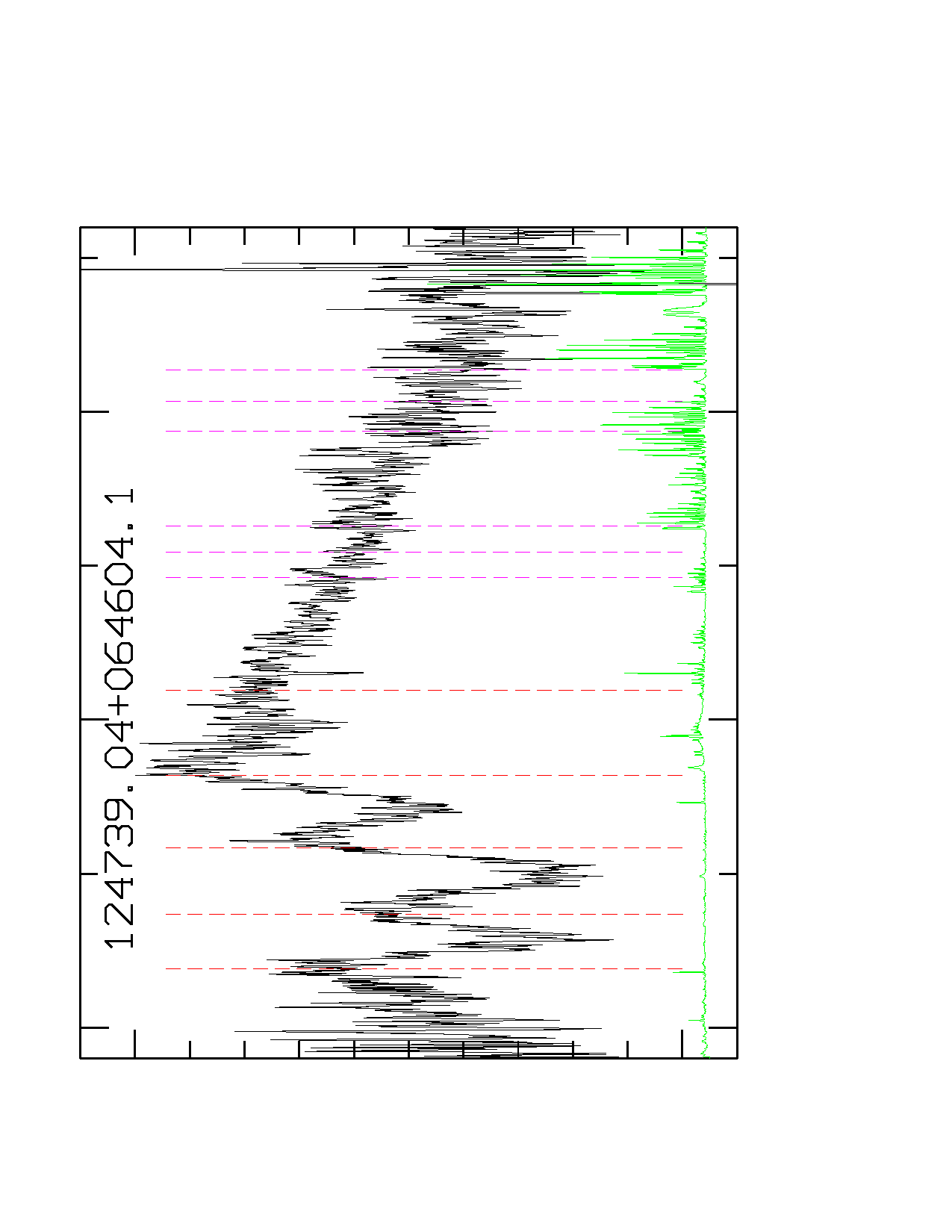}\hfill \=
\includegraphics[viewport=50 70 490 650,angle=270,width=4.5cm,clip]{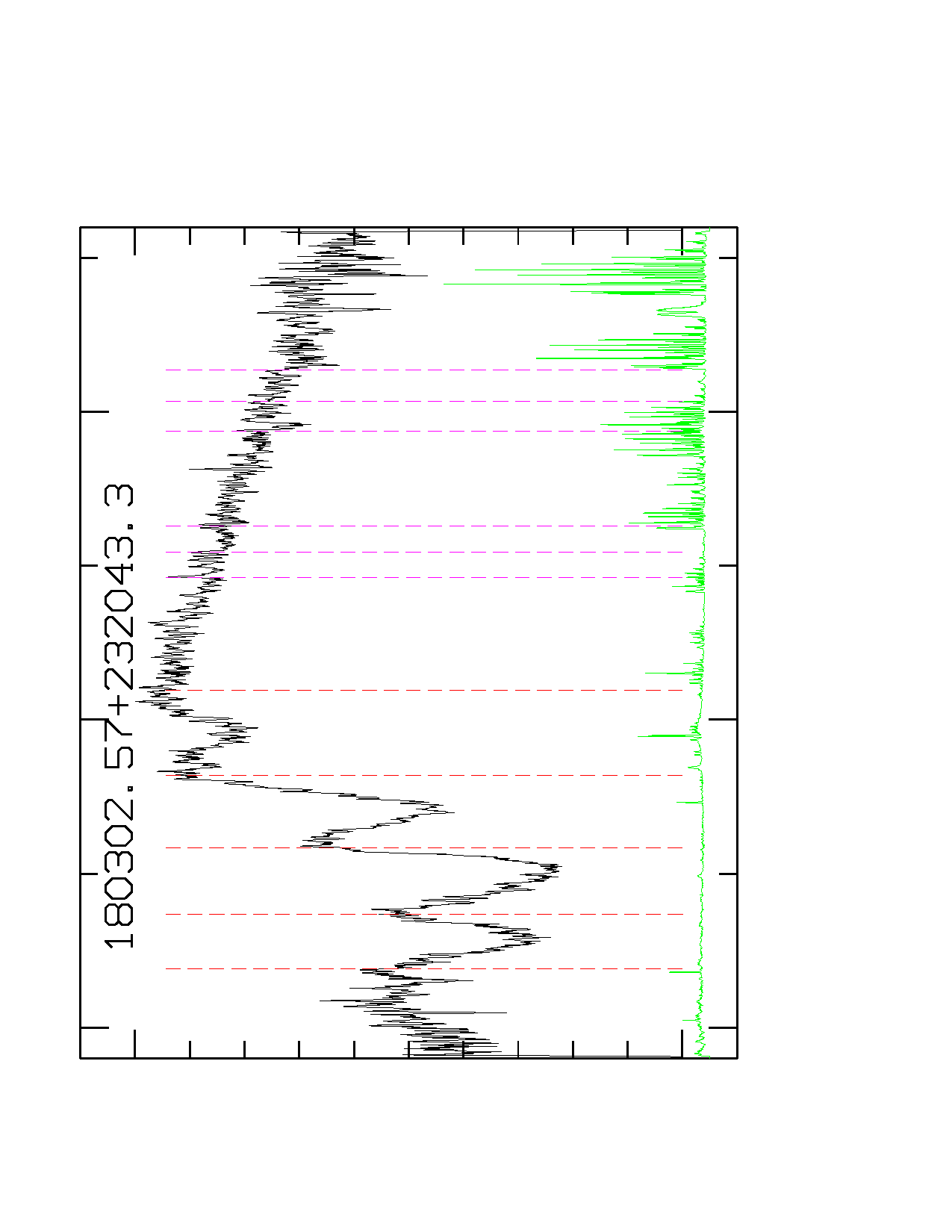}\hfill \\
\includegraphics[viewport=50 20 560 650,angle=270,width=4.9cm,clip]{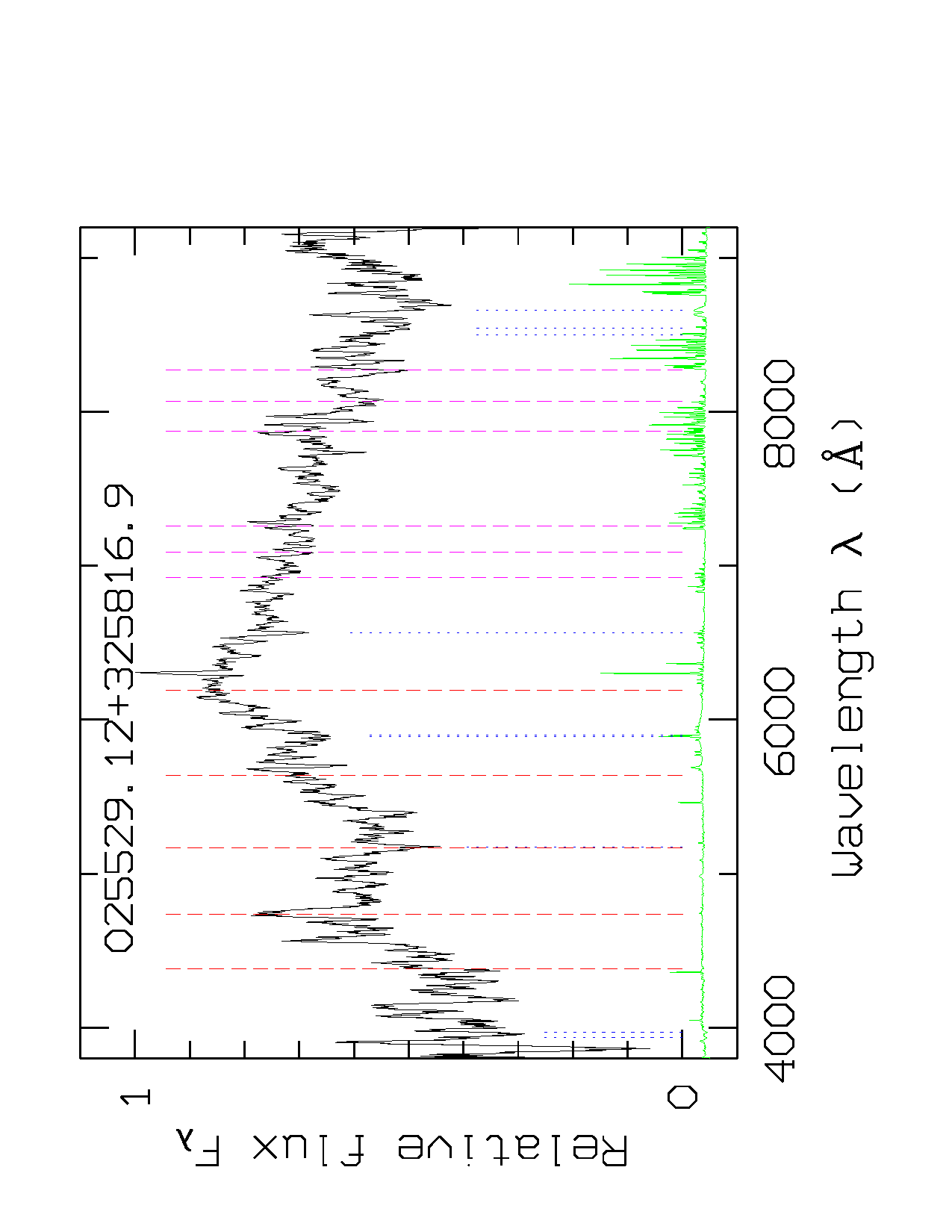}\hfill \=
\includegraphics[viewport=50 70 560 650,angle=270,width=4.5cm,clip]{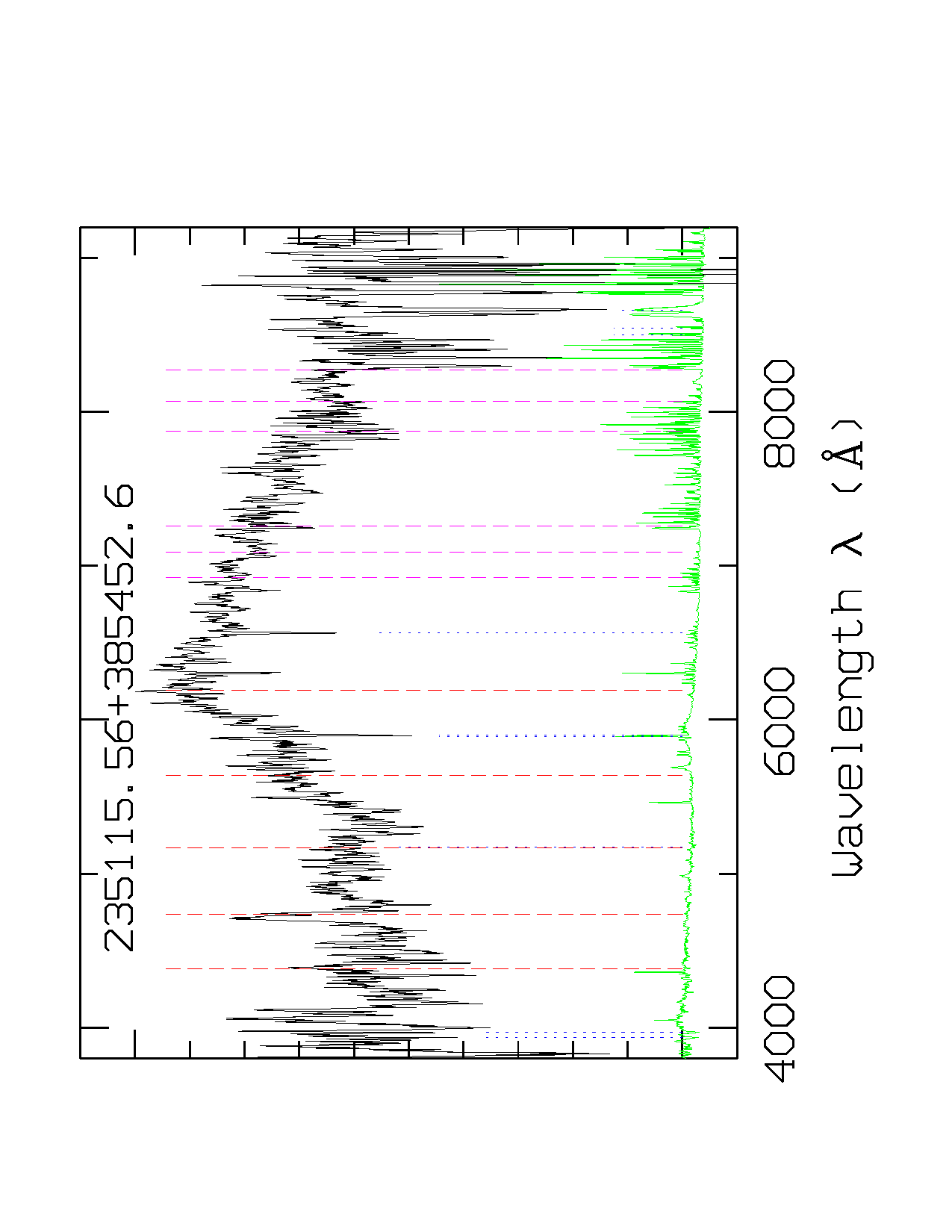}\hfill \=
\includegraphics[viewport=50 70 560 650,angle=270,width=4.5cm,clip]{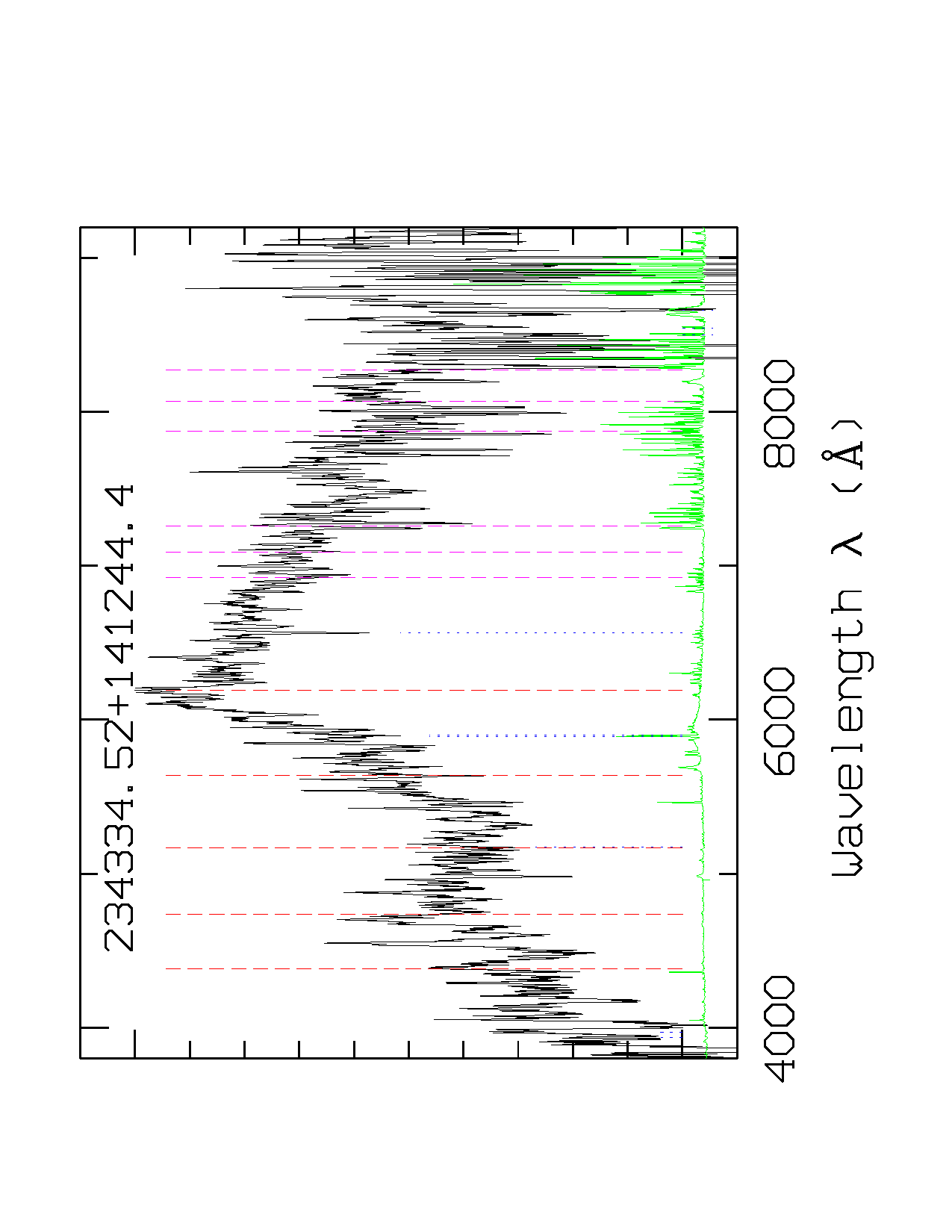}\hfill \=
\includegraphics[viewport=50 70 560 650,angle=270,width=4.5cm,clip]{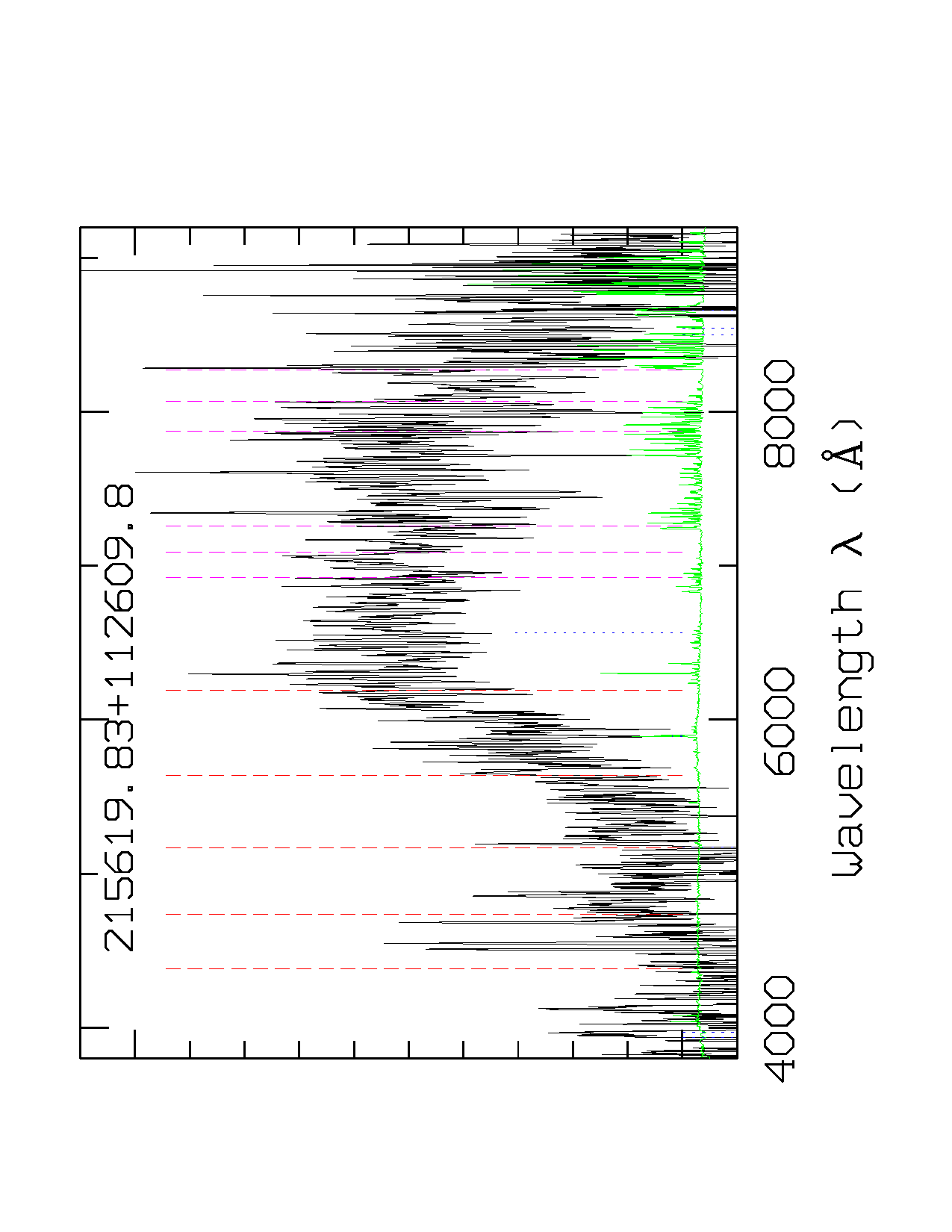}\hfill
\end{tabbing}
\caption{DQ WDs (top) and uncertain spectral types (bottom).}
\label{fig:uncertain_and_cont}
\end{figure*}

There is a certain danger in the selection process of wrongly classifying peculiar stellar types as 3000 \AA\ break quasars.
This holds in particular for peculiar DQ white dwarfs (DQ WDs; Pelletier et al. \cite{Pelletier1986}; 
Koester \& Knist \cite{Koester2006}; Hall \& Maxwell \cite{Hall2008}) with their characteristic absorption 
troughs resulting from from sequences of Swan bands of the C$_2$ molecules. The absorption troughs can be deep, widely 
overlapping, and sometimes apparently distorted (Kowalsky \cite{Kowalsky2010}). 
In combination with a blue continuum at longer wavelengths, the abrupt decline in the spectrum around 
$\lambda \approx 6000$ \AA\ towards shorter wavelengths makes such spectra similar,
in some respects, to 3000 \AA\ break quasars at 
$z \approx 1$. 

At the top of Fig.\,\ref{fig:uncertain_and_cont}, the SDSS spectra of four peculiar DQ WDs are shown to illustrate the effect. The bandheads of the usually strongest sequences of C$_2$ bands (Tanabashi et al. \cite{Tanabashi2007})
were over-plotted as vertical red dashed lines and those from the red CN bands (Davis et al. \cite{Davis1987}) as magenta lines. The downscaled sky spectrum is over-plotted in green at the bottom of each panel.
\object{J133359.83+001655.0} is a rare peculiar cool magnetic DQ (DQpec) with broad molecular bands 
(Schmidt et al. \cite{Schmidt2003}; Vornanen et al. \cite{Vornanen2013}). 
\object{J095051.70+122758.5} looks quite similar, with pronounced band heads at roughly 5600\AA, 5200\AA, and 4700\AA, and 
is obviously a newly discovered peculiar DQ.\footnote{At least, it is listed neither in the SDSS DR7 white dwarf catalogue 
from Kleinman et al. (\cite{Kleinman2013}) nor in the catalogue of carbon stars and DQ white dwarfs from Green 
(\cite{Green2013}).}
\object{J124739.04+064604.5} and \object{J180302.57+232043.3}, also classified as peculiar DQs (Kilic et al. \cite{Kilic2010}; Limoges et al. \cite{Limoges2013}; Kleinman et al. \cite{Kleinman2013}), look similar to the former two spectra, but display a declining flux density towards the shortest wavelengths. All four stars have significant proper motions between 200 and 400 mas yr$^{.1}$.

The interpretation of the spectra in the bottom row of Fig.\,\ref{fig:uncertain_and_cont} is much more uncertain. 
None of them was found in the SIMBAD database\footnote{http://simbad.u-strasbg.fr/simbad/}. The 
spectra show a break at $\sim 6200$\AA\ that could be interpreted as a redshifted 3000 \AA\ break of a quasar. 
As an alternative, it can be identified with the bandhead of the reddest of the five marked C$_2$ troughs in C stars, 
where all other other Swan bands are, however, only weakly indicated, if at all. 
It is very unusual for DQs that the $\sim 4700$\AA\ trough is missing, while the $\sim 6200$\AA\ trough is so strong.
The proper motions are essentially zero for 
\object{J235115.56+385452.6} ($9 \pm 13$ mas yr$^{-1}$) and \object{J234334.52+141244.4} ($0 \pm 0$ mas yr$^{-1}$), 
not measured for \object{J215619.83+112609.8}, and only marginally significant for \object{J025529.12+325816.9} ($34 \pm 12$ mas yr$^{-1}$). On the other hand, a stellar origin
is indicated by the presence of stellar absorption lines at $z=0$
indicated by the blue dotted vertical lines (from left to right: \ion{Ca}{ii}$\lambda\lambda$ 3935,3970, 
\ion{Mg}{i} $\lambda$ 5175, \ion{Na}{i} $\lambda\lambda$ 5892,5898, H$\alpha$, and the \ion{Ca}{ii} infrared triplet). 
Though a definitive classification remains unclear, we conclude that the first three, 
\object{J025529.12+325816.9}, \object{J235115.56+385452.6}, \object{J234334.52+141244.4} are most likely stars,
perhaps peculiar carbon stars. The spectrum of \object{J215619.83+112609.8} is simply too noisy. 
We emphasise that we paid particular attention in the selection of 3000 \AA\ break quasar candidates to 
minimise the risk of contamination by objects with such uncertain classification as for those in the bottom row of 
Fig.\,\ref{fig:uncertain_and_cont}.

We checked our final sample from Table\,\ref{tab:sample} also against the catalogue of optically selected BL Lac objects 
from SDSS-DR7 (Plotkin et al. \cite{Plotkin2010}). Only one source, \object{J075437.85+422115.3}, was classified there as a 
high-confidence BL Lac candidate.

\section{Spectral energy distributions}\label{sec:SED}

\subsection{Observational data}\label{sec:obsdat}

The present study takes advantage of existing archival data. 
The major surveys exploited here, apart from the SDSS, 
are the {\it Two Micron All-Sky Survey} (2MASS; Skrutskie et al. \cite{Skrutskie2006}), the 
{\it UKIRT Infrared Deep Sky Survey} (UKIDSS; Lawrence et al. \cite{Lawrence2007}; Hambly et al. \cite{Hambly2008}) 
in the near infrared (NIR), the {\it Wide-Field Infrared Survey Explorer} 
(WISE; Wright et al. \cite{Wright2010}) in NIR and mid infrared (MIR), the {\it Faint Images of the Radio Sky at 
Twenty Centimeters} (FIRST; White et al. \cite{White2000}), and the data products from from the 
{\it Galaxy Evolution Explorer} (GALEX; Morrissey et al. \cite{Morissey2007}) in the near and far 
ultraviolet (NUV, FUV). 

The VizieR Service at CDS Strasbourg\footnote{http://vizier.u-strasbg.fr/viz-bin/VizieR} was used to 
extract the data from the {\it SDSS Photometric Catalog, Release 7} (Abazajian et al. \cite{Abazajian2009}), 
the {\it GALEX-DR5 (GR5) sources from AIS and MIS} (Bianchi et al. \cite{Bianchi2011}), and 
the {\it UKIDSS-DR9 Large Area Surveys} (Lawrence et al. \cite{Lawrence2007}). 
The NASA/IPAC Infrared Science Archive 
(IRSA)\footnote{http://irsa.ipac.caltech.edu/applications/Gator/index.html} was used to identify the SDSS 
sources in the 
{\it 2MASS All-Sky Point Source Catalogue} and in the {\it AllWISE Source Catalog} from WISE. 
IRSA was also used to search for counterparts in the {\it IRAS Faint Source Catalogue} (Moshir et al. \cite{Moshir1992}) 
in the far infrared (FIR) and in the catalogues from the {\it Spitzer Space Observatory} (Werner et al. 
\cite{Werner2004}) in MIR and FIR. 
The {\it Herschel Stripe 82 Survey} (Viero \cite{Viero2014}) was checked for sub-mm detections 
using VizieR. No source was detected by IRAS, one source by {\it Herschel}, and only a few sources 
by {\it Spitzer}. A summary of the number of detections in the various photometric bands is given in 
Table\,\ref{tab:detections}.

\begin{table}[hbpt]
\caption{Photometric bands, wavelengths, and numbers of detections $N$ per band for samples A, B, and C, and total number.}
\begin{tabular}{lrrrrr}
\hline
Band         &$\lambda$\,[$\mu$m]& $N_A$&$N_B$&$N_C$& $N_{\rm tot}$\\
\hline
GALEX FUV    &   0.1539     &  1 &  0 &  0 &  1 \\   
GALEX NUV    &   0.2316     & 11 &  5 &  3 & 19 \\   
SDSS u       &   0.3543     & 22 & 15 & 21 & 58 \\   
SDSS g       &   0.4770     & 23 & 15 & 25 & 63 \\   
SDSS r       &   0.6231     & 23 & 15 & 26 & 64 \\   
SDSS i       &   0.7625     & 23 & 15 & 26 & 64 \\   
SDSS z       &   0.9134     & 23 & 15 & 25 & 63 \\   
UKIDSS Y     &   1.0305     & 10 &  3 & 10 & 23 \\   
UKIDSS J     &   1.2483     & 10 &  3 & 10 & 23 \\   
UKIDSS H     &   1.6316     & 10 &  4 & 11 & 25 \\   
UKIDSS K     &   2.2010     & 10 &  4 & 11 & 25 \\   
2MASS J      &   1.235      &  7 &  4 & 10 & 21 \\   
2MASS H      &   1.662      &  7 &  3 & 10 & 20 \\   
2MASS K      &   2.159      &  7 &  3 & 10 & 20 \\   
WISE W1      &   3.4        & 23 & 15 & 26 & 64 \\   
WISE W2      &   4.6        & 23 & 15 & 26 & 64 \\   
WISE W3      &  12          & 23 & 15 & 25 & 63 \\   
WISE W4      &  22          & 20 & 13 & 19 & 52 \\   
IRAC F1      &   3.6        &  2 &  0 &  0 &  2 \\   
IRAC F2      &   4.5        &  2 &  1 &  0 &  3 \\   
IRAC F3      &   5.8        &  2 &  1 &  0 &  3 \\   
IRAC F4      &   8.0        &  2 &  1 &  0 &  3 \\   
MIPS F1      &  24          &  2 &  1 &  2 &  5 \\   
MIPS F2      &  70          &  0 &  0 &  2 &  2 \\   
MIPS F3      & 160          & C 0 &  0 &  1 &  1 \\  
Herschel 250 & 250          &  1 &  0 &  0 &  1 \\   
Herschel 350 & 350          &  1 &  0 &  0 &  1 \\   
Herschel 500 & 500          &  1 &  0 &  0 &  1 \\   
FIRST        & $2\cdot10^5$ & 17 &  9 &  4 & 30 \\   
\hline
\end{tabular}
\label{tab:detections}
\end{table}

\begin{figure}[h]
\includegraphics[viewport=30 40 580 800,angle=270,width=9.4cm,clip]{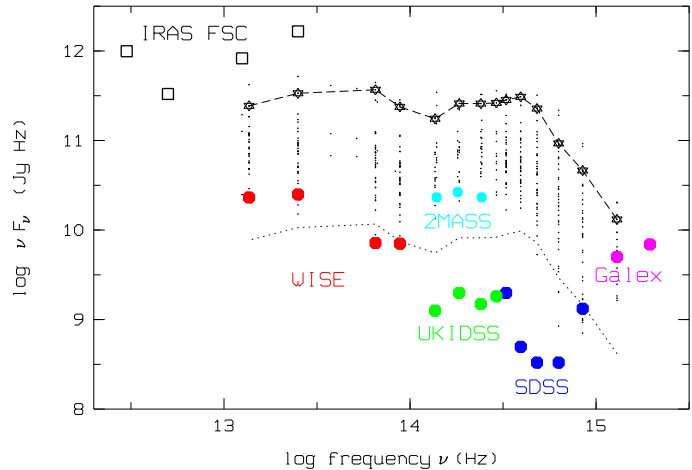}
\caption{Observed fluxes for the quasars from Table\,\ref{tab:sample} (small black dots) and sensitivity limits of
the major surveys used for the present study (filled symbols) and for the IRAS Faint Source Catalogue (open squares).
Star symbols and dashed curve: \object{J220445.27+003141.8};
dotted curve: \object{J220445.27+003141.8} shifted by 1.5 dex downwards.
}
\label{fig:survey_limits}
\end{figure}

Magnitudes were transformed into fluxes and the fluxes were corrected for Galactic foreground 
extinction using $E(B-V)$ from Schlafly \& Finkbeiner 
(\cite{Schlafly2011})\footnote{http://irsa.ipac.caltech.edu/applications/DUST}.
Figure\,\ref{fig:survey_limits} shows the observed UV, optical, and IR fluxes of the quasars from our sample 
after foreground extinction correction. Also plotted are the flux limits of the corresponding major surveys. 
The GALEX limits were adopted from Bianchi et al. (\cite{Bianchi2007}) for GALEX/SDSS matched sources with
photometric errors $< 0.3$ mag, but the limits are a function of the field position and can thus be lower.
We used the Mikulski Archive for Space Telescopes (MAST)\footnote{https://galex.stsci.edu/GR6/}
to select the faintest GALEX sources in fields of 10 arcmin size around our quasars. In nearly all fields, GALEX
sources were found that are several tenths of a magnitude fainter than the above-mentioned limits.
The survey limits given for the four WISE bands are point source sensitivities in unconfused regions on the
ecliptic\footnote{http://wise2.ipac.caltech.edu/docs/release/allsky/}. 
The WISE sensitivity improves towards the ecliptic poles due to denser coverage and lower zodiacal background. 

For illustration purposes, we over-plotted in Fig.\,\ref{fig:survey_limits} the prototypical 3000 \AA\ 
break quasar \object{J220445.27+003141.8}. Provided that the SED of our quasars does not increase
much more strongly towards the FIR than for \object{J220445.27+003141.8}, the IRAS Faint Source Catalogue
(Moshir et al. \cite{Moshir1992}) is not deep enough to contain the brightest quasars from our sample.
On the other hand, a substantial part of our sample is above the NUV limit, but not above the FUV limit.
Nevertheless, upper GALEX limits provide useful information on the shape of the SED
at the shortest wavelengths. The SED of \object{J220445.27+003141.8} shifted towards fainter fluxes by 1.5 dex is
indicated by the dotted curve. It demonstrates that even the faintest quasars should be detected by UKIDSS and in the
WISE bands w1 and w2 if their SED is similar to that  of \object{J220445.27+003141.8}.

Among the 64 quasars from from Table\,\ref{tab:sample}, 28 were found to have radio counterparts 
in the {\it FIRST Survey Catalog, Version 2014Dec17} (Helfand et al. \cite{Helfand2015}).
The fields of the remaining 36 objects were inspected on the FIRST image  
cutouts\footnote{http://third.ucllnl.org/cgi-bin/firstcutout} to search for possible sources slightly below the
hard catalogue limit of 1 mJy. Only one additional radio source was found, \object{J115436.60+030006.3} in addition 
to \object{J134246.24+284027.5}, which was already known as a sub-mJy source. 
The FIRST radio fluxes were not directly involved in the study of the SED, but were used to estimate the
fraction of radio-loud quasars in our samples (Sect.\,\ref{sec:RL}).

The optical spectra were downloaded from the SDSS DR12 (Alam et al. \cite{Alam2015}), the spectrum of
\object{J134246.24+284027.5} is from Meusinger et al. (\cite{Meusinger2005}). For 33 quasars, BOSS spectra are available that have a wider wavelength coverage than the spectra from the SDSS spectrograph. If a quasar had both a SDSS and a BOSS spectrum, we always downloaded the latter. All spectra were corrected for Galactic foreground extinction, again using $E(B-V)$ from Schlafly \& Finkbeiner (\cite{Schlafly2011}).

\subsection{Comparison of the individual SEDs with the SED of ordinary quasars}

\subsubsection{Fitting the photometric data}

In general, the wavelength coverage in the IR is too poor to fit model SEDs 
to the observational data in a same way as done, for examplem by Farrah et al. (\cite{Farrah2012}) and 
Lazarova et al. (\cite{Lazarova2012}). Instead we restricted our analysis of the individual quasar SEDs 
to a qualitative comparison with the composite SED of normal quasars. 
We assumed that the SED is composed of an UV/optical part and an IR part, 
roughly separated by the $1\mu$m inflection. According to the AGN standard model, the accretion 
disk is responsible for the UV and optical continuum in the big blue bump (BBB), 
whereas the IR bump is attributed to the putative dusty torus and a possible starburst component in the 
quasar host galaxy. 

\begin{figure}[h]
\includegraphics[viewport=30 40 580 800,angle=270,width=9.2cm,clip]{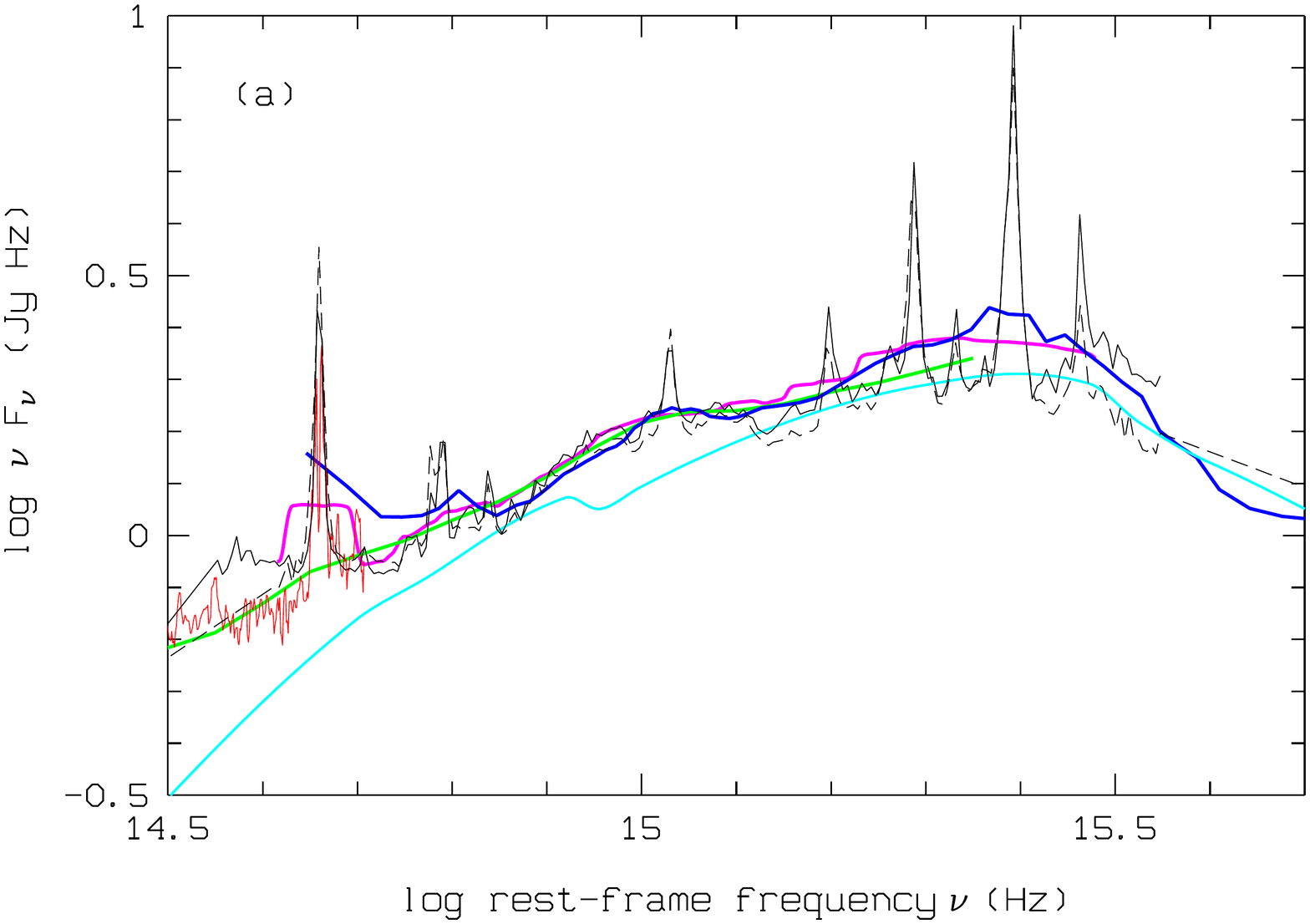}
\includegraphics[viewport=30 40 580 800,angle=270,width=9.2cm,clip]{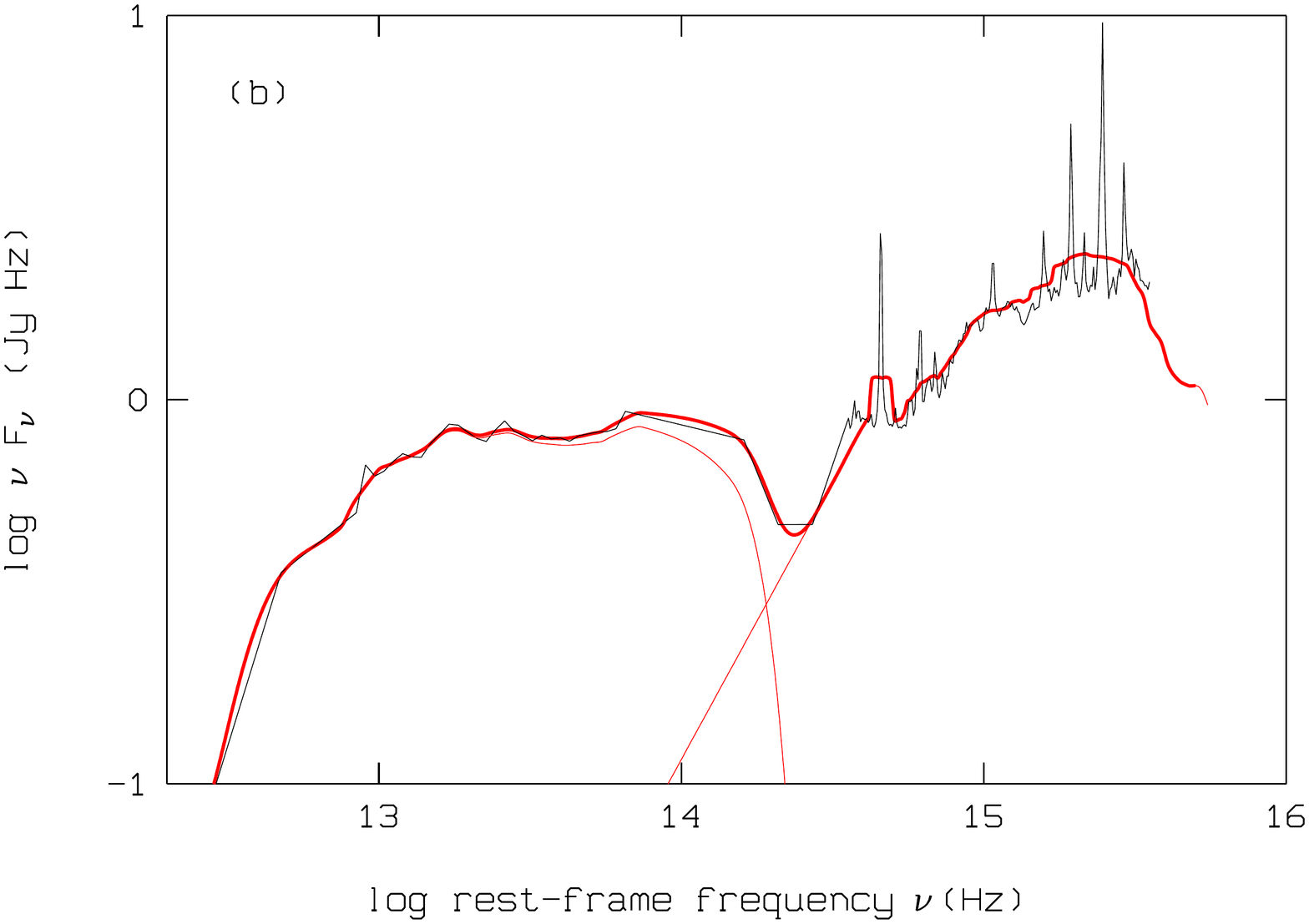}
\caption{
SED of normal quasars. (a) Big blue bump. 
(b) Broad-band SED from the far infrarad to the extreme ultraviolet. Thin red curves: adopted SED for the torus (+ host) and for the
central engine, respectively. Thick red curve: total flux. Black curve: composite spectrum from 
Shang et al. (\cite{Shang2011}).
} 
\label{fig:SED_Shang}
\end{figure}

Shang et al. ({\cite{Shang2011}) constructed composite (median) SEDs both for radio-quiet and 
radio-loud quasars from high-quality single-epoch multi-wavelength data for 85 
optically bright non-blazar quasars. Although there is a large variation from quasar to quasar, 
these composite SEDs are considered representative for UV-optically bright quasars.
Accordingly we used the Shang et al. composite to construct two components of the SED starting with 
the BBB. Figure\,\ref{fig:SED_Shang}a shows the corresponding part
from the $1\,\mu$m  inflection to the Lyman break. The black solid 
curve is the composite for the radio-quiet quasars, the black dashed curve is for the radio-loud ones. 
The SED derived from broadband photometry should be compared with a smoothed version of the composite spectrum, 
shown here in magenta for only the radio-quiet quasars. The blue curve is the composite
SED derived from SDSS and GALEX photometric data (Trammell et al. \cite{Trammell2007}), and it
extends to shorter wavelengths than in Shang et al. ({\cite{Shang2011}). The green curve is the 
composite from SDSS and 2MASS photometry (Labita et al. \cite{Labita2008}), the red one is 
the NIR spectral template from Glikman et al. (\cite{Glikman2006}). For comparison, the cyan curve is 
the continuum of a simple multi-temperature blackbody accretion disk model (Shakura \& Sunyaev \cite{Shakura1973}) 
with a disk temperature parameter 
$T^\ast = 9\cdot10^4$\,K roughly corrected for effects that have been considered by more elaborated disk 
models (see Meusinger \& Weiss \cite{Meusinger2013}). 

To construct the blue component of the SED, i.e. BBB plus emission lines, we used a smoothed version of the
radio-quiet composite\footnote{The composite for radio-quiet quasars was used because the radio loudness of 
our sources is generally low, though the percentage of FIRST radio sources is relatively high (see Sect.\,\ref{sec:RL}).} 
from Shang et al. ({\cite{Shang2011}) at intermediate wavelengths between H$\alpha$ and Ly$\alpha$.
Towards longer wavelengths, the SED is combined with a power law with $\alpha_\nu = 0.44$ 
(Kishimoto et al. \cite{Kishimoto2008}). On the other hand, the data from Trammell et al. (\cite{Trammell2007}) 
is used to continue the SED shortwards of Ly$\alpha$. Subtracting the BBB component from the Shang et al composite yields the IR component. Both components are shown in Fig.\,\ref{fig:SED_Shang}b, together with the entire composite
spectrum.

\subsubsection{Intrinsic dust reddening}\label{sec:intr_dust}

The general shape of the spectra of 3000 \AA\ break quasars suggests that reddening by dust may play a role. 
In general, thermal dust emission of quasars in the infrared exhibits substantial dust masses. 
The dust is assumed to be concentrated in a torus-like configuration around the central engine 
(Antonucci et al. \cite{Antonucci1993}) in a circum-nuclear starburst or distributed on a galaxy-wide scale 
over the host galaxy.   There are tentative arguments supporting the 
view that quasars spend a substantial part of their lifetimes in a dust-enshrouded environment, especially in 
young evolutionary stages (e.g., Sanders et al. \cite{Sanders1988}; Hopkins et al. \cite{Hopkins2004}). 
FeLoBAL quasars are invariably reddened AGNs with high IR luminosities (e.g., Farrah et al. \cite{Farrah2007}, 
\cite{Farrah2012}). Excess reddening by foreground dusty absorbers seem to be rare (Fynbo et al. \cite{Fynbo2013}), 
though there are convincing detections in a few cases. 
Here we simply assume that the dust is located at the redshift of the quasars and that 
the blue component of the quasar SED may need to be corrected for intrinsic reddening, while reddening of 
the IR component is negligible in our context.

There has been a long-standing debate over the reddening law for quasars. The lack of a strong
$\lambda2175$ extinction bump and of a substantial curvature of the UV continuum led to the assumption of 
only modest reddening (e.g., Pitman et al. \cite{Pitman2000}; Cheng et al \cite{Cheng1991}). 
From the analysis of a large sample of SDSS quasars, Richards et al. (\cite{Richards2003}) concluded that 
the reddening law in quasars at $z<2.2$ can be described by the reddening curve from the Small Magellanic 
Cloud (SMC), but not by the reddening in the Large Magellanic Cloud or the Milky Way (MW) galaxy. A similar 
conclusion was drawn by Hopkins et al. (\cite{Hopkins2004}). The SMC extinction is characterised by a 
stronger increase in the reddening in the UV compared to the MW. 

Gaskell et al. (\cite{Gaskell2004}) analysed a large sample of AGNs and concluded that the UV extinction 
curves are very flat, probably caused by a relative lack of small dust grains in the environment 
of the quasar.  Similarly, ``grey'' extinction curves were also predicted for ultra-luminous infrared galaxies
(ULIRGs) at $z \sim 1$ and were attributed to the lack of small grains as predicted by supernova dust models 
(Shimizu et al. \cite{Shimizu2011}).  

\begin{figure}[h]
\includegraphics[viewport=125 30 570 790,angle=270,width=9.0cm,clip]{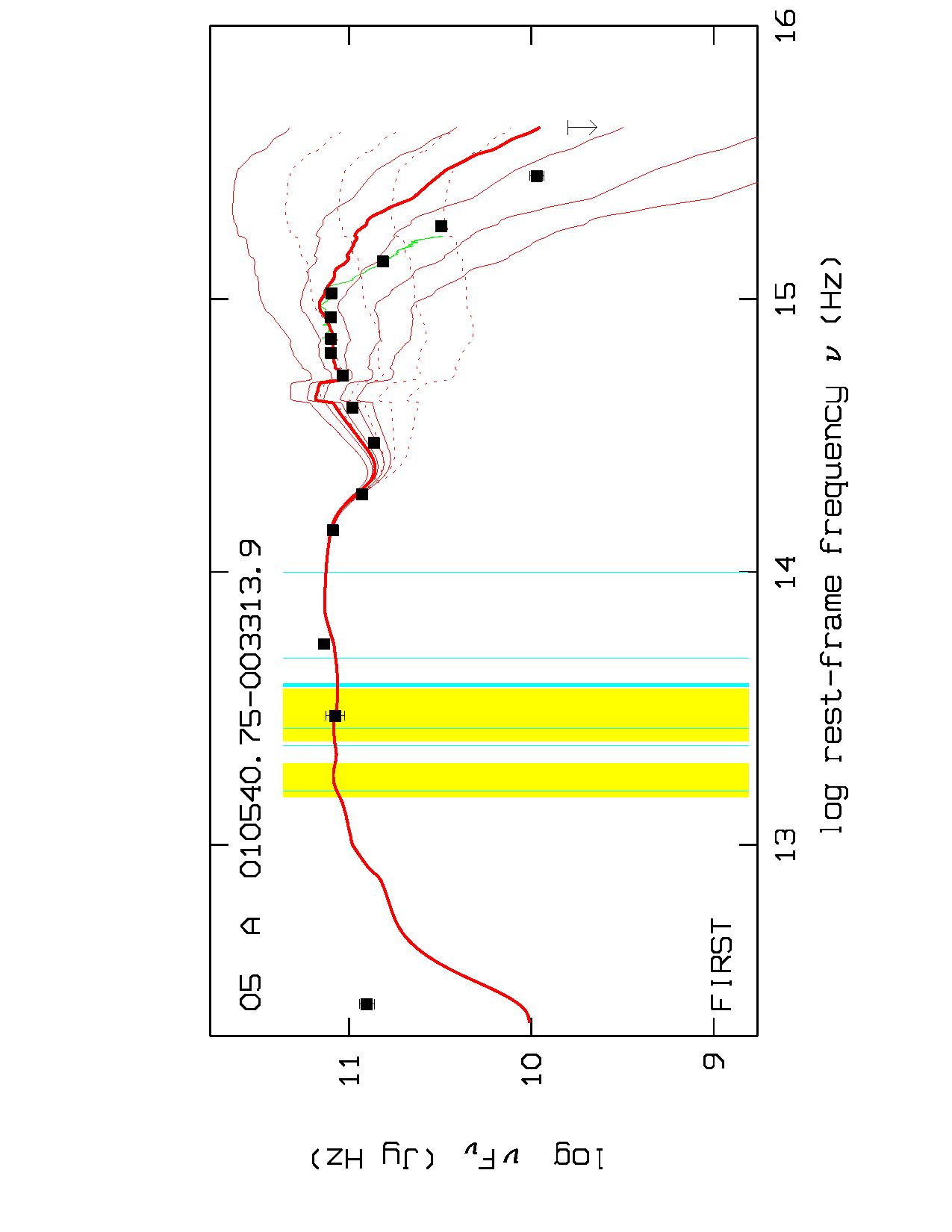}
\includegraphics[viewport=125 30 570 790,angle=270,width=9.0cm,clip]{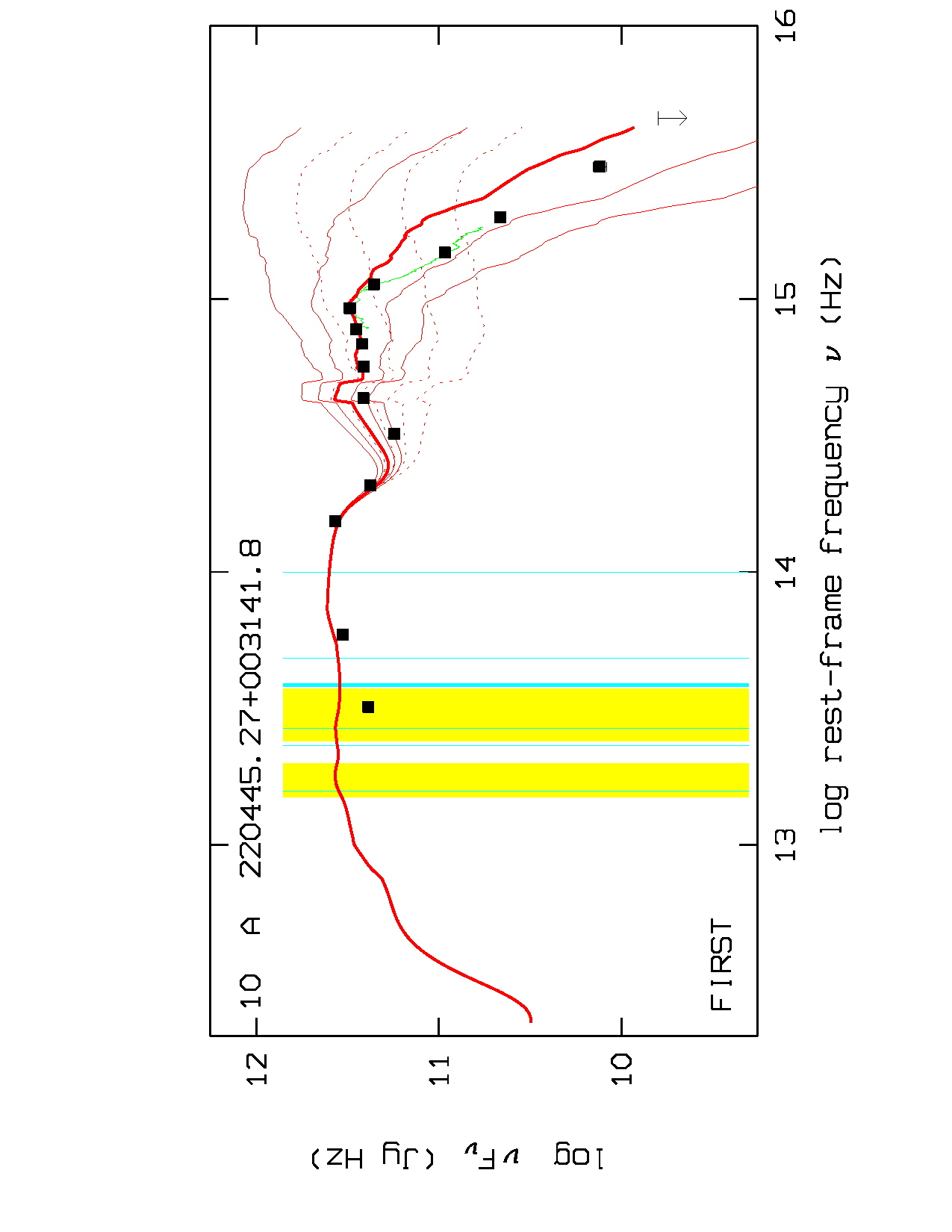}
\caption{
Broad-band SEDs of the two prototypical 3000 \AA\ break quasars.
For comparison, the quasar composite from Fig.\,\ref{fig:SED_Shang} is plotted for different
reddenings.}
\label{fig:SED_ec_5_examples}
\end{figure}

Here we estimated the intrinsic reddening by comparing the individual quasar SED with the reddened 
composite from Fig.\,\ref{fig:SED_Shang}b. First the BBB component was reddened, and then 
the sum of the two components of the quasar composite SED was fitted to the observed quasar SED. 
The SMC reddening curve (Pei \cite{Pei1992}) was adopted, but the Gaskell curve was used for comparison as well.
The fitting procedure has two free parameters, the colour excess $E(B-V)$ and the relative shift 
between the BBB and the IR component. Owing to the highly peculiar SEDs, the procedure is subject 
to substantial uncertainties. Because of the 3000 \AA\ break in the spectra from subsamples A and B 
and the usually strong effect of BALs shortwards of 3000 \AA\ for sample C, the fit has to be restricted 
essentially to the spectral range between the $1\mu$m deflection and $\sim$3000 \AA.

For illustration, Fig.\,\ref{fig:SED_ec_5_examples} shows the two prototypical 3000 \AA\ break quasars \object{J010540.75-003313.9} and \object{J220445.27+003141.8}. The black boxes indicate the fluxes from the photometric
data in the various bands, downward arrows indicate upper limits. The BOSS spectra are over-plotted in green.
The thin red curves show the composite SED reddened with either the SMC (solid curves) 
or the Gaskell (dotted curves) extinction law for five different values of $E(B-V)$ in equal steps
from 0 to 0.4 mag for SMC and 0 to 0.2 mag for Gaskell dust (top to bottom). The positions
of the silicate absorption troughs in the MIR are indicated by the vertical yellow stripes. The vertical 
green lines indicate the positions of the PAH emission features.  The best-fitting SMC extinction-corrected 
SED is highlighted by the thick red curve. 

The best-matching intrinsic reddening $E(B-V)$ is listed in Table\,\ref{tab:sample}. In the individual case,
the inferred values should only to be considered indicative for several reasons (see below). 
The mean values for the groups A, B, and C are 
0.17$+/-$0.10, 0.12$+/-$0.12, and 0.10$+/-$0.09 for SMC reddening and 
0.11$+/-$0.06, 0.08$+/-$0.08, and 0.07$+/-$0.07 for the Gaskell law. 
Assuming SMC reddening, 19\% of the quasar SEDs are matched with $E(B-V)_{\rm smc} = 0$ mag and
62\% with $E(B-V)_{\rm smc} > 0.1$ mag. 
This is a much higher percentage than for SDSS quasar in general. 
Hopkins et al. (\cite{Hopkins2004}) estimated that 2\% of the quasars from the main SDSS targeting procedure
have $E(B-V)_{\rm smc} > 0.1$ mag. Quasars with stronger reddening of $E(B-V)_{\rm smc} > 0.4$ mag 
(e.g., Hall et al. \cite{Hall2002a}) are exceptionally rare in SDSS. The strongest reddening in our sample
is $E(B-V)=0.41$ mag.
The corresponding values for the Gaskell law are
35\% and 30\% with a maximum reddening 0.27 mag.

The spectra of at least some quasars from our sample exhibit strong broad iron emission features
in the optical and in the UV that complicate the comparison with the standard quasar SED.
Stronger-than-usual iron emission can lead to underestimating the amount of reddening. In principle, 
contamination by host galaxy light provides another potential 
source of uncertainty. However, because of the relatively high luminosities of our quasars 
(see Sect.\,\ref{sec:L3000}; Paper I), 
we assume that the host contribution should be negligible on average.

\subsubsection{Individual quasar broad-band SEDs}\label{sec:individual_SEDs}

The individual SEDs for all quasars are shown in Fig.\,\ref{fig:spectra_and_seds}.
The observed optical spectra, expressed by log\,$F_\lambda$ $(\lambda)$, are given on the left-hand side.
The panels on the right-hand side show the broad-band SEDs log\,$\nu F_\nu$ (log $\nu$) in the quasar's rest frame.
More detailed information on Fig.\,\ref{fig:spectra_and_seds} is given in Appendix B.

\subsection{Sample-averaged SEDs}

\begin{figure*}[htbp]
\begin{tabbing}
\includegraphics[viewport=60 0 570 790,angle=270,width=8.5cm,clip]{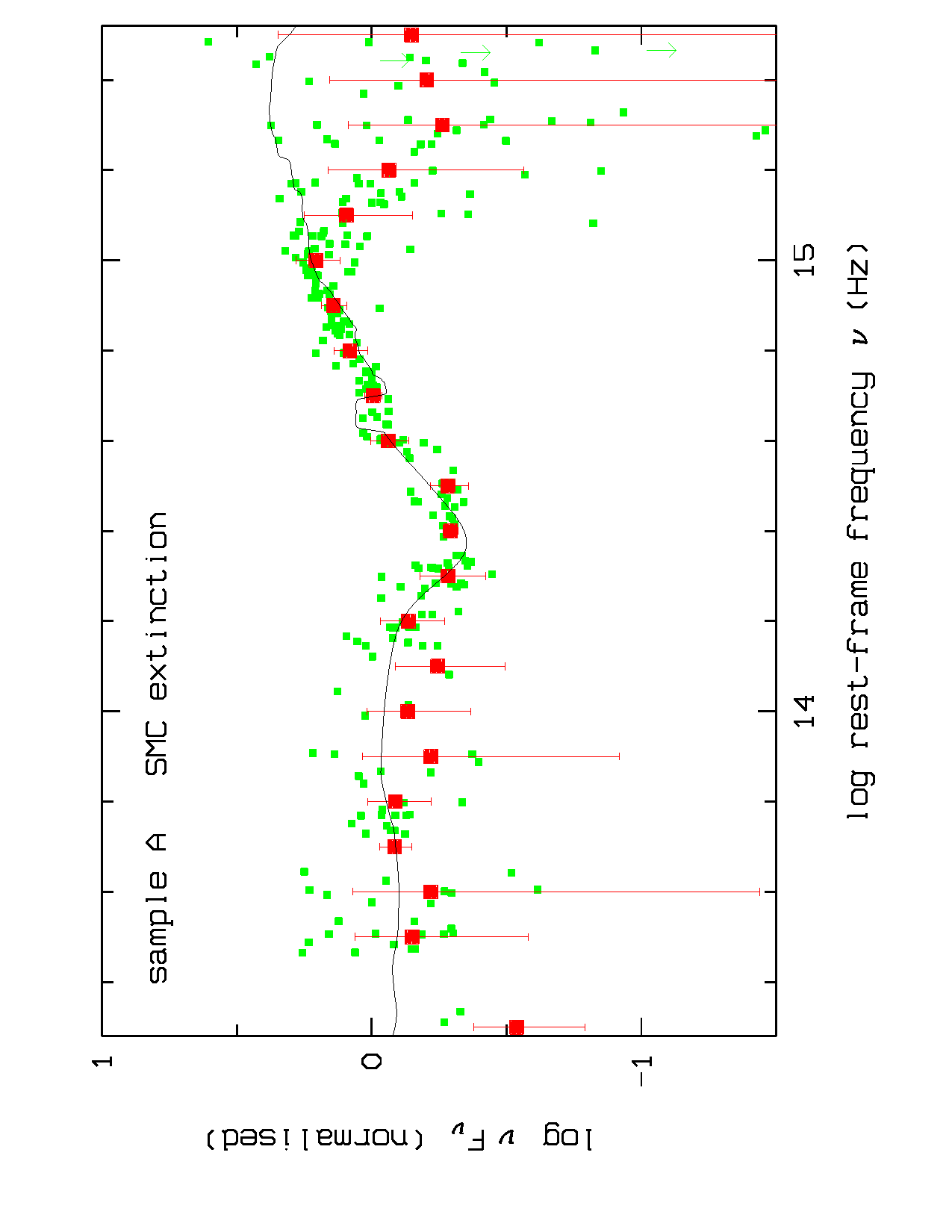}\hfill \=
\includegraphics[viewport=60 0 570 790,angle=270,width=8.5cm,clip]{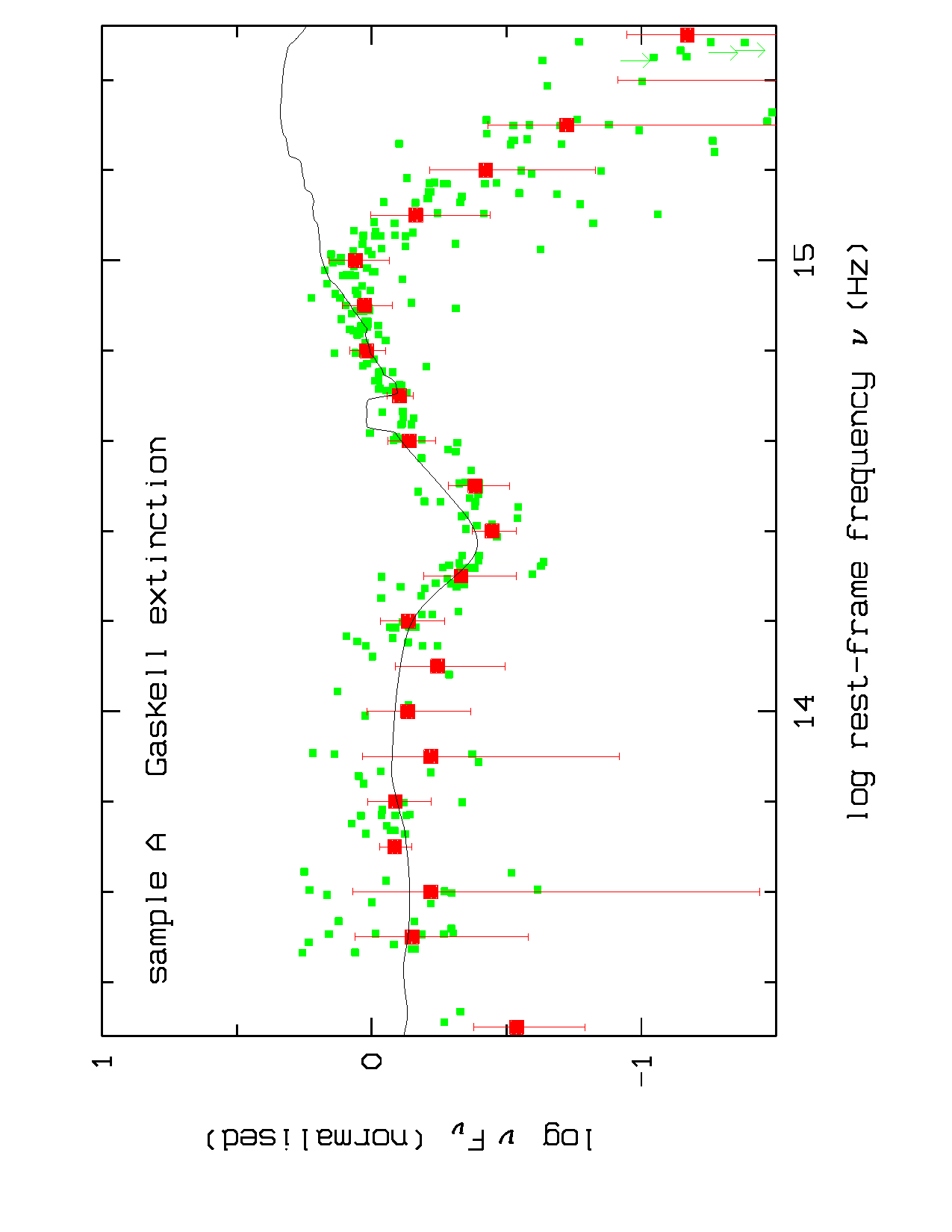}\hfill  \\
\includegraphics[viewport=60 0 570 790,angle=270,width=8.5cm,clip]{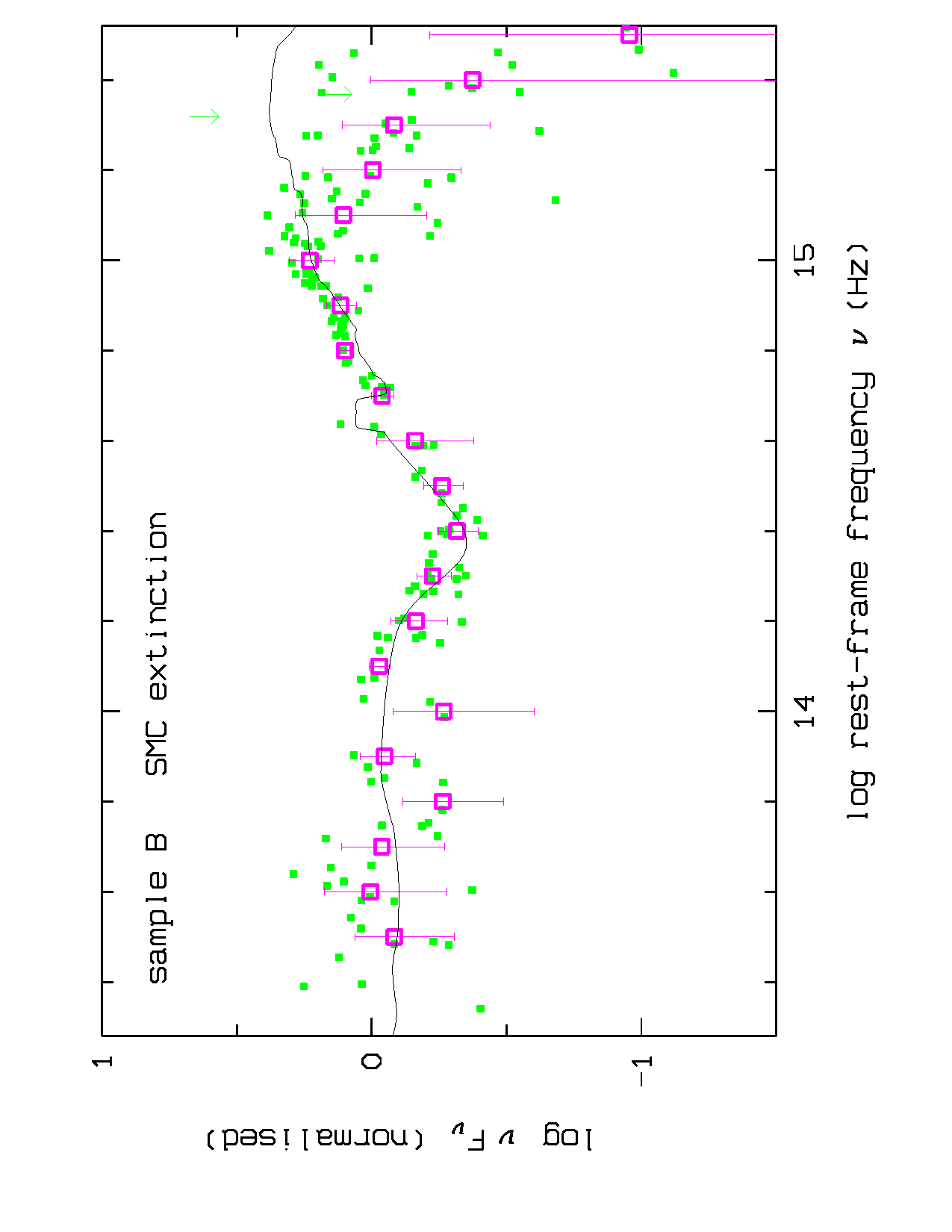}\hfill \=
\includegraphics[viewport=60 0 570 790,angle=270,width=8.5cm,clip]{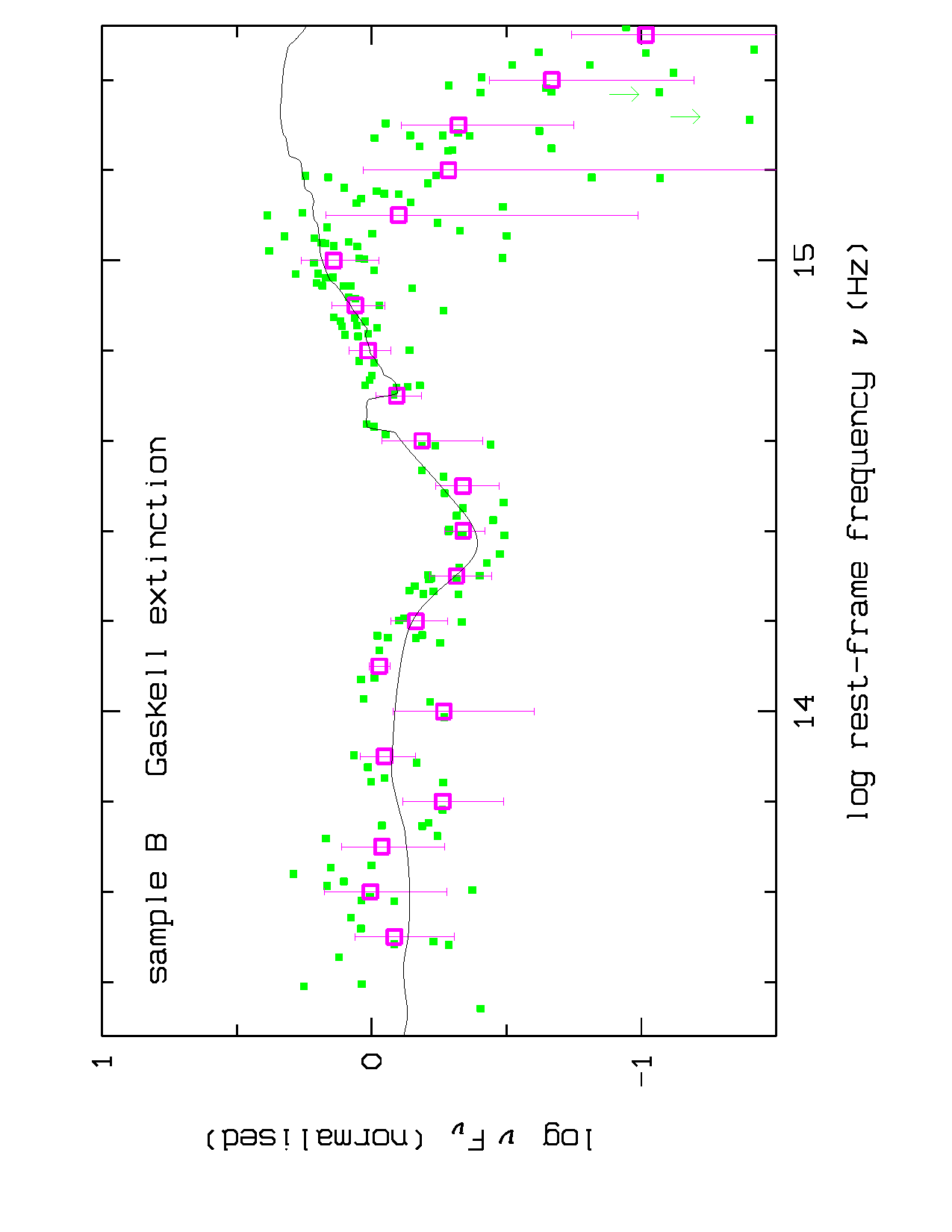}\hfill  \\
\includegraphics[viewport=60 0 570 790,angle=270,width=8.5cm,clip]{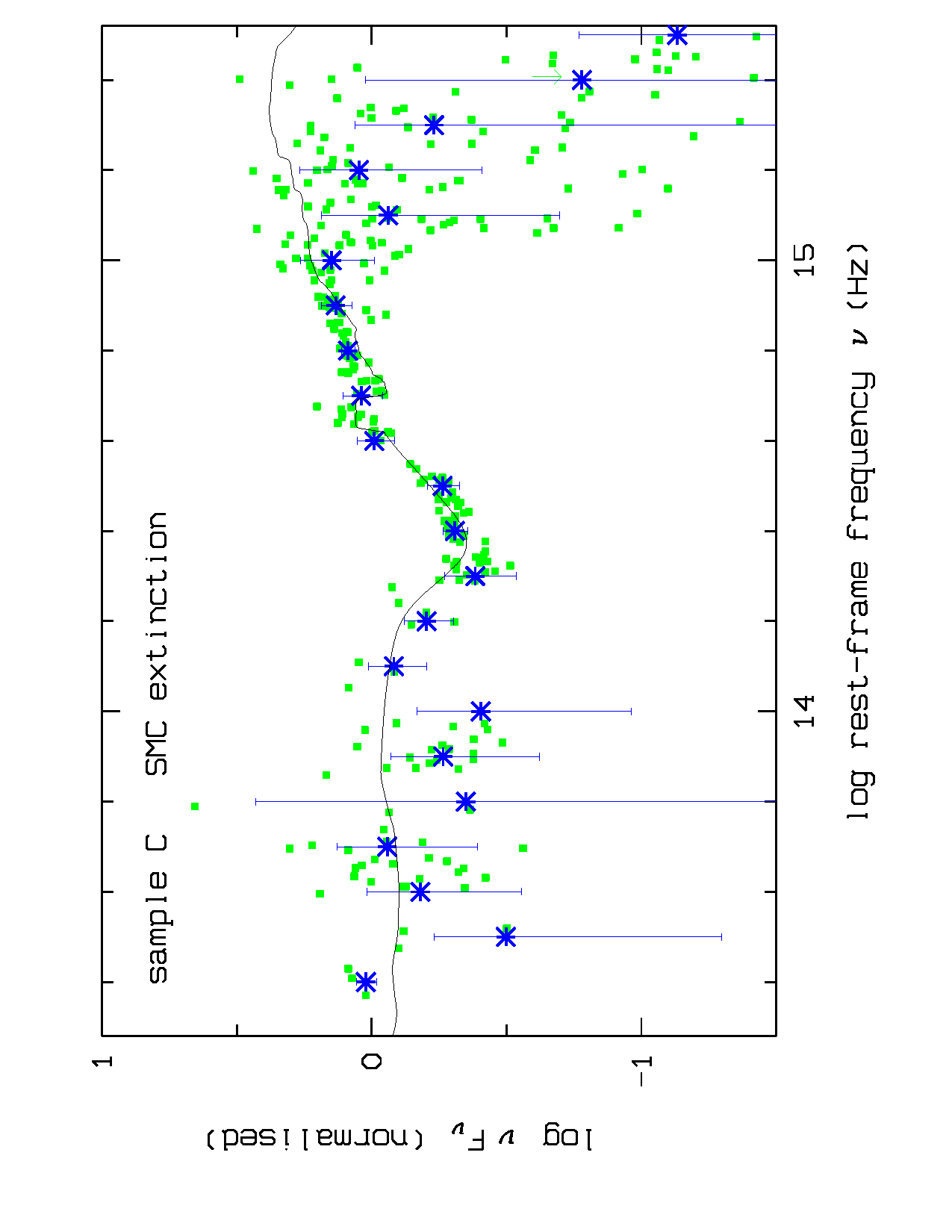}\hfill \=
\includegraphics[viewport=60 0 570 790,angle=270,width=8.5cm,clip]{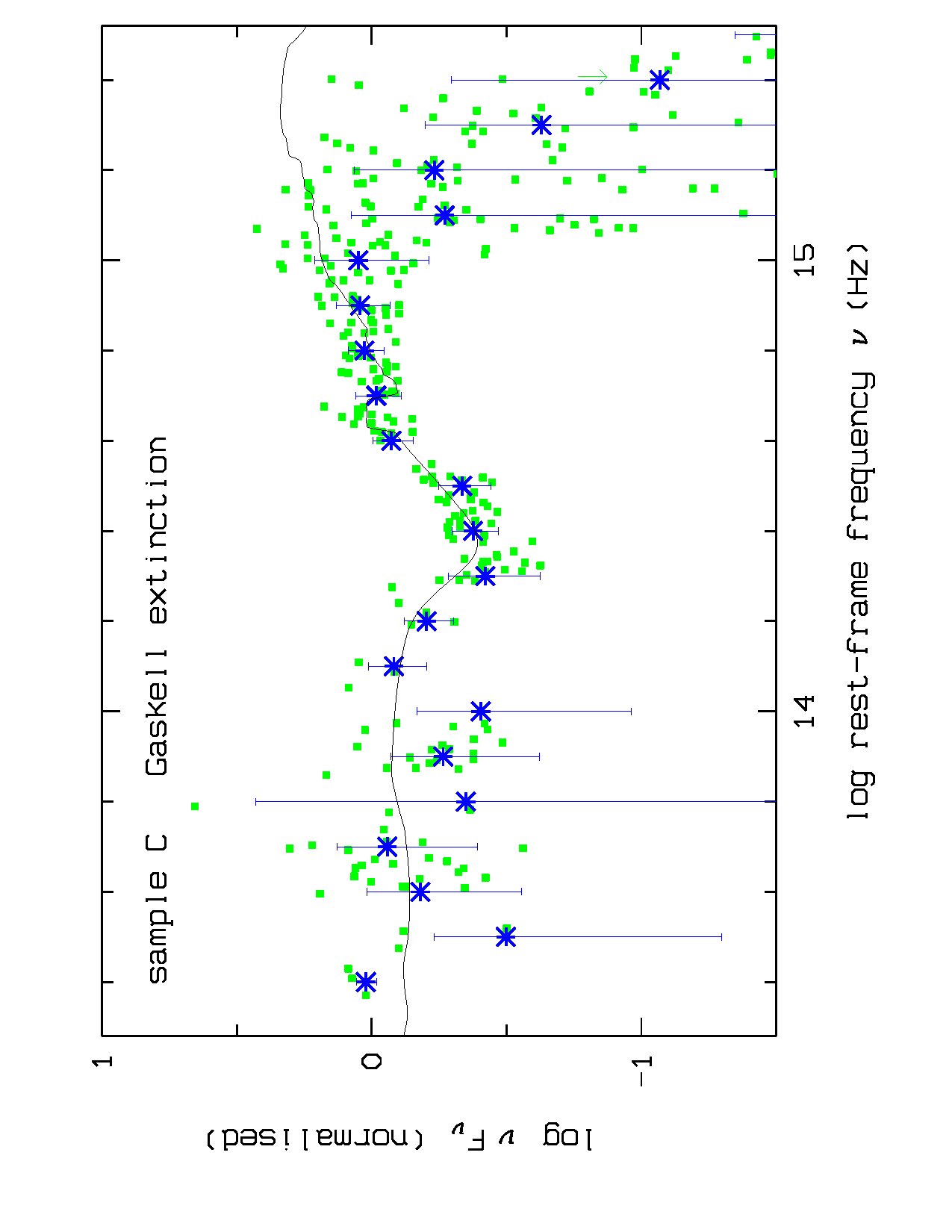}\hfill  \\
\includegraphics[viewport=60 0 570 790,angle=270,width=8.5cm,clip]{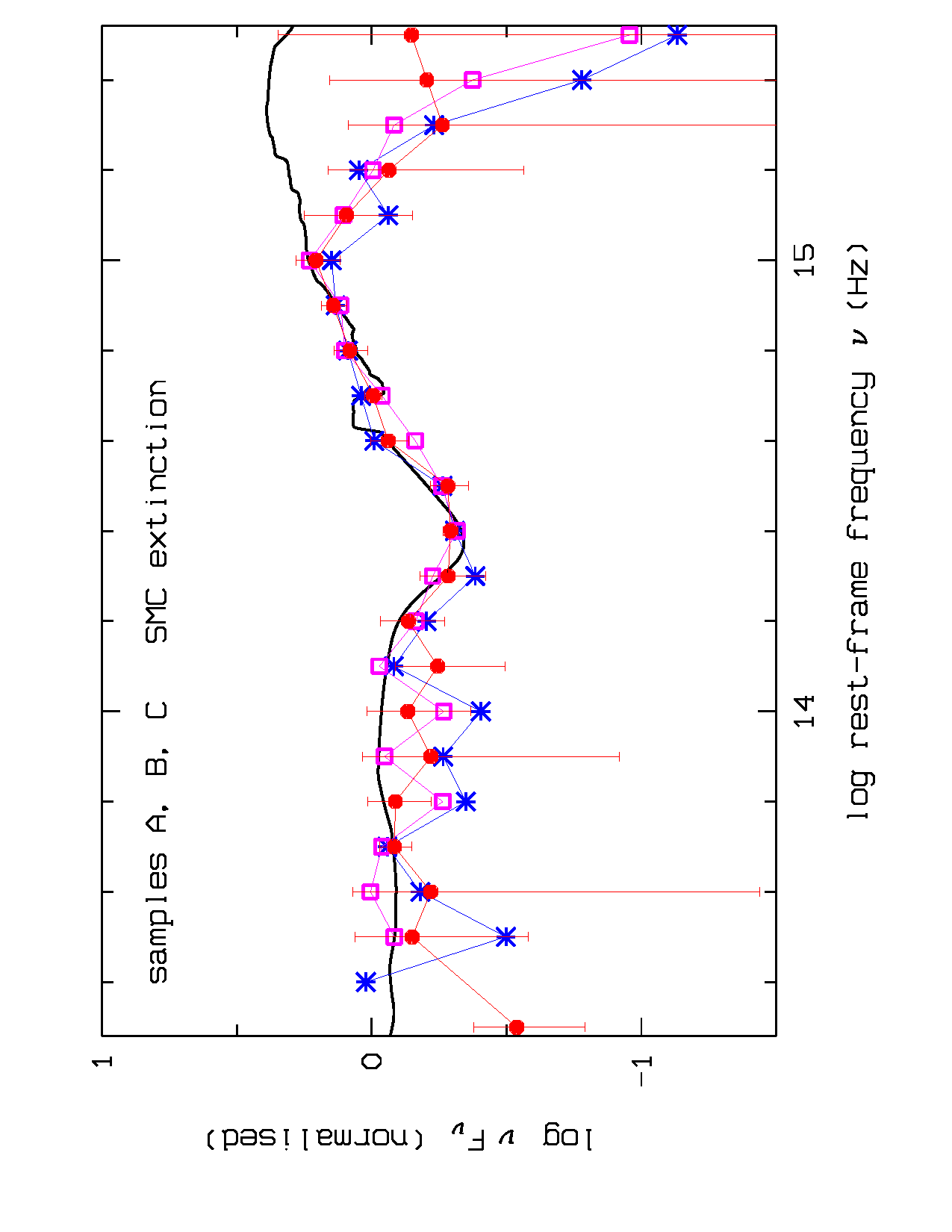}\hfill \=
\includegraphics[viewport=60 0 570 790,angle=270,width=8.5cm,clip]{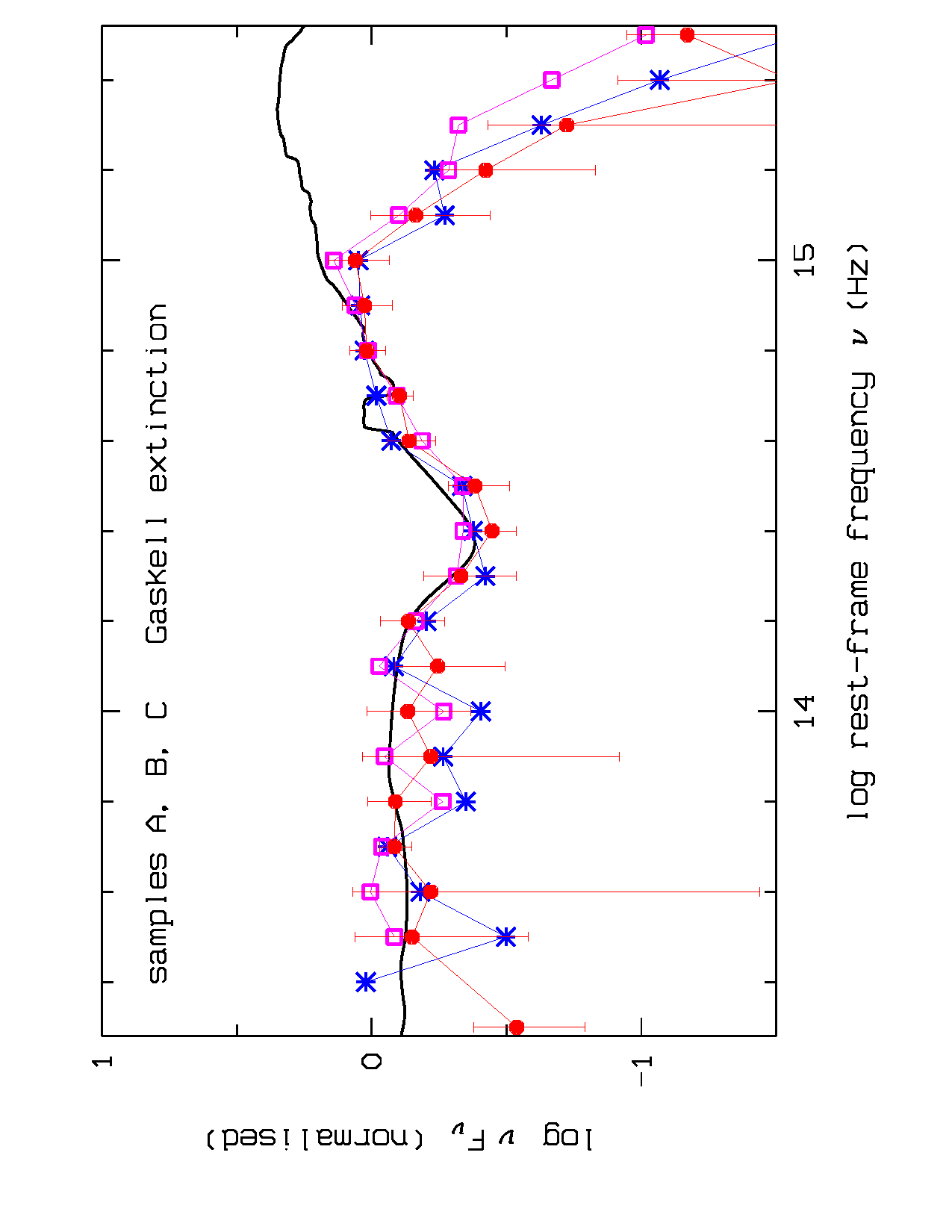}\hfill
\end{tabbing}
\caption{
Composite SEDs from photometric data corrected for Galactic foreground reddening and intrinsic dust reddening (green filled squares) for the quasars in samples A, B, and C by adopting SMC reddening (left) and Gaskell reddening (right), respectively: median values (symbols as in Fig.\,\ref{fig:z_here_z_sdss}) and standard deviations (vertical error bars). The normal quasar composite from Fig.\,\ref{fig:SED_Shang} is over-plotted in black.
}
\label{fig:mean_SED_ec_ic}
\end{figure*}

The comparison of single quasar SEDs with a composite spectrum should be considered with care.
Shang et al. ({\cite{Shang2011}) caution that there is a large variation in their individual quasar SEDs. 
Given the effects of absorption by dust and gas, an even larger object-to-object variation is
expected for the unusual quasars from our sample.  Temporal flux variability adds a further complication, 
especially at short wavelengths, i.e. when connecting SDSS with GALEX data taken at different epochs. 
These difficulties are substantially reduced in a statistical approach where ensemble averaged SEDs are used.

We normalised the extinction-corrected fluxes for each single quasar in such a way that the normal 
quasar composite SED from Fig.\,\ref{fig:SED_Shang}b is best matched both around $4000$ \AA\ and around the 
1\,$\mu$m inflection. The normalised fluxes are plotted as green squares in Fig.\,\ref{fig:mean_SED_ec_ic} 
for the samples A to C in the first three rows from top. SMC extinction correction was applied on the
left-hand side, and the Gaskell extinction curve was used at the right-hand side. 
The symbols from Fig.\,\ref{fig:z_here_z_sdss} with error bars represent arithmetic median values and standard
deviations in frequency bins of 0.1 dex width. The normal quasar composite is over-plotted in black.
The same symbols were used in the panels at the bottom of Fig.\,\ref{fig:mean_SED_ec_ic} to compare the
medians from the three samples with each other. The symbols were interconnected just to guide the eye.

The following inferences are made from Fig.\,\ref{fig:mean_SED_ec_ic}. In most cases, the
decline at log\,$\nu \mbox{[Hz]} \ga 15$, defining the 3000 \AA\ break quasars from sample A, 
does not disappear after the correction for intrinsic dust reddening. Exceptions are
\object{J160827.08+075811.5} (\# 6), \object{J111541.01+263328.6} (\# 11), and perhaps \object{J161836.09+153313.5} (\# 12) for SMC dust. We caution, however, that the reddening correction might be over-estimated for these spectra. The SMC dust-corrected data of sample A seem to indicate a slight change of the slope of the average SED at highest frequencies. However, given the small number of data points and the huge scatter, this trend should not be considered significant. The effect
is much weaker for the data corrected with the Gaskell reddening law. 
Surprisingly, there are no significant differences between the averaged SEDs of 
the three samples. In particular, the average SED of the unusual BAL quasars from sample C shows
the same trend in decreasing flux at $\log\,\nu \ga 15$ as the 3000 \AA\ break quasars from sample A.  
The ensemble-averaged SEDs of the three samples seem to be indistinguishable from each other and from the
SED of normal quasars. We only notice a trend towards slightly lower MIR fluxes or alternatively slightly 
higher UV fluxes for our quasars.

\subsection{Luminosities}

\subsubsection{Infrared luminosity}

Quasars are known to be invariably luminous in the IR with luminosities comparable 
to ULIRGs with $L_{\rm IR} > 10^{12} L_\odot$. Haas et al. (\cite{Haas2003}) derived 
3 -- 1000 $\mu$m IR luminosities for a sample of 64 Palomar-Green quasars. All of their quasars at 
$0.5 \la z \la 2$ detected in the IR have $L_{\rm IR} > 10^{12} L_\odot$ with $L_{\rm IR} > 10^{13} L_\odot$ 
for the majority. Farrah et al. (\cite{Farrah2012}) investigated the optical to FIR SEDs of a sample of reddened 
FeLoBAL quasars with $0.8 < z < 1.8$ and found 1 -- 1000 $\mu$m IR luminosities at rest-frame $L_{\rm IR} > 10^{\rm 12} L_\odot$ with $L_{\rm IR} > 10^{\rm 13} L_\odot$ for one third of the sample.

Under the assumption of isotropy, the IR luminosity is $L_{\rm IR} = 4\pi\, d_{\rm L}^2\, f_{\rm \nu,IR}$, where 
$d_{\rm L}$ is the luminosity distance and $f_{\rm \nu,IR}$ is the flux $F_\nu$ integrated over the IR spectral range. 
To estimate the IR luminosities for the quasars from our sample, we simply integrated the best matching 
quasar composite  from 1 -- 1000 $\mu$m in the rest frame. The results are listed in Table\,\ref{tab:sample}.
With the exception of J140025-012957 at $z=0.585$, all quasars have $L_{\rm IR} > 10^{\rm 12} L_\odot$.
In line with Farrah et al. (\cite{Farrah2012}), we found that 39\% and  30\% of the quasars from samples A and B, respectively, have $L_{\rm IR} > 10^{\rm 13} L_\odot$, whereas the percentage is higher (77\%) for sample C.
The $L_{\rm IR}$ histograms are similar for all three samples. The mean luminosities and standard deviations are 
$\overline{L}_{\rm IR} = 1.9\pm2.8, 1.1\pm1.6$, and $1.7\pm1.2 \ 10^{13} L_{\odot}$ for samples A, B, and C.

\subsubsection{Monochromatic 3000 \AA\ luminosity}\label{sec:L3000}

We computed monochromatic luminosities at 3000 \AA\ rest frame as
$ L_{\rm 3000} = 4\pi\, d_{\rm L}^2\, f_{\rm \nu,3000}$ where $f_{\rm \nu,3000} = \nu\,F_{\nu}$ at $\lambda = 3000$\AA.
As for the IR luminosities, the flux was estimated from the best-matching normal quasar composite. 
That means, the 3000 \AA\ break is assumed to be caused by absorption by dust or gas and is corrected for in that way.
In other words, it is implicitly assumed that the intrinsic emission is
represented by the SED of normal quasars. This is in principle correct for the reddened BAL quasars in sample C, but it is
only a working hypothesis for samples A and B (see Sect.\,\ref{sec:disc}).  
Figure\,\ref{fig:L3000_z} displays the comparison of those 3000 \AA\ 
luminosities with the $L_{\rm 3000}$  given by Shen et al. (\cite{Shen2011}) for the quasars from the SDSS DR7 quasar catalogue (Schneider et al. \cite{Schneider2010}).\footnote{The catalogue from Shen et al. (\cite{Shen2011}) gives $L_{3000}$ only for $z \la 2.25$.} Independent of the redshift, the corrected 3000 \AA\ luminosities for the quasars from our samples tend to be higher than those of typical SDSS quasars.

\begin{figure}[htbp]
\includegraphics[viewport=10 40 560 780,angle=270,width=9.1cm,clip]{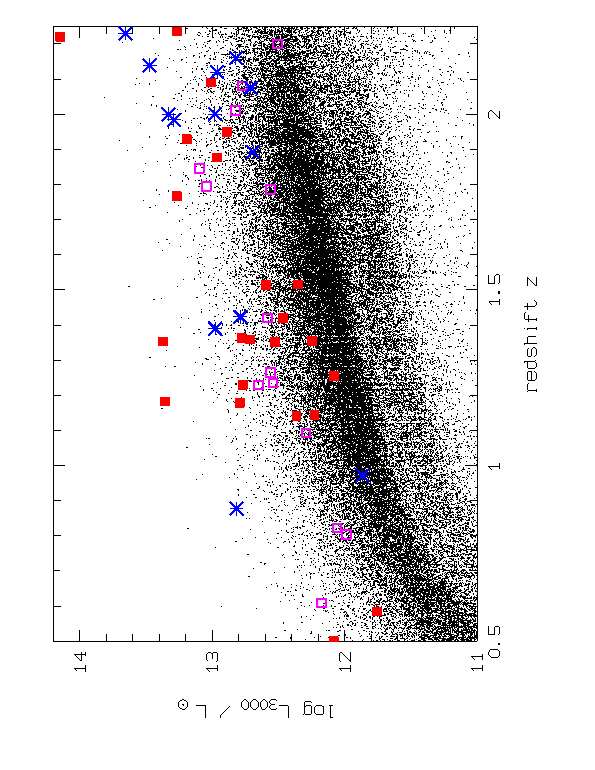}
\caption{Monochromatic 3000 \AA\ luminosities, $L_{3000}$, versus redshift for 80\,847 quasars from the 
Shen catalogue (small black dots) and for the present samples. Symbols as in Fig.\,\ref{fig:z_here_z_sdss}.
}
\label{fig:L3000_z}
\end{figure}

Figure \ref{fig:LIR_L3000} shows the distribution of the ratios of the IR luminosity to the 
monochromatic luminosity at 3000 \AA\ for samples A, B, and C. The histograms are similar for
all three samples, except for a trend towards lower ratios in sample C. The mean ratios are 
$1.73 \pm 0.51, 1.60 \pm 0.39$, and $1.37 \pm 0.59$ for samples A, B, and C, in good agreement 
with the ratio $L_{\rm IR}/_{3000} \approx 1.8$ derived from the composite spectrum for ordinary 
radio-quiet quasars. 

\begin{figure}[h]
\includegraphics[viewport=10 0 560 780,angle=0,width=8.5cm,clip]{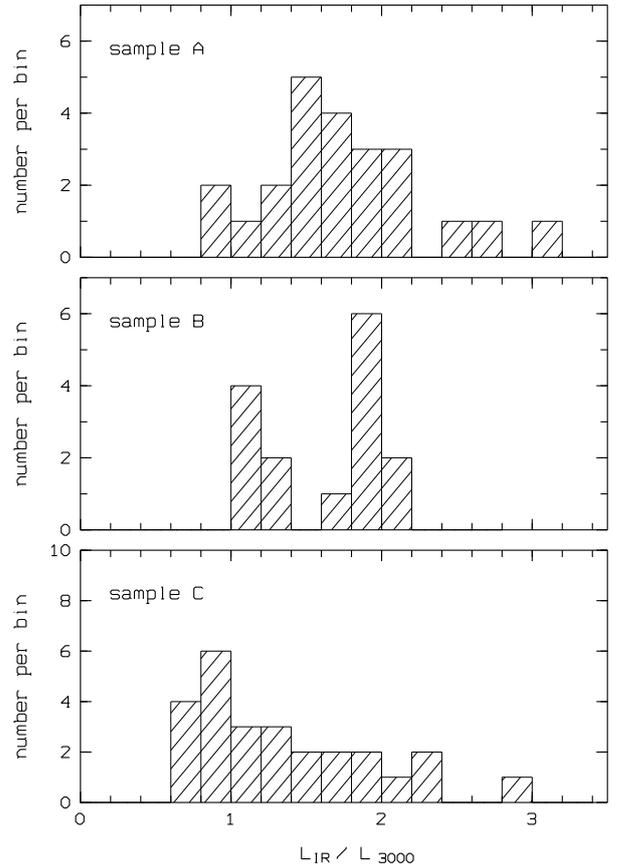}
\caption{
Histograms of the ratio of the IR luminosity $L_{\rm IR}$ to the 3000 \AA\ monochromatic 
luminosity $L_{\rm 3000}$ for samples A, B, and C.
}
\label{fig:LIR_L3000}
\end{figure}

To illustrate the effect of the absorption correction, the left-hand panel of Fig.\,\ref{fig:L3000_shen} shows the comparison of $L_{\rm 3000}$ from our study with the monochromatic 3000 \AA\ luminosities given by Shen et al. (\cite{Shen2011}), where we corrected the latter for the intrinsic dust reddening with the values of $E(B-V)$ from Table\,\ref{tab:sample}. The scatter around the 1:1 relation (dotted line) reflects that the dust reddening correction procedure
corrects only for a part of the decline of the SED at $\lambda \la 3000$ \AA. It is thus not surprising that the scatter 
is greatest for quasars from sample C with their strong absorption troughs. The largest differences are measured for  
\object{J094317.59+541705.1} (\# 51) and \object{J173049.10+585059.5} (\# 46), two extreme FeLoBAL quasars where the continuum drops rapidly at $\lambda \la 3000$ \AA\ to a level close to the detection threshold down to the shortest wavelengths covered by the optical spectra (see also Urrutia et al. \cite{Urrutia2009}; Hall et al. \cite{Hall2002a}; Paper I). In the middle panel of Fig.\,\ref{fig:L3000_shen}, the IR luminosity is plotted versus the (uncorrected) 3000 \AA\ luminosity from Shen et al. (\cite{Shen2011}). For comparison we over-plotted those FeLoBAL quasars from Farrah et al. (\cite{Farrah2012}) for which $L_{\rm 3000}$ is available in the Shen catalogue (black symbols). The dotted line in this diagram represents the quasar composite spectrum. The right-hand panel of Fig.\,\ref{fig:L3000_shen} shows $L_{\rm IR}$ versus $L_{\rm 3000}$. In agreement with Figs.\,\ref{fig:mean_SED_ec_ic} and \ref{fig:LIR_L3000}, the majority of our quasars are fitted by the relation for normal quasars with a trend towards either slightly lower $L_{\rm IR}$ or slightly higher $L_{3000}$ for some quasars, in particular from sample C.

\begin{figure*}[hbtp]
\begin{tabbing}
\includegraphics[viewport=20 150 560 680,angle=270,width=6.2cm,clip]{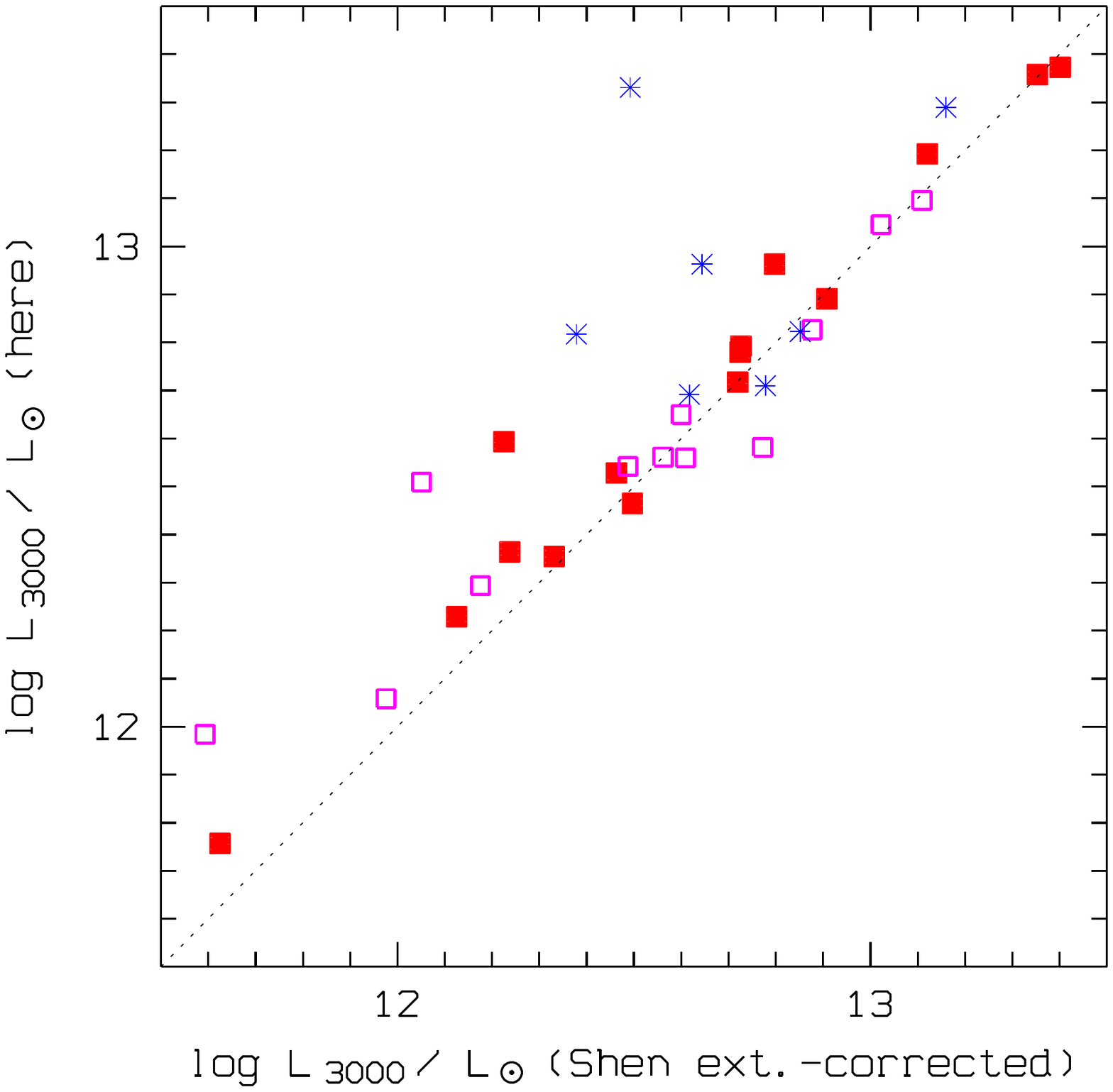}\hfill \=
\includegraphics[viewport=20 150 560 680,angle=270,width=6.2cm,clip]{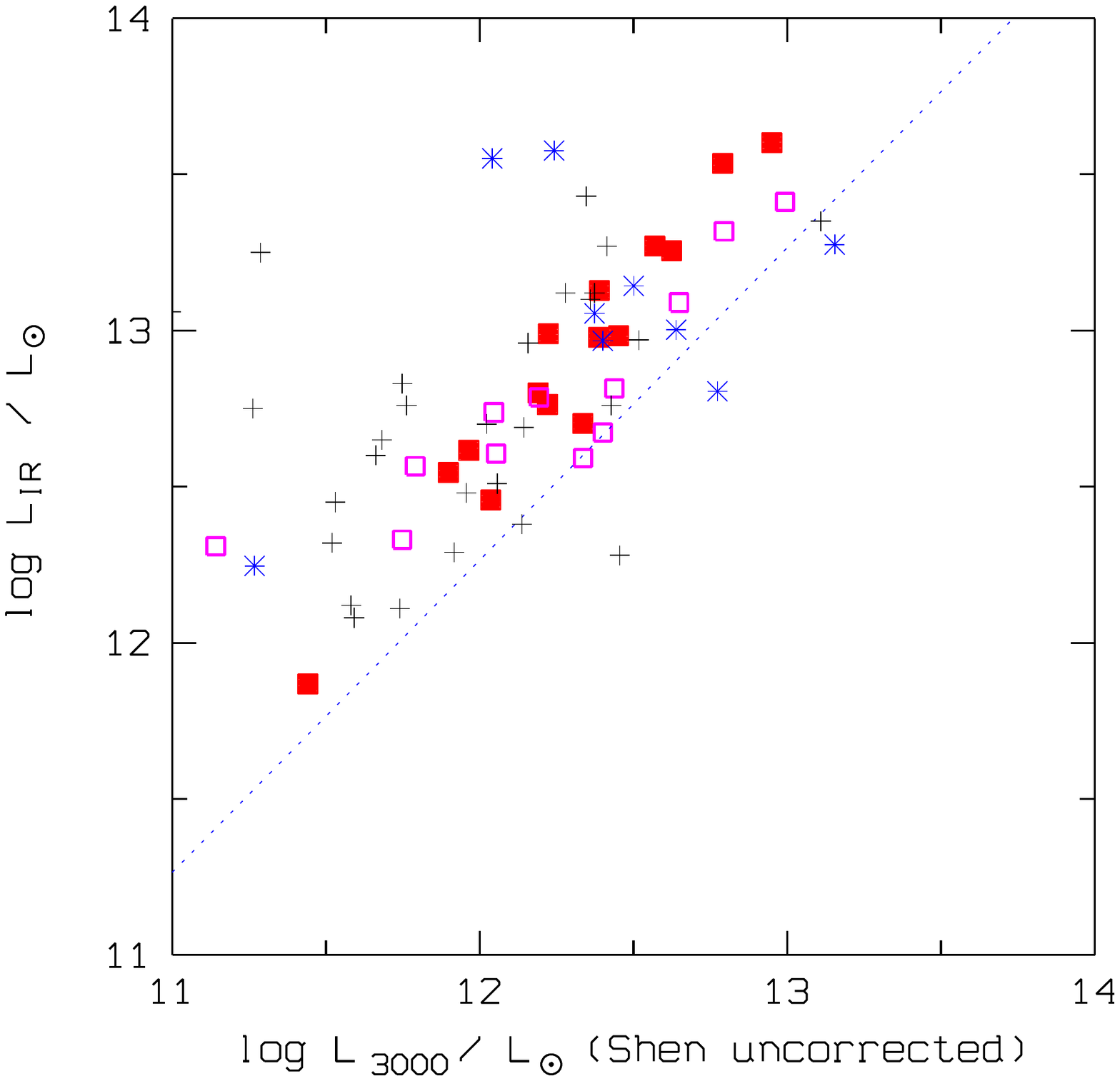}\hfill \=
\includegraphics[viewport=20 150 560 680,angle=270,width=6.2cm,clip]{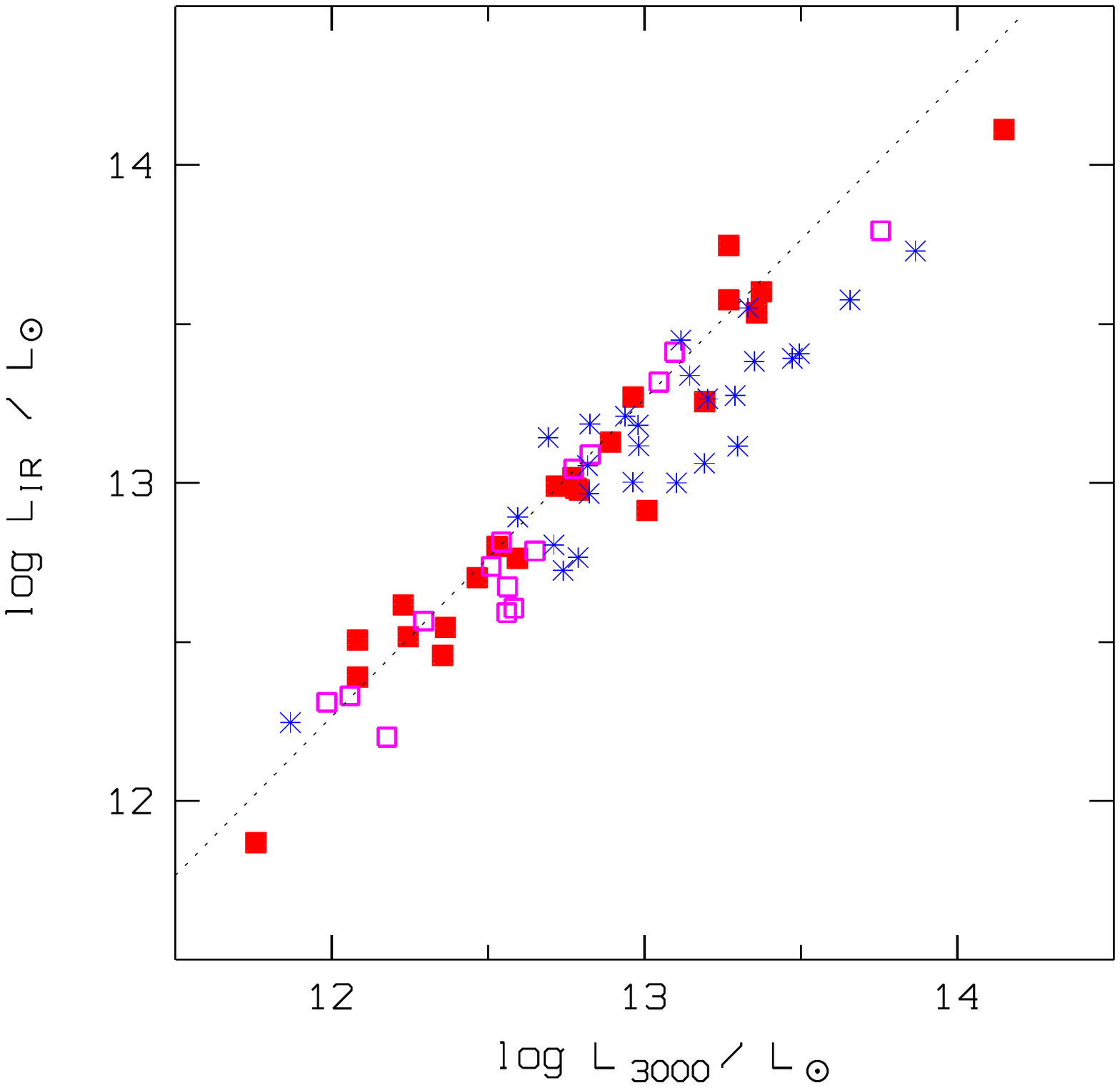}
\end{tabbing}
\caption{{\it Left}:
Monochromatic luminosity $L_{3000}$ derived here from the fit of the AGN composite SED to the
extinction-corrected photometric data compared with the values from the Shen catalogue after correcting for
the same amount of extinction. Symbols as in Fig.\,\ref{fig:z_here_z_sdss}.
{\it Middle}:
Infrared luminosity $L_{\rm IR}$ versus the monochromatic 3000 \AA\ luminosity $L_{3000}$ from the Shen catalogue
without extinction corrections. The black plus signs are for the quasars from Farrah et al. (\cite{Farrah2012}).
{\it Right}:
Infrared luminosity $L_{\rm IR}$ versus the extinction-corrected luminosity $L_{3000}$ from the present study.
}
\label{fig:L3000_shen}
\end{figure*}

\subsection{Radio loudness}\label{sec:RL}

Thirty quasars (47\%) from our sample were detected by FIRST, 33 quasars are below the FIRST level, and
one quasar, \object{J005041.59+143755.9}, is not in the FIRST area.
The percentage of FIRST detected quasars decreases from 74\% for sample A to 60\% for sample B 
and 16\% for sample C. It was shown in Paper I that the radio detection fraction strongly correlates with
the degree of peculiarity of the spectrum and that this trend can be fully explained as a selection effect.
Of the 22 SDSS quasars in sample A, 16 (73\%) were selected for SDSS spectroscopy because they were 
detected as FIRST radio sources. The ratio of the number of quasars targeted as FIRST sources to the 
number of quasars targeted not as such is 2, compared to 0.01 for the normal quasar population in the 
SDSS DR7 quasar catalogue (see Paper I).  

Quasars are usually classified as radio-loud based on the radio-to-optical flux ratio.
We use the K-corrected ratio $R = F_{\rm 5 GHz}/F_{\rm 2500\AA}$ of the 5 GHz radio flux density to the 
2500 \AA\ UV flux density in the quasar rest frame as a measure of radio loudness (Stocke et al. \cite{Stocke1992}).
The 5 GHz flux is extrapolated from the FIRST 1.4 GHz flux adopting a power law $F \propto \nu^{\ \alpha_{\rm R}}$ with 
$\alpha_{\rm R} = -0.5$. The UV flux density at 2500 \AA\ is taken from the best-matching SED after correction for
intrinsic reddening. That is, the $R$ parameter is corrected not only for intrinsic dust reddening but also 
for the strong BAL absorption effects. The results are listed in the last column of Table\,\ref{tab:sample}, where upper limits are given for FIRST non-detections assuming an upper limit of 0.7 mJy for the detection on the FIRST image cutouts (see Sect.\,\ref{sec:obsdat}).

The value $R=10$ is commonly used to distinguish radio-loud and radio-quiet quasars
(e.g., 
Stocke et al. \cite{Stocke1992};
Francis et al. \cite{Francis1993};
Urry \& Padovani \cite{Urry1995};
Richards et al. \cite{Richards2011}),
although this value is to some degree rather arbitrary 
(e.g., Falcke et al. \cite{Falcke1996}; 
Wang et al. \cite{Wang2006}).
In general, the distribution of the radio loudness of quasars appears to
be bimodal with about 10\% being radio loud 
(Kellermann et al. \cite{Kellermann1989}; 
White et al. \cite{White2000};
Ivezi\'c et al. \cite{Ivezic2002}). 
The mean value of the radio-loudness parameter for the 63 quasars from Table\,\ref{tab:sample} in the FIRST area
is $\bar{R} = 5.6$. Only three quasars (5\%) have $10 < R < 100$. This range of the radio-loudness parameter 
is sometimes called radio-intermediate (Falcke et al. \cite{Falcke1996}; Wang et al. \cite{Wang2006}). All three 
quasars belong to samples A and B.

\section{Discussion}\label{sec:disc}

\begin{figure*}[htbp]
\begin{tabbing}
\includegraphics[viewport=80 20 580 780,angle=270,width=8.8cm,clip]{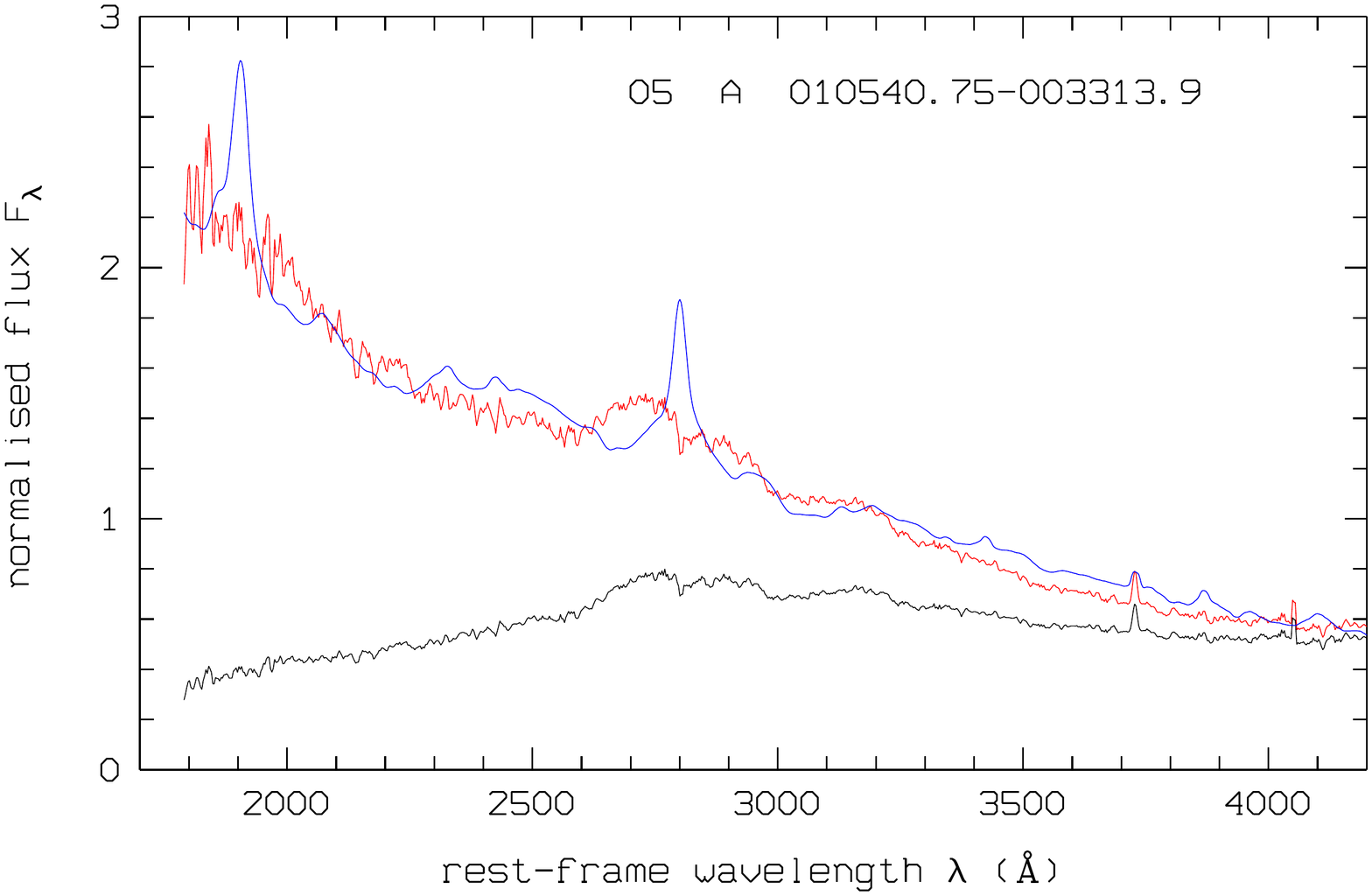}\hfill \=
\includegraphics[viewport=80 20 580 780,angle=270,width=8.8cm,clip]{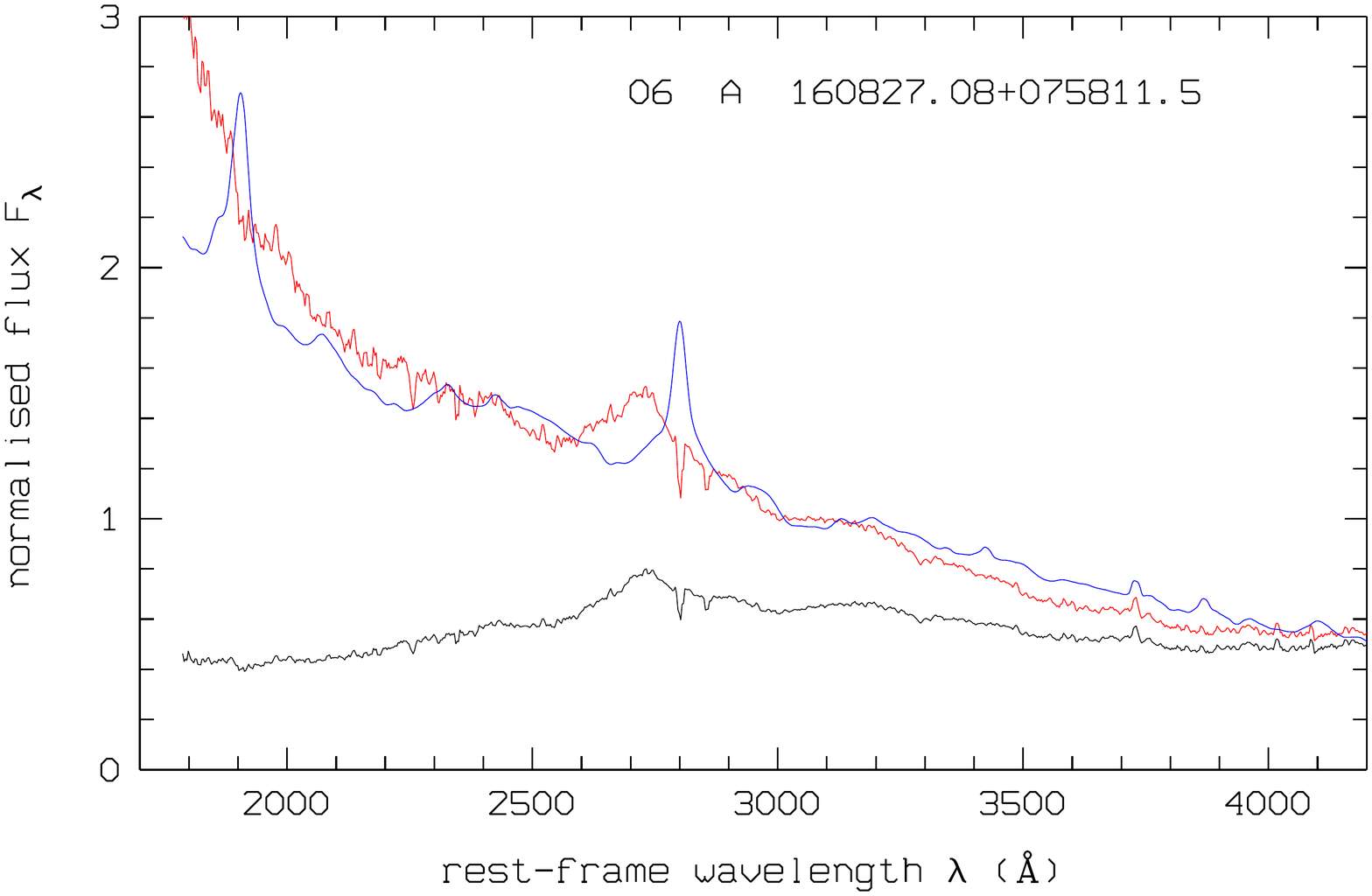}\hfill \\
\includegraphics[viewport=80 20 580 780,angle=270,width=8.8cm,clip]{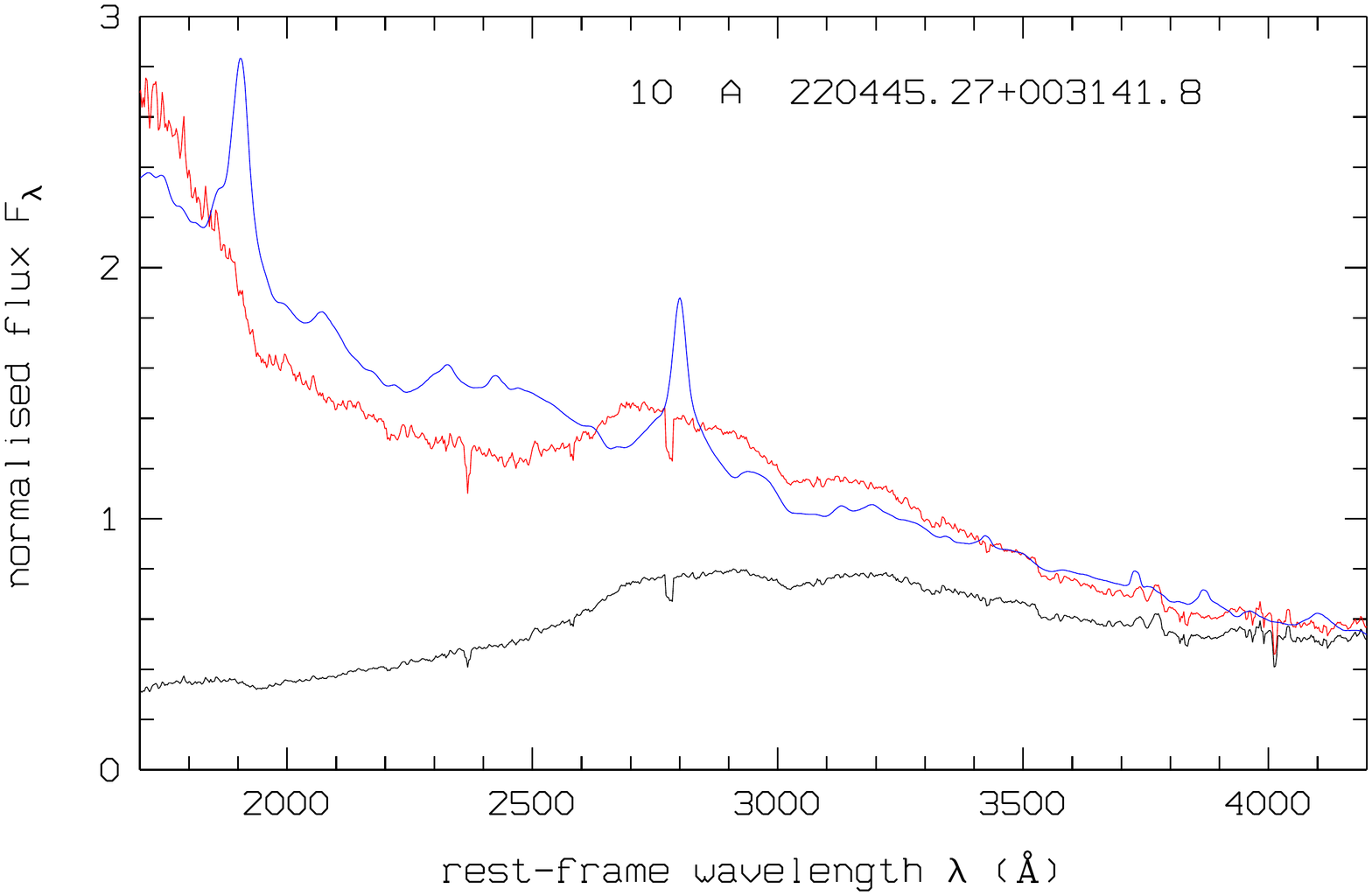}\hfill \=
\includegraphics[viewport=80 20 580 780,angle=270,width=8.8cm,clip]{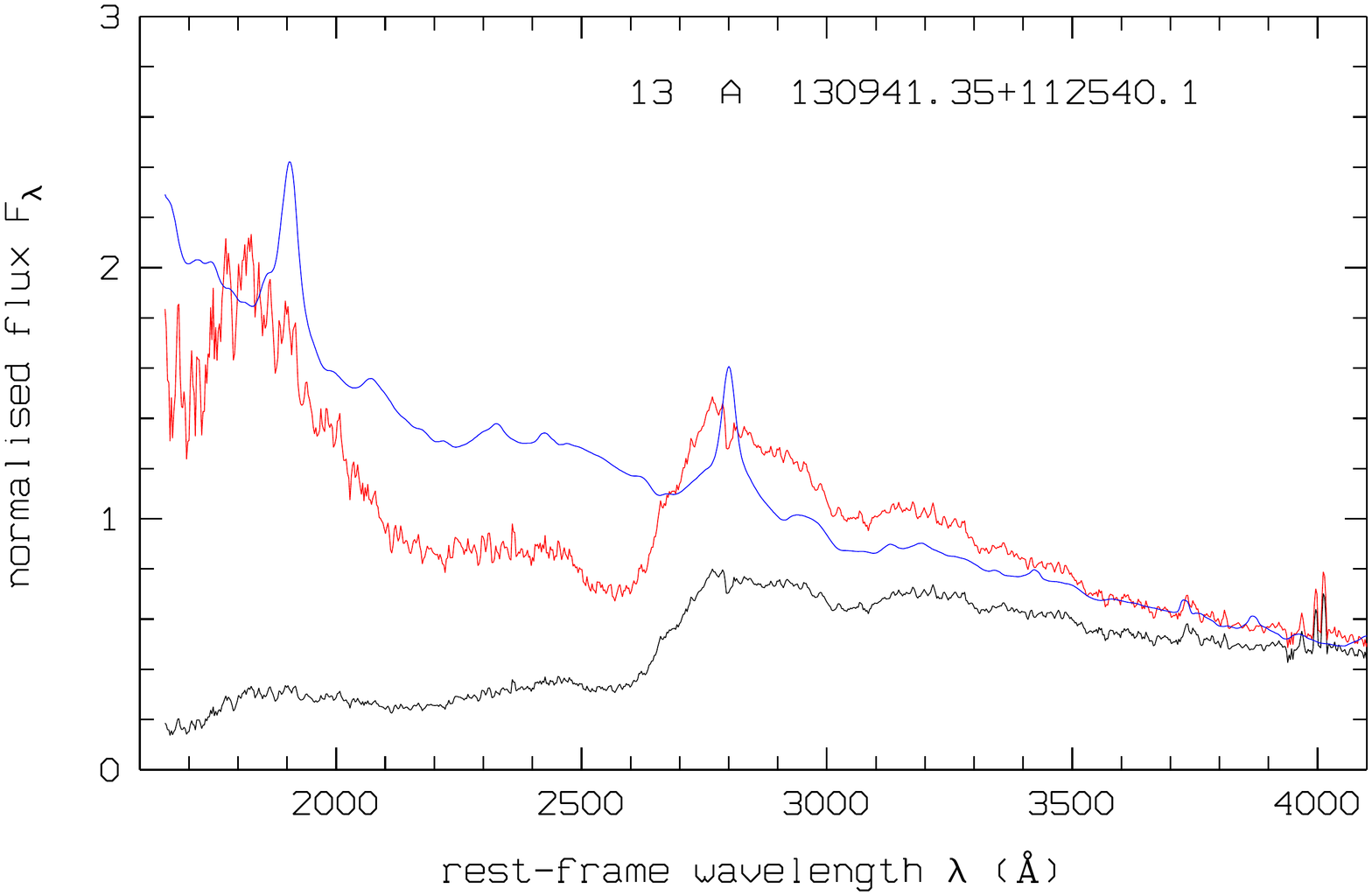}\hfill \\
\includegraphics[viewport=80 20 580 780,angle=270,width=8.8cm,clip]{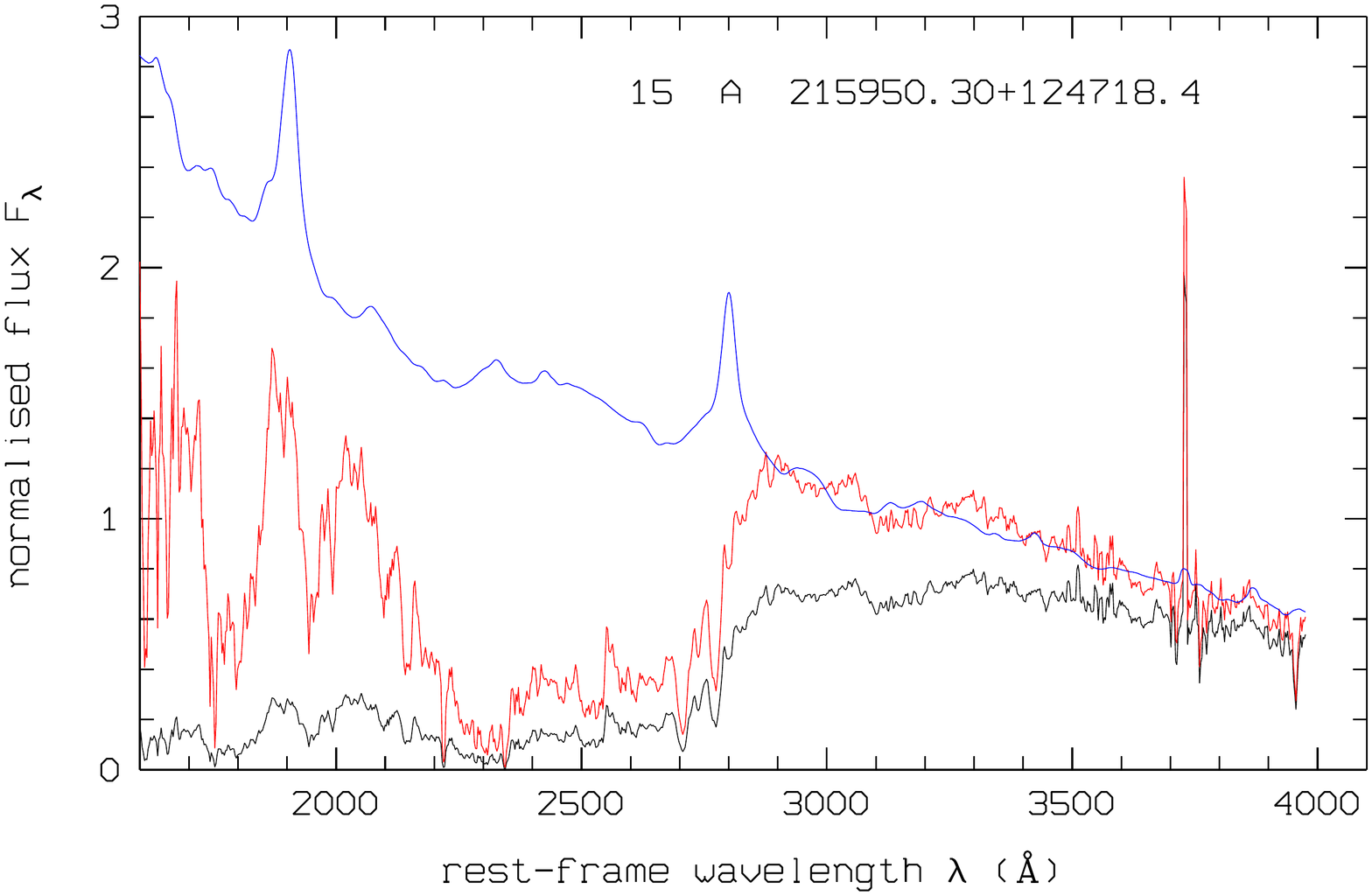}\hfill \=
\includegraphics[viewport=80 20 580 780,angle=270,width=8.8cm,clip]{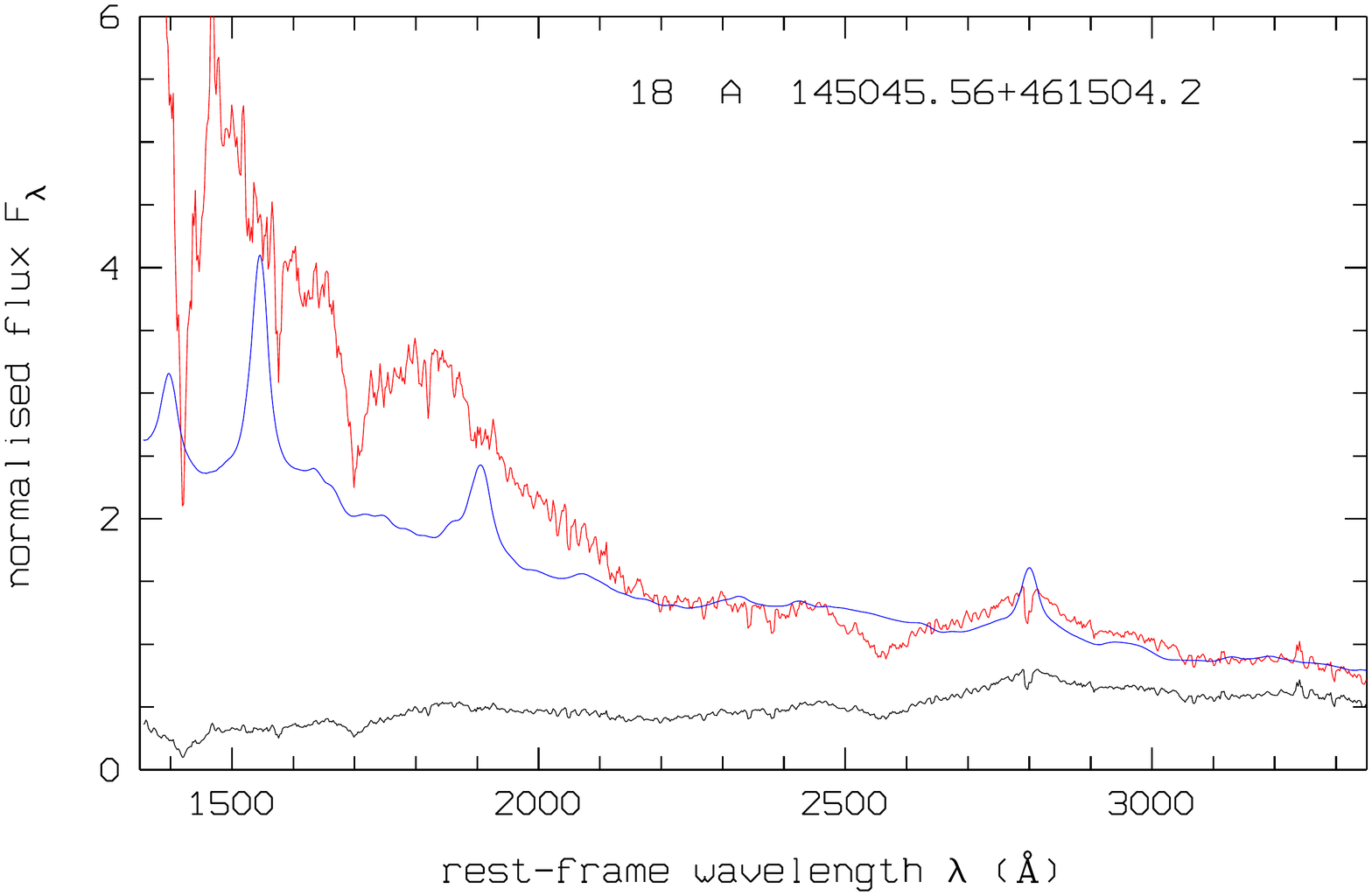}\hfill
\end{tabbing}
\caption{
Spectra of six 3000 \AA\ break quasars from sample A, de-reddened with the reddening law from
Jiang et al. (\cite{Jiang2013}; red). The observed spectra (after correction for Galactic foreground extinction) are shown in black, the SDSS quasar composite (Vanden Berk et al. \cite{VandenBerk2001}) in blue.
}
\label{fig:6examples_Jiang_dered}
\end{figure*}

\subsection{Extreme overlapping absorption troughs}\label{sec:disc_a}

The photometric data of the prototypical quasar J$010540.75-003313.9$ indicate 
enhanced FIR emission (Fig.\,\ref{fig:SED_ec_5_examples}). For the vast majority of our objects, however, 
the FIR is not covered by our photometric data, and no conclusion can be drawn on a possible starburst component.
From MIR to UV, the SED is dominated by the AGN. The results from the previous section demonstrate that the 
SED of 3000 \AA\ break quasars from MIR to NUV is indistinguishable from the SED of unusual BAL quasars.
All three quasar samples have SEDs that are similar to that one of normal quasars. This agrees with 
previous studies (see Sect.\,\ref{sec:intro}) showing that BAL quasars are not significantly different 
from non-BAL quasars in terms of their MIR-UV SED and IR luminosities. 
The similarity of the mean SEDs of samples A, B, and C from MIR to FUV supports the explanation of 
3000 \AA\ quasars as LoBAL quasars. 

It seems reasonable to assume that the main difference between samples A and C 
is unusually wide, overlapping absorption troughs for sample A quasars. Green (\cite{Green2006}) proposes 
a warm (ionised), probably filamentary relativistic wind that is smooth in velocity space with a
profile ``more like a shallow bowl than a trough''. The proposed absorption structure is then not identical to
typical BALs, but its velocity profile is broader and smoother. 
Single absorbers with high outflow velocities are indicated for some objects from our sample.
For example, the absorber system at $z_{\rm abs} = 1.158$ in the $z_{\rm em} = 1.359$ quasar J161836.09+153313.5 (\# 12) 
with a relative velocity $\sim 0.07 c$ (Sect.\,\ref{sect:individual}) makes it plausible that high-velocity BALs 
contribute to the absorption. The onset of a broad absorption structure at about 3000 \AA\ (towards shorter wavelengths) 
is likely due to overlapping absorption troughs from \ion{Mg}{ii} and \ion{Fe}{ii}, possibly in combination with 
iron emission at $\lambda \ga 3000$ \AA. It is well known that LoBAL quasars have enhanced Fe II emission line blends, and it is evident that this is also the case for many quasars from the comparison sample C. Blends by UV iron emission lines are likely to contribute also to the observed continuum structures in the spectra from sample A, but this is much less evident.
The understanding of their UV spectra would certainly benefit from a detailed quantitative analysis of the 
iron emission, but this is clearly beyond the scope of the present paper.

Figure\,\ref{fig:two_spectra} indicates spectral variability for both \object{J010540.75-003313.9} and
\object{J220445.27+003141.8} (Sect\,\ref{sec:sample}). The difference between the two spectra of \object{J220445.27+003141.8} shows the characteristic variability behaviour of the quasar continuum, namely stronger variability at shorter wavelengths (Wilhite et al. \cite{Wilhite2005}; Meusinger et al. \cite{Meusinger2011}).
The addition of BALs can strongly modulate the wavelength dependence of quasar variability, and in fact, 
BAL trough variability has been observed in a number of cases (Hall et al. \cite{Hall2002a} and references therein;
Sect.\,1). The difference of the two spectra of \object{J010540.75-003313.9} at $\lambda_{\rm obs} \la 5400$\AA\ is
clearly different from that of \object{J220445.27+003141.8} and might hint at variable absorption troughs. If true, the observed spectral variability of \object{J010540.75-003313.9} further supports the interpretation of the spectral shape at $\lambda < 3000$ \AA\ as due to unusually broad absorption line structures.

\subsection{Other, hypothetical explanations}\label{sec:alternative}

\subsubsection{Reddening with a steeper than SMC extinction curve}\label{sec:disc_b}

Although most of the statistical studies on dust reddening of quasars suggest an extinction curve like
in the SMC or flatter (Sect.\,\ref{sec:intr_dust}), reddening curves steeper than in the SMC were suggested
for some samples or individual quasars
(Hall et al. \cite{Hall2002a};
Zhou et al. \cite{Zhou2006};
Fynbo et al. \cite{Fynbo2013};
Jiang et al. \cite{Jiang2013}; 
Leighly et al. \cite{Leighly2014}).
Presuming that the continuum of the de-reddened broad-band SED is similar to that of the 
average quasar SED, Jiang et al. (\cite{Jiang2013}) have recently derived a reddening curve for the local 
3000 \AA\ break quasar \object{IRAS 14026+4341} that steeply rises towards short wavelengths at 
$\lambda < 3000$ \AA, which is very different from the reddening usually observed in nearby galaxies. 
These authors argue that such an unusual reddening law can be caused by the speculative assumption of
an exotic dust grain size distribution that lacks large grains.

\begin{figure}[h]
\includegraphics[viewport=60 20 580 800,angle=270,width=9.0cm,clip]{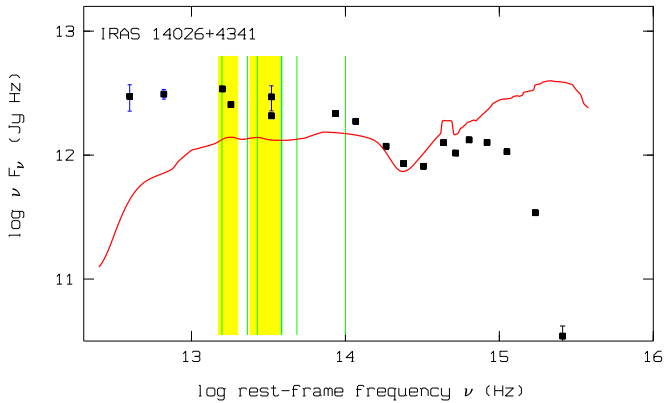}
\caption{Broad-band SED of \object{IRAS 14026+4341} (black squares) compared with the quasar composite SED from Fig.\,\ref{fig:SED_Shang} 
(red).
}
\label{fig:IRAS14026_broad_band}
\end{figure} 

Is the spectral shape of the 3000 \AA\ break quasars from sample A explained by 
the Jiang reddening curve? Figure\,\ref{fig:6examples_Jiang_dered} shows six representative 
examples. The quasars \# 10, 13, 15, and 18 clearly indicate absorption structures 
shortwards of the \ion{Mg}{ii} line that are better matched by a more or less flat trough or bowl 
than by the Jiang reddening curve. In the case of \object{J215950.30+124718.4} (\# 15), 
the de-reddened spectrum displays characteristic absorption structures from \ion{Fe}{ii}. 
\object{J145045.56+461504.2} (\# 18) is obviously over-corrected below 2200 \AA. 
The de-reddened spectrum of  \object{J160827.08+075811.5} (\# 6) resembles a weak-line quasar, but
with an unusual broad emission structure shortwards of \ion{Mg}{ii}. One might speculate that this
feature is an extremely broad wing of the \ion{Mg}{ii} (see also Hall et al. \cite{Hall2002a}), 
but this would require an unrealistically low systemic redshift of $z_{\rm em} = 1.13$.\footnote{There 
is an absorber system at $z_{\rm abs} = 1.182$. Since the absorber must be between the source and 
the observer, the only explanation would be an associated absorber moving towards the quasar with an 
unrealistic velocity $v = c [R^2 - 1] / [R^2 + 1] \approx - 7300$ km s$^{-1}$, where $R = (1+z_{\rm em})/(1+z_{\rm abs})$ 
(Foltz et al. \cite{Foltz1986}).} The structure in the de-reddened spectra therefore remains unexplained. 
The same applies to \object{J010540.75-003313.9} (\# 5).

We conclude that the reddening curve derived by Jiang et al. (\cite{Jiang2013}) for IRAS 14026+4341 does
not provide a reasonable explanation of the characteristic features of the 3000 \AA\ break quasars from our sample. 
There are two arguments that \object{IRAS 14026+4341} is different from our 3000 \AA\ break quasars. 
First, spectropolarimetric observations of \object{IRAS 14026+4341} revealed high polarisation that rises rapidly towards 
the blue, peaks at $ \sim3000$ \AA\ (rest-frame), and remains nearly constant at shorter wavelengths 
(Hines et al. \cite{Hines2001}). However, spectropolarimetric observations of the two prototypical 3000 \AA\ quasars \object{J010540.75-003313.9} and \object{J220445.27+003141.8} showed that their polarisation
is very weak (DiPompeo et al. \cite{DiPompeo2011}). 
Second, \object{IRAS 14026+4341} is brighter in the MIR and FIR than typical quasars
(Fig.\,\ref{fig:IRAS14026_broad_band}), but the quasars from our sample are not 
(see Figs.\,\ref{fig:mean_SED_ec_ic} and \ref{fig:L3000_shen}).

\subsubsection{Truncated accretion disks}\label{sect:disc_c}

Finally, we briefly address the question of whether the 3000 \AA\ break can be explained by an accretion disk 
with a strongly suppressed innermost part. An interesting candidate is the extremely X-ray weak BAL quasar 
\object{PG 0043+039} whose UV spectrum peaks at $\sim2500$\AA\ and shows a set of broad humps that 
can be explained as cyclotron line emission from a strongly magnetised hot plasma close to the supermassive
black hole (SMBH; Kollatschny et al. \cite{Kollatschny2015a},\cite{Kollatschny2015b}).

\begin{figure}[htbp]
\includegraphics[viewport=25 0 570 790,angle=270,width=8.5cm,clip]{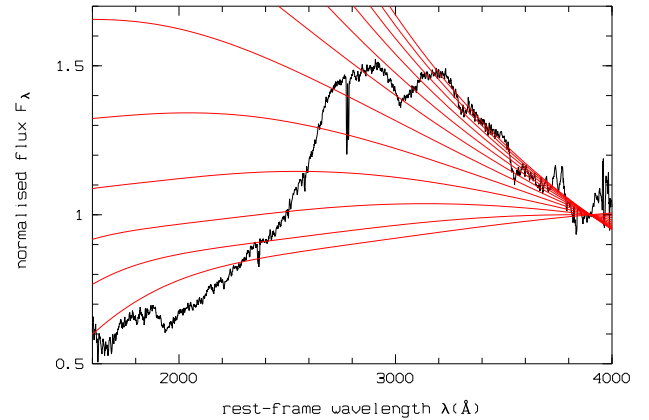}
\caption{BOSS spectrum of \object{J220445.27+003141.8} (black) compared with the spectra of an accretion disks
with a temperature parameter $T^\ast = 9\cdot 10^5$\,K (Meusinger \& Weiss \cite{Meusinger2013}) and an
inner hole of radius $r_{\rm in}$ from 3 to 38 (red curves, top to bottom) in logarithmic steps of 0.1 dex, 
arbitrarily normalised at 3900\AA.}
\label{fig:AD_with_hole}
\end{figure}

\begin{figure}[hhhh]
\includegraphics[viewport=30 0 600 790,angle=270,width=8.5cm,clip]{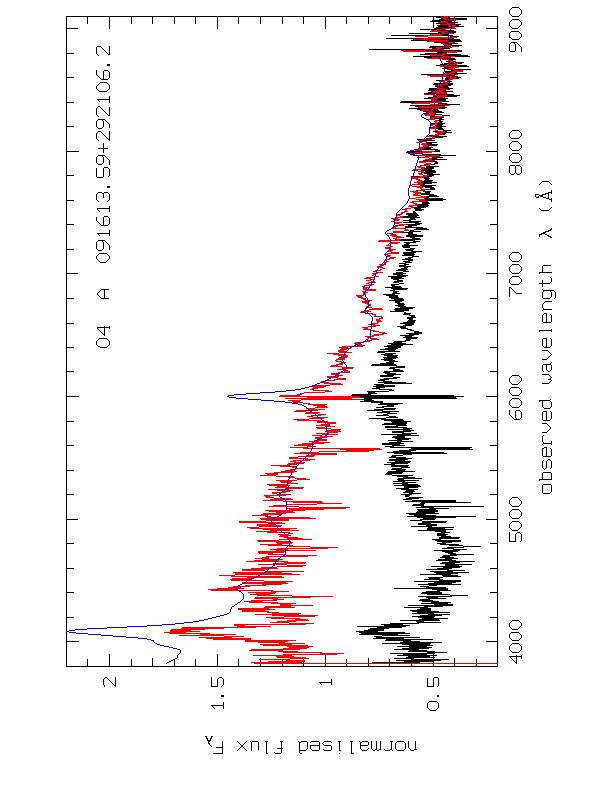}
\includegraphics[viewport=50 0 600 790,angle=270,width=8.5cm,clip]{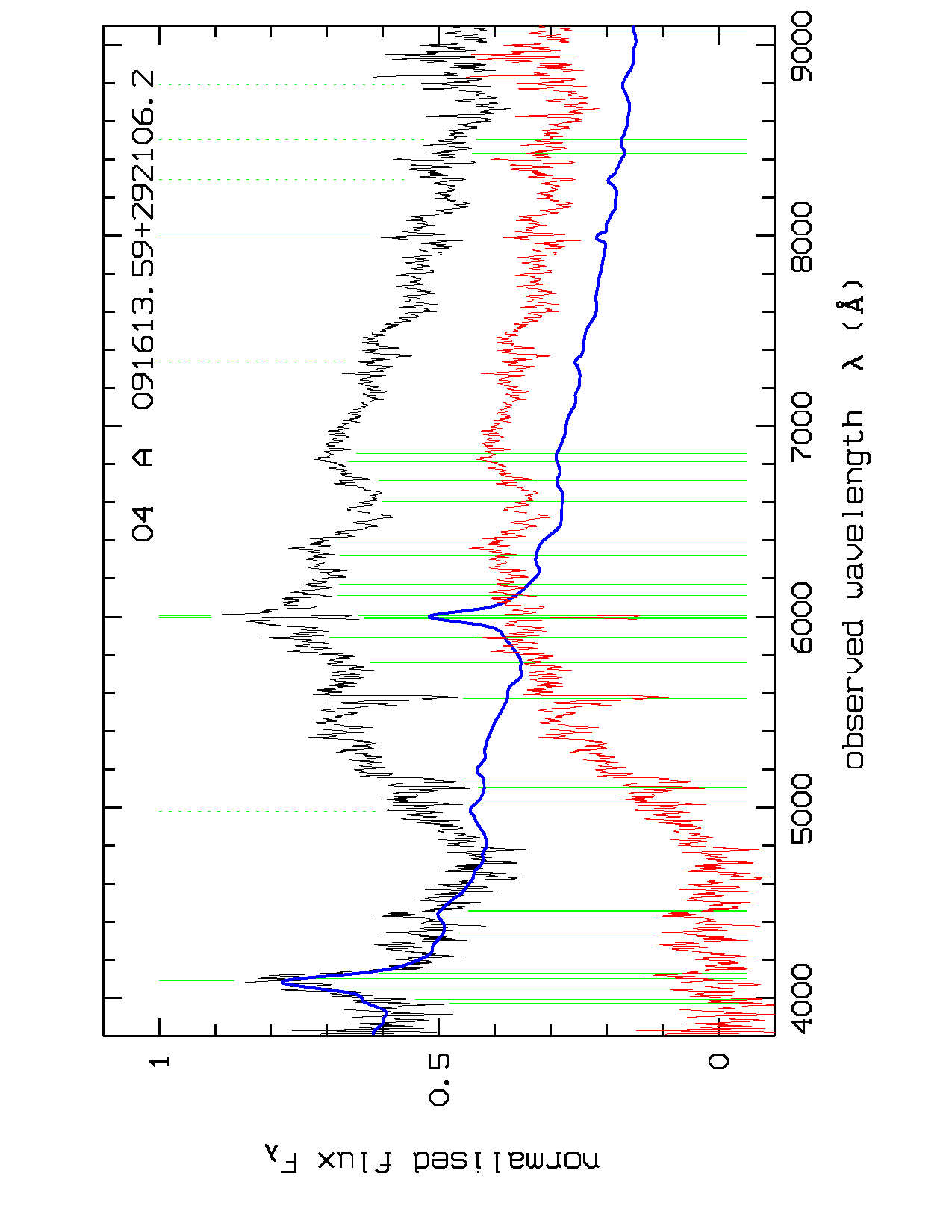}
\caption{Observed SDSS spectrum of J091613.59+292106.2 (black).
Top: De-reddened by the Milky Way extinction curve with $E(B-V) = 0.2$ mag (red) and compared with the
SDSS quasar composite (blue).
Bottom: SDSS quasar composite fitted at $\lambda < 4600$ \AA. The subtraction of the fitted composite
yields the red spectrum.}
\label{fig:091613_diff}
\end{figure}

A natural reason for a distortion of the accretion flow could be an inspiralling secondary SMBH in a close supermassive 
binary black hole (SMBBH; G\"ultekin \& Miller \cite{Gultekin2012}). 
The SMBHs in a galaxy merger are expected to sink into the potential well 
of the remnant via dynamical friction and to form a SMBBH (Begelman et al. \cite{Begelman1980}). 
The first luminous, spatially resolved binary quasar that clearly inhabits an ongoing galaxy merger 
was presented by Green et al. (\cite{Green2010}). The first FeLoBAL quasar in a relatively small 
separation binary was discovered by Gregg et al. (\cite{Gregg2002}). At a small separation and a sufficiently large ratio of the mass of the secondary SMBH to the binary mass, almost the entire region inside of the orbit of the secondary will be dynamically unstable. The major effect will be an inner hole in the accretion disk (AD) and thus a substantial suppression of the emission at short wavelengths, leaving a red SED (G\"ultekin \& Miller \cite{Gultekin2012}). Direct observational evidence of close SMBBHs is sparse, but there are promising candidates
(e.g., Valtonen et al. \cite{Valtonen2011}; 
Dotti et al. \cite{Dotti2012}; 
Graham et al. \cite{Graham2015}). 

A circum-binary AD with a central hole was recently proposed to explain the optical-to-UV continuum of
the nearby quasar Mrk 231 (Yan et al. \cite{Yan2015}). Its optical and UV spectrum  shows a maximum 
at $\sim 4000$ \AA\ followed by a decline at a shorter wavelength that is, however, much shallower than  
for the 3000 \AA\ break quasars. We computed spectra of model ADs assuming the Shakura \& Sunyaev (\cite{Shakura1973}) 
approximation for a set of disk temperature parameters with the inner radius as a free parameter.
As expected, the decline in the observed spectra is much steeper than
predicted by the model (Fig.\,\ref{fig:AD_with_hole}).

\subsubsection{Further peculiarities}\label{sec:special}

In a few cases, peculiarities in the spectra of very unusual quasars might stem
from the superimposition with other objects on the line of sight (Paper I). 
\object{J091613.59+292106.2} (\# 4) is the only quasar in sample A where the decrease in the SDSS spectrum 
shortwards of the peak turns into an increase at short wavelengths. In Paper I, we argued that this feature 
is perhaps due to reddening by Galactic dust\footnote{with its characteristic 2175 \AA\ trough} at
the redshift of the quasar. However, the agreement between the de-reddened spectrum and the quasar
composite is rather poor at $\lambda_{\rm obs} \la 4400$ \AA\ (Fig.\,\ref{fig:091613_diff}, top).
Alternatively, the observed spectrum might be the superimposition of the 
spectra of two quasars at approximately the same redshift (Fig.\,\ref{fig:091613_diff}, bottom).
The part at $\lambda_{\rm obs} \la 4600$ is well matched by the unreddened
SDSS quasar composite and might thus represent an ordinary quasar. 
The residual from subtracting this component from the observed spectrum of \object{J091613.59+292106.2} 
is similar to a typical 3000 \AA\ break quasar spectrum.

\section{Summary and conclusions}\label{sec:summary}

The 3000 \AA\ break quasars constitute a rare type of luminous AGNs. The defining spectral features are
a usual blue continuum at $\lambda \ga 3000$ \AA,\ in combination with a drop-off around 3000 \AA\ that does
not clearly show the typical structure of broad absorption lines, and a lack of substantial typical quasar emission lines.
The prototypes are \object{J010540.75-003313.9} and \object{J220445.27+003141.8}
from Hall et al. (\cite{Hall2002a}). 
We searched for 3000 \AA\ break quasars in the spectra database of the SDSS DR7 and DR10 using
a number of large Kohonen self-organising maps.
We compiled a sample of 23 quasars of the type of 3000 \AA\
break quasars (sample A). In addition, we compiled both a comparison sample of
other unusual quasars that is dominated by LoBAL quasars with overlapping
absorption BAL troughs or with many absorption lines (sample C) and a sample
with quasars showing properties between those of samples A and C (sample B).
Particular attention was paid to avoiding contamination by unusual DQ white dwarfs that were found to mimic the spectra of 3000 \AA\ break quasars.

We constructed broad-band SEDs for all 64 quasars using archive data from GALEX, 2MASS, UKIDSS, WISE, {\it Spitzer}, 
and {\it Herschel}, in addition to the photometric and spectroscopic data from the SDSS. We corrected for dust reddening at 
the redshift of the quasars by the comparison with the mean broad-band SED of normal quasars for standard assumptions
on the reddening law (SMC reddening and alternatively a flat reddening curve). The derived mean values of the colour excess
$E(B-V) \approx 0.1 - 0.2$ mag are higher than for normal quasars. However, as for the entire SDSS quasar population,  
strongly reddened quasars with $E(B-V) > 0.4$ mag are very rare.
Provided the intrinsic SED is similar to the standard quasar SED, the quasars from all three samples
are intrinsically relatively bright in the UV. The absorption-corrected monochromatic
luminosities at 3000 \AA\ are, on average, higher than for normal quasars at the same redshift.

At wavelengths $\ga 3000$ \AA, the de-reddened arithmetic median composite SEDs from MIR to NUV are
very similar for the three samples and are indistinguishable from the SED of normal quasars.
The 3000 \AA\ break in the continuum does not vanish after de-reddening with standard reddening laws. 
The similarity of the broad-band SEDs of samples A, B, and C supports the interpretation of 3000 \AA\ break quasars
as extreme versions of LoBAL quasars. We conclude that 3000 \AA\ break quasars are most likely extreme versions of LoBAL quasars. A definitive proof could be provided by high-resolution spectra with sufficiently high signal-to-noise
ratio and/or by further multi-epoch spectroscopy. In either case, it is quite possible that individual blue-shifted absorption features might be separately identified.

The percentage of FIRST-detected radio sources (74\%, 60\%, and 16\%  for samples A, B, and C, respectively)  
is much higher than for the entire population of SDSS quasars. This is most likely a selection bias caused by the 
fact that the majority of the quasars with spectra similar to 3000 \AA\ break quasars were targeted by the SDSS, 
not because of their UV properties but because they were detected as radio sources. If this is true, it seems 
likely that a much larger population of such quasars exists that can be found by a dedicated search program focused on the UV properties when making use of the photometric data from IR surveys in combination with proper motion data from ESA's Gaia mission to reduce stellar contamination.

%
\begin{acknowledgements}

We are grateful to Martin Haas and the anonymous referee for helpful comments on an
early version of this paper. We also owe thanks to Aick in der Au and J\"org Br\"unecke for
their improvements of the computation and manageability of the SOMs.

This research made use of data products from the Sloan
Digital Sky Survey (SDSS). Funding for the SDSS and SDSS-II has been provided by
the Alfred P. Sloan Foundation, the Participating Institutions
(see below), the National Science Foundation, the National
Aeronautics and Space Administration, the U.S. Department
of Energy, the Japanese Monbukagakusho, the Max Planck
Society, and the Higher Education Funding Council for
England. The SDSS Web site is http://www.sdss.org/.
The SDSS is managed by the Astrophysical Research
Consortium (ARC) for the Participating Institutions.
The Participating Institutions are the American
Museum of Natural History, Astrophysical Institute
Potsdam, University of Basel, University of Cambridge
(Cambridge University), Case Western Reserve University,
the University of Chicago, the Fermi National
Accelerator Laboratory (Fermilab), the Institute
for Advanced Study, the Japan Participation Group,
the Johns Hopkins University, the Joint Institute
for Nuclear Astrophysics, the Kavli Institute for
Particle Astrophysics and Cosmology, the Korean
Scientist Group, the Los Alamos National Laboratory,
the Max-Planck-Institute for Astronomy (MPIA),
the Max-Planck-Institute for Astrophysics (MPA),
the New Mexico State University, the Ohio State
University, the University of Pittsburgh, University
of Portsmouth, Princeton University, the United
States Naval Observatory, and the University of
Washington. 

This publication has made extensive use of the VizieR catalogue access
tool, CDS, Strasbourg, France and of data obtained from 
the NASA/IPAC Infrared Science Archive (IRSA), operated by the 
Jet Propulsion Laboratories/California Institute of Technology, founded
by the National Aeronautic and Space Administration. In particular  
we made use of the data products from 2MASS, WISE, Spitzer, and Herschel.
This work is also based in part on the data obtained as part of the 
UKIRT Infrared Deep Sky Survey.
Some of the data presented in this paper were obtained from the 
Mikulski Archive for Space Telescopes (MAST). STScI is operated by the 
Association of Universities for Research in Astronomy, Inc., under 
NASA contract NAS5-26555. Support for MAST for non-HST data is
provided by the NASA Office of Space Science via grant NNX09AF08G and
by other grants and contracts.

\end{acknowledgements}
%

%
{}

%

\newpage

\appendix
\section{Remarks on individual quasars}\label{sect:individual}      
\subsection{Sample A}

(1)\ {\it  J105528.80+312411.3}\\ 
Originally discovered by the FIRST Bright Quasar Survey (FBQS; White et al. \cite{White2000}).
Appears to be a low-$z$ version of the prototypical 3000 \AA\ break quasars (Hall et al. \cite{Hall2002a}).
Strong broad iron emission and associated \ion{Mg}{ii} absorption at $z_{\rm abs} =0.492$. Emission redshift only roughly estimated to $z_{\rm em} \approx 0.5$. Comparison of the SDSS spectra from three epochs indicates strong spectral variability at $\lambda_{\rm rest} < 3400$ \AA\ (Zhang et al. \cite{Zhang2015}).\\

\noindent
(2) \ {\it J140025.53-012957.0}\\
Quite similar to J105528.80+312411.3 (\# 1). Emission redshift from [\ion{O}{ii}] $\lambda$3728, H$\beta$,
and [\ion{O}{iii}] $\lambda\lambda$4960.3, 5008.2 in agreement with the absorption redshift from
\ion{Mg}{ii}. Belongs to the LoBAL quasar sample studied by Lazarova et al. (\cite{Lazarova2012}) in the MIR.\\

\noindent
(3) \ {\it J134050.80+152138.7}\\
Several strong narrow emission lines. Emission redshift in good agreement with the redshift of the associated
absorption from \ion{Mg}{ii} and \ion{Fe}{ii}.\\

\noindent
(4) \ {\it J091613.59+292106.2}\\
Similar to J134050.80 +152138.7 (\# 3) at $\lambda \ga 4600$ \AA, but peculiar at shorter wavelengths
(Sect.\,\ref{sec:special}).\\ 

\noindent
(5) \ {\it \object{J010540.75-003313.9}}\\
One of the two prototypical 3000 \AA\ break quasars discovered by Hall et al. (\cite{Hall2002a}).
BOSS spectrum displays broad iron emission features near 4000 \AA\ and 4200 \AA\  and the
[\ion{O}{ii}] $\lambda$3728 emission line. Emission redshift $z = 1.179$
in good agreement with the redshift of the resolved associated Mg II absorption. Deviation of
the BOSS spectrum from the SDSS spectrum at $\lambda_{\rm obs} \la 5400$ \AA\  probably indicates
variable absorption troughs (Fig.\,\ref{fig:two_spectra}, Sect.\,\ref{sec:disc_a}).\\ 

\noindent
(6) \ {\it J160827.08+075811.5}\\
Discovered by Plotkin et al. \cite{Plotkin2010} (and independently in Paper I).
If the narrow line \ion{Mg}{ii} and \ion{Fe}{ii} absorber system and the 
[\ion{O}{ii}] $\lambda$3730  emitter indicate the systemic redshift of the quasar, the
pronounced peak in the observed spectrum is shortward of the \ion{Mg}{ii} line 
at $\sim 2700$\AA\ rest-frame.\\  

\noindent
(7) \ {\it J152928.53+133426.6}\\
Targeted by SDSS as a high-$z$ quasar and wrongly classified by the pipeline of SDSS DR12
as a galaxy at $z=0.578$. WISE colour index $w1-w2 = 1.4$\,mag typical for quasars.
Two peaks inside the deep absorption trough at $\sim 6200$ \AA\ separated by 16 \AA\ can be identified with the
\ion{Mg}{ii} $\lambda\lambda$ 2796.3, 2803.5 doublet at $z = 1.23$. 
Weak [\ion{Ne}{iii}] $\lambda$3870 emission is indicated at the same redshift, but at low S/N. \\ 

\noindent
(8) \ {\it J134246.24+284027.5}\\
In paper I, $z = 1.3$ was given. Here we assume $z=1.255$ as a better match, but the redshift 
remains fairly uncertain.\\

\noindent
(9) \ {\it J091940.97+064459.9}\\
Associated absorption from \ion{Mg}{ii} and \ion{Fe}{ii} at $z = 1.352$ and [\ion{O}{ii}] emission at the 
same redshift. Classified as a possible lower confidence BL Lac object by Plotkin et al. (\cite{Plotkin2008}).\\

\noindent
(10) \ {\it \object{J220445.27+003141.8}}\\
One of the two prototypical 3000 \AA\ break quasars
discovered by Hall et al. (\cite{Hall2002a}).
Continuum similar to \object{J010540.75-003313.9} (\# 5).
Associated \ion{Mg}{ii} and \ion{Fe}{ii} absorption at z = 1.335.
Emission redshift uncertain. Hall et al. (\cite{Hall2002a}) derived
$z_{\rm em} = 1.353$ from cross-correlation with \object{J010540.75-003313.9}. BOSS spectrum displays
two faint peaks at $\sim 8800$\AA, where the blue peak was identified here with 
[\ion{O}{ii}] $\lambda$3728 at $z = 1.353$.\\

\noindent
(11) \ {\it J111541.01+263328.6}\\
The (noisy) SDSS spectrum displays two absorption troughs from \ion{Mg}{ii} and \ion{Al}{iii} at 
$z_{\rm abs} = 1.327$. Identification of the double emission peak at the red edge of the 6500\AA\ trough
with \ion{Mg}{ii} $\lambda\lambda$ 2796.3, 2803.5 yields $z_{\rm em} = 1.355$. Emission redshift
uncertain.\\

\noindent
(12) \ {\it J161836.09+153313.5}\\
Another source discovered by Plotkin et al. \cite{Plotkin2010} and independently in Paper I. In earlier SDSS 
data releases identified as high-$z$ quasar due to a confusion of the peak with Lyman $\alpha$.  Carballo et al. 
(\cite{Carballo2008}) give a revised value $z = 1.322$ based on the identification of the peak with \ion{Mg}{ii}.
SDSS spectrum similar to J160827.08+075811.5 (\# 6).
If the narrow emission line at $\lambda_{\rm obs} \approx 8800$\AA\ is [\ion{O}{ii}] $\lambda$3730,                       the systemic redshift is $z_{\rm em} = 1.359$ and the peak of the spectrum is at shorter
wavelengths than \ion{Mg}{ii}. The system of narrow \ion{Mg}{ii} and \ion{Fe}{ii} absorbers at
$z_{\rm abs} = 1.158$ make it plausible that high-velocity BALs contribute to the absorption.\\

\noindent
(13) \ {\it J130941.35+112540.1}\\
Another source discovered by Plotkin et al. \cite{Plotkin2010} and independently in Paper I.
Similar to J134246.24+284027.5 (\# 8), though the break is at the blue side of the \ion{Mg}{ii} absorption.\\

\noindent
(14) \ {\it J152423.18+115312.0}\\
Similar to J130941.35+112540.1 (\# 14). No clearly identified emission or absorption line.
Redshift estimated from fitting the SDSS composite at the red side of the peak and by the comparison
with J130941.35+112540.1 (\# 13) at the blue side.\\

\noindent
(15) \ {\it J215950.30+124718.4}\\
Discovered by Plotkin et al. \cite{Plotkin2010} and independently in Paper I.
BOSS spectrum  reveals strong [\ion{O}{ii}] $\lambda3730$ and probably
weak [\ion{Ne}{iii}] $\lambda$ 3870  emission at $z_{\rm em} = 1.514$ as well as 
slightly redshifted Ca H+K absorption. Similar to J134246.24+284027.5 (\# 8) but 
stronger partial coverage of the continuum source by overlapping trough absorption.\\

\noindent
(16) \ {\it J085502.20+280219.6}\\
Similar to J091940.97+064459.9 (\# 9). Absorber system at $z_{\rm abs} = 1.515$, no emission lines
(the increase shortward of 3900 \AA\ is most likely an artifact).\\

\noindent
(17) \ {\it J134408.32+283932.0}\\
Similar to J085502.20+280219.6 (\# 16), strong narrow \ion{Mg}{ii} and \ion{Fe}{ii} absorption
at $z_{\rm abs} = 1.767$. At this redshift, the [\ion{O}{ii}] $\lambda$3730 is not covered by the BOSS spectrum.
No other emission line is seen. Known as a 2MASS quasar with radio spectral properties suggestive
of young radio jets (Geogakakis et al. \cite{Georgakakis2012}).\\

\noindent
(18) \ {\it J145045.56+461504.2}\\ 
Discovered by Plotkin et al. \cite{Plotkin2010} and independently in Paper I. Redshift $z_{\rm abs} = 1.877$
estimated from the identification of the observed absorption lines at 8043\AA\ and 8063\AA\ with
the \ion{Mg}{ii} $\lambda\lambda$ 2796.3, 2803.5 doublet.\\     

\noindent
(19) \ {\it J110511.15+530806.5}\\
Strongly reddened narrow-line quasar. BOSS spectrum displays strong narrow emission lines from
\ion{Si}{iv}, \ion{C}{iv}, \ion{Al}{iii}, \ion{C}{iii}, and  \ion{Mg}{ii} and an associated \ion{Mg}{ii} 
absorption line. Reddening steeper than in the SMC, perhaps even steeper than
for IRAS 14026+4341 (Sect.\,\ref{sec:disc_b}).\\

\noindent
(20) \ {\it J075437.85+422115.3}\\
Continuum is similar to J110511.15+530806.5 (\# 19), \ion{Mg}{ii} emission is indicated.
Classified as a high-confidence BL Lac candidate by Plotkin et al. (\cite{Plotkin2010}).\\

\noindent
(21) \ {\it J092602.98+162809.2}\\
Red quasar with a weak \ion{C}{iii}] emission line and only moderate \ion{Mg}{ii} and \ion{Al}{iii}
absorption troughs at the same redshift. Redward of \ion{Mg}{ii} $\lambda$ 2800 the spectral slope seems 
to turn blue.\\   

\noindent
(22) \ {\it J013435.66-093103.0}\\
Peculiar in various respects: very red (Urrutia et al. \cite{Urrutia2005}),
highest radio-loudness in our sample ($R \approx 65$; Sect.\,\ref{sec:RL}), and 
a gravitationally lensed quasar with complex radio morphology (Winn et al. \cite{Winn2002}, \cite{Winn2003}). 
Explained by a model wherein lensing and reddening is caused by the absorber system at $z\approx0.76$
(Hall et al. \cite{Hall2002b}).\\

\noindent
(23) \ {\it J100933.22+255901.1}\\
Another quasar where the continuum appears red shortward of \ion{Mg}{ii} and blue at longer wavelengths. 
HiBALs from Lyman $\alpha$, \ion{Si}{iv}, and \ion{C}{iv} and associated absorption lines
from \ion{Mg}{ii} and \ion{Fe}{ii} at approximately the same redshift.

\subsection{Samples B and C}

In the following we make short notes on only a few remarkable objects from samples B and C.\\

\noindent
(24) \ {\it J141428.30+185646.0}\\
One of the few known FeLoBALs at $z<0.7$. The spectrum displays strong \ion{Fe}{ii} emission.\\ 

\noindent
(29) \ {\it J120337.91+153006.6}\\
Absorption line variability found by McGraw et al. (\cite{McGraw2015}).\\ 

\noindent
(30) \ {\it J164941.87+401455.9}\\
Similar to SDSSJ033810.85+005617.6 from Hall et al. (\cite{Hall2002a}).\\  

\noindent
(31) \ {\it J153341.88+150059.5}\\
Behind the outer disk of the spiral galaxy NGC 5951 at $z = 0.0059$.\\ 

\noindent
(32-35) \ {\it J165117.31+240836.3, J101723.04+230322.1, J151627.40+305219.7, J073816.91+314437.0}\\
High-velocity BALs. \\ 

\noindent
(39) \ {\it J030000.57+004828.0}\\
Intensively studied overlapping-trough FeLoBAL 
(Lacy et al. \cite{Lacy2002}; 
Gregg et al. \cite{Gregg2002};
Morabito et al. \cite{Morabito2011};
Farrah et al. \cite{Farrah2012}). 
Detailed analysis of the spectral properties
by Hall et al. (\cite{Hall2003}) based on VLT+UVES spectroscopy. 
No line variability indicated (McGraw et al. \cite{McGraw2015}).\\

\noindent
(41) \ {\it J115436.60+030006.3}\\
Another well-studied overlapping-trough FeLoBAL, similar to SDSSJ030000.57+004828.0 (\# 39)
(Hall et al. \cite{Hall2002a}; Farrah et al. \cite{Farrah2012}).
No line variability indicated (McGraw et al. \cite{McGraw2015}).\\

\noindent
(46 and 51) \ {\it J173049.10+585059.5, J094317.59+541705.1}\\
Two extreme overlapping FeLoBAL quasars where there is always no flux shortward of the onset of
\ion{Fe}{ii} UV2 at 7000 \AA\ observed. J173049.10+585059.5 was discussed by Hall et al. 
(\cite{Hall2002a}), J094317.59+541705.1 was discovered by Urrutia et al. (\cite{Urrutia2009}).\\

\noindent
(52 and 59) \ {\it J162527.73+093332.8, J100424.89+122922.2}\\
Similar to \# 46 and \# 51, but at slightly higher $z$. 100424.89+122922.2 was classified as one of the 
most heavily reddened quasars with $R-K = 6.26$ mag (Glikman et al. \cite{Glikman2004}).\\

\noindent
(50, 53, 55, and 59) are examples of LoBAL quasars showing many narrow troughs in their
spectra.

\newpage

\section{Individual quasar SEDs}

Figure\,\ref{fig:spectra_and_seds} displays the individual SEDs for all quasars from our samples A, B, and C 
in the same order as in Table\,\ref{tab:sample}. On the left-hand side, the observed optical spectra
expressed by log\,$F_\lambda$ $(\lambda)$ are shown in black, over-plotted with a slightly smoothed version in red.
Each panel is labelled with the running number from Table\,\ref{tab:sample}, the name, and the redshifts
estimated from emission and/or absorption features. For comparison, the SDSS quasar composite spectrum
from Vanden Berk et al. (\cite{VandenBerk2001}) is over-plotted in blue. The vertical solid green
lines above the quasar spectrum indicate the positions of the strongest emission lines in the composite,
the dotted lines indicate typical weaker emission lines. The vertical green lines below the quasar spectrum
indicate the positions of strong absorption lines seen in the spectra of BAL quasars (taken from
Table\,1 in Hall et al. \cite{Hall2002a}).

The panels on the right-hand side of Fig.\,\ref{fig:spectra_and_seds} show the broad-band SEDs as
log\,$\nu F_\nu$ (log $\nu$) in the quasar's rest frame. The transformation into the rest frame
is based on the emission redshift, if available, and  on the absorption redshift in all other cases.
The original fluxes are plotted as red squares with error bars without correction for intrinsic
dust reddening. For the majority of the data points, the error bars are smaller than the symbol size.
Upper limits (for the data from GALEX and WISE) are indicated by downward arrows.
FIRST radio sources are labelled as such in the lower left-hand corner of each panel.
For quasars measured in the NIR bands J, H, and K both by 2MASS and UKIDSS, only the UKIDSS fluxes
were plotted.  The blue squares are the fluxes corrected for intrinsic SMC
reddening, the magenta squares for Gaskell dust. The data points are interconnected by dotted
lines in order to guide the eye. The SMC extinction-corrected optical spectra are over-plotted
in cyan. The vertical yellow stripes indicate the positions of the silicate absorption troughs in
the MIR, the vertical green lines mark the positions of the PAH emission features.

\begin{figure*}[h]
\begin{tabbing}
\includegraphics[viewport=125 0 570 790,angle=270,width=8.0cm,clip]{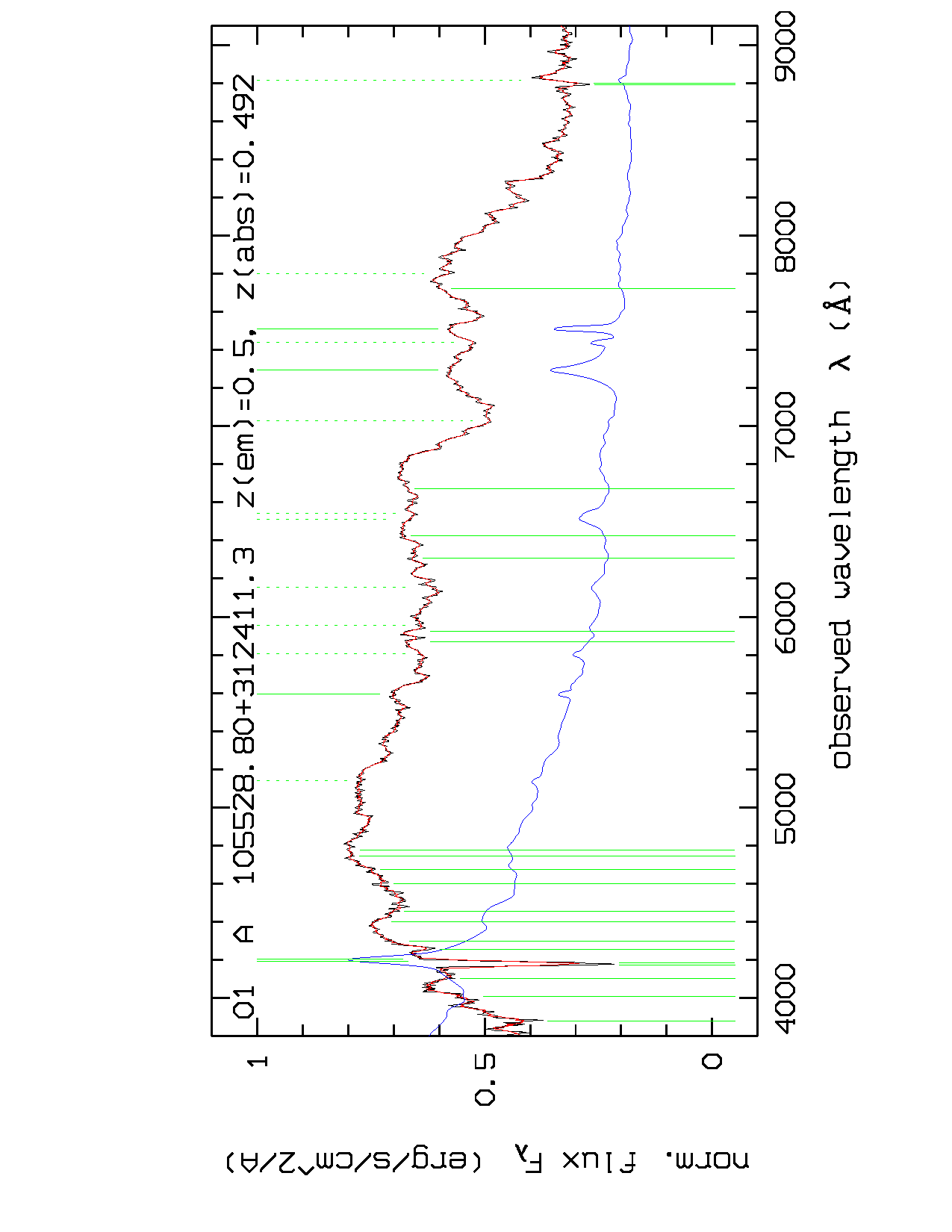}\hfill \=
\includegraphics[viewport=125 0 570 790,angle=270,width=8.0cm,clip]{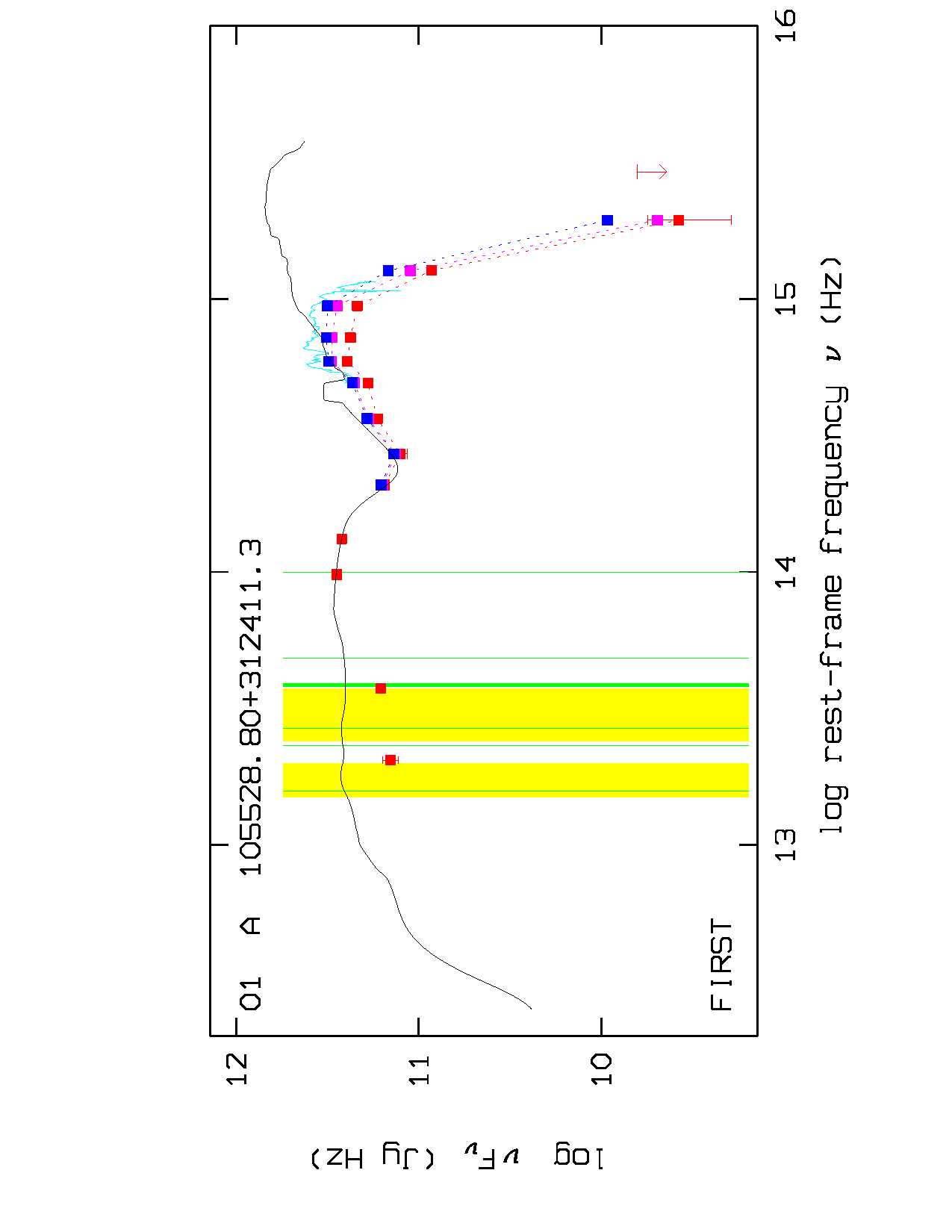}\hfill  \\
\includegraphics[viewport=125 0 570 790,angle=270,width=8.0cm,clip]{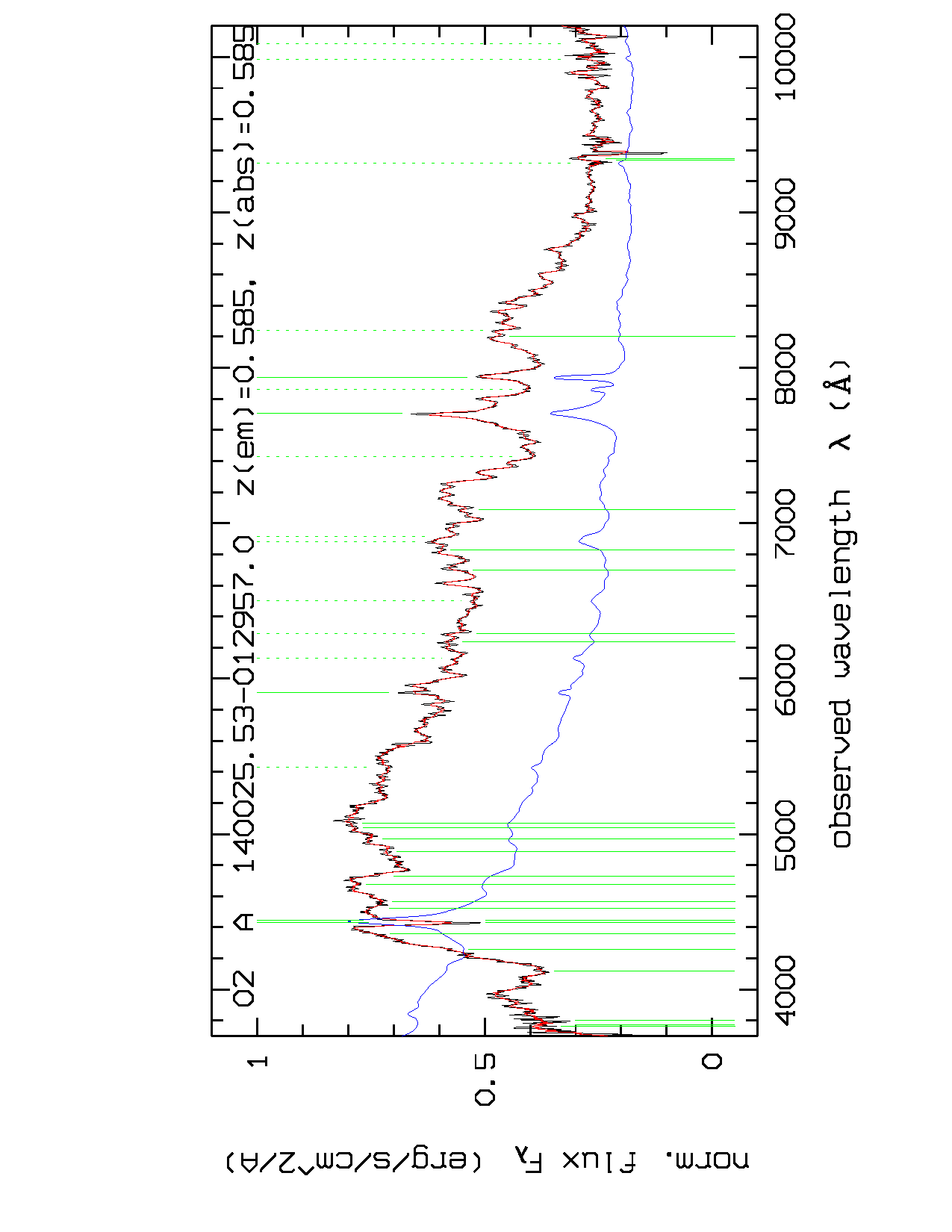}\hfill \=
\includegraphics[viewport=125 0 570 790,angle=270,width=8.0cm,clip]{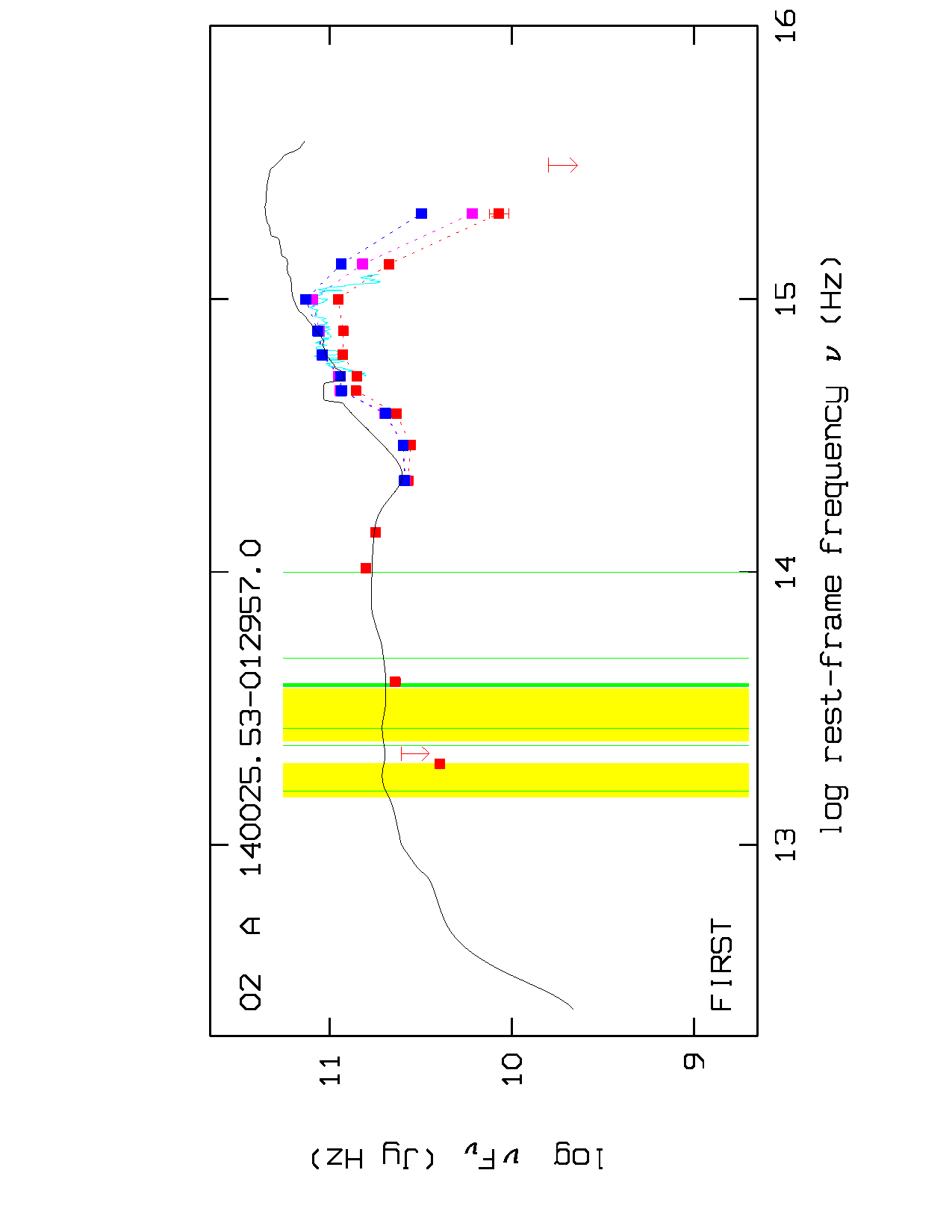}\hfill  \\
\includegraphics[viewport=125 0 570 790,angle=270,width=8.0cm,clip]{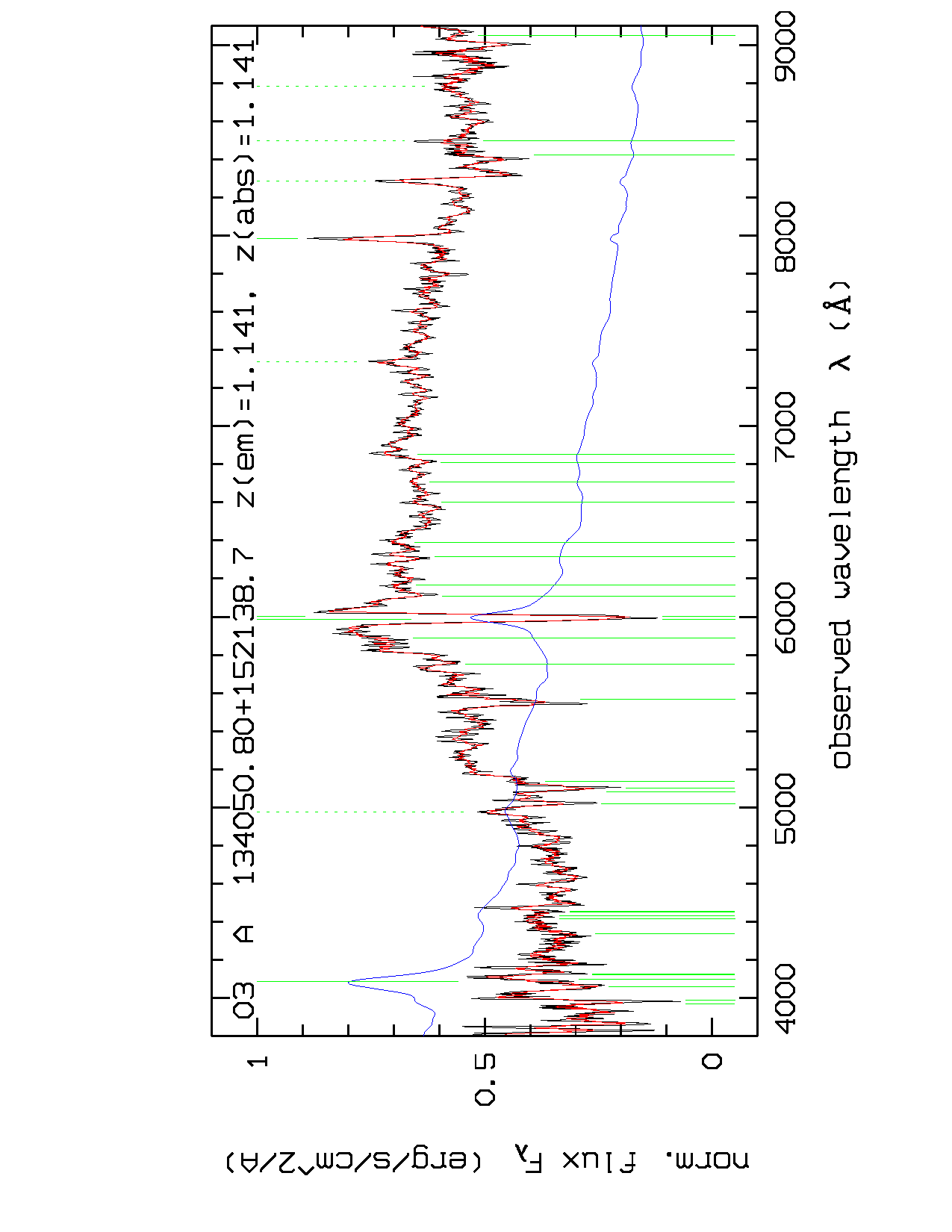}\hfill \=
\includegraphics[viewport=125 0 570 790,angle=270,width=8.0cm,clip]{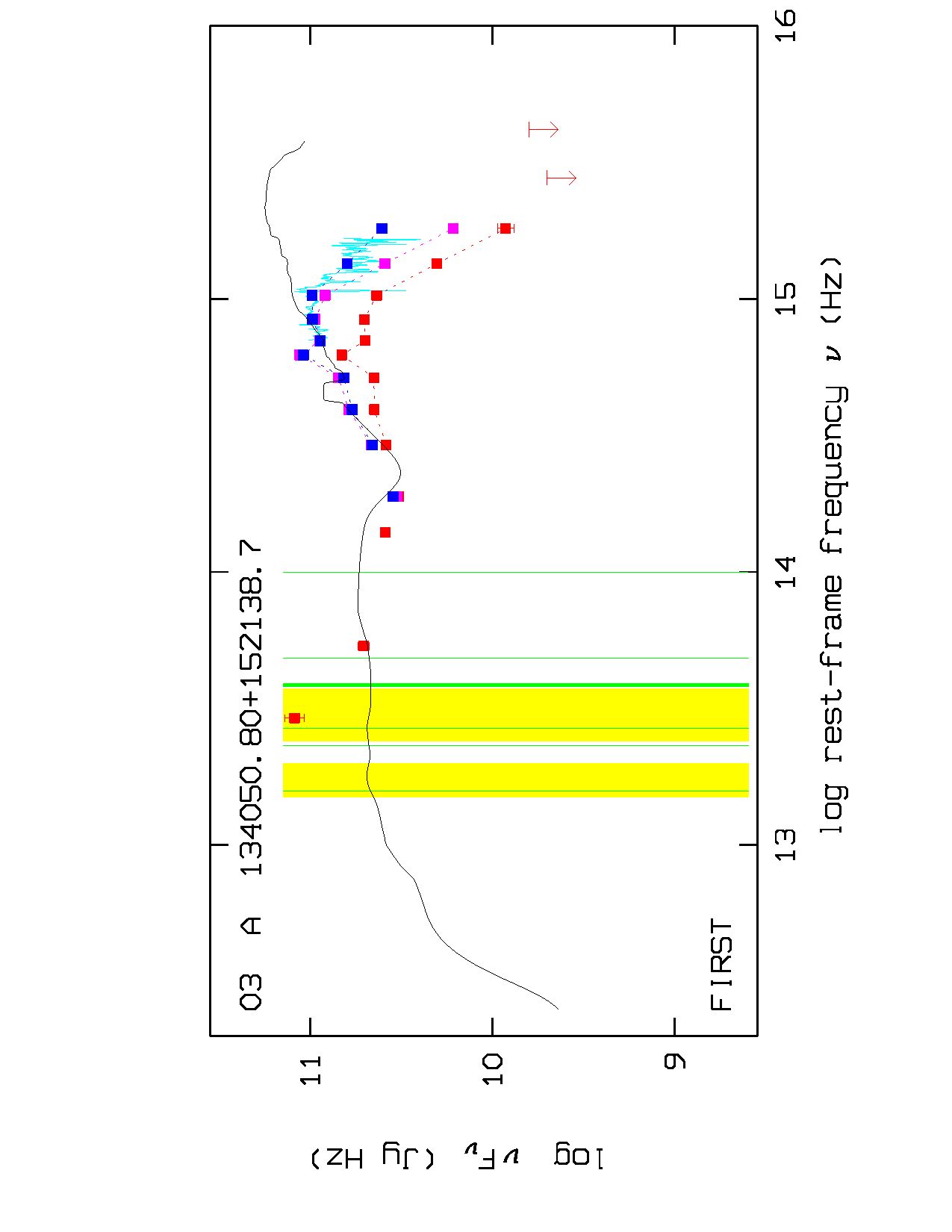}\hfill \\
\includegraphics[viewport=125 0 570 790,angle=270,width=8.0cm,clip]{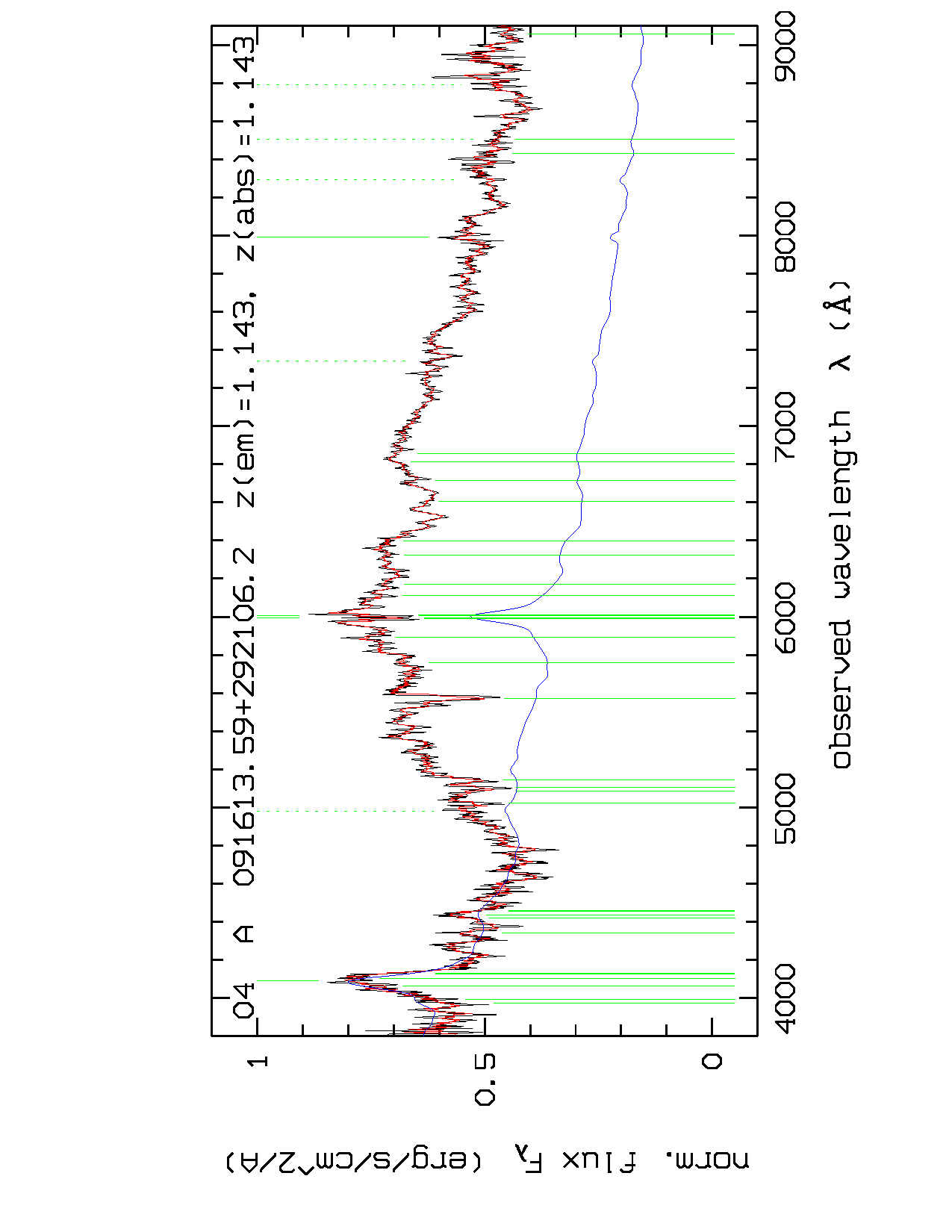}\hfill \=
\includegraphics[viewport=125 0 570 790,angle=270,width=8.0cm,clip]{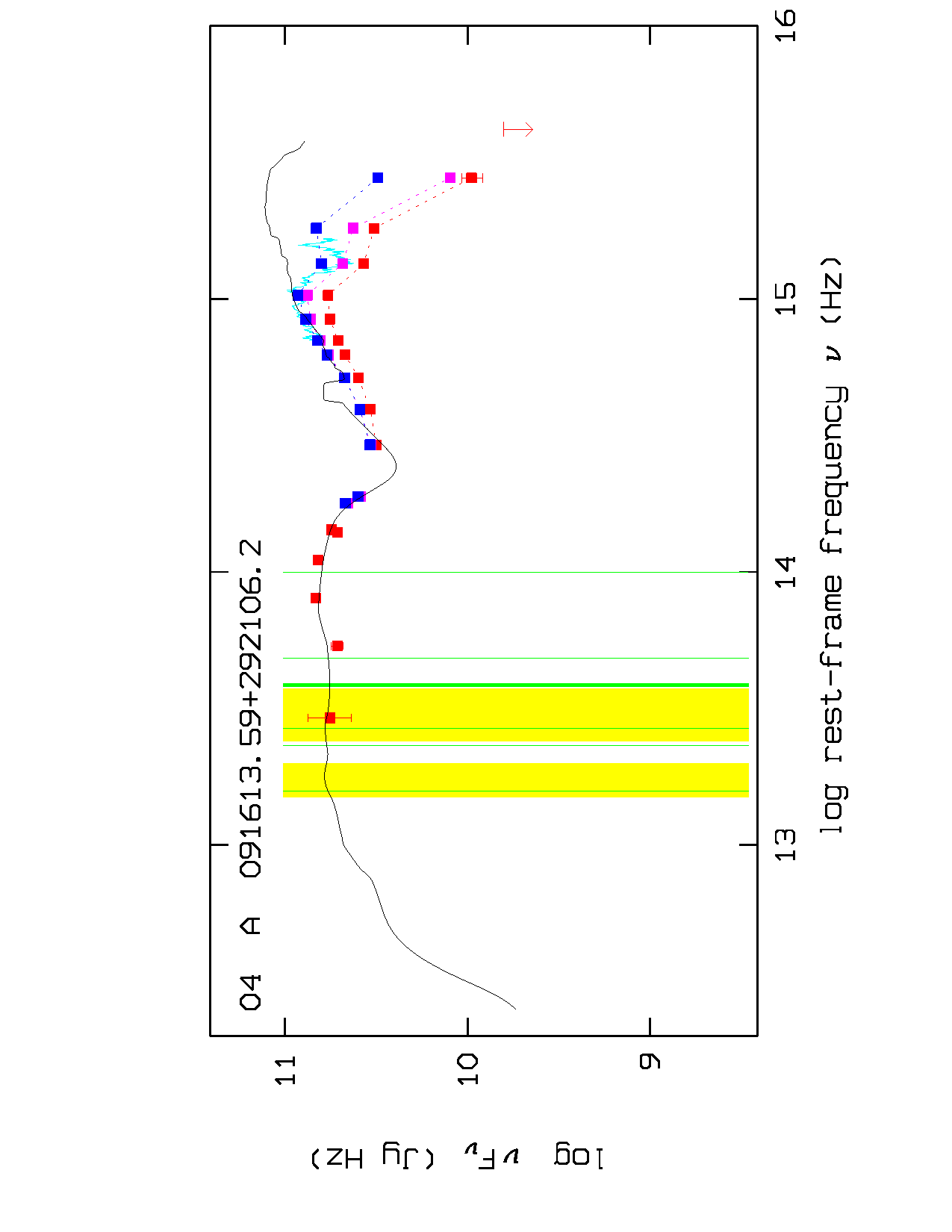}\hfill \\
\includegraphics[viewport=125 0 570 790,angle=270,width=8.0cm,clip]{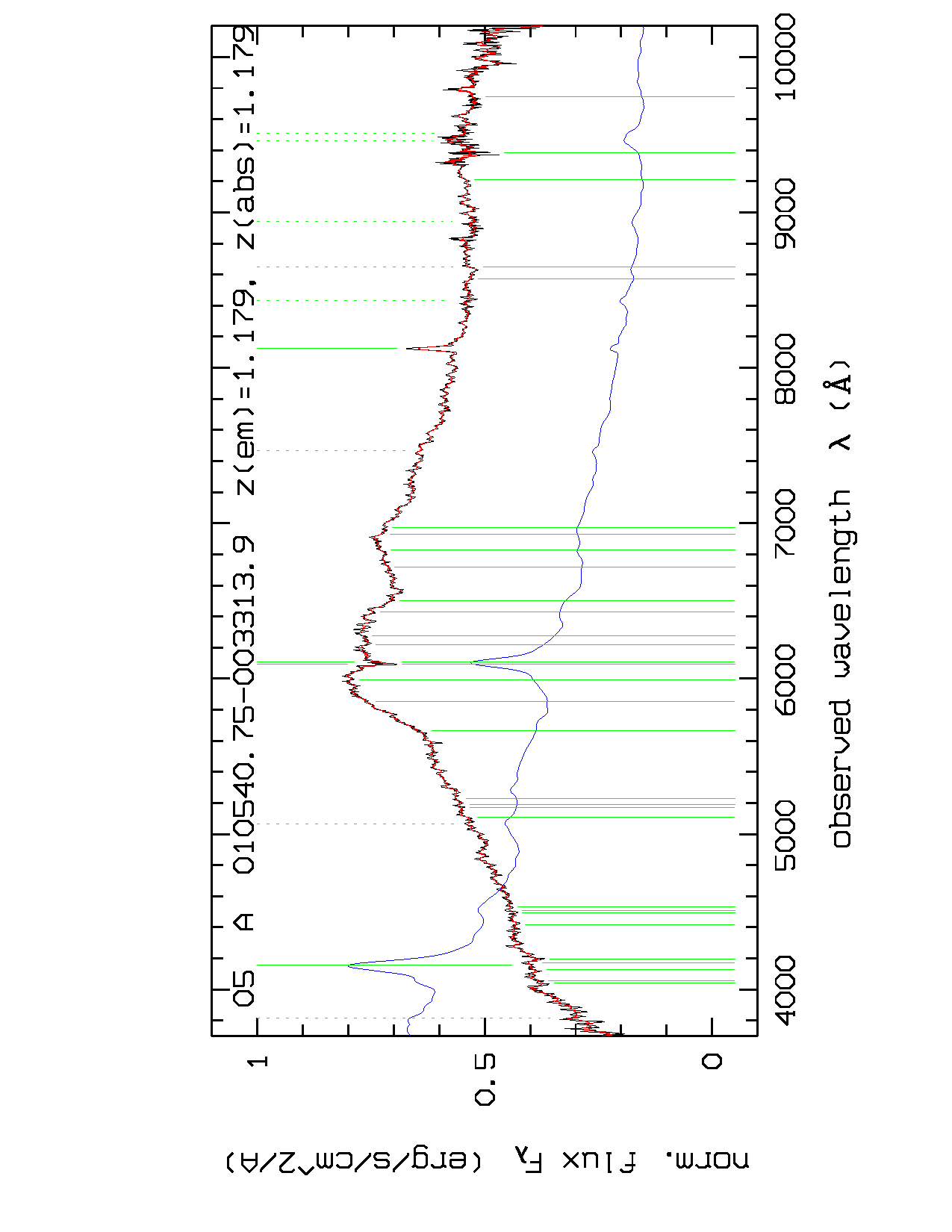}\hfill \=
\includegraphics[viewport=125 0 570 790,angle=270,width=8.0cm,clip]{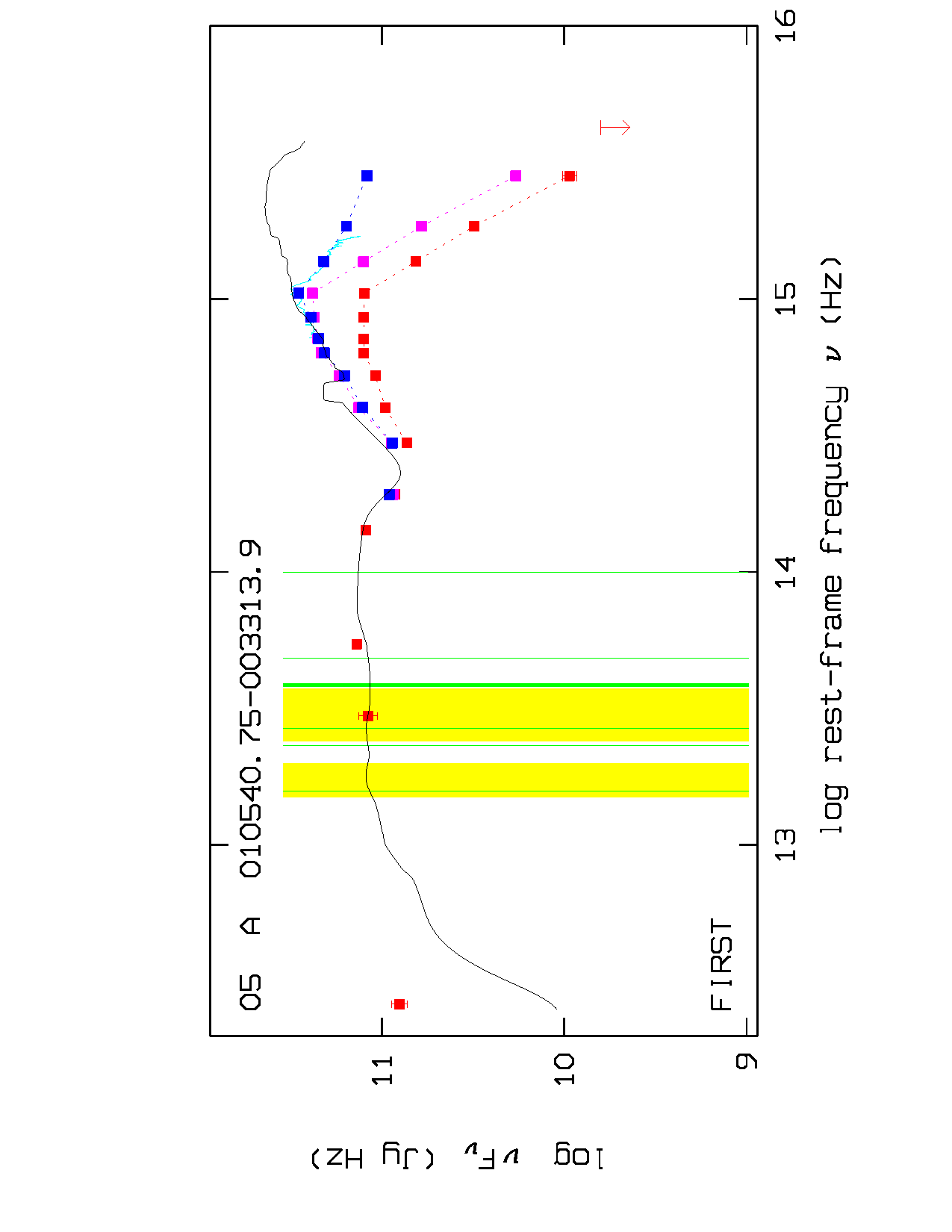}\hfill
\end{tabbing}
\caption{Sample A.}
\label{fig:spectra_and_seds}
\end{figure*}
\clearpage

\begin{figure*}[h]
\begin{tabbing}
\includegraphics[viewport=125 0 570 790,angle=270,width=8.0cm,clip]{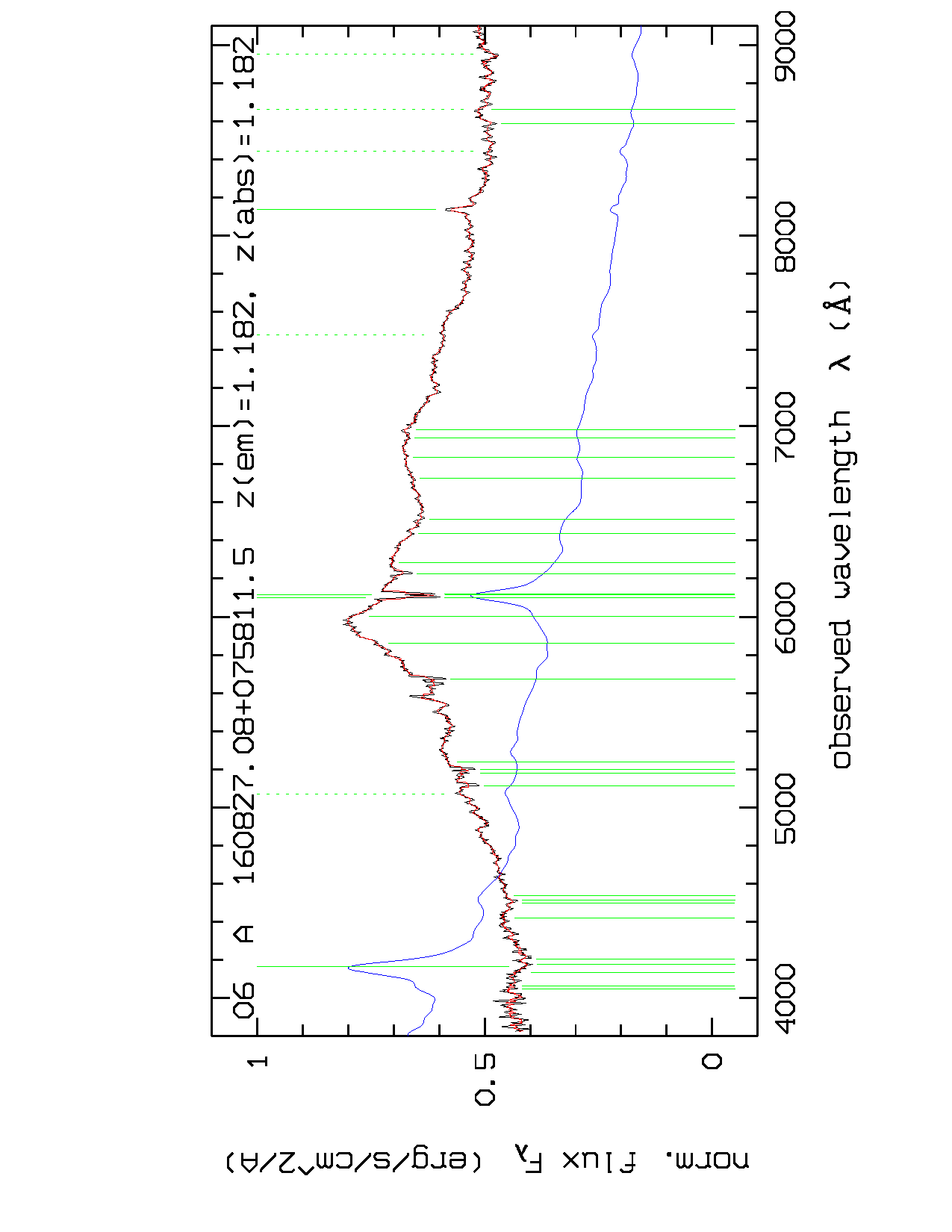}\hfill \=
\includegraphics[viewport=125 0 570 790,angle=270,width=8.0cm,clip]{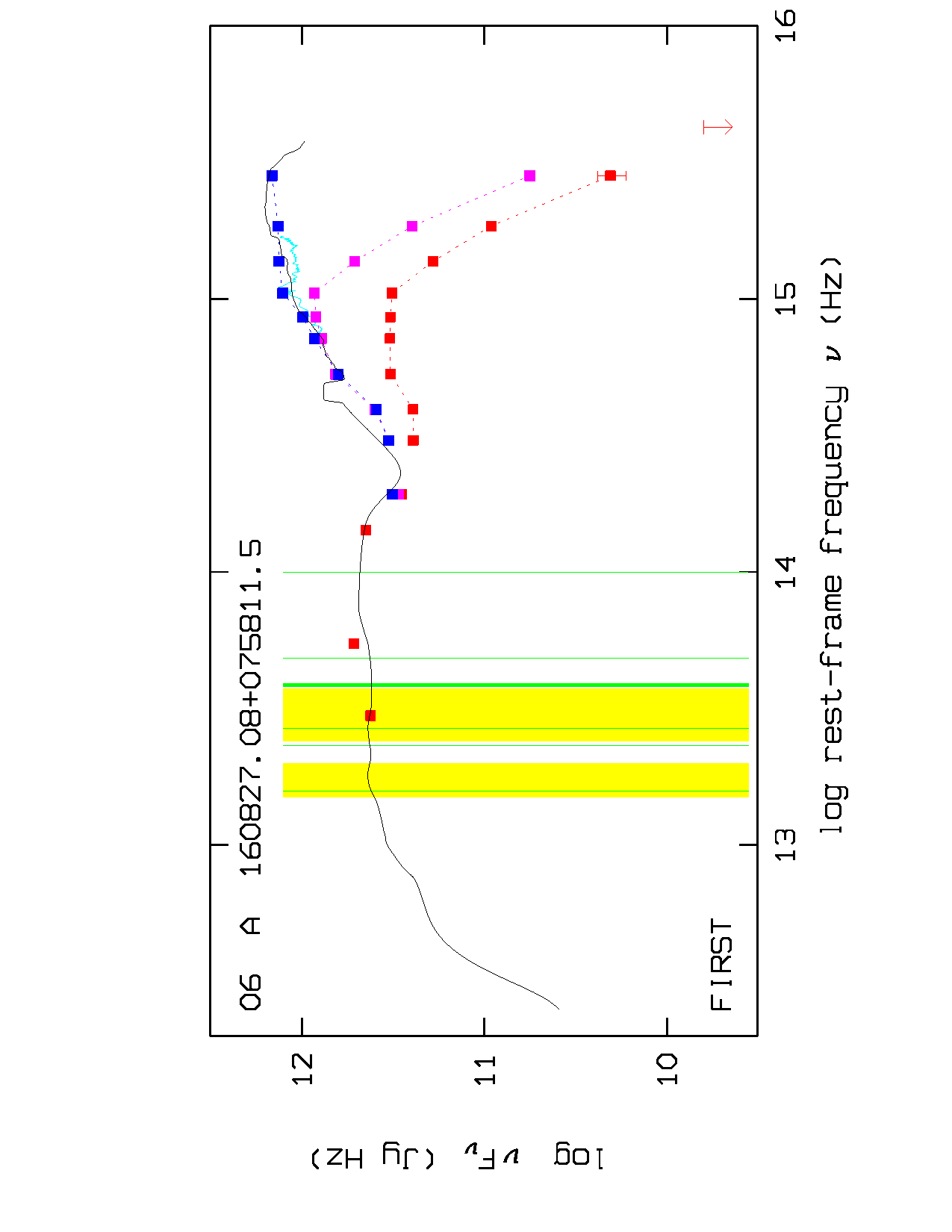}\hfill  \\
\includegraphics[viewport=125 0 570 790,angle=270,width=8.0cm,clip]{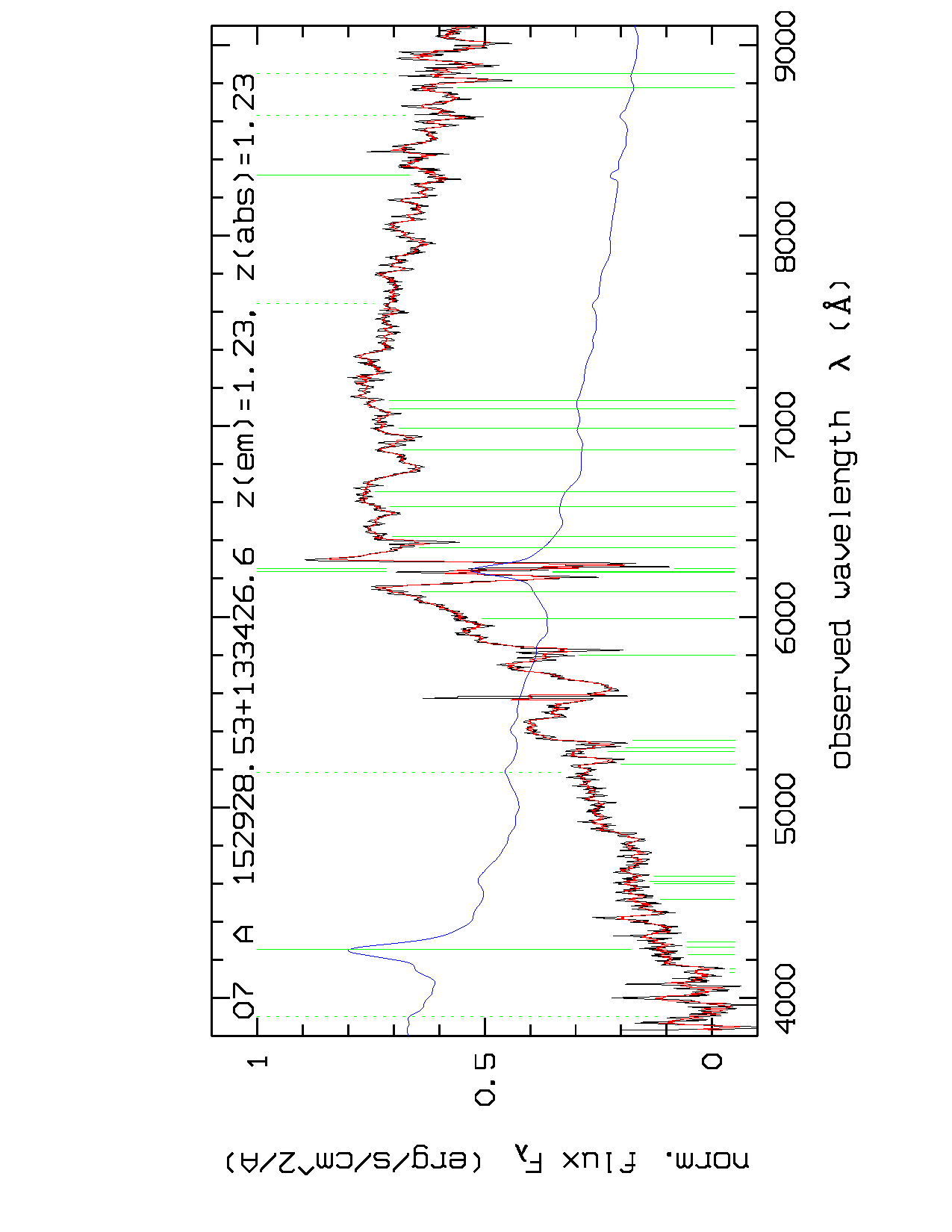}\hfill \=
\includegraphics[viewport=125 0 570 790,angle=270,width=8.0cm,clip]{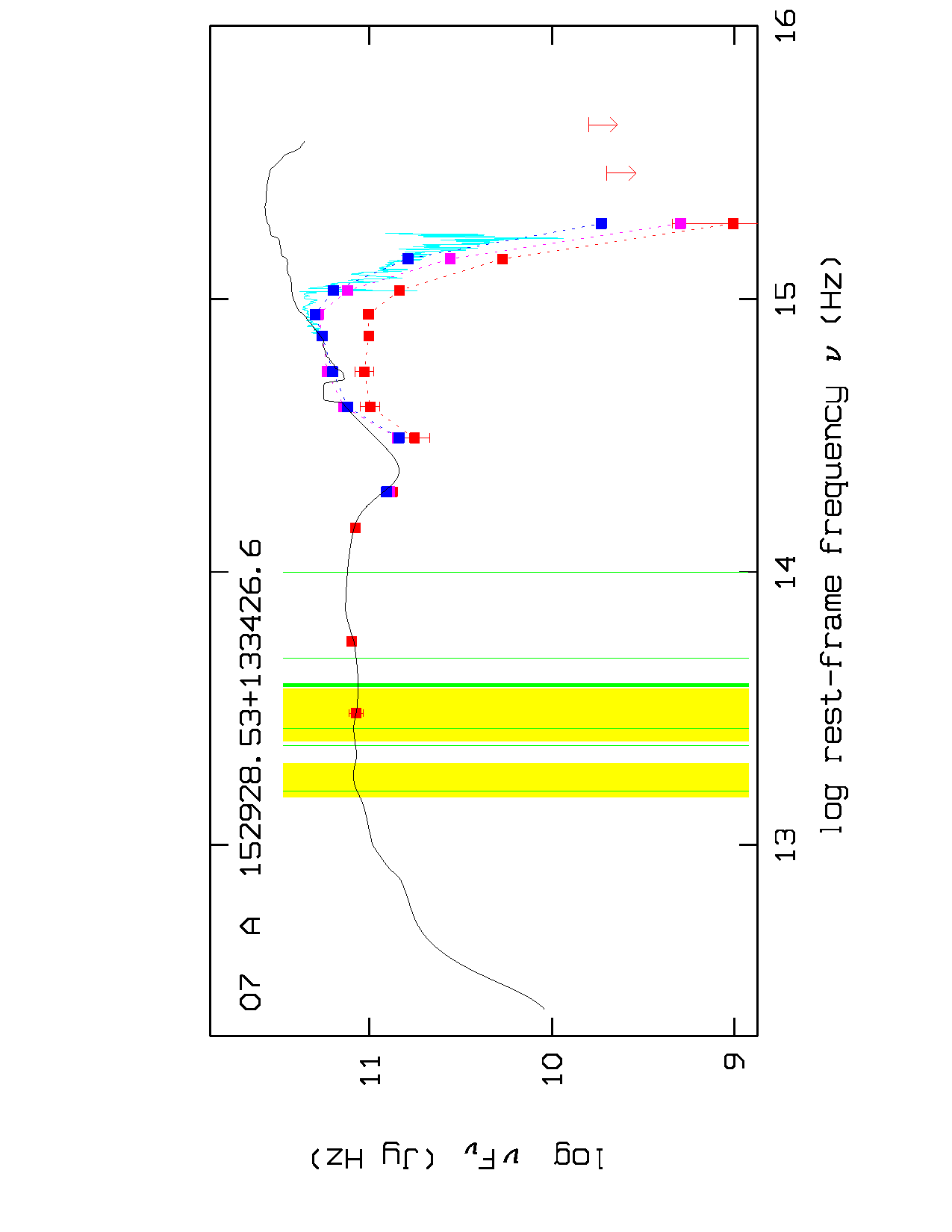}\hfill \\
\includegraphics[viewport=125 0 570 790,angle=270,width=8.0cm,clip]{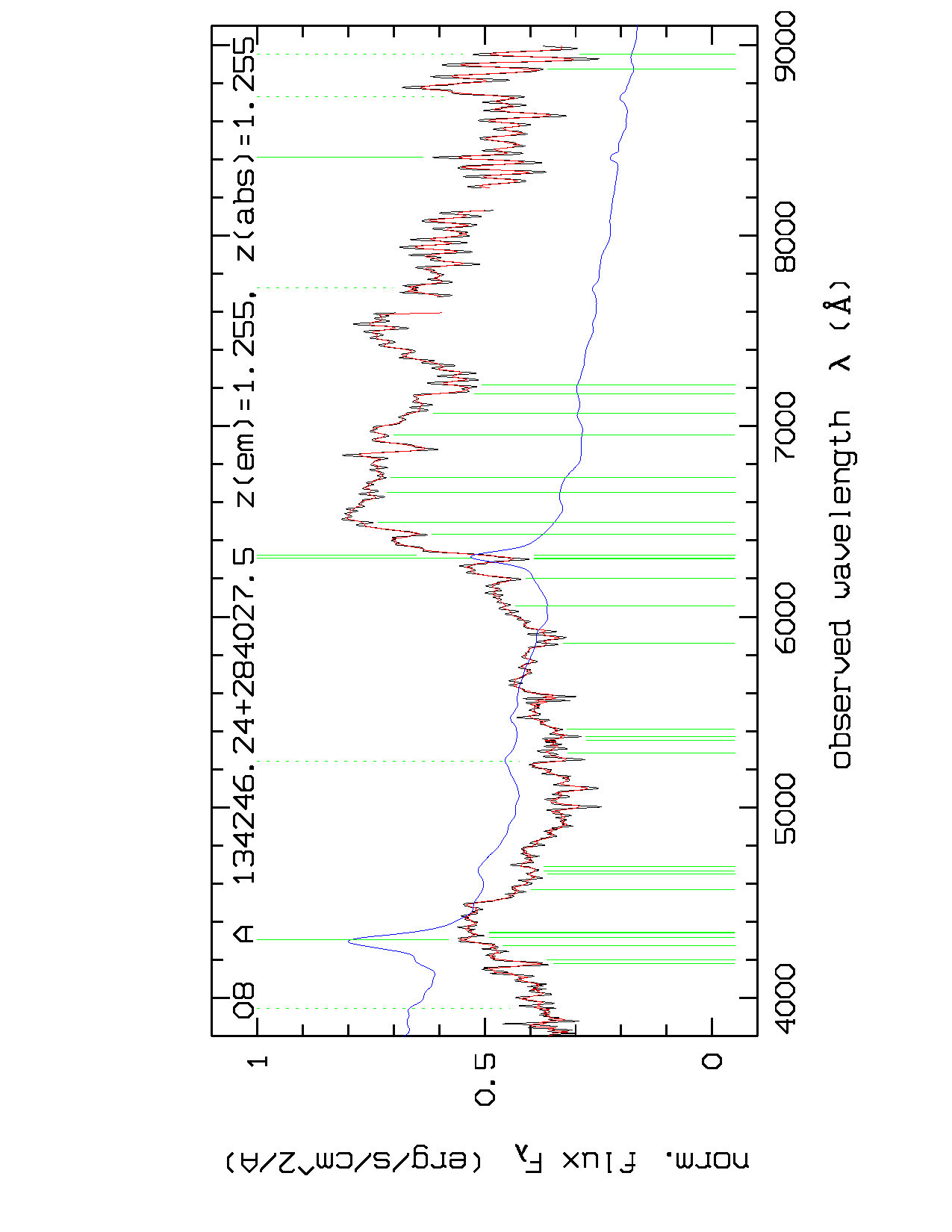}\hfill \=
\includegraphics[viewport=125 0 570 790,angle=270,width=8.0cm,clip]{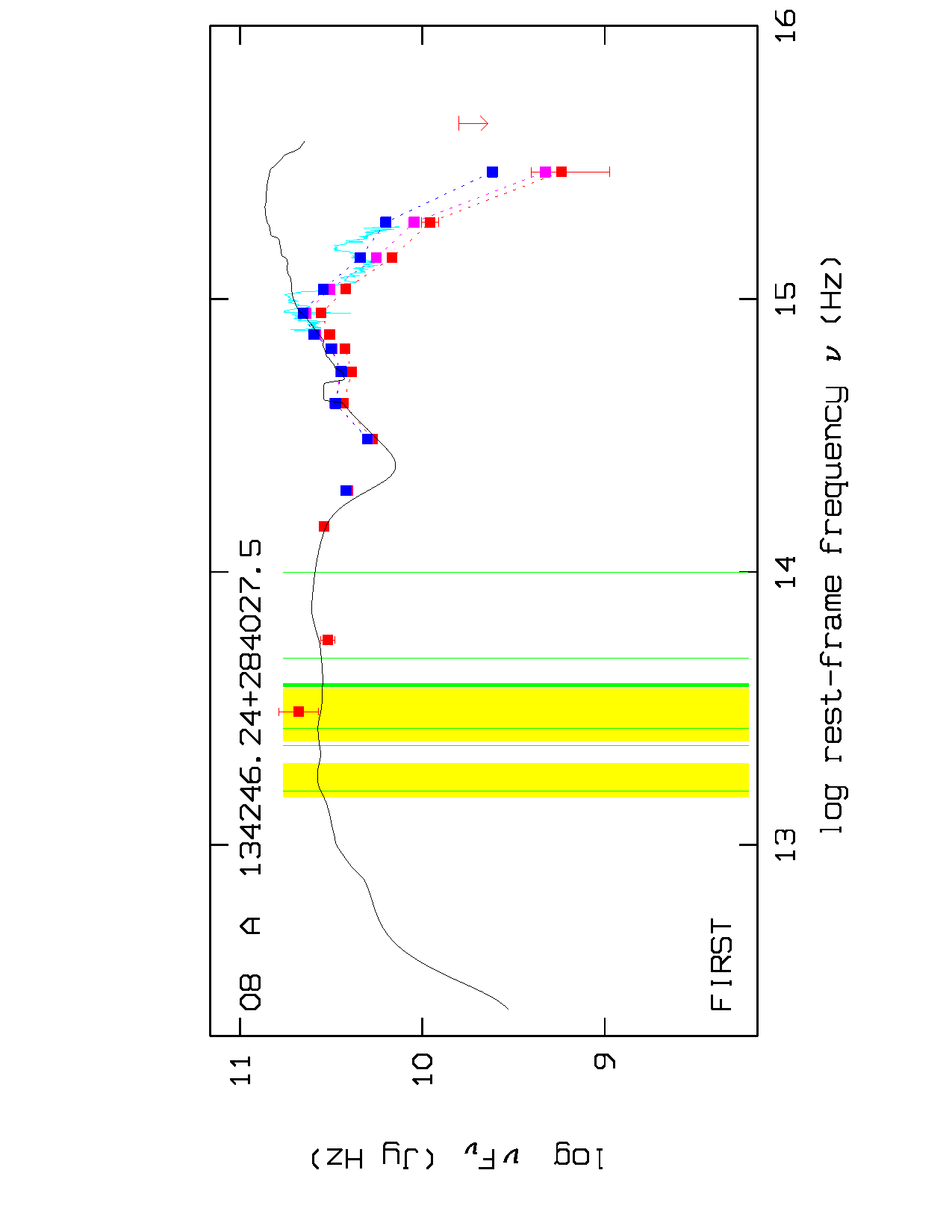}\hfill \\
\includegraphics[viewport=125 0 570 790,angle=270,width=8.0cm,clip]{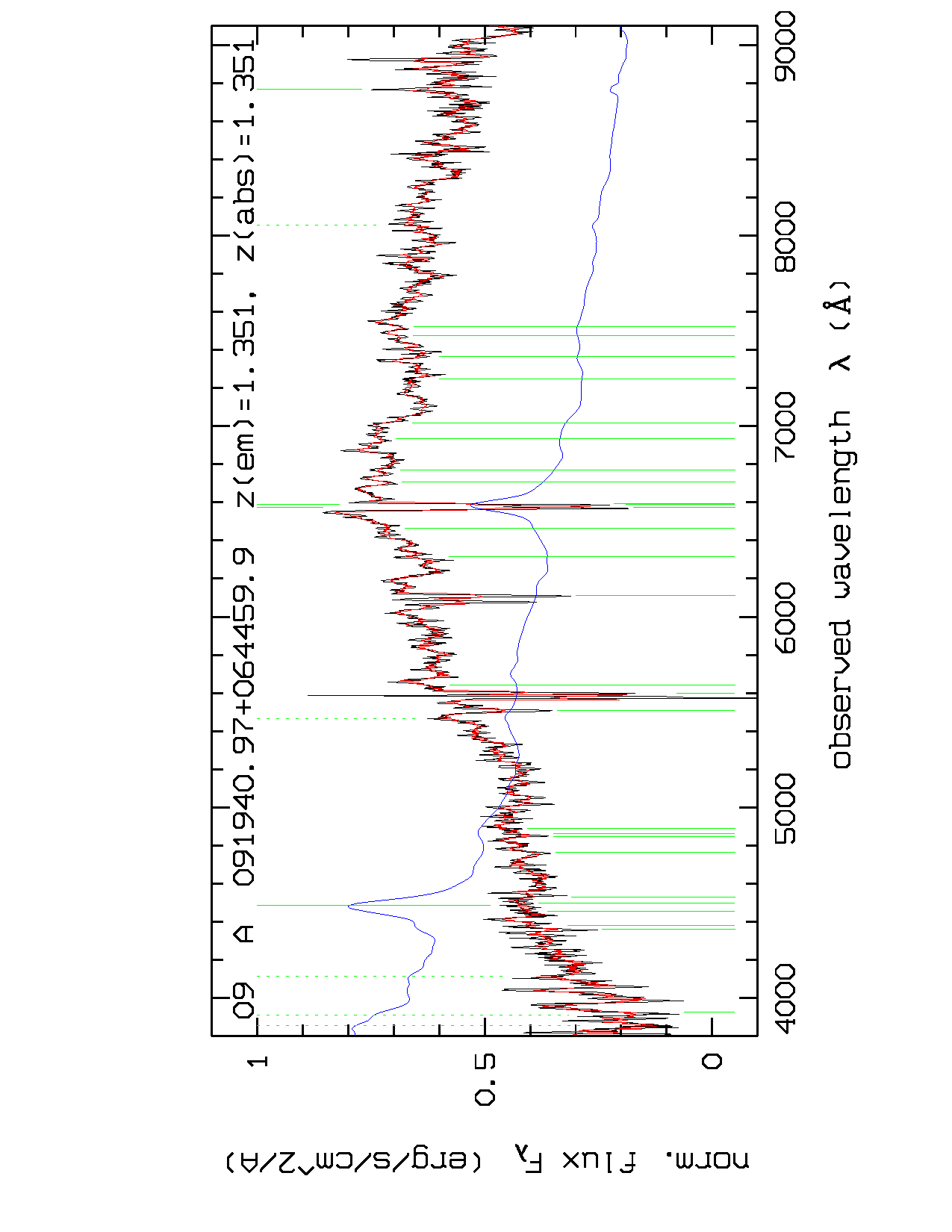}\hfill \=
\includegraphics[viewport=125 0 570 790,angle=270,width=8.0cm,clip]{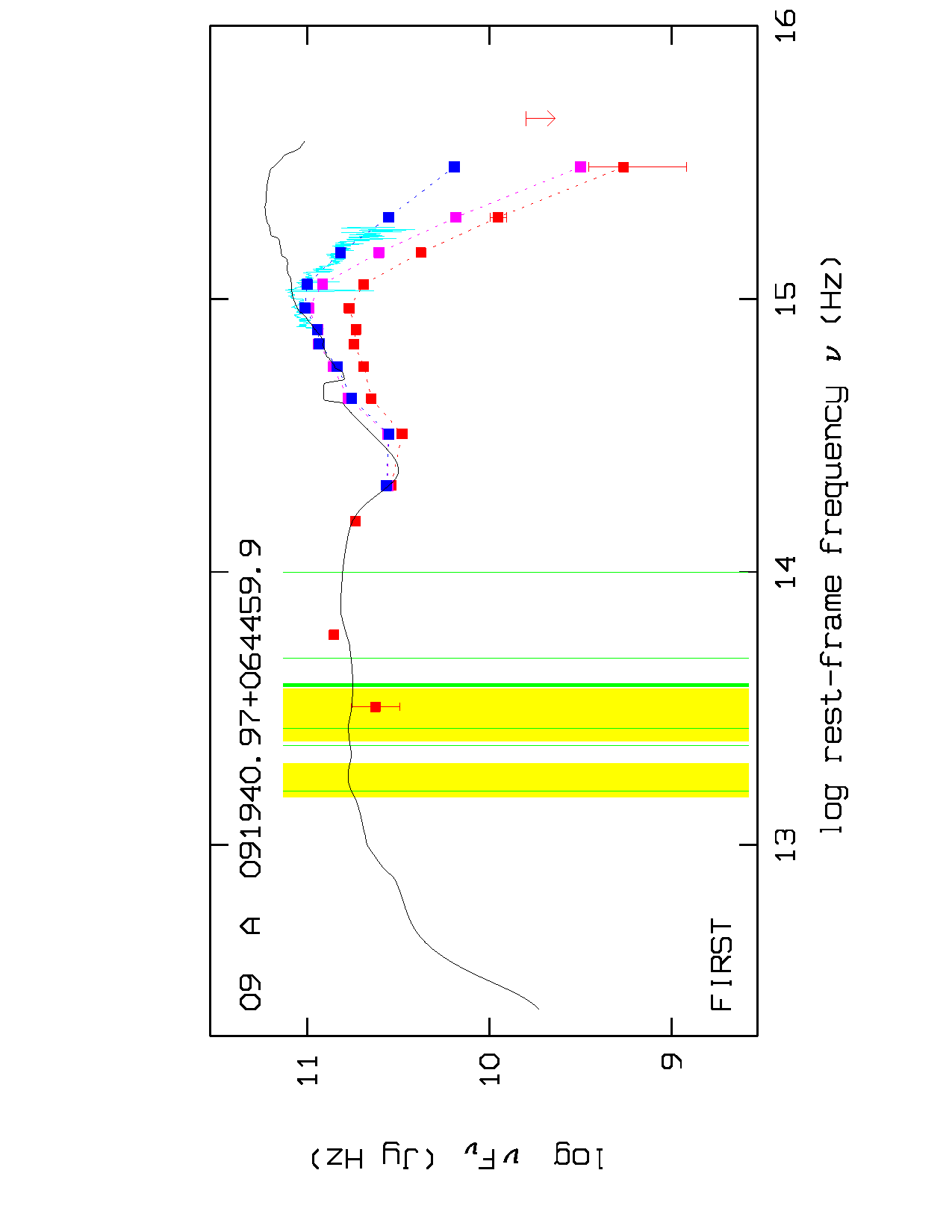}\hfill \\
\includegraphics[viewport=125 0 570 790,angle=270,width=8.0cm,clip]{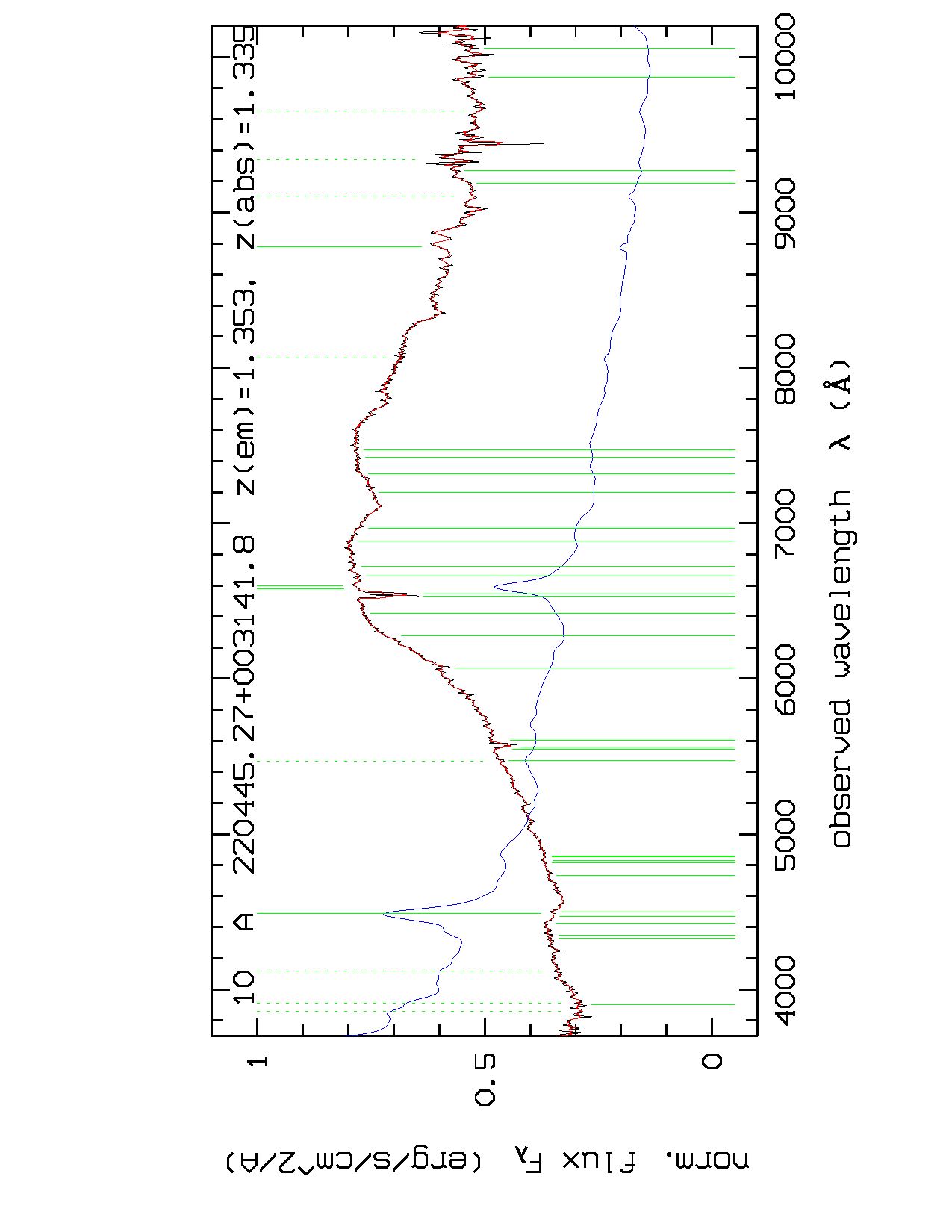}\hfill \=
\includegraphics[viewport=125 0 570 790,angle=270,width=8.0cm,clip]{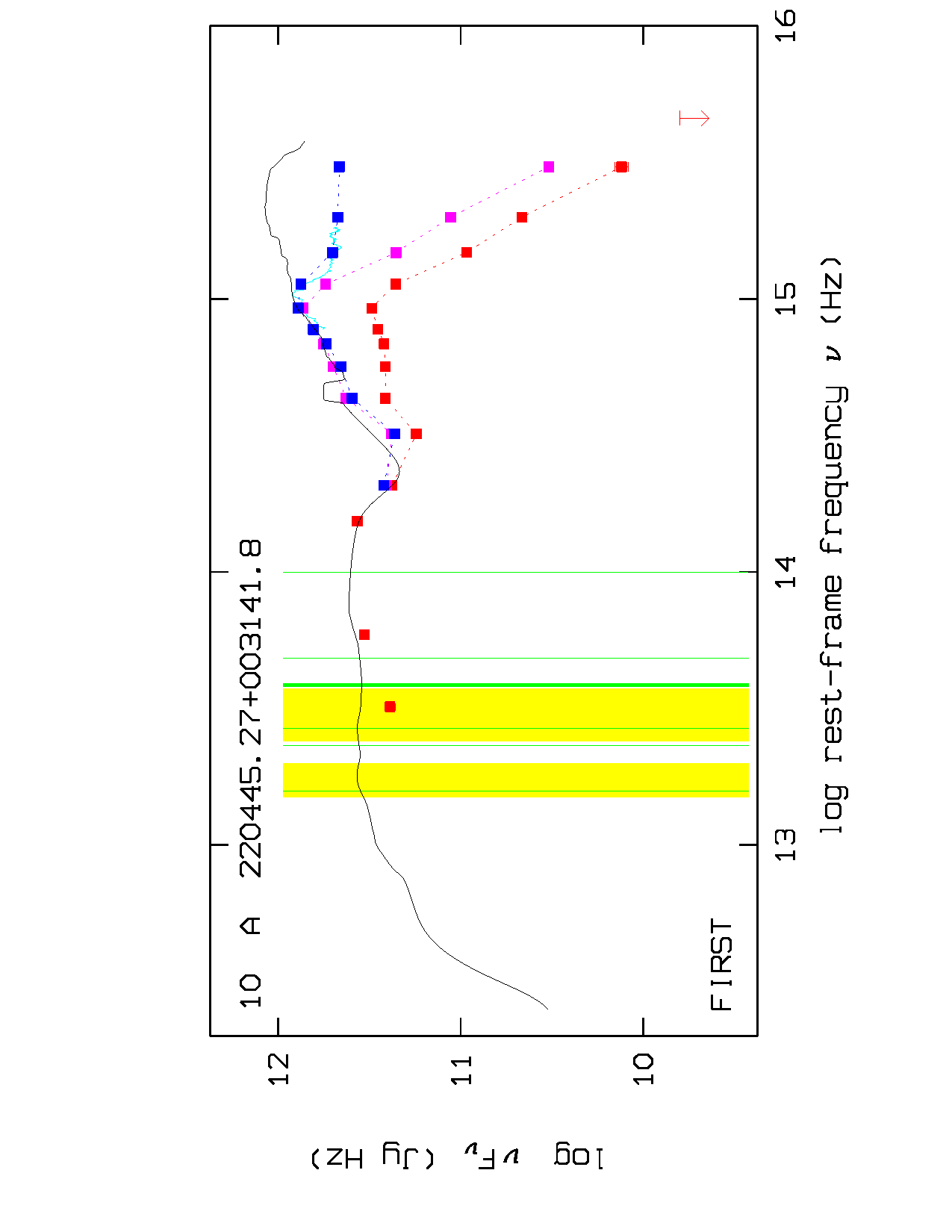}\hfill
\end{tabbing}
\caption{Sample A - continued (1).}
\end{figure*}\clearpage

\begin{figure*}[h]
\begin{tabbing}
\includegraphics[viewport=125 0 570 790,angle=270,width=8.0cm,clip]{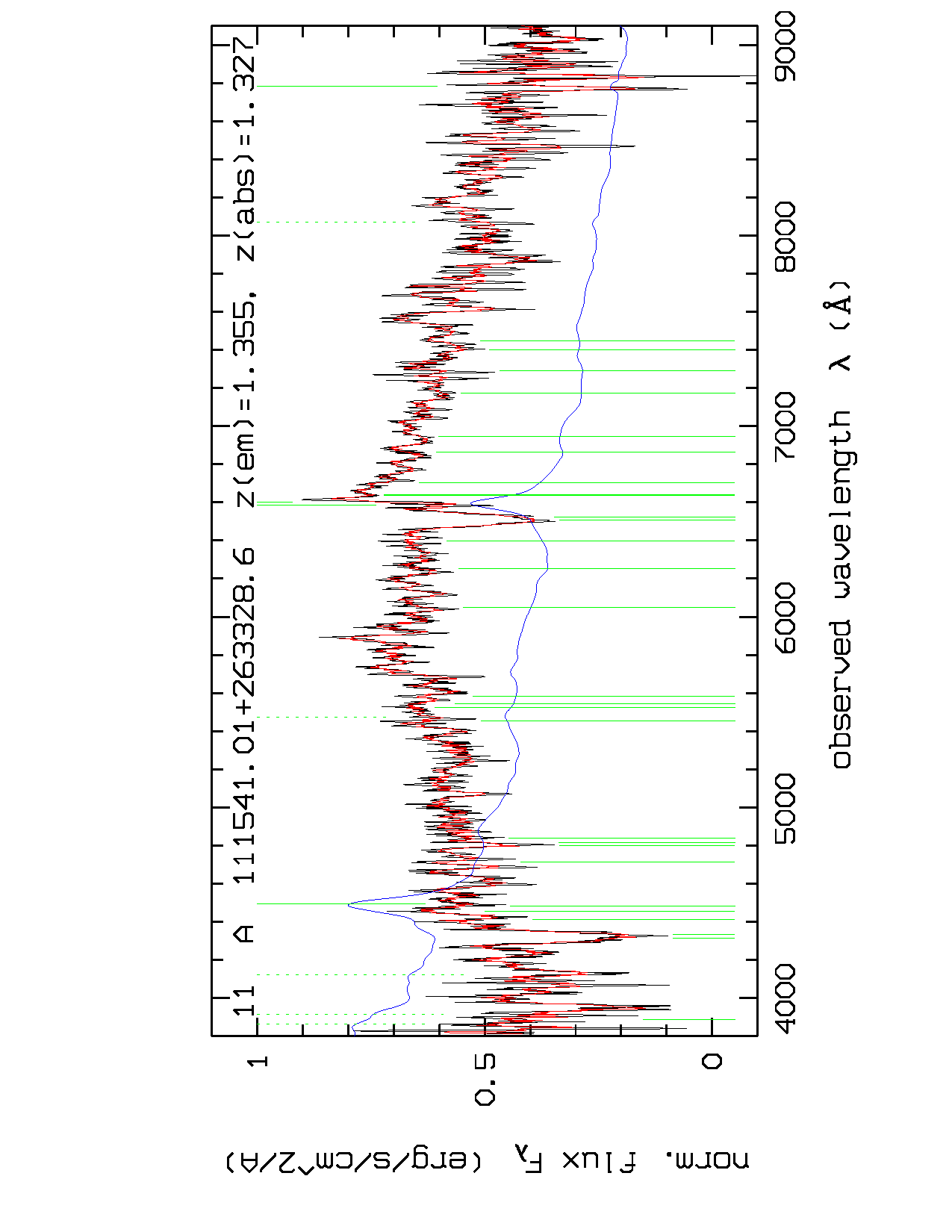}\hfill \=
\includegraphics[viewport=125 0 570 790,angle=270,width=8.0cm,clip]{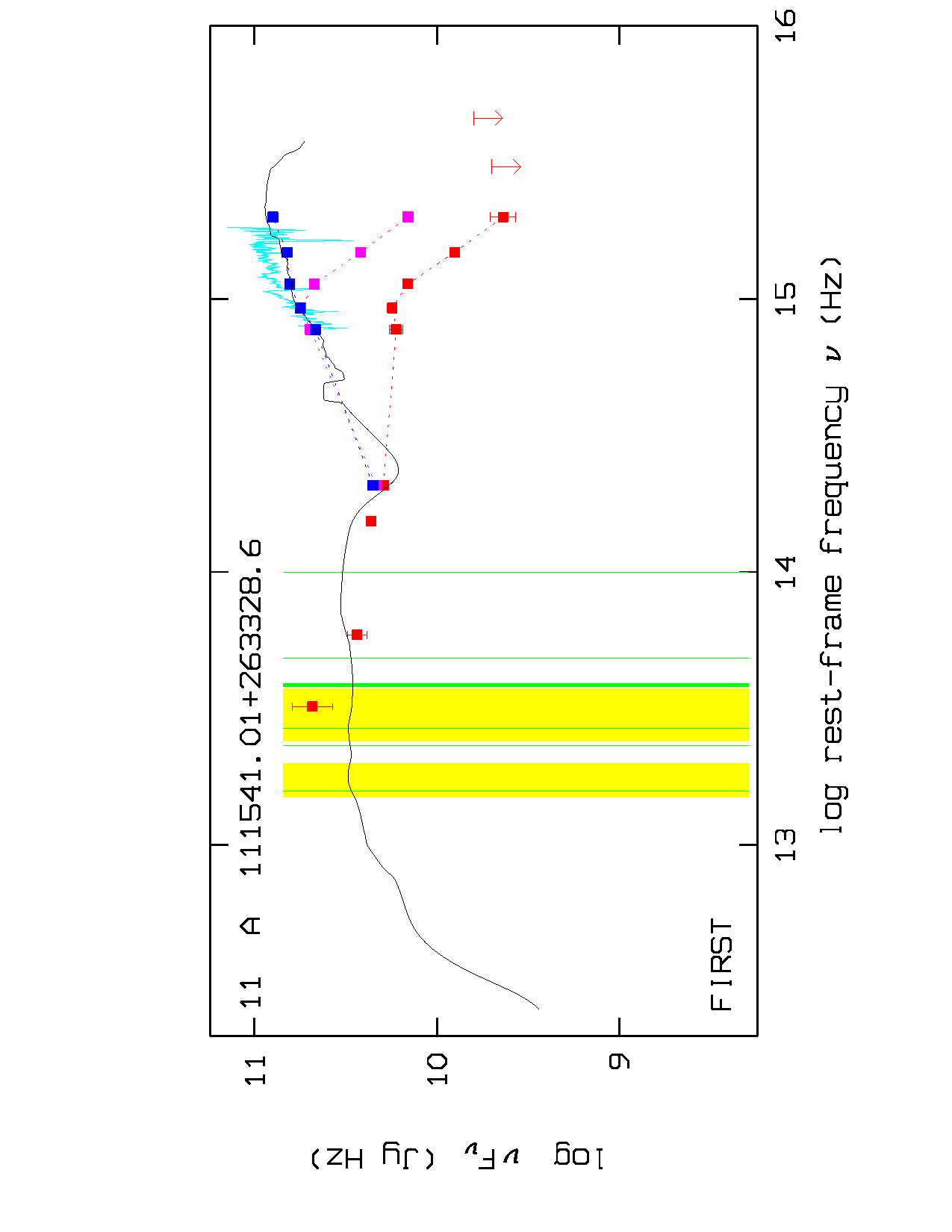}\hfill \\
\includegraphics[viewport=125 0 570 790,angle=270,width=8.0cm,clip]{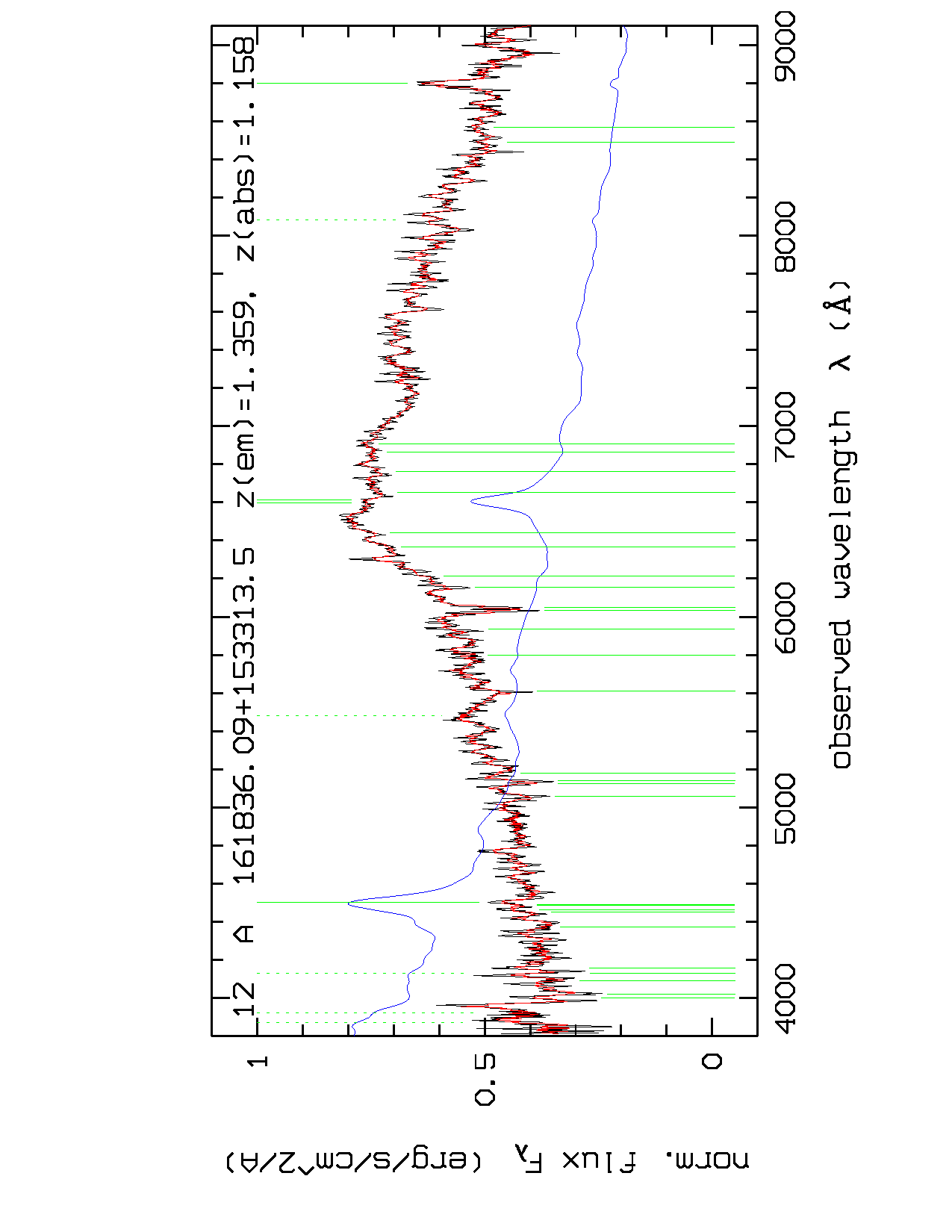}\hfill \=
\includegraphics[viewport=125 0 570 790,angle=270,width=8.0cm,clip]{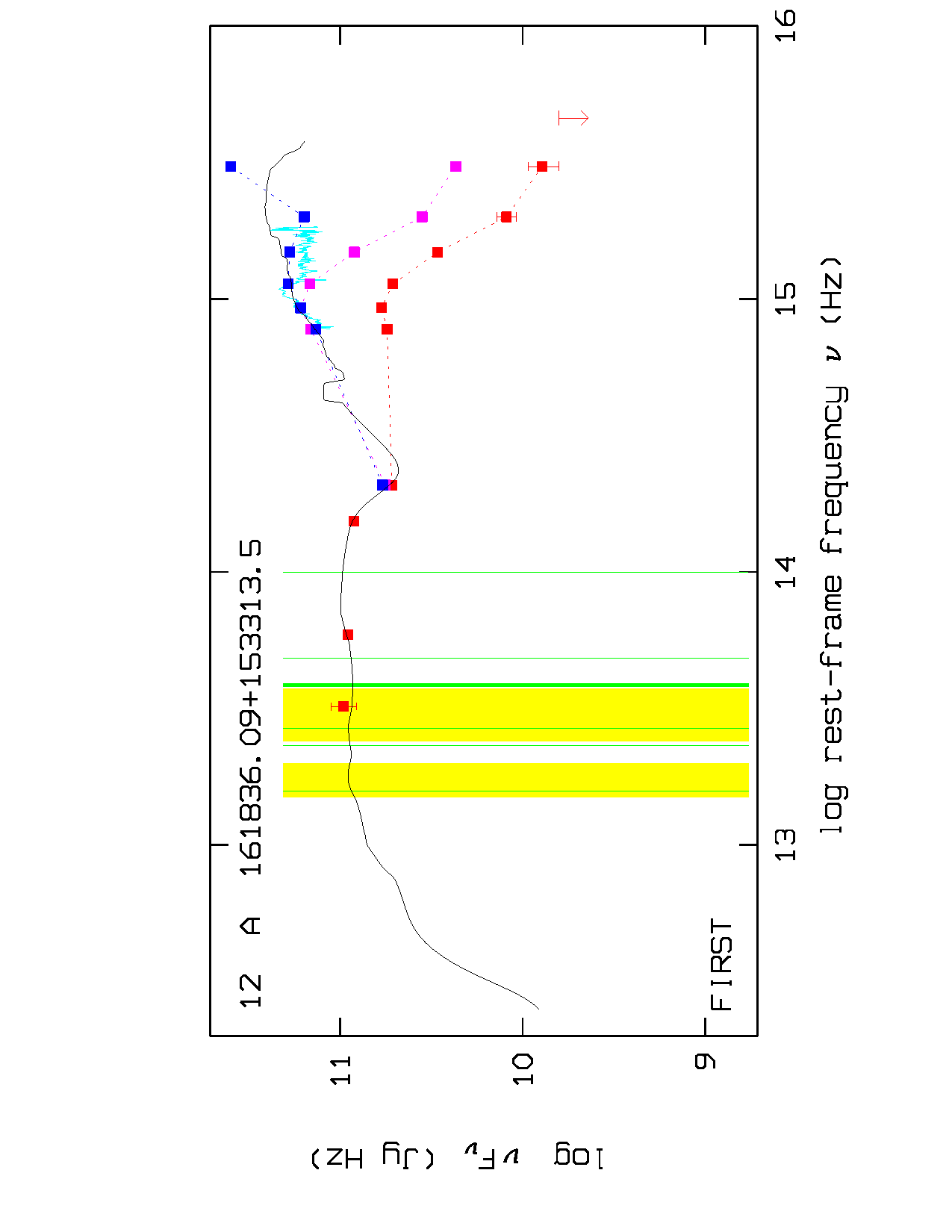}\hfill \\
\includegraphics[viewport=125 0 570 790,angle=270,width=8.0cm,clip]{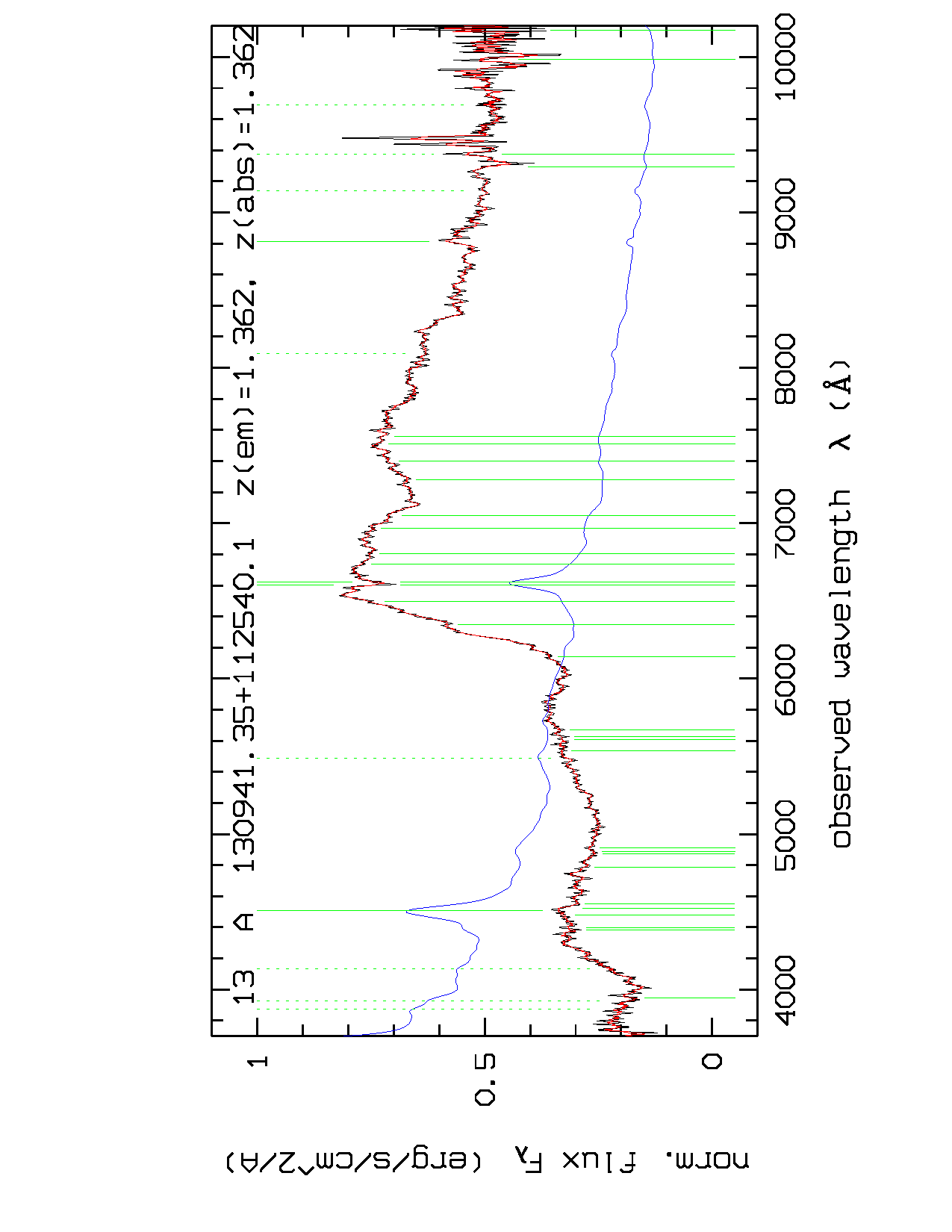}\hfill \=
\includegraphics[viewport=125 0 570 790,angle=270,width=8.0cm,clip]{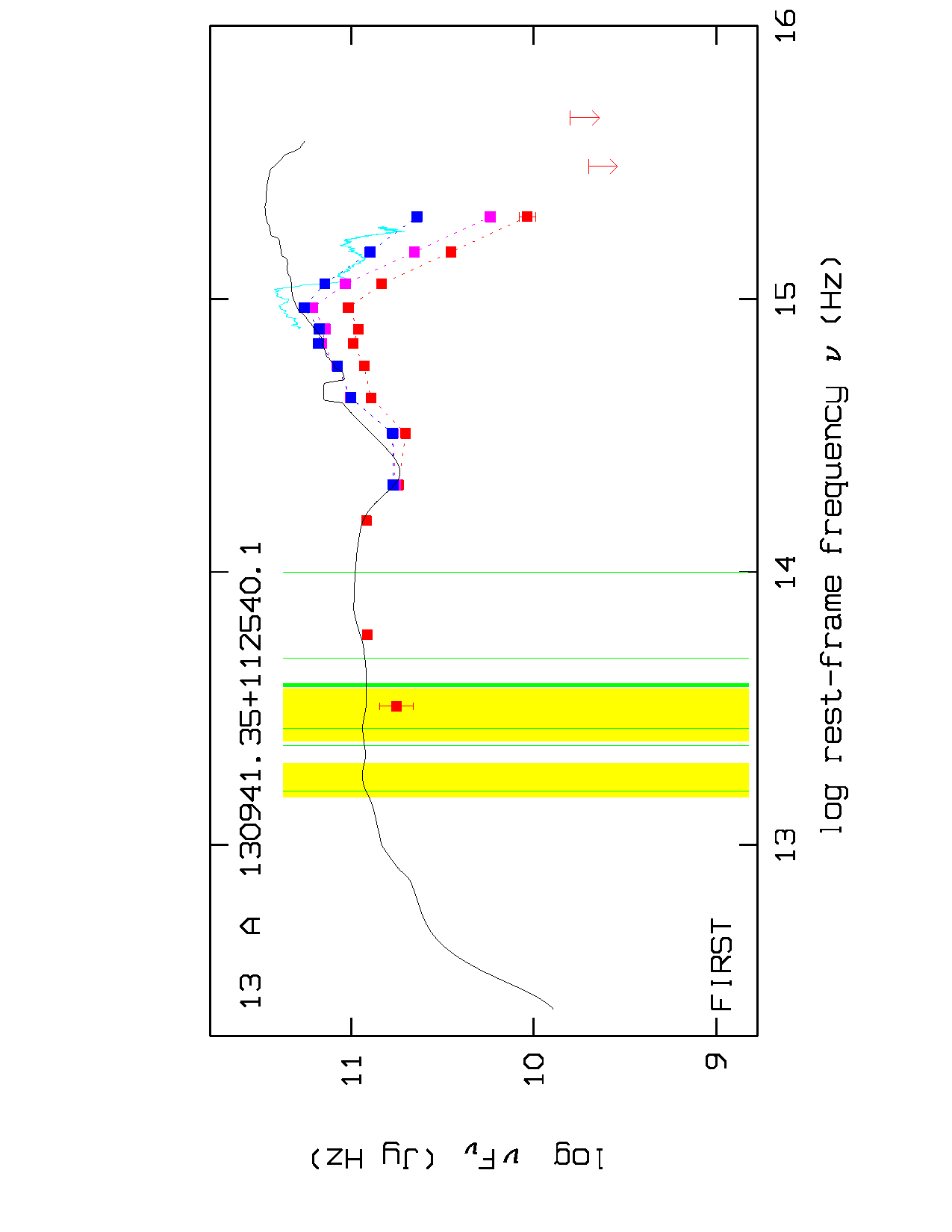}\hfill \\
\includegraphics[viewport=125 0 570 790,angle=270,width=8.0cm,clip]{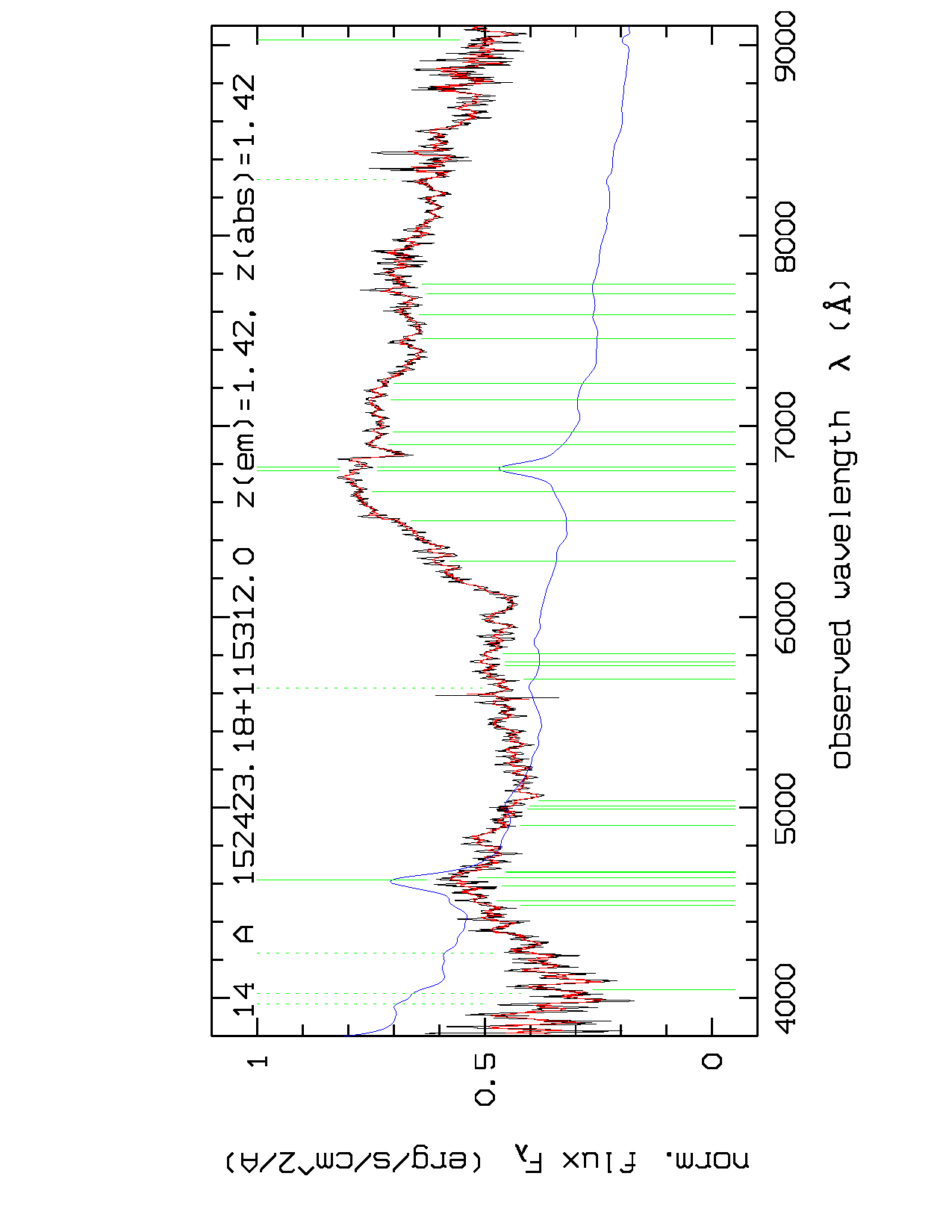}\hfill \=
\includegraphics[viewport=125 0 570 790,angle=270,width=8.0cm,clip]{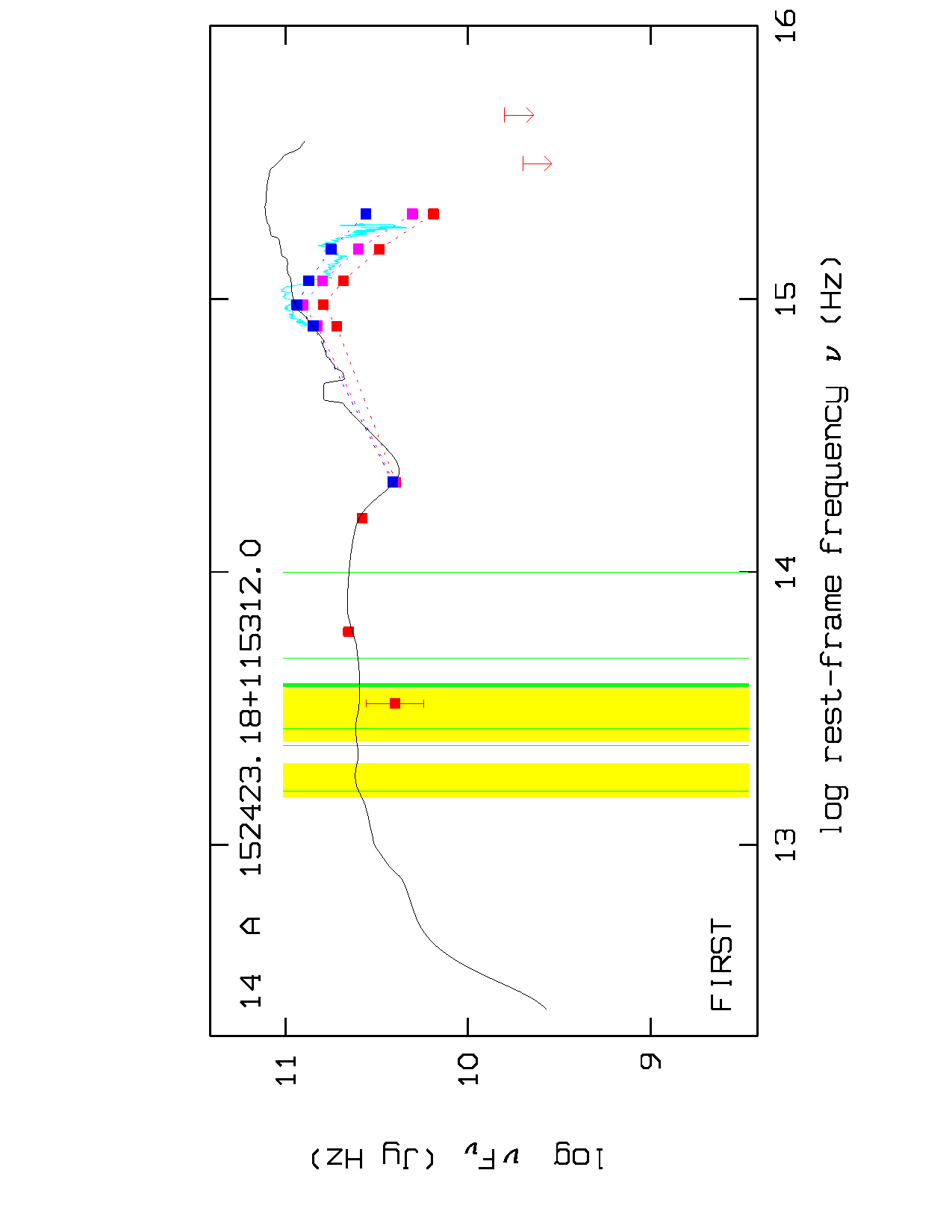}\hfill \\
\includegraphics[viewport=125 0 570 790,angle=270,width=8.0cm,clip]{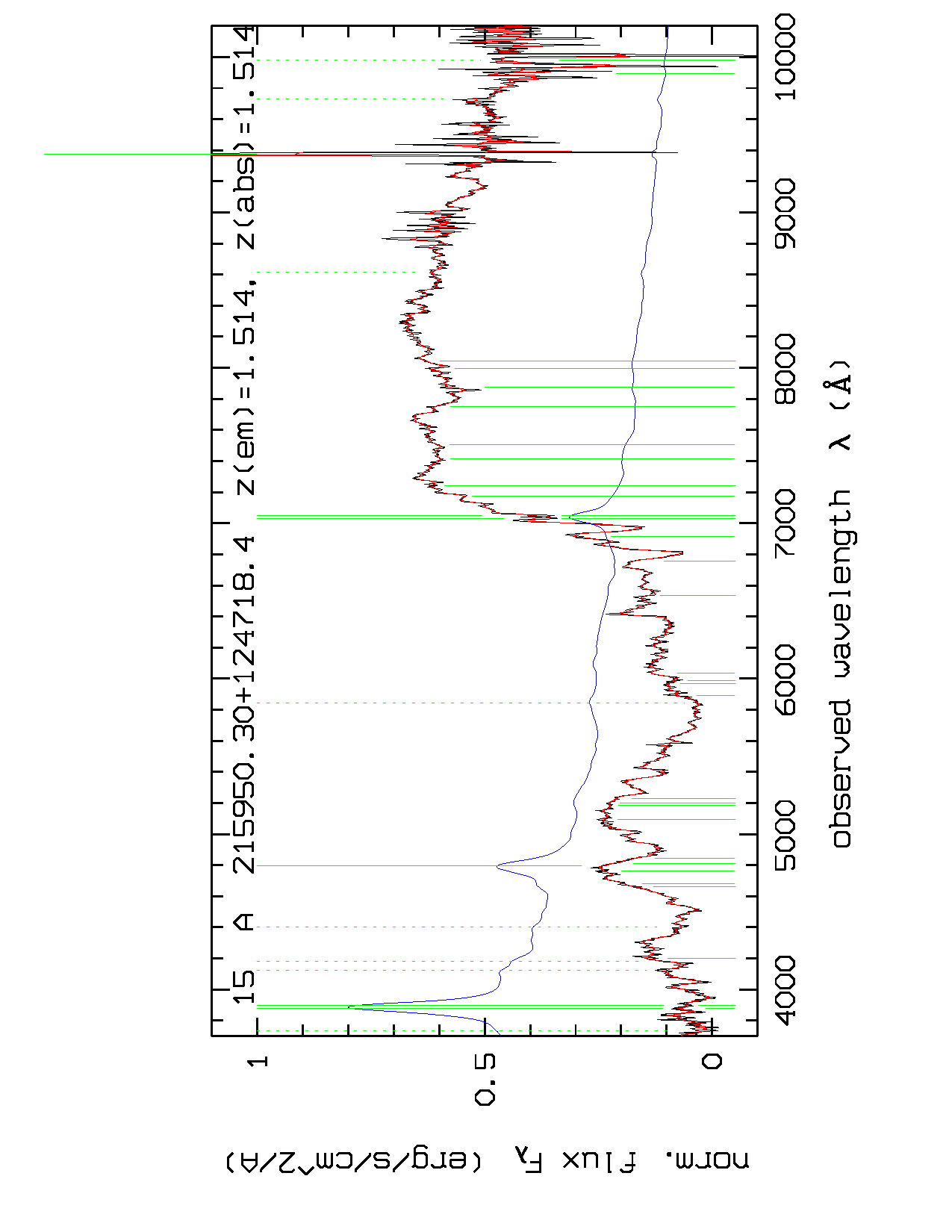}\hfill \=
\includegraphics[viewport=125 0 570 790,angle=270,width=8.0cm,clip]{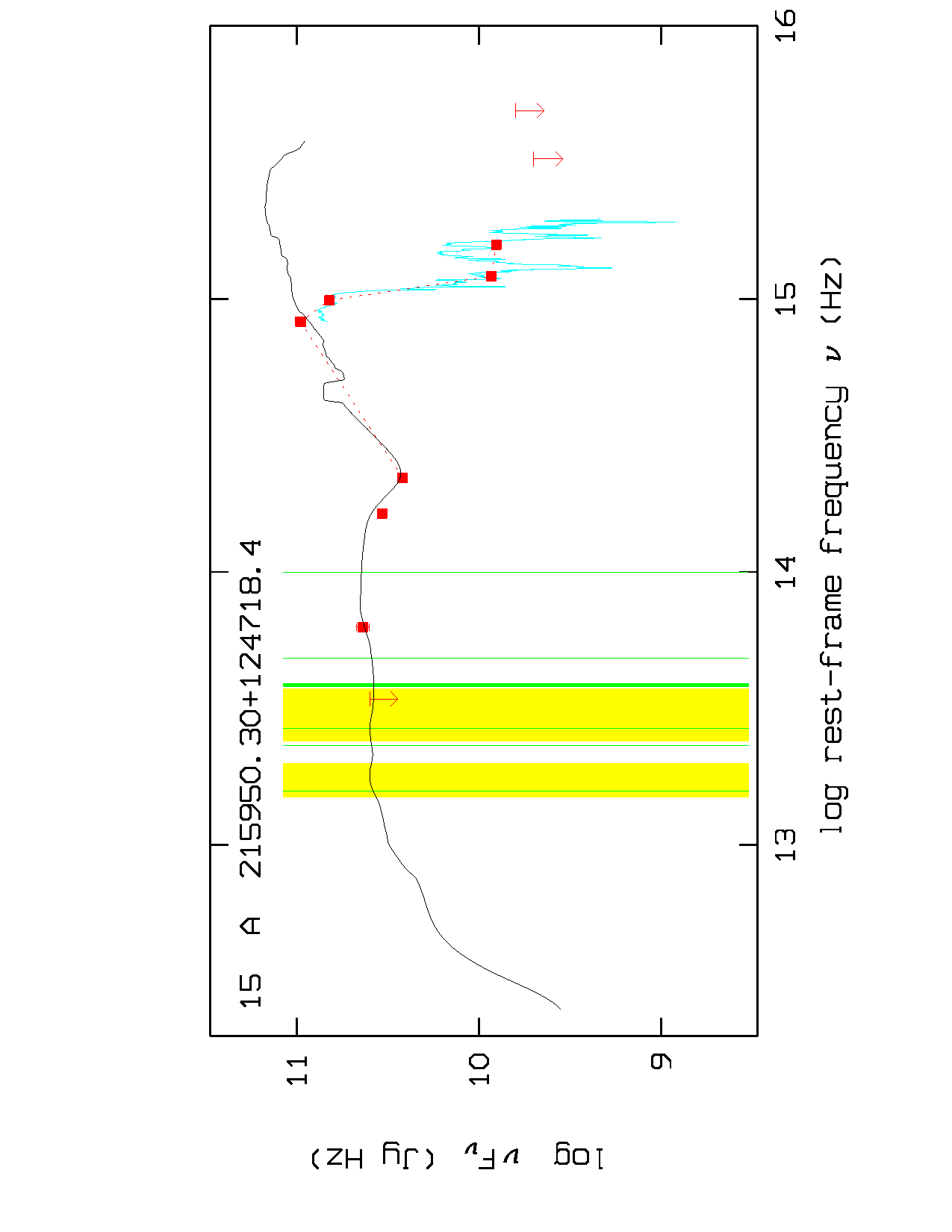}\hfill
\end{tabbing}
\caption{Sample A - continued (2).}
\end{figure*}\clearpage

\begin{figure*}[h]
\begin{tabbing}
\includegraphics[viewport=125 0 570 790,angle=270,width=8.0cm,clip]{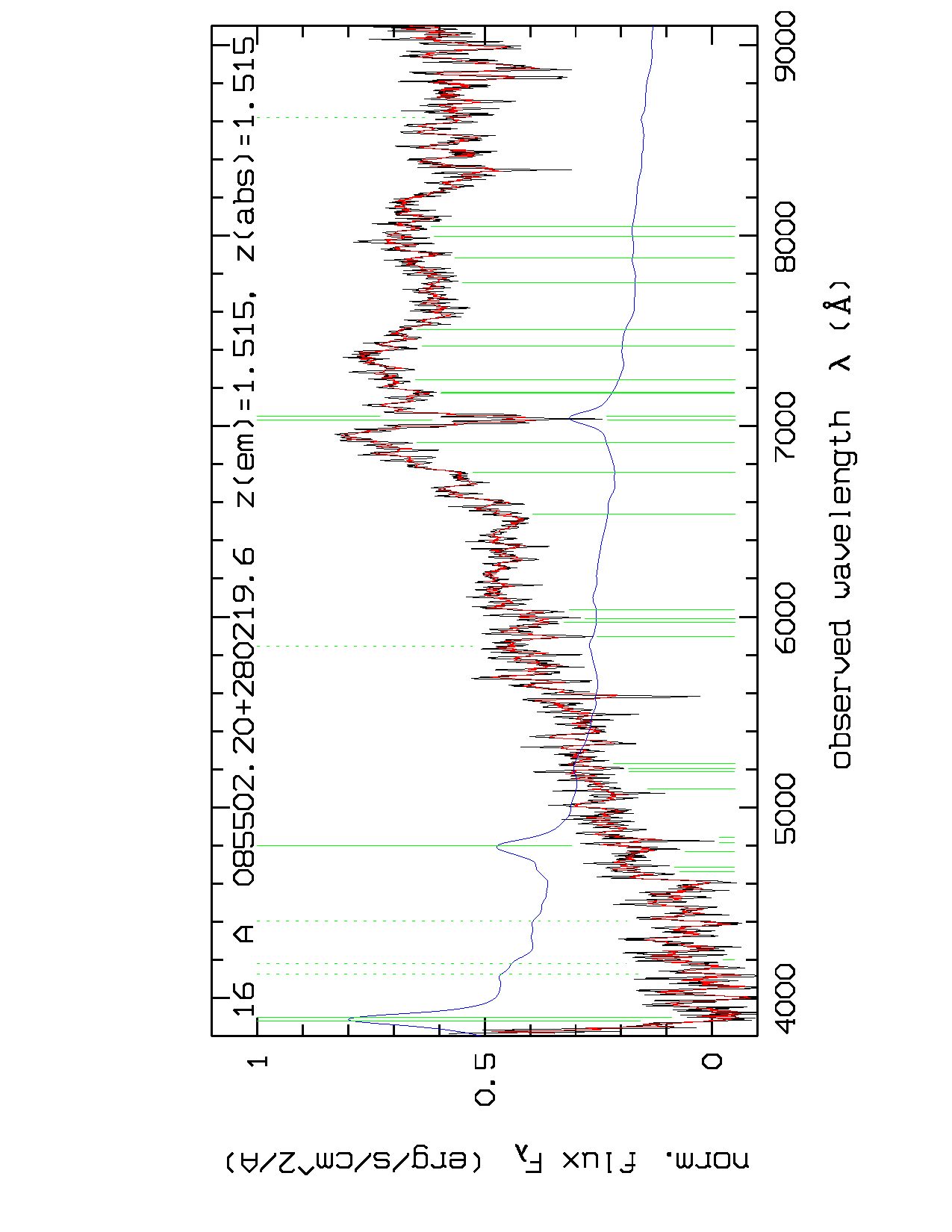}\hfill \=
\includegraphics[viewport=125 0 570 790,angle=270,width=8.0cm,clip]{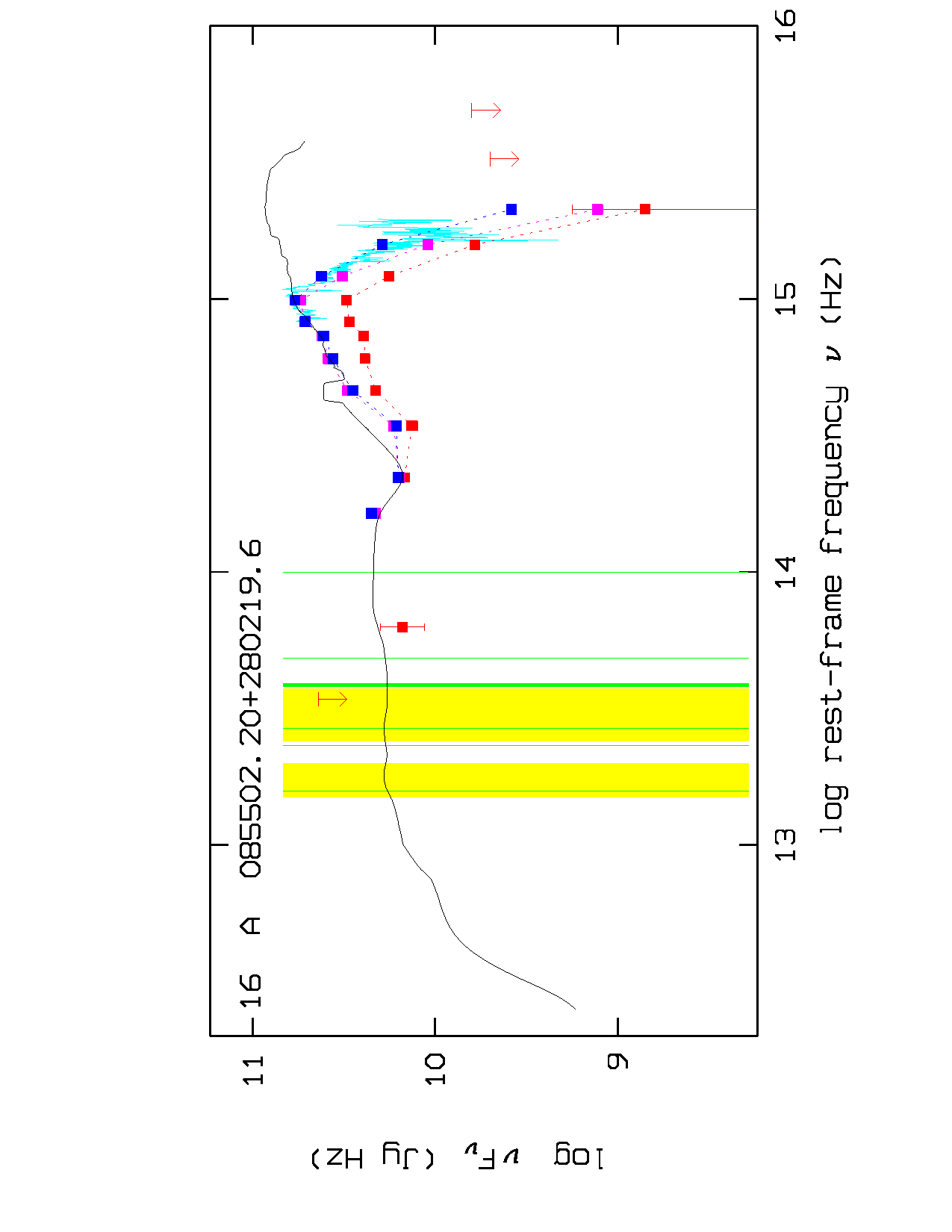}\hfill \\
\includegraphics[viewport=125 0 570 790,angle=270,width=8.0cm,clip]{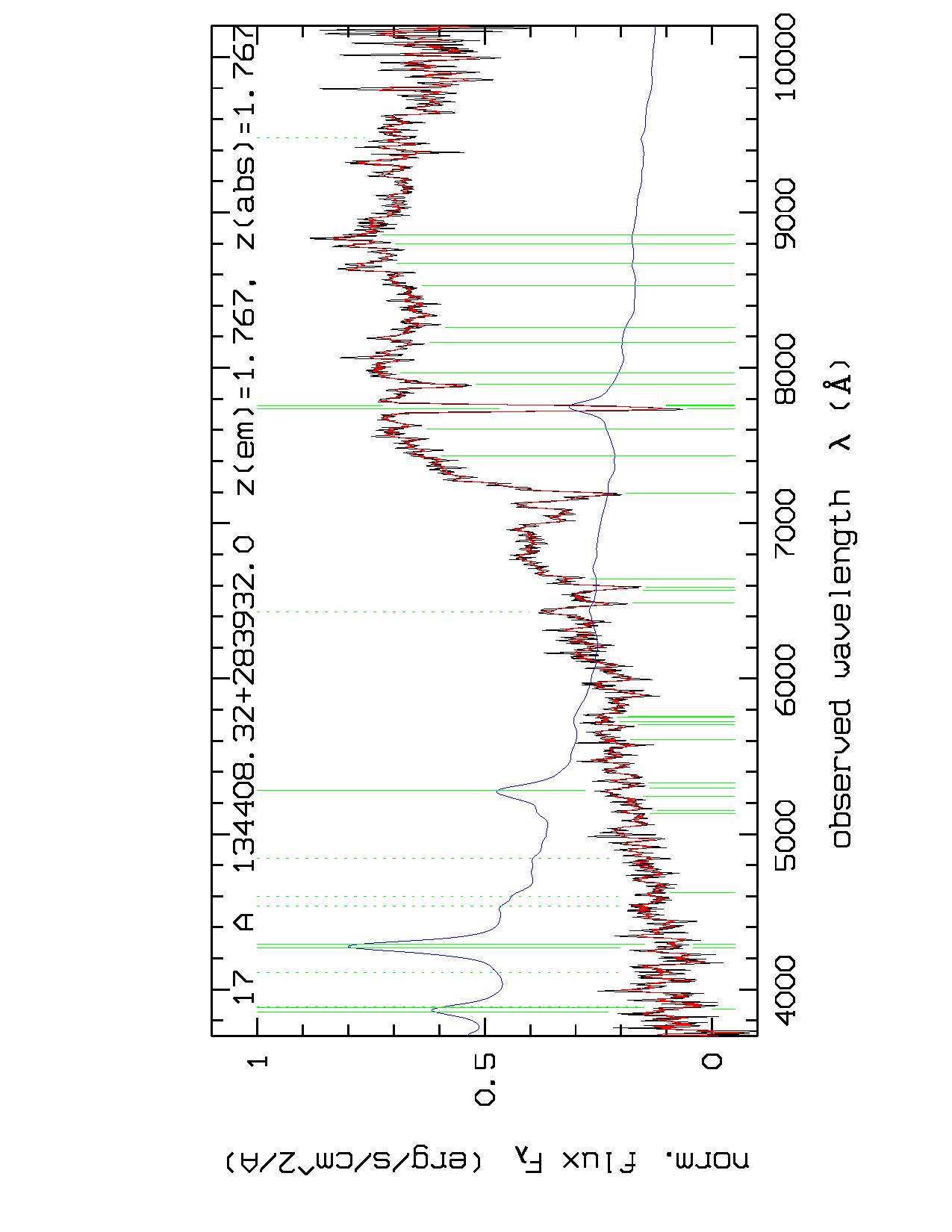}\hfill \=
\includegraphics[viewport=125 0 570 790,angle=270,width=8.0cm,clip]{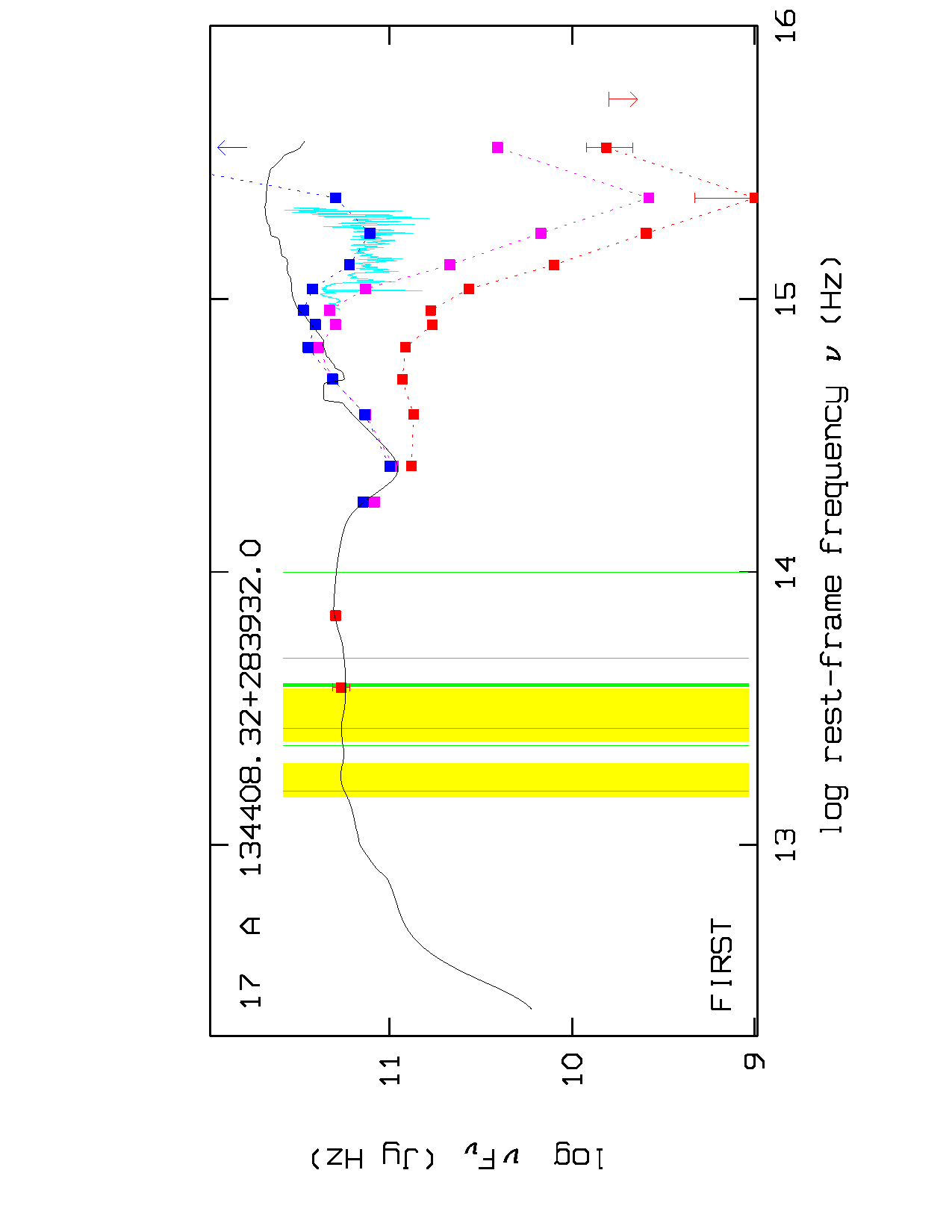}\hfill \\
\includegraphics[viewport=125 0 570 790,angle=270,width=8.0cm,clip]{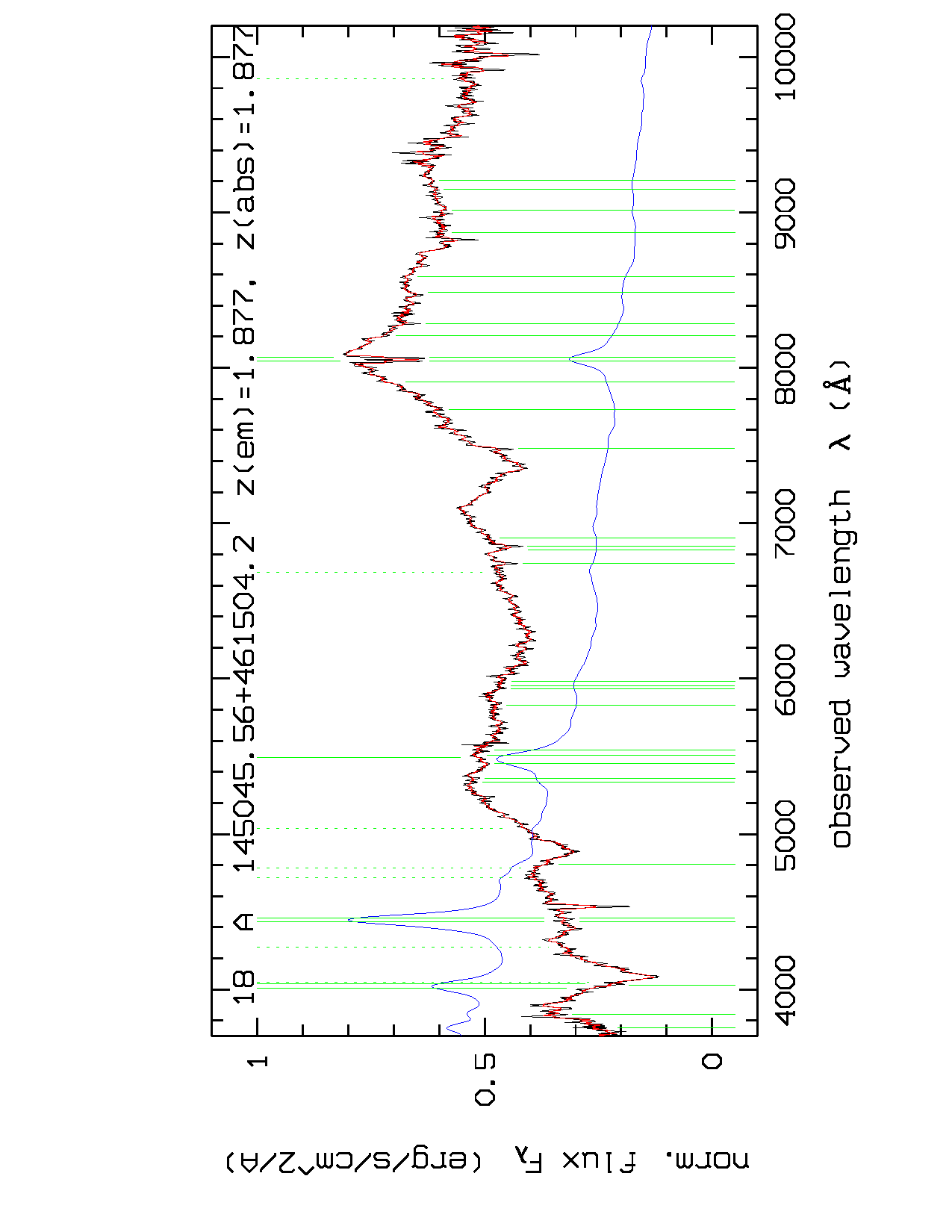}\hfill \=
\includegraphics[viewport=125 0 570 790,angle=270,width=8.0cm,clip]{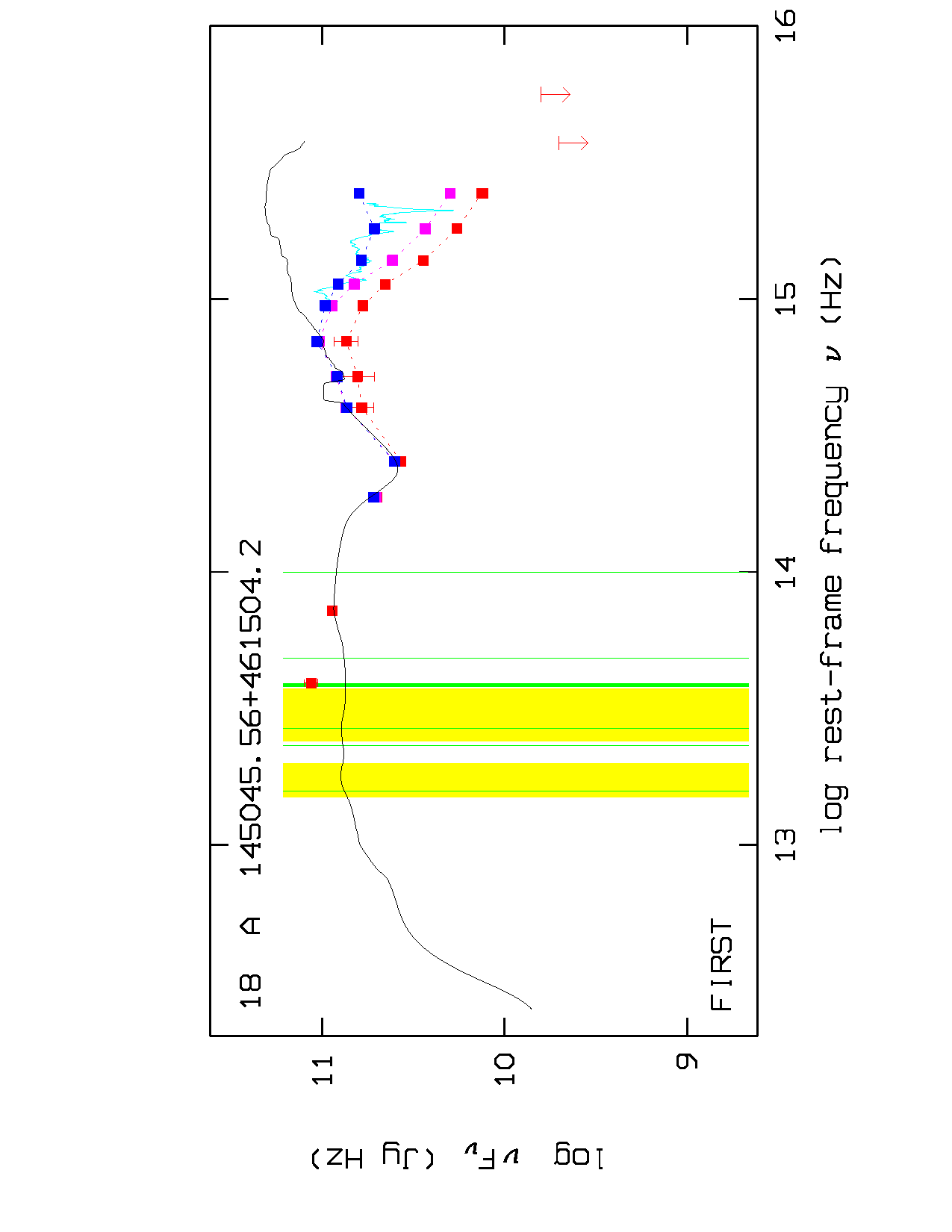}\hfill \\
\includegraphics[viewport=125 0 570 790,angle=270,width=8.0cm,clip]{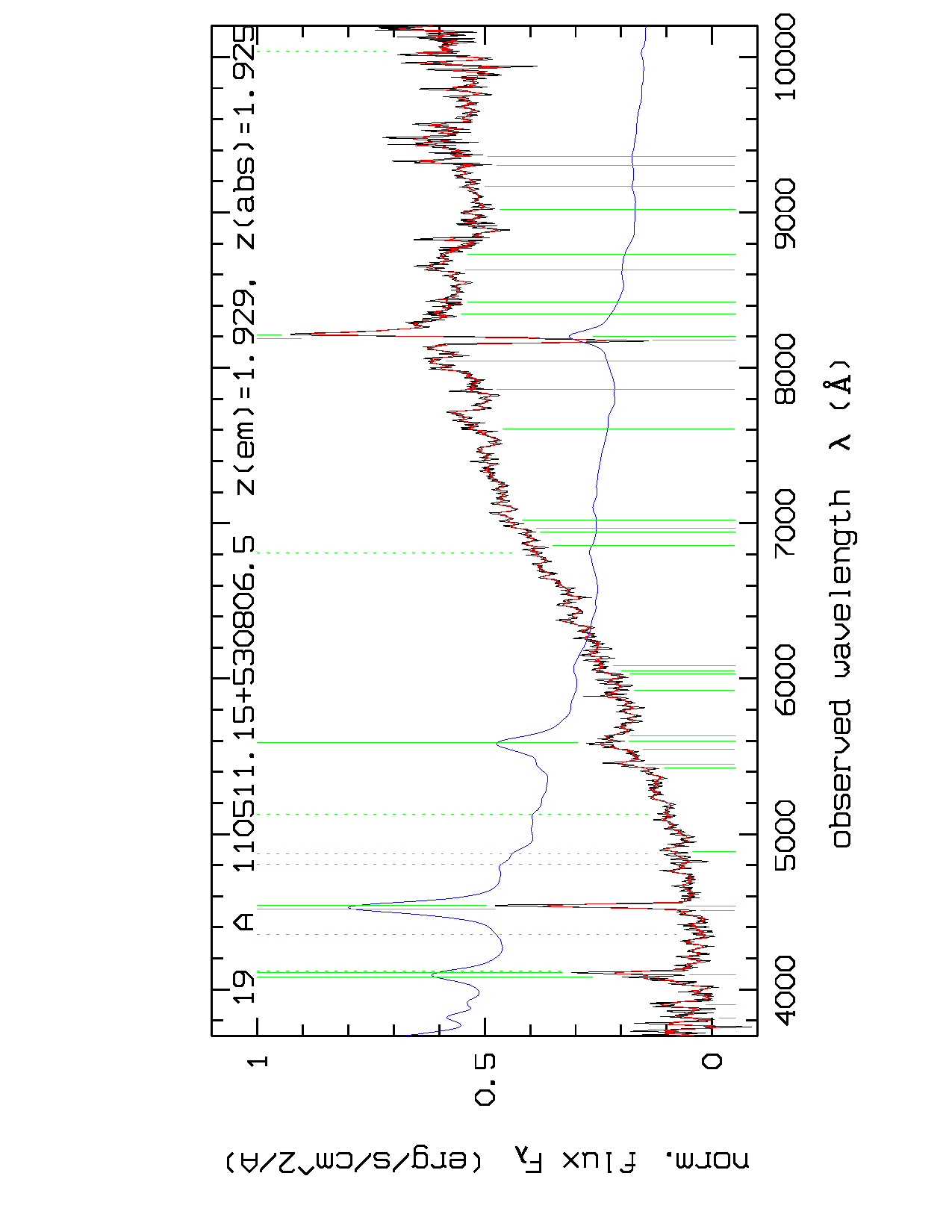}\hfill \=
\includegraphics[viewport=125 0 570 790,angle=270,width=8.0cm,clip]{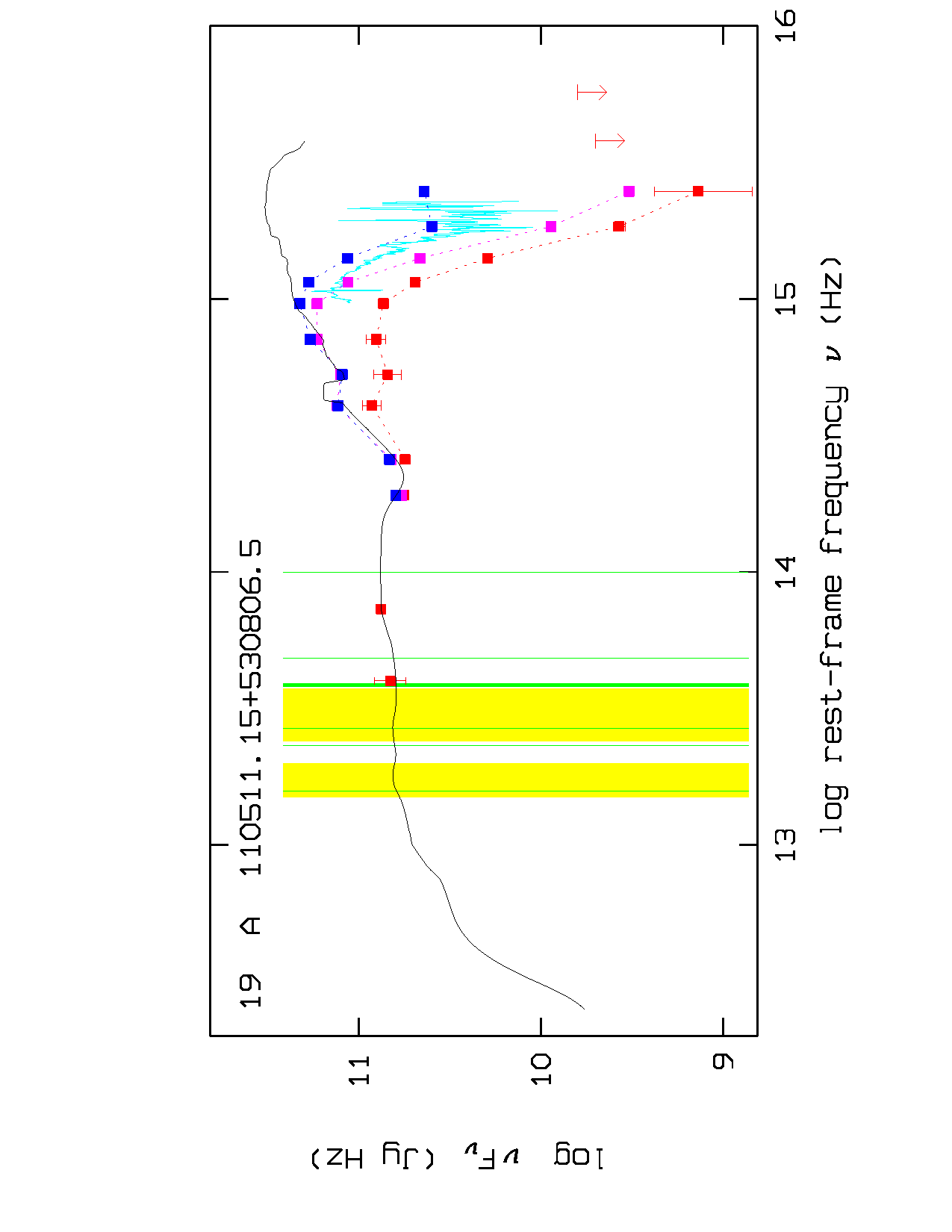}\hfill \\
\includegraphics[viewport=125 0 570 790,angle=270,width=8.0cm,clip]{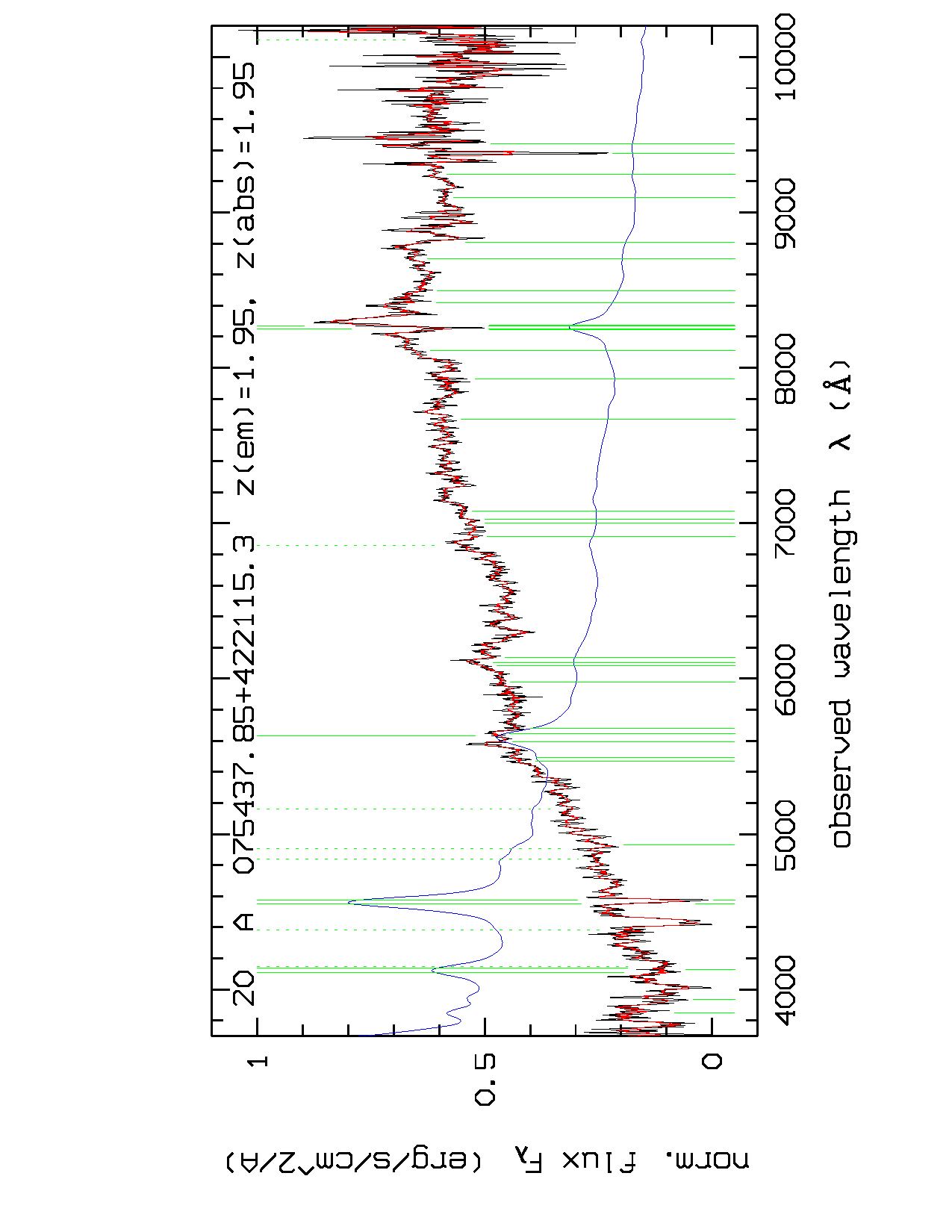}\hfill \=
\includegraphics[viewport=125 0 570 790,angle=270,width=8.0cm,clip]{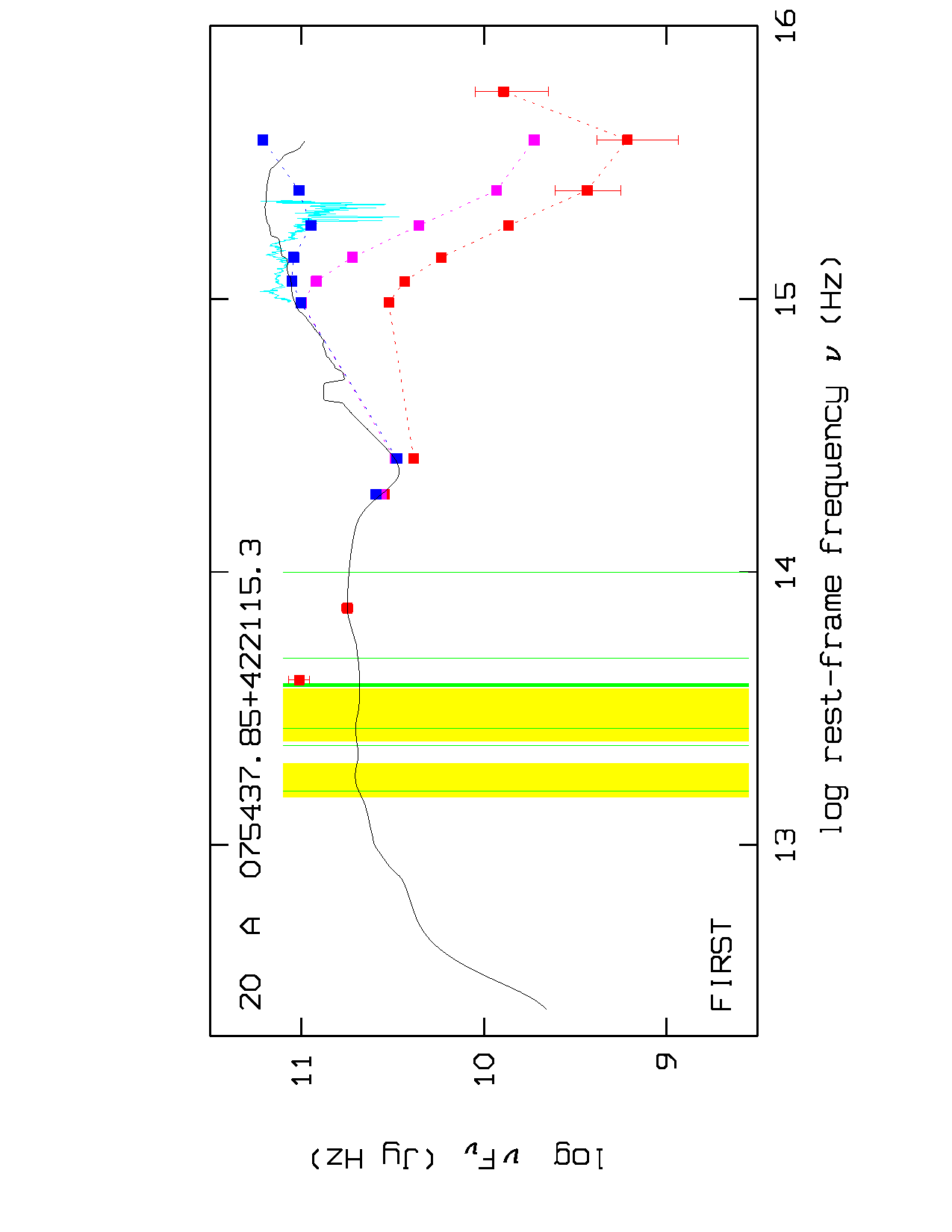}\hfill
\end{tabbing}
\caption{Sample A - continued (3).}
\end{figure*}\clearpage

\begin{figure*}[h]
\begin{tabbing}\\
\includegraphics[viewport=125 0 570 790,angle=270,width=8.0cm,clip]{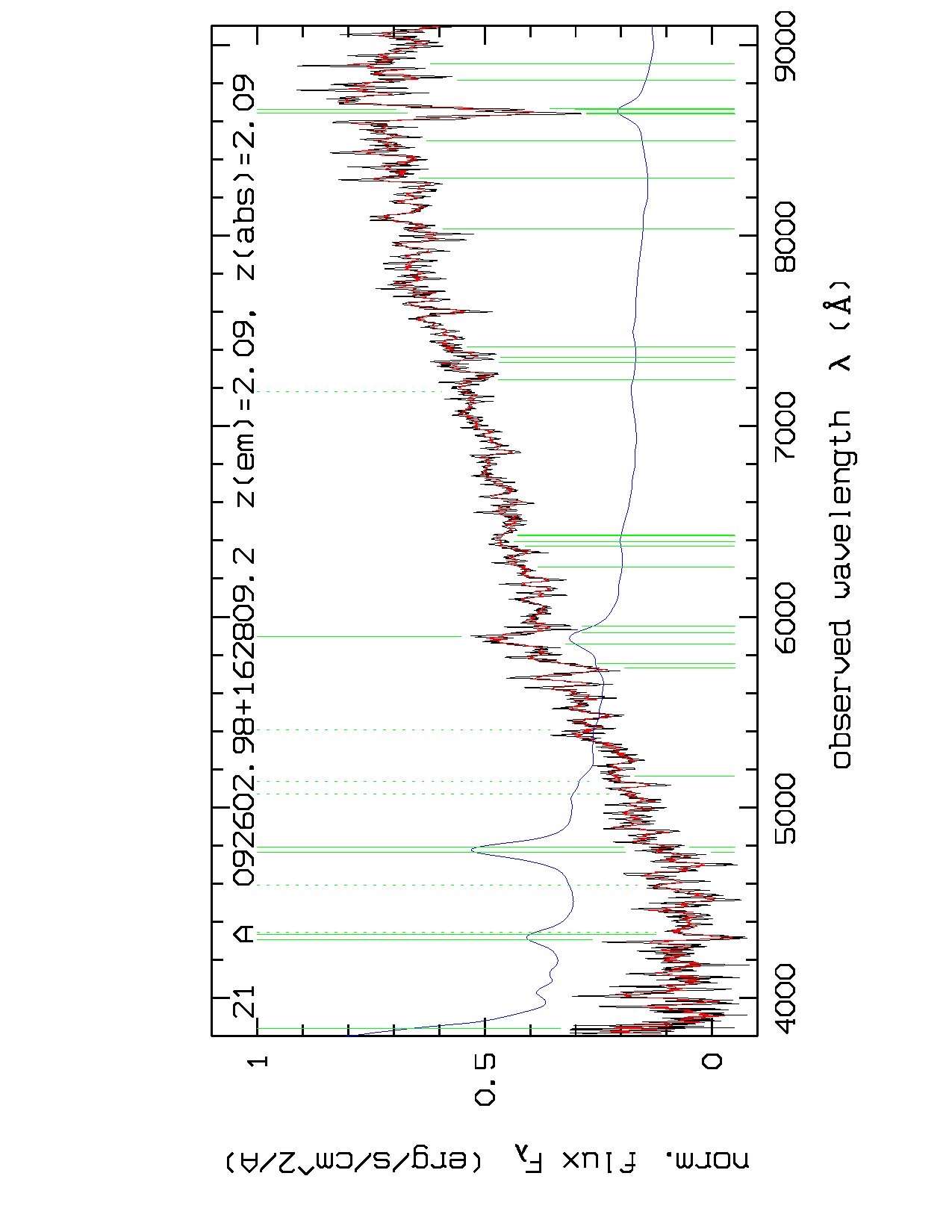}\hfill \=
\includegraphics[viewport=125 0 570 790,angle=270,width=8.0cm,clip]{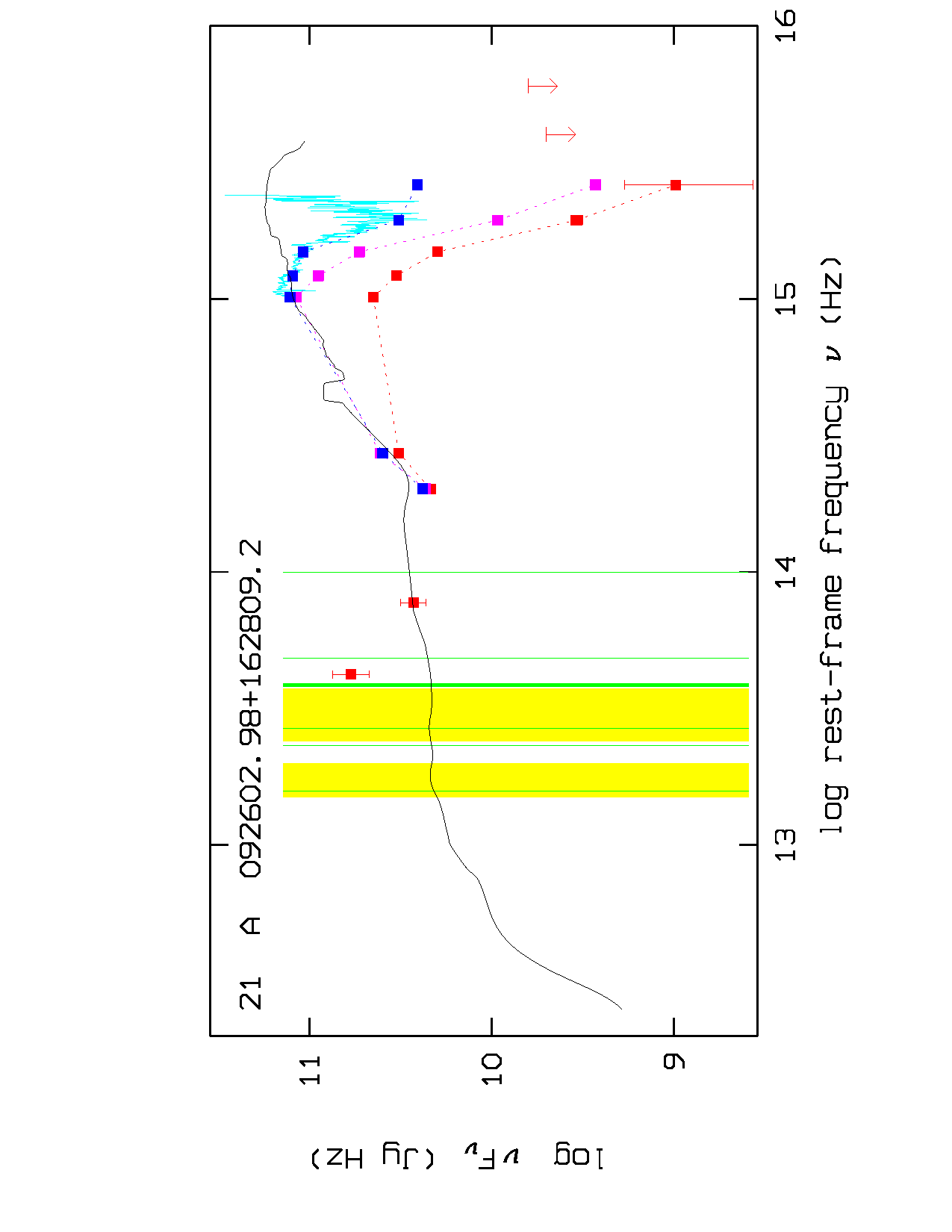}\hfill \\
\includegraphics[viewport=125 0 570 790,angle=270,width=8.0cm,clip]{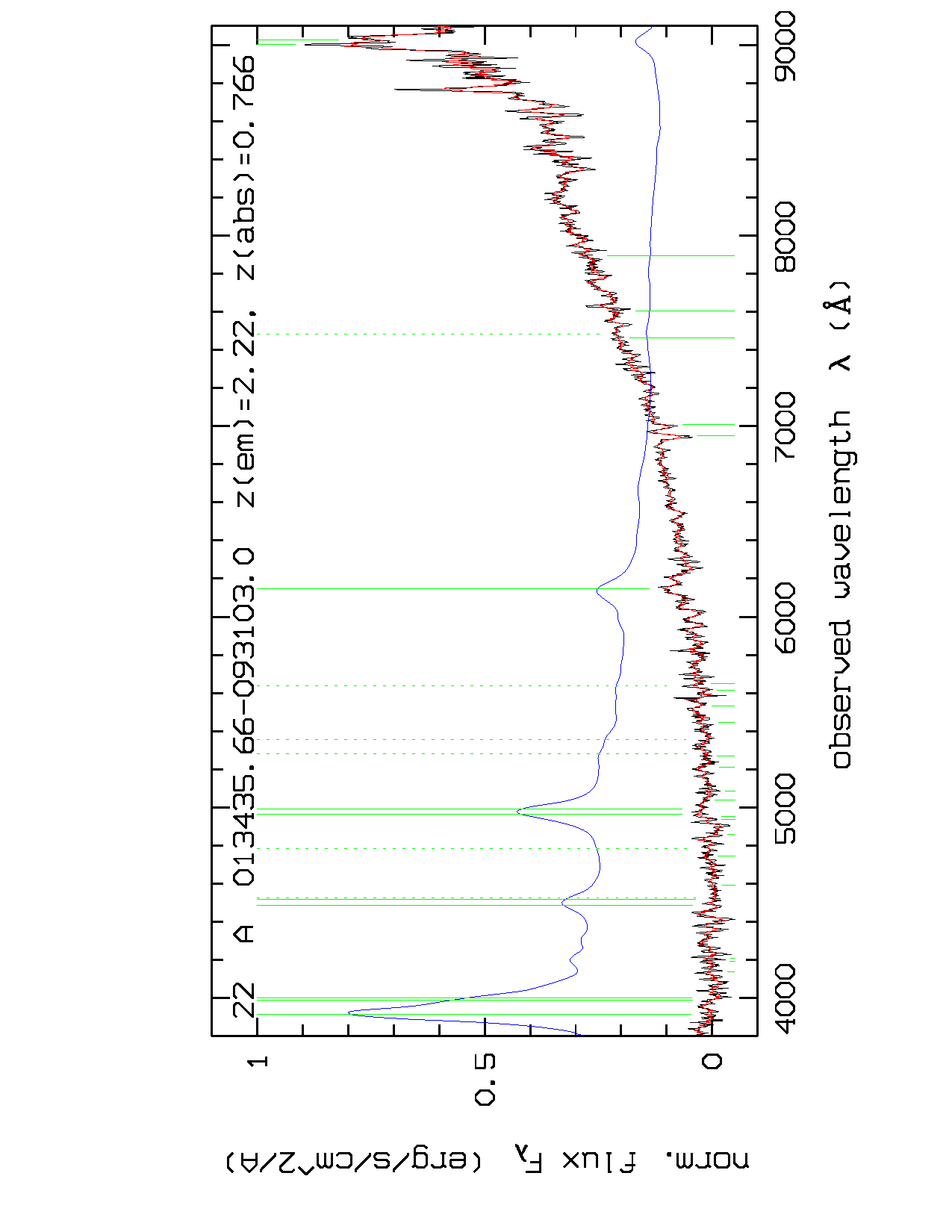}\hfill \=
\includegraphics[viewport=125 0 570 790,angle=270,width=8.0cm,clip]{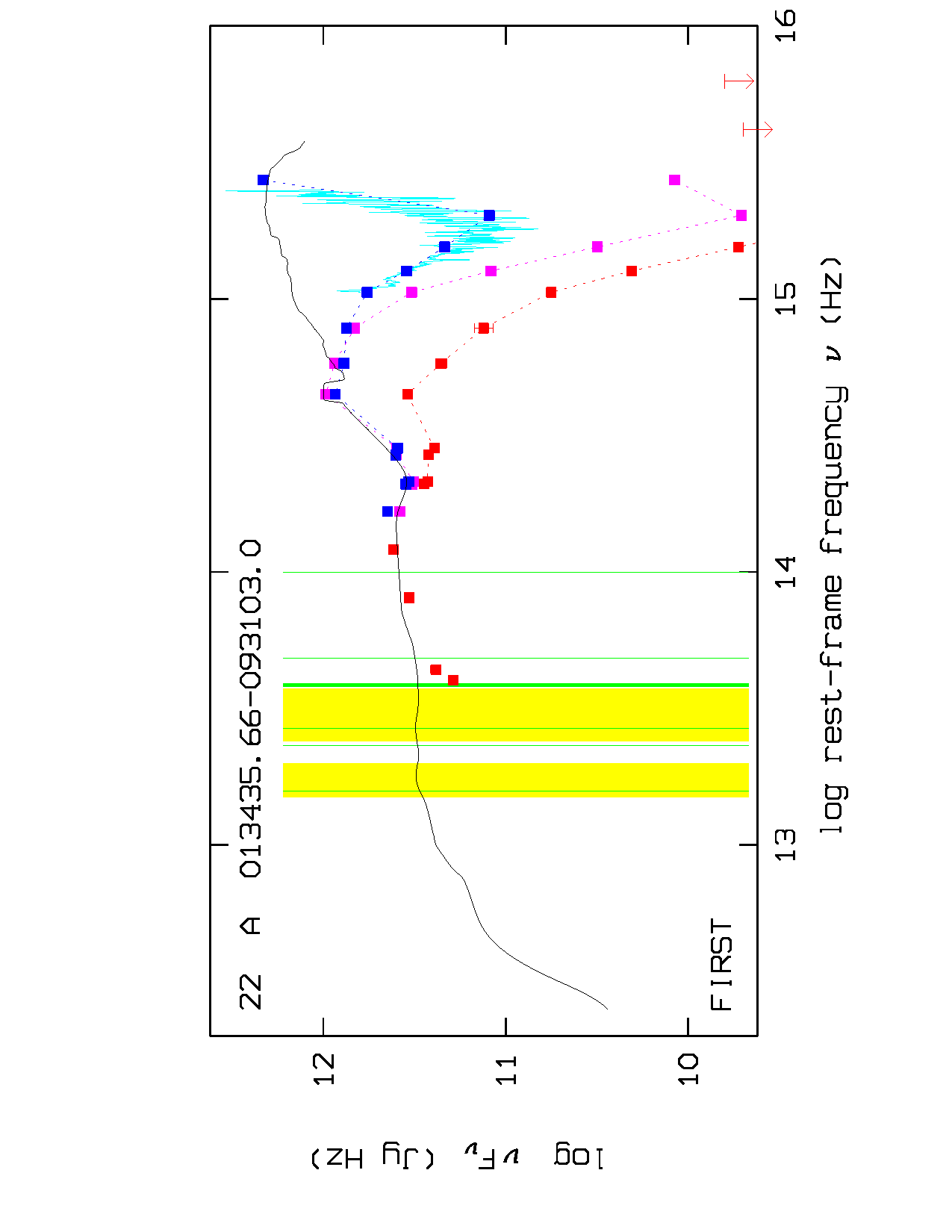}\hfill \\
\includegraphics[viewport=125 0 570 790,angle=270,width=8.0cm,clip]{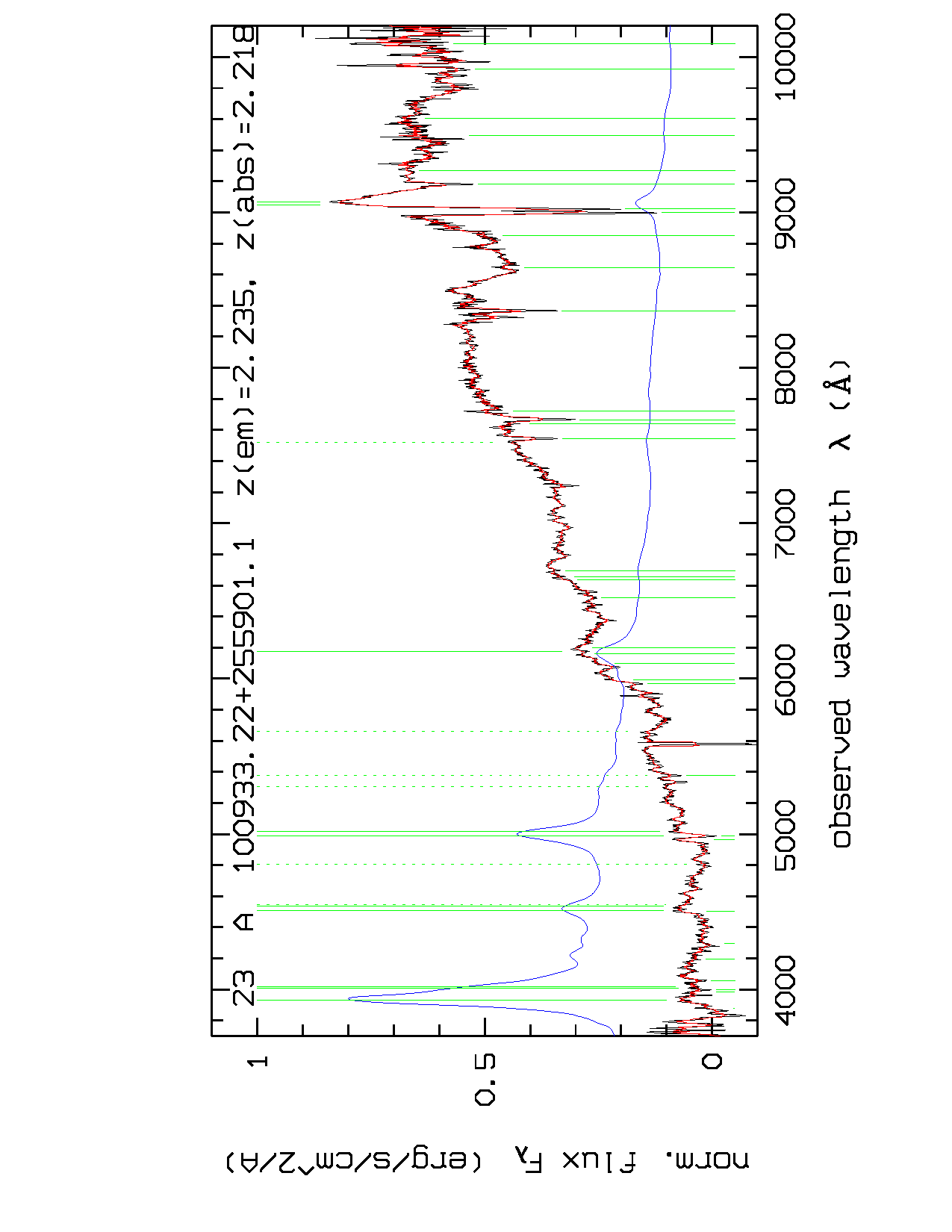}\hfill \=
\includegraphics[viewport=125 0 570 790,angle=270,width=8.0cm,clip]{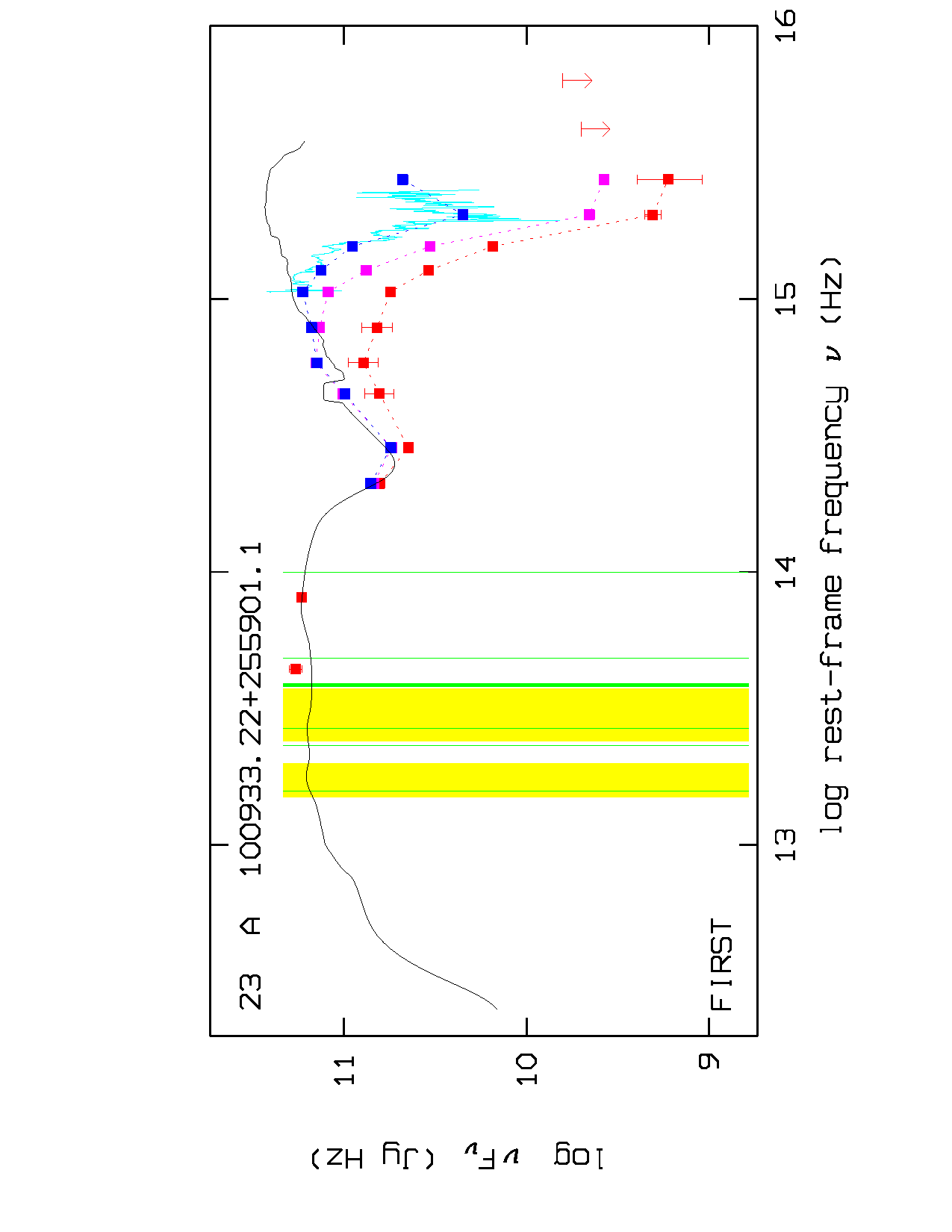}\hfill
\end{tabbing}
\caption{Sample A - continued (4).}
\end{figure*}\clearpage

\begin{figure*}[h]
\begin{tabbing}
\includegraphics[viewport=125 0 570 790,angle=270,width=8.0cm,clip]{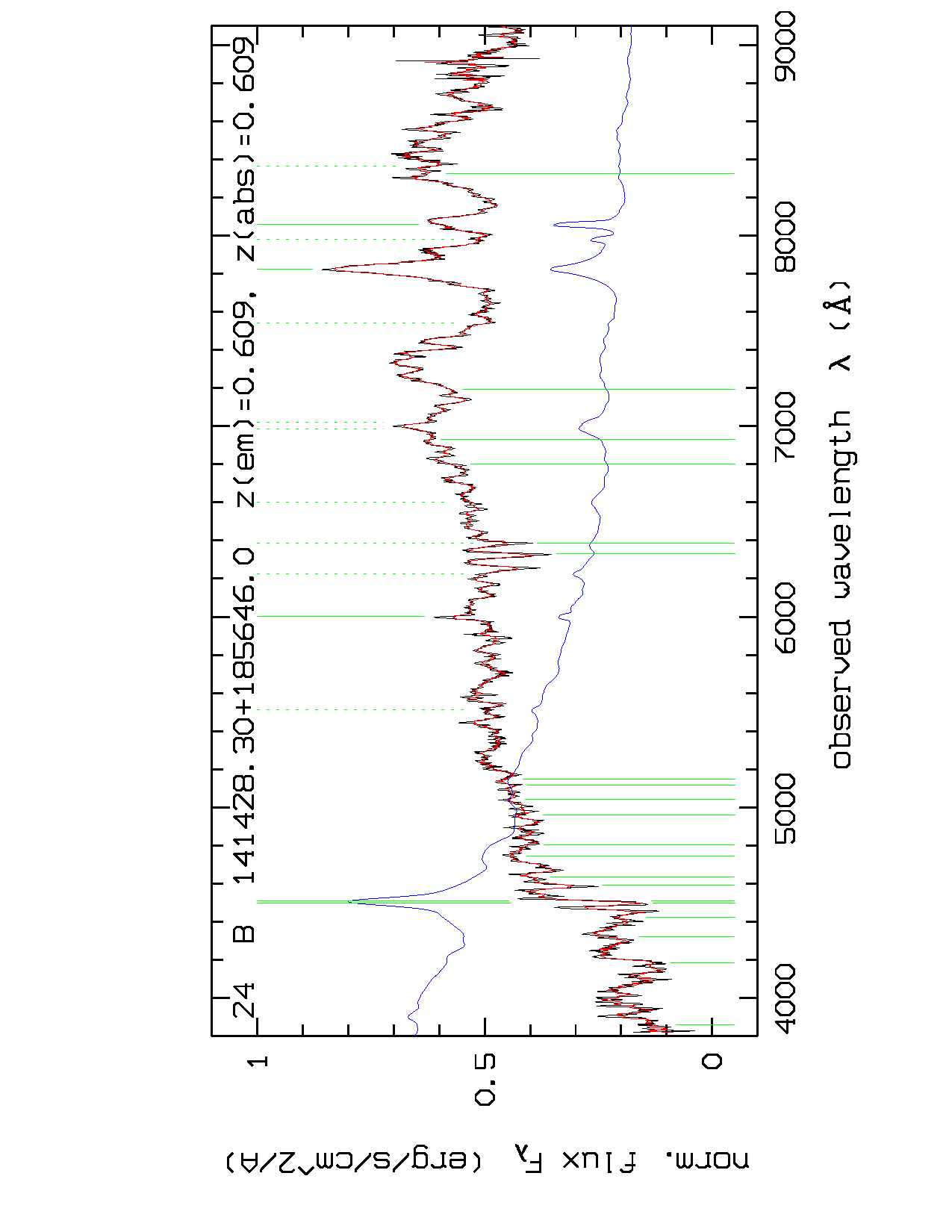}\hfill \=
\includegraphics[viewport=125 0 570 790,angle=270,width=8.0cm,clip]{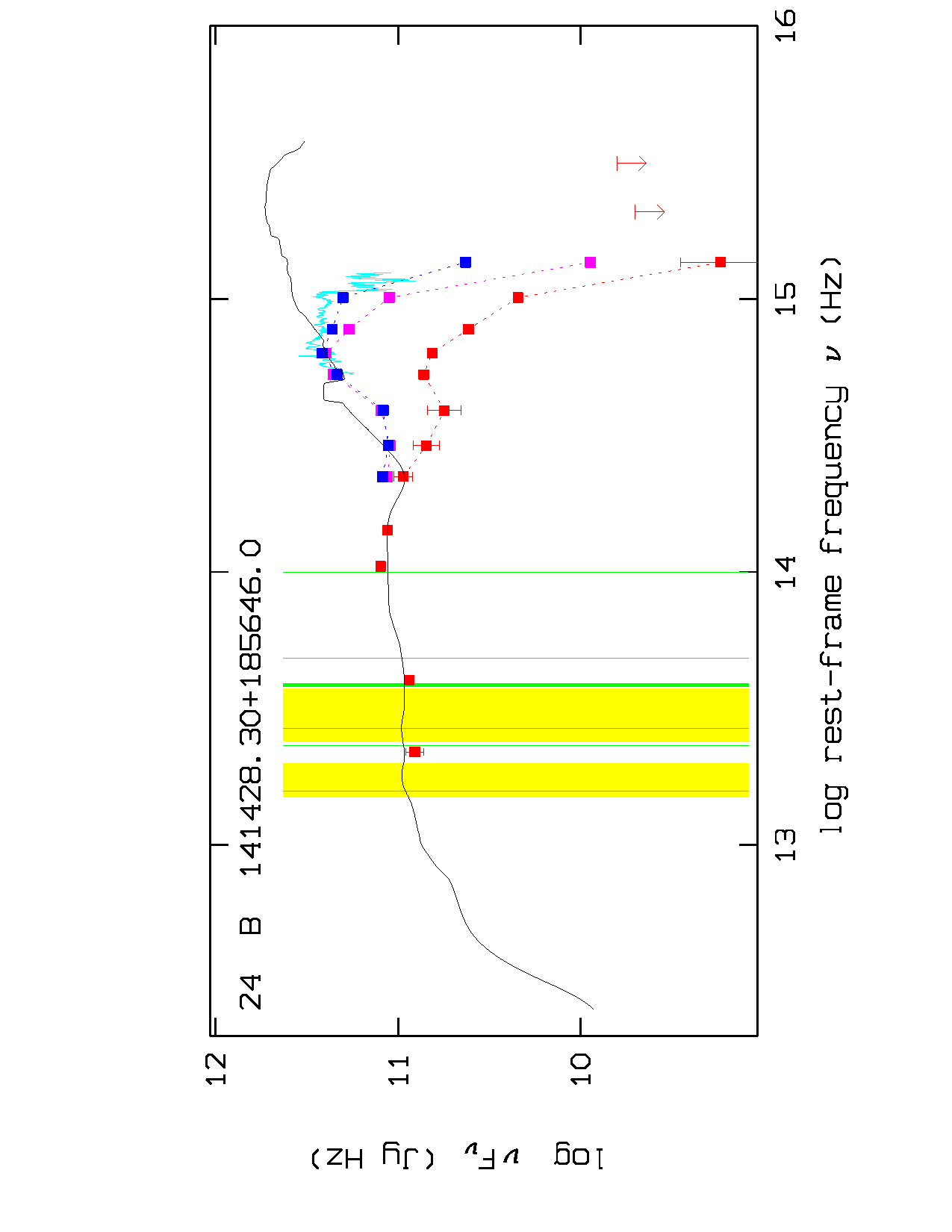}\hfill \\
\includegraphics[viewport=125 0 570 790,angle=270,width=8.0cm,clip]{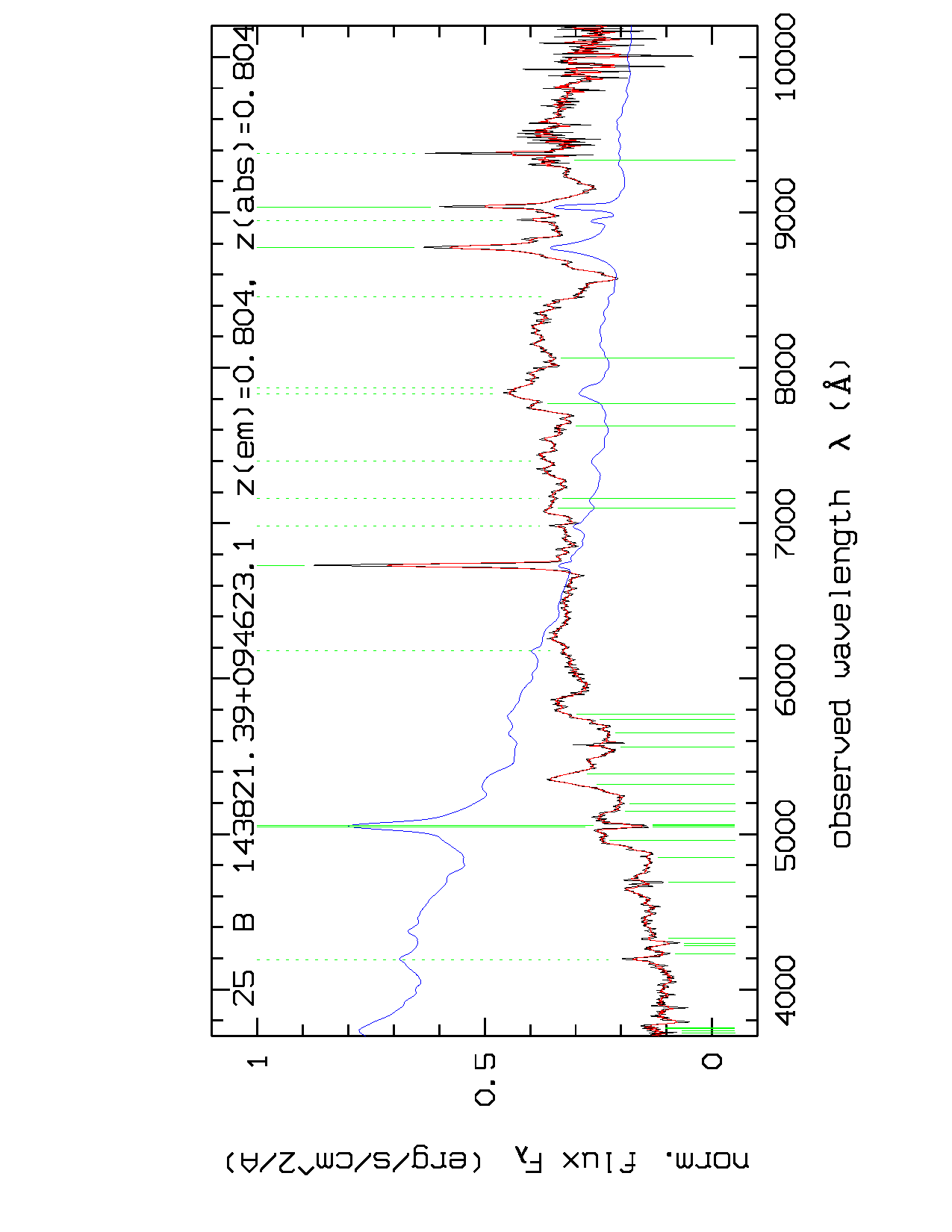}\hfill \=
\includegraphics[viewport=125 0 570 790,angle=270,width=8.0cm,clip]{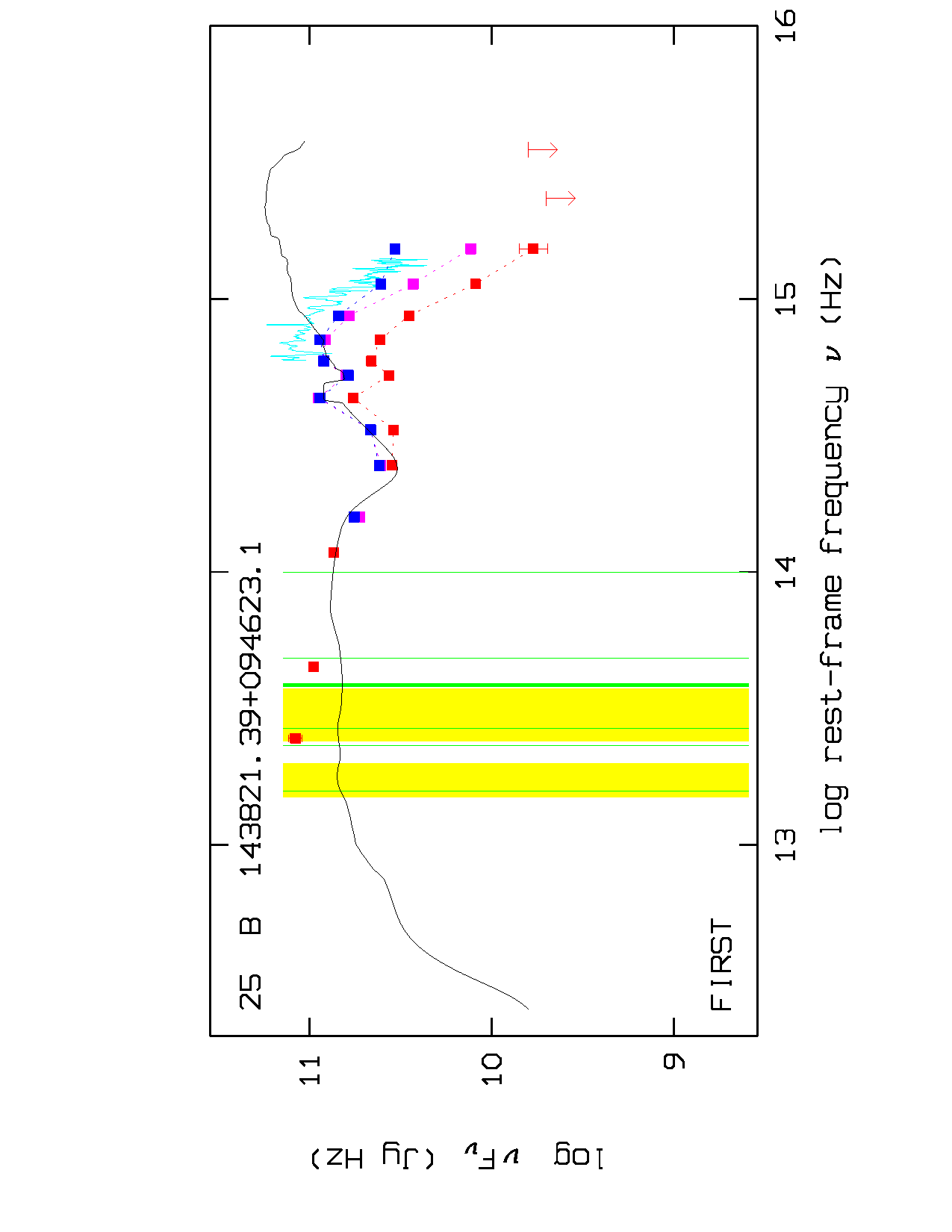}\hfill \\
\includegraphics[viewport=125 0 570 790,angle=270,width=8.0cm,clip]{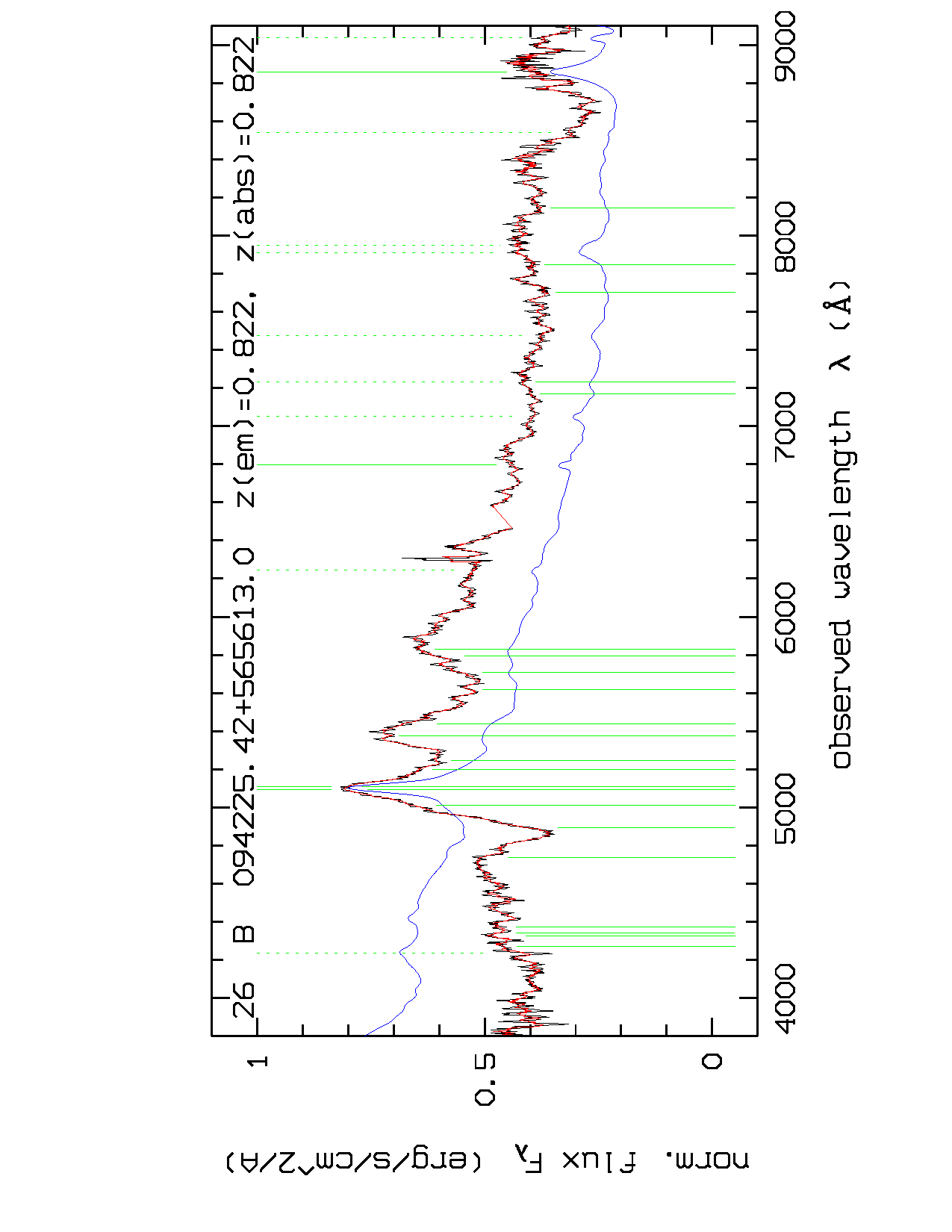}\hfill \=
\includegraphics[viewport=125 0 570 790,angle=270,width=8.0cm,clip]{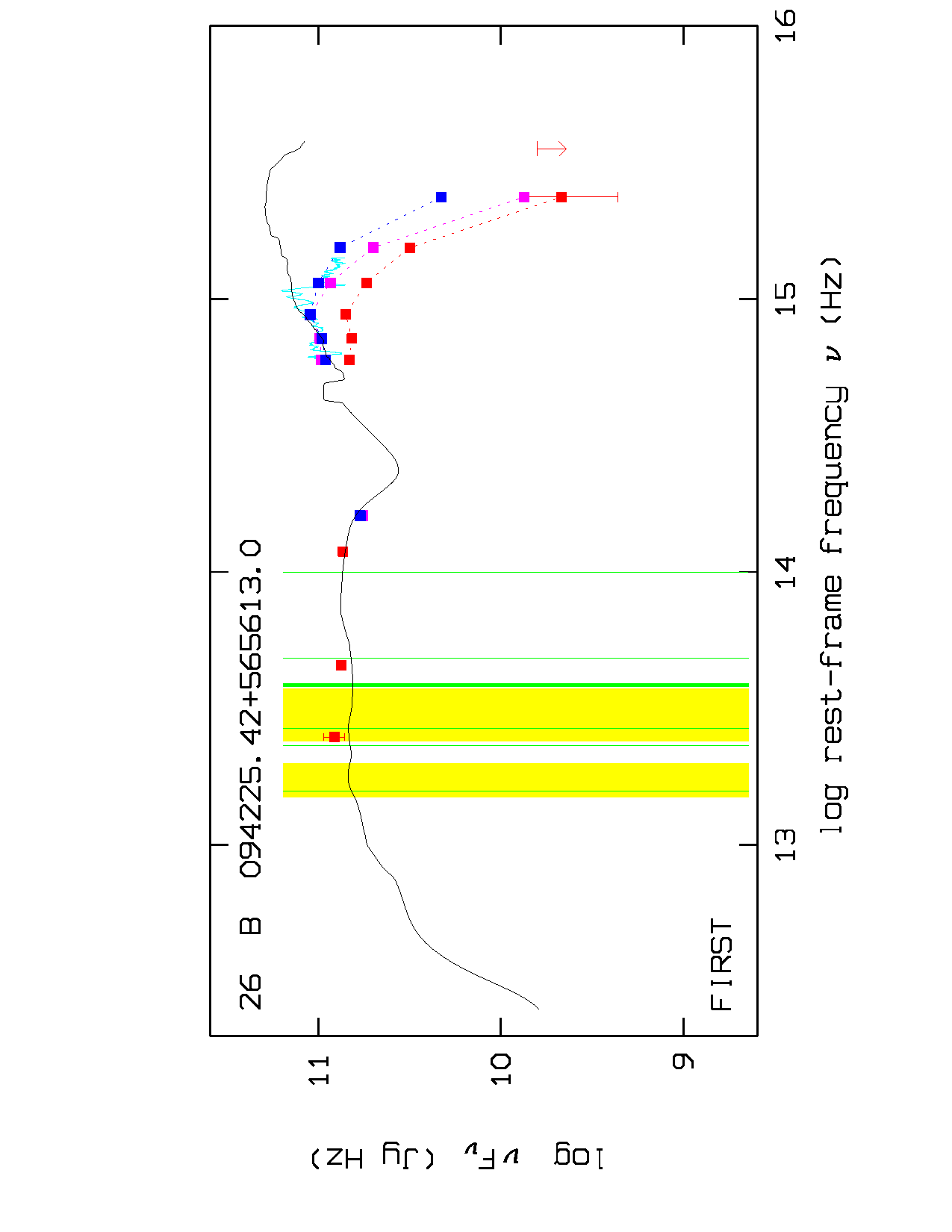}\hfill \\
\includegraphics[viewport=125 0 570 790,angle=270,width=8.0cm,clip]{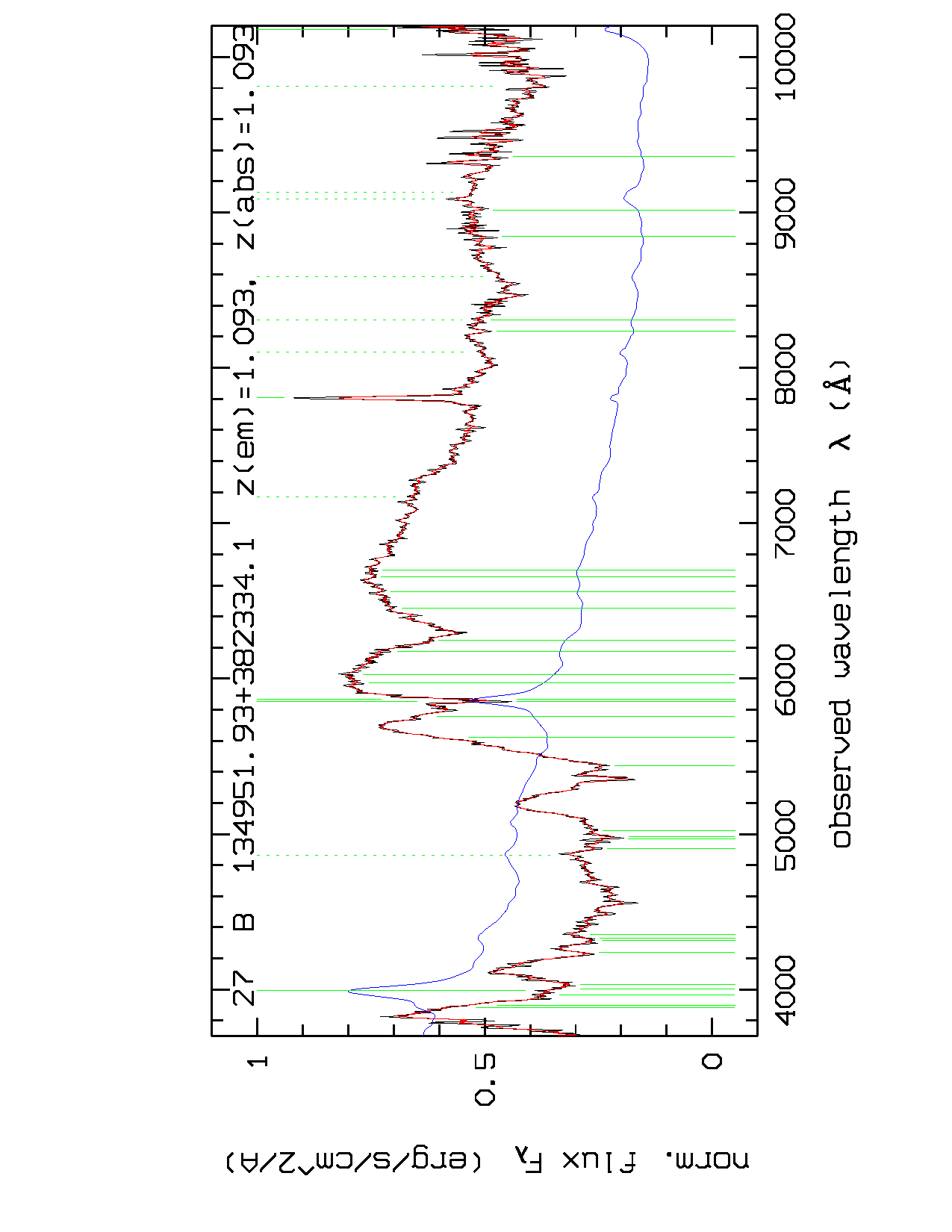}\hfill \=
\includegraphics[viewport=125 0 570 790,angle=270,width=8.0cm,clip]{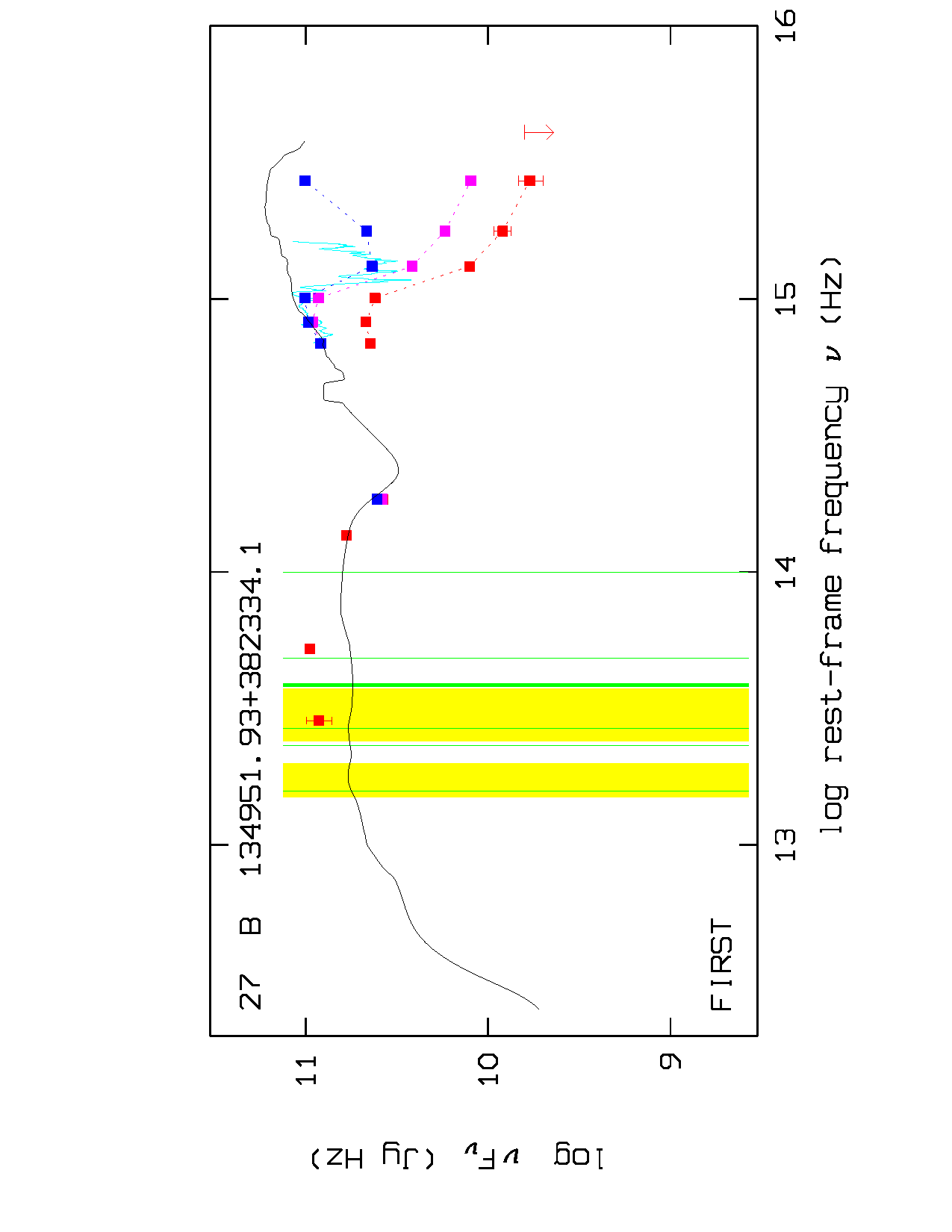}\hfill \\
\includegraphics[viewport=125 0 570 790,angle=270,width=8.0cm,clip]{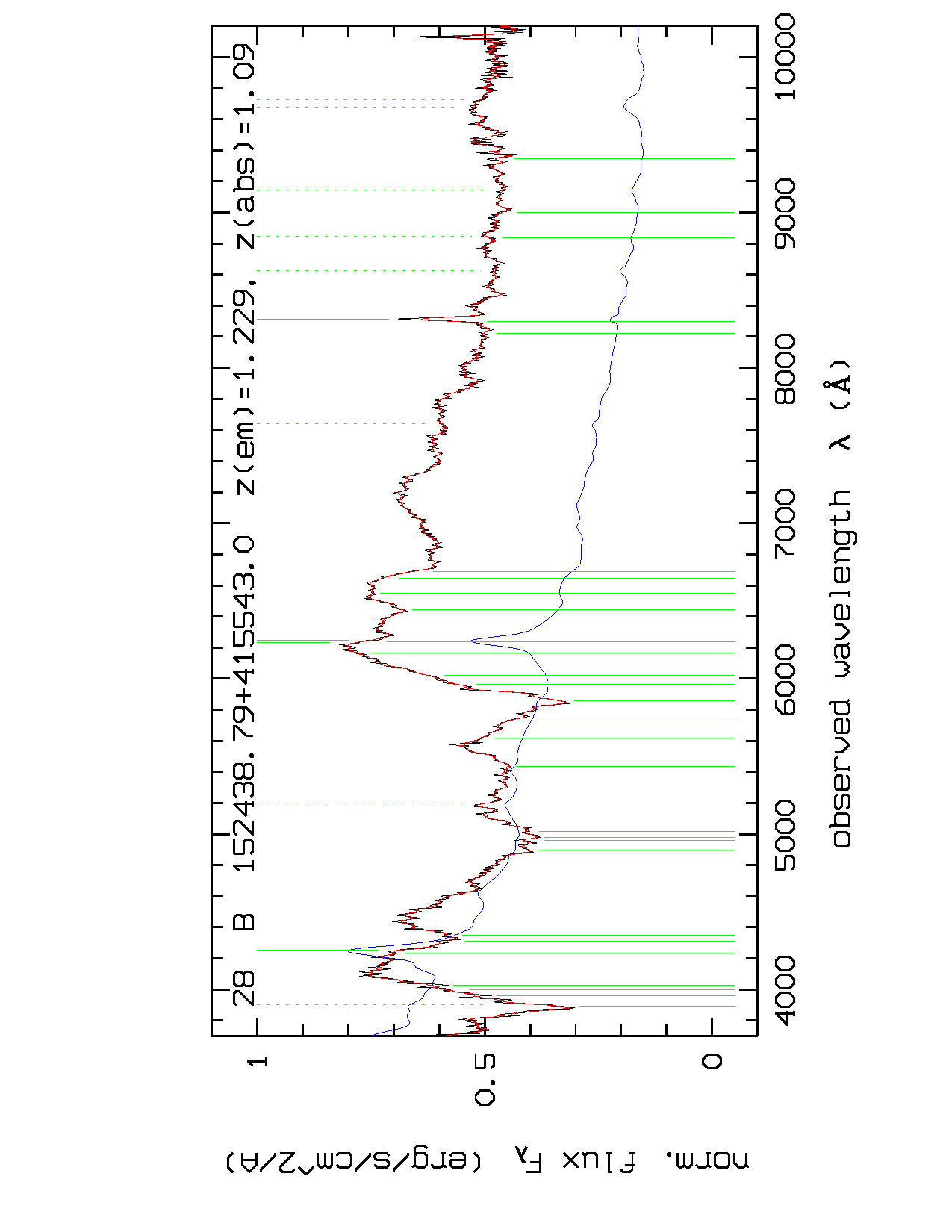}\hfill \=
\includegraphics[viewport=125 0 570 790,angle=270,width=8.0cm,clip]{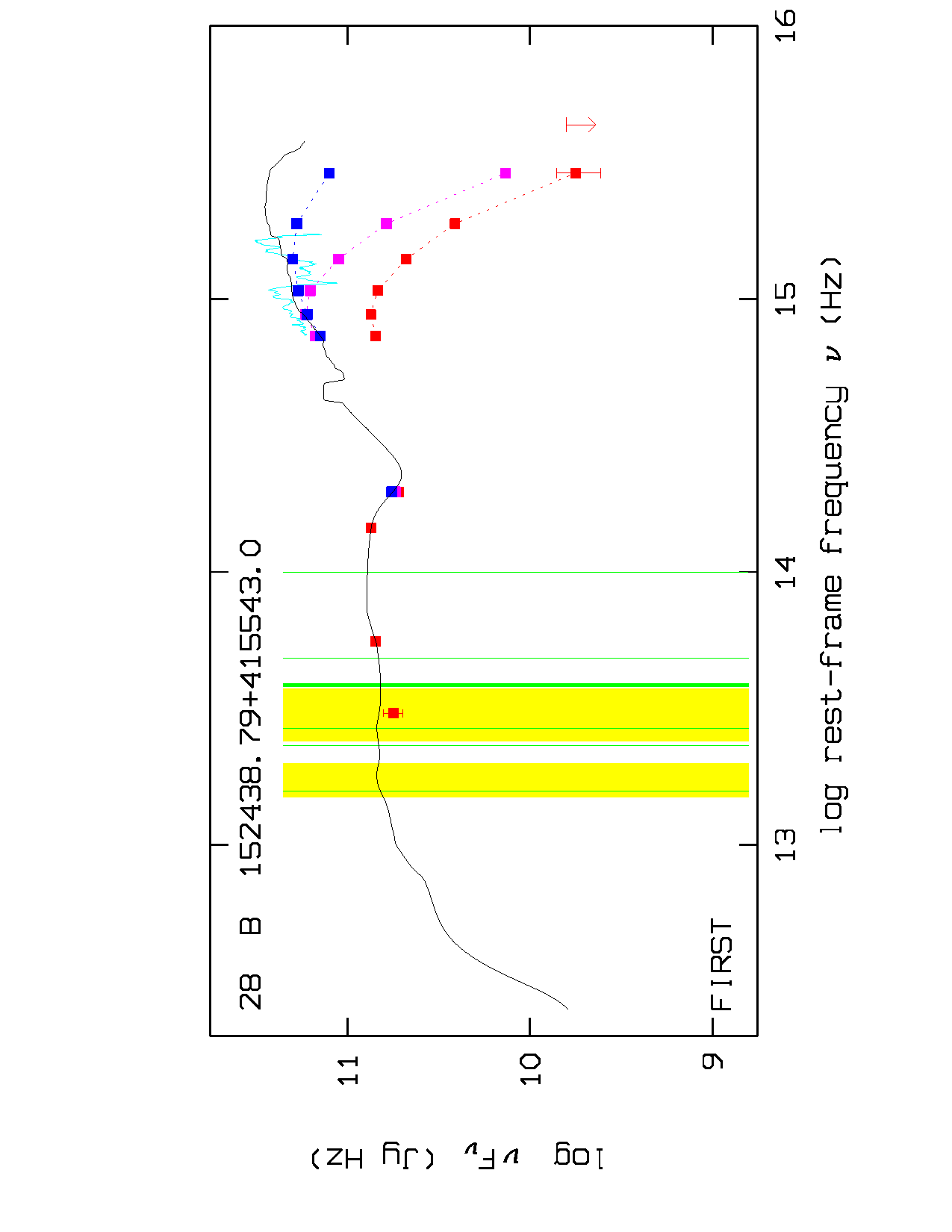}\hfill
\end{tabbing}
\caption{Sample B.}
\end{figure*}\clearpage

\begin{figure*}[h]
\begin{tabbing}
\includegraphics[viewport=125 0 570 790,angle=270,width=8.0cm,clip]{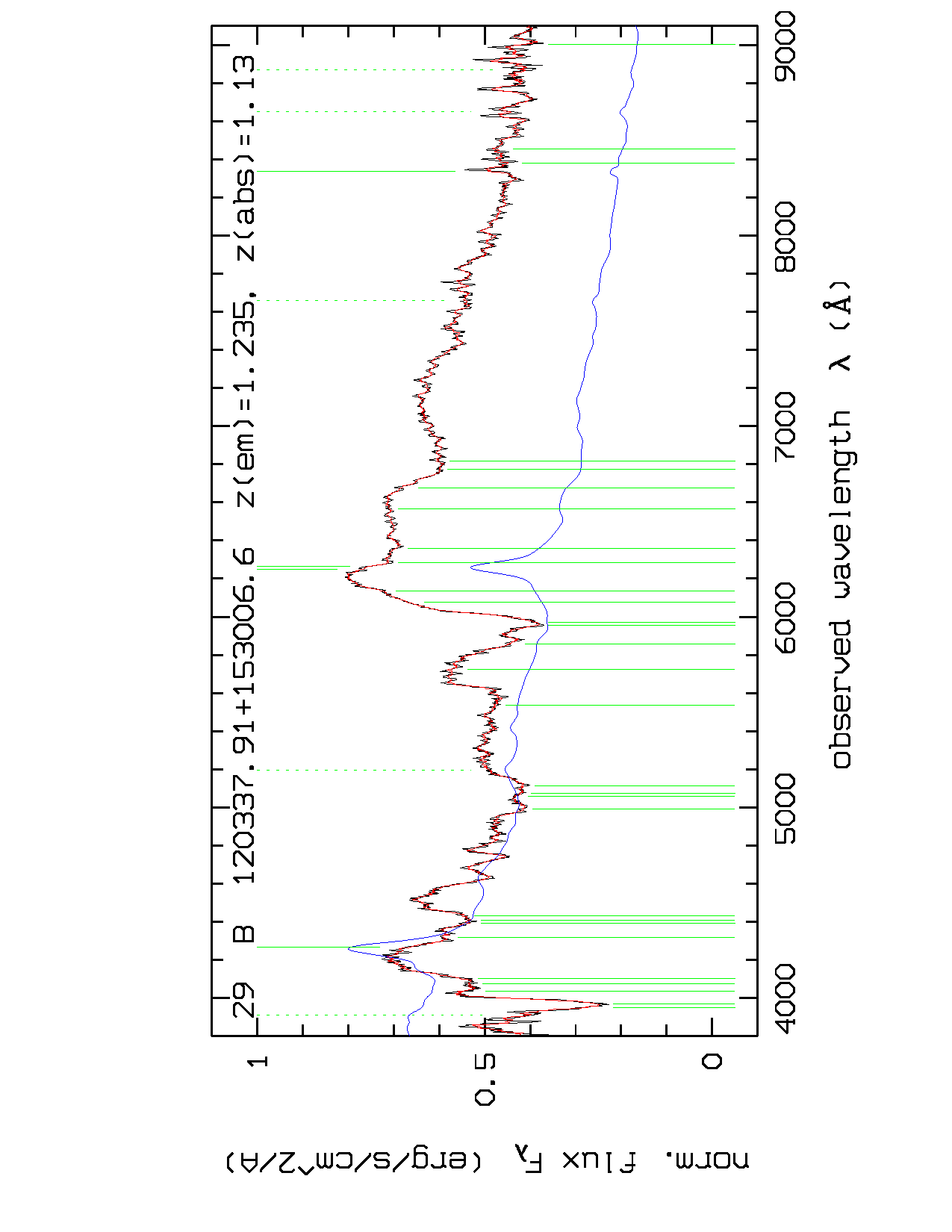}\hfill \=
\includegraphics[viewport=125 0 570 790,angle=270,width=8.0cm,clip]{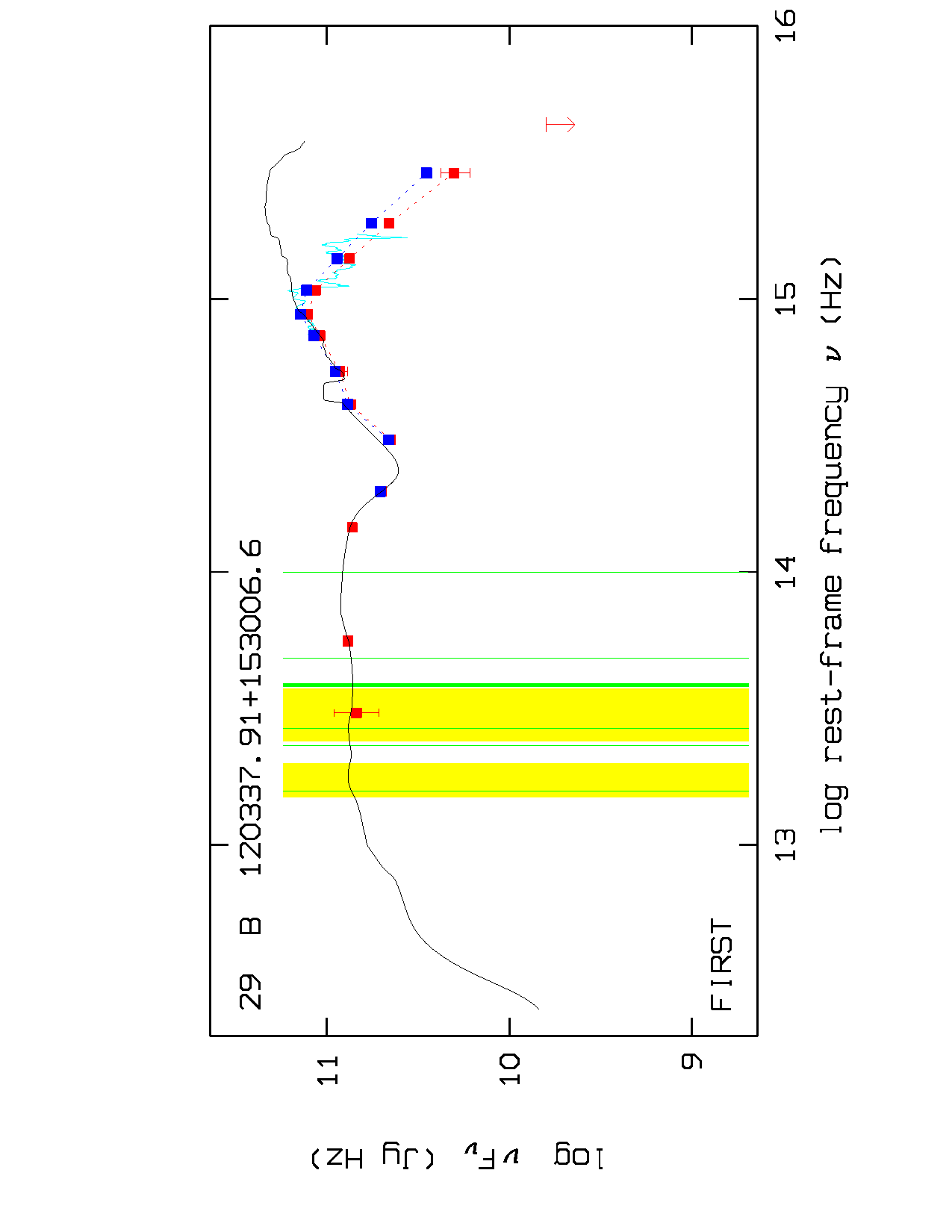}\hfill \\
\includegraphics[viewport=125 0 570 790,angle=270,width=8.0cm,clip]{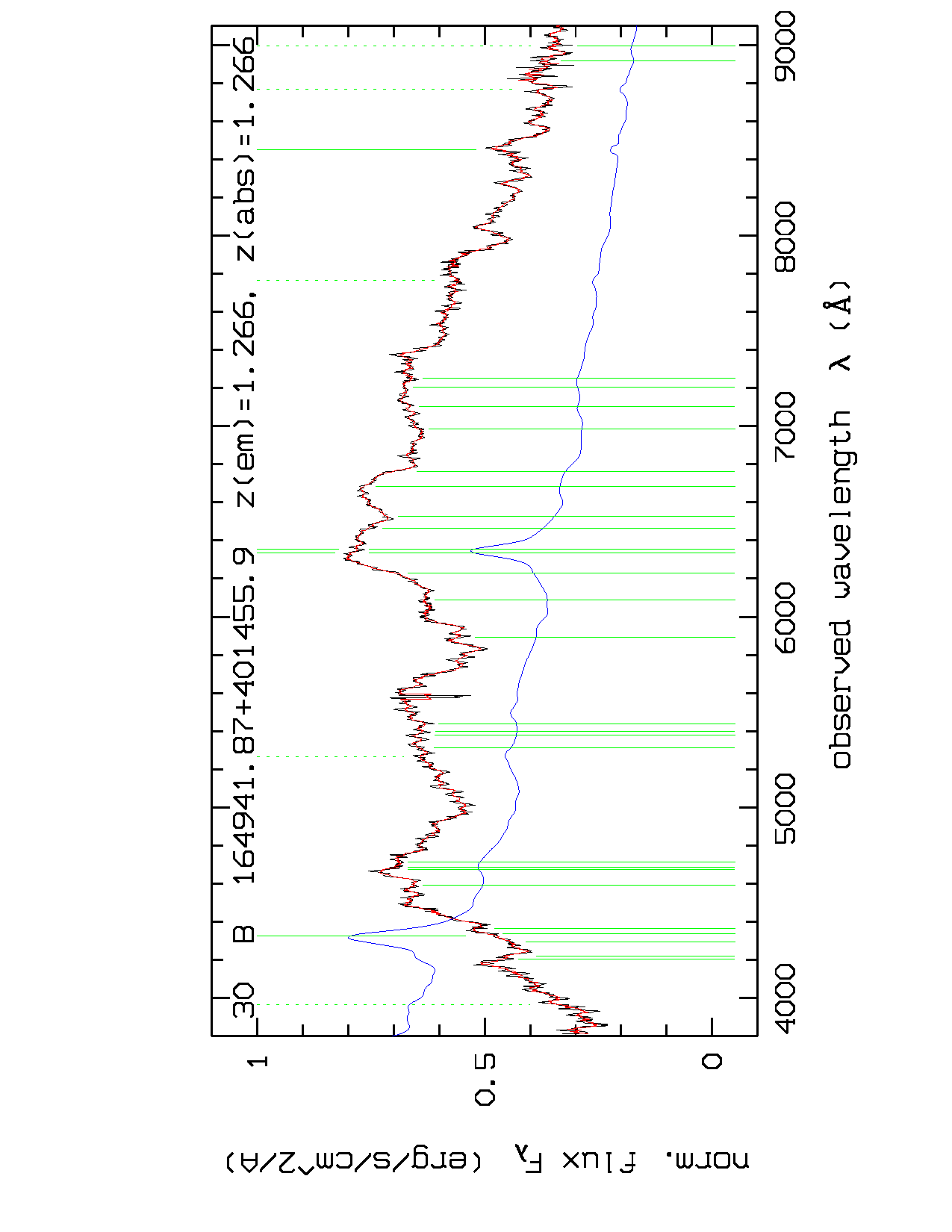}\hfill \=
\includegraphics[viewport=125 0 570 790,angle=270,width=8.0cm,clip]{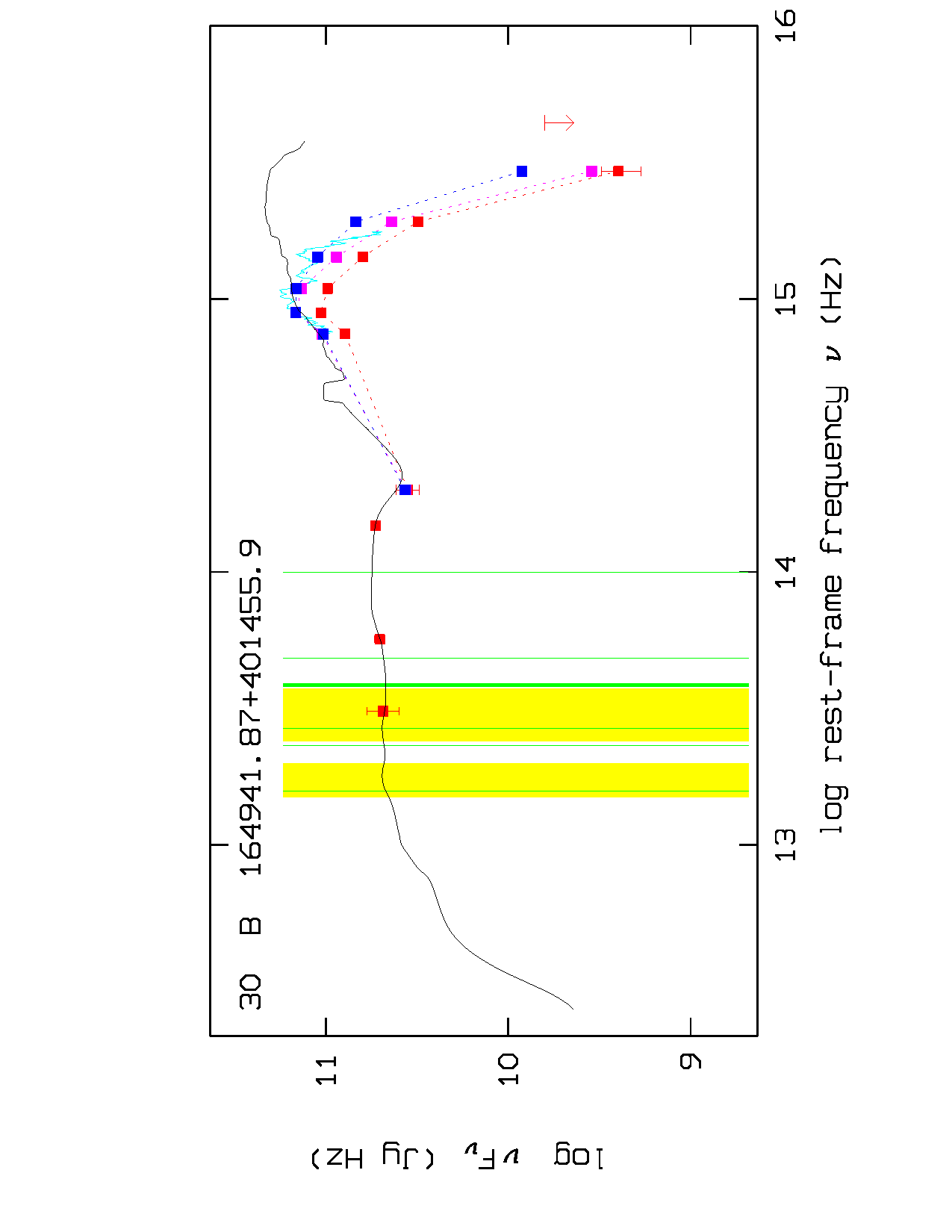}\hfill \\
\includegraphics[viewport=125 0 570 790,angle=270,width=8.0cm,clip]{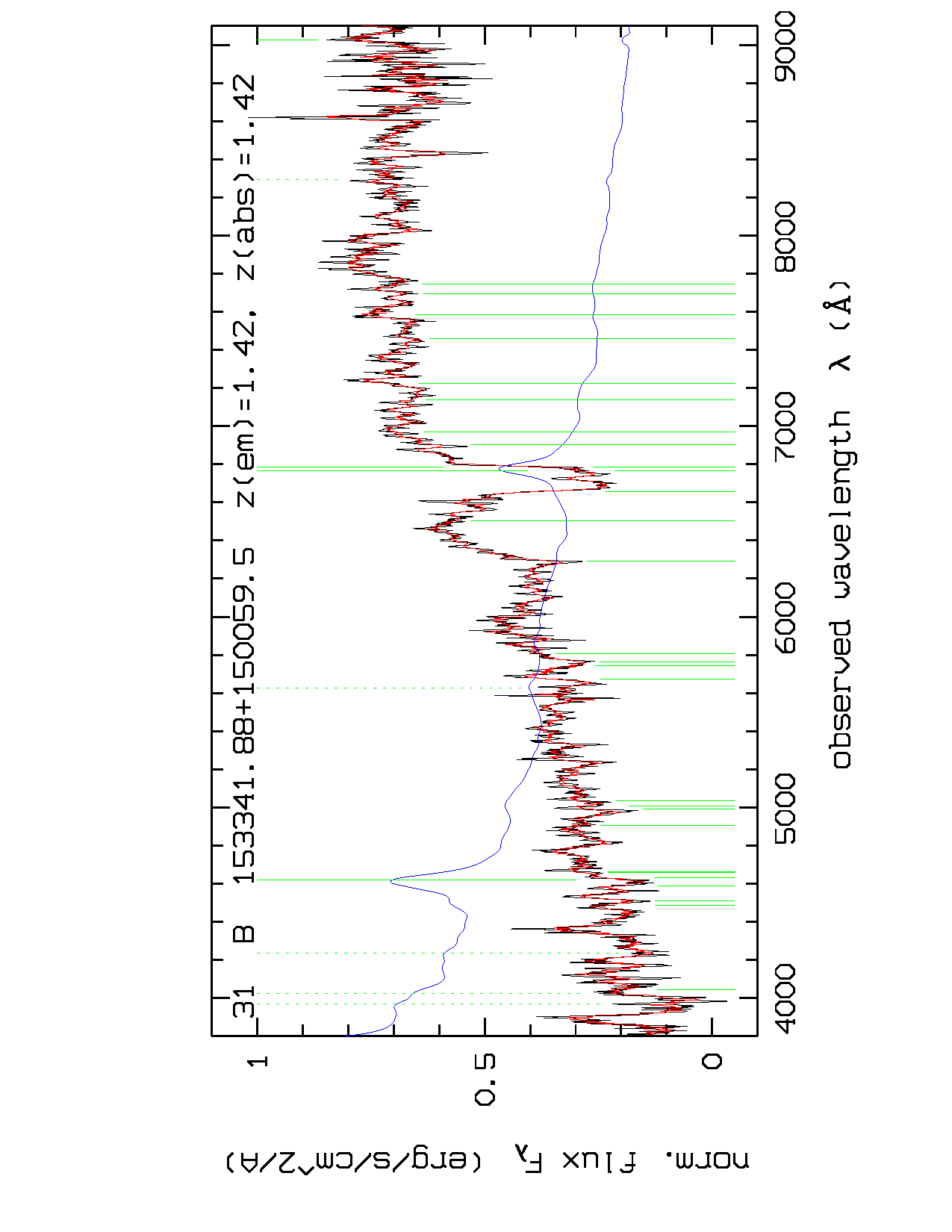}\hfill \=
\includegraphics[viewport=125 0 570 790,angle=270,width=8.0cm,clip]{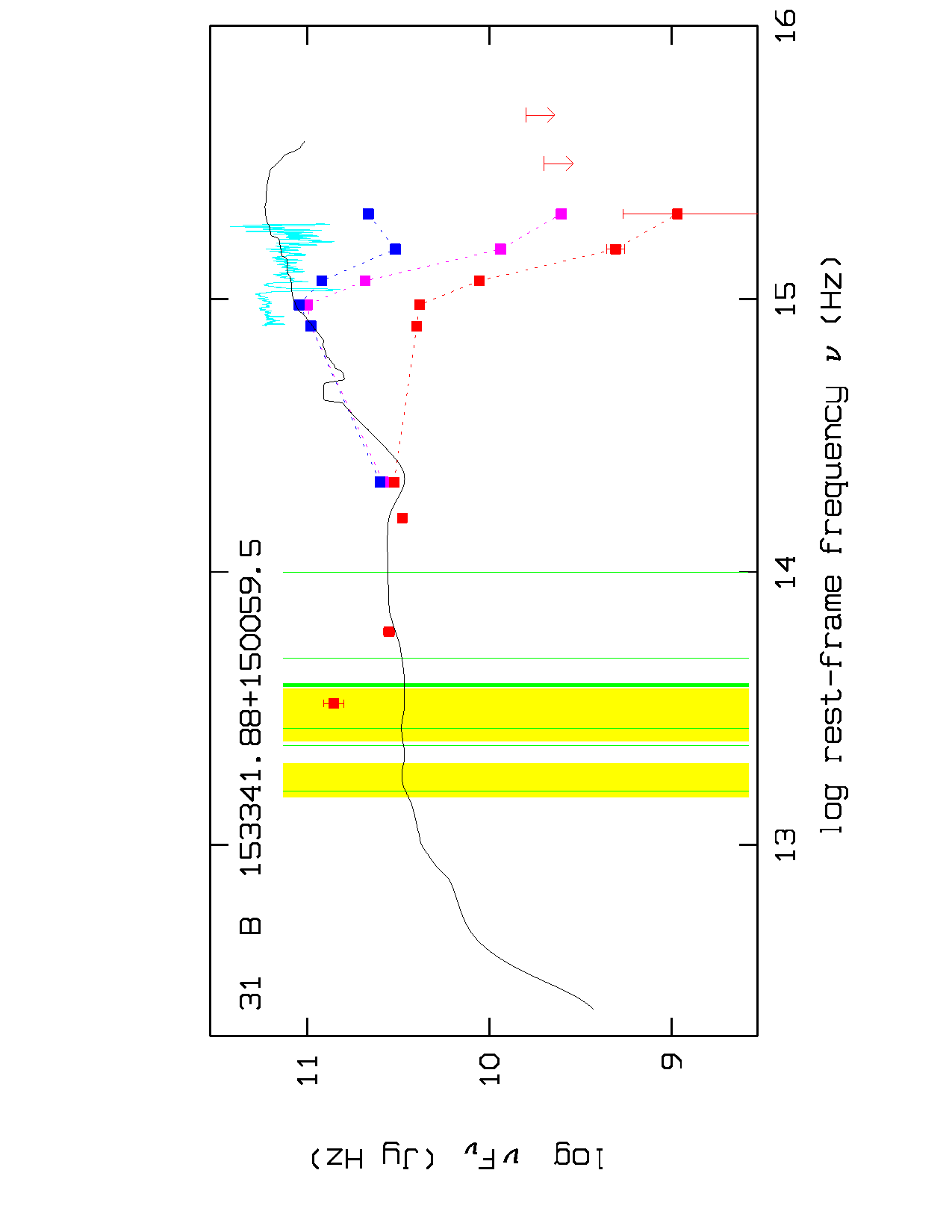}\hfill \\
\includegraphics[viewport=125 0 570 790,angle=270,width=8.0cm,clip]{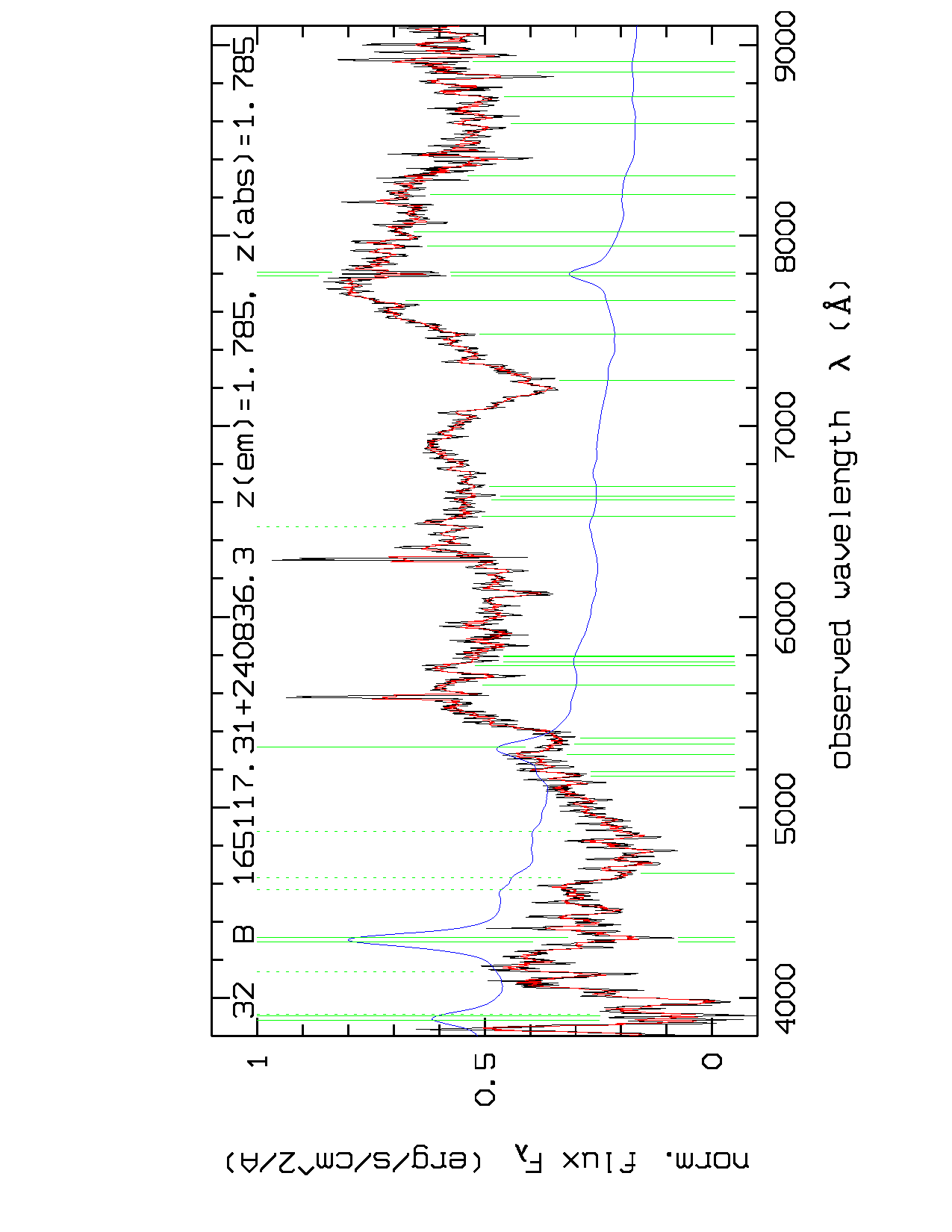}\hfill \=
\includegraphics[viewport=125 0 570 790,angle=270,width=8.0cm,clip]{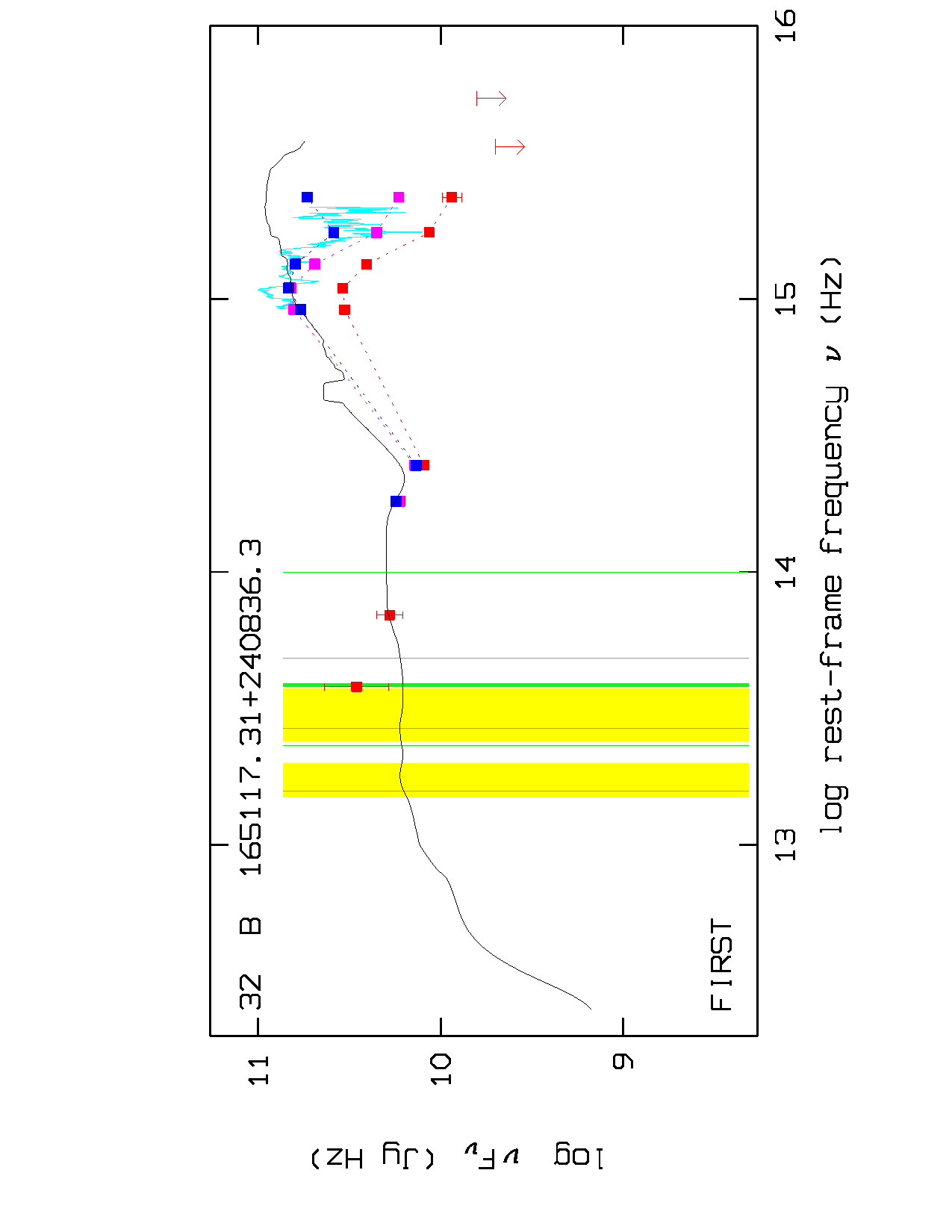}\hfill \\
\includegraphics[viewport=125 0 570 790,angle=270,width=8.0cm,clip]{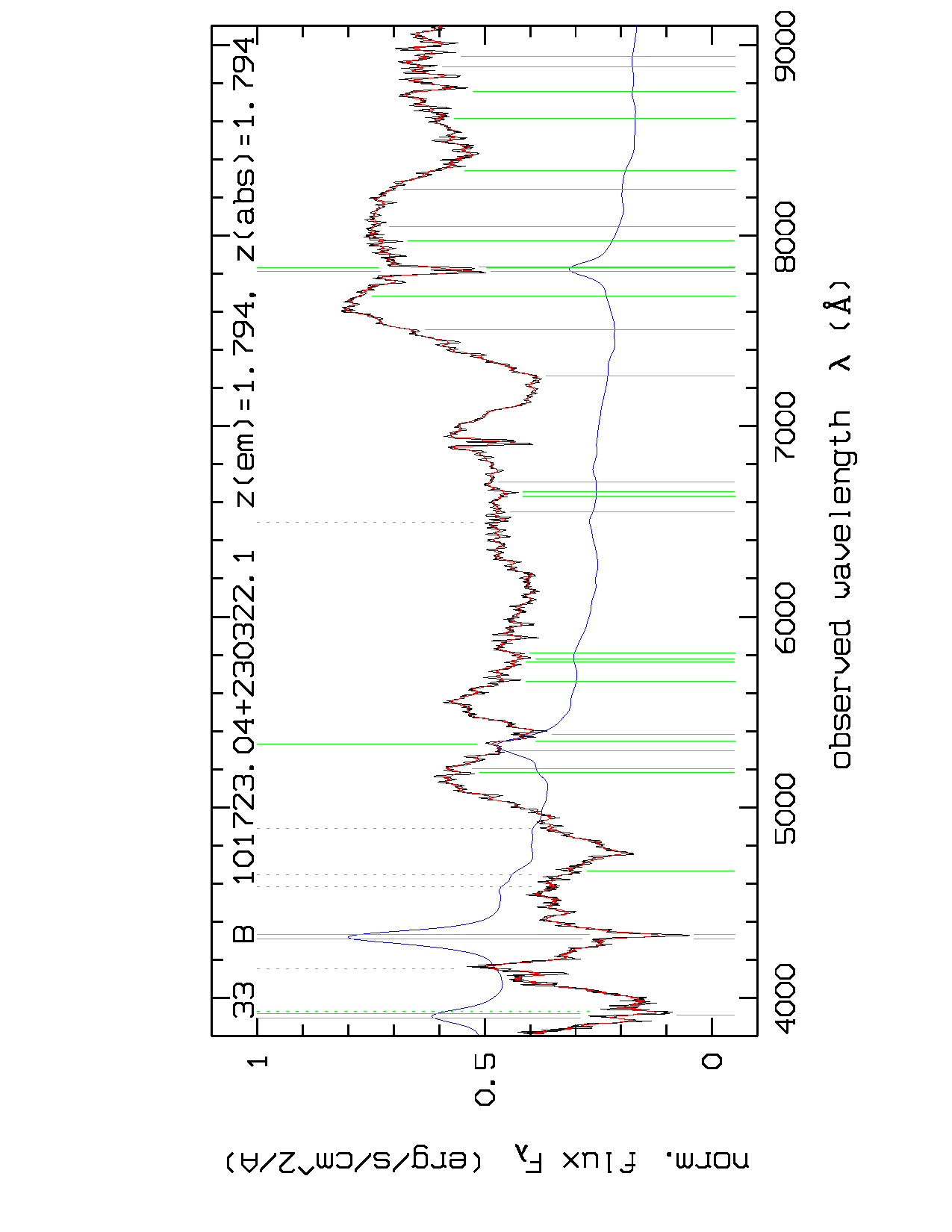}\hfill \=
\includegraphics[viewport=125 0 570 790,angle=270,width=8.0cm,clip]{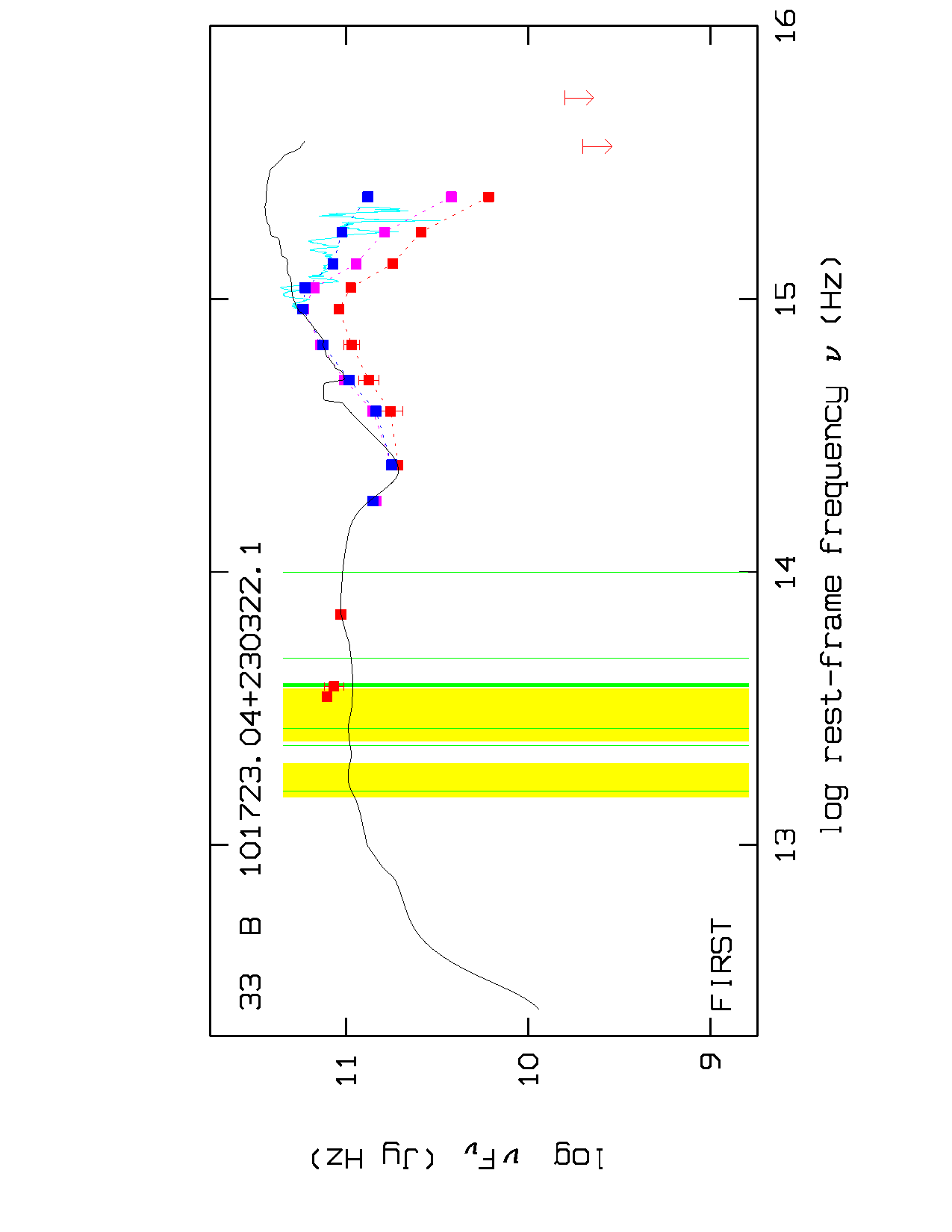}\hfill
\end{tabbing}
\caption{Sample B - continued (1).}
\end{figure*}\clearpage

\begin{figure*}[h]
\begin{tabbing}
\includegraphics[viewport=125 0 570 790,angle=270,width=8.0cm,clip]{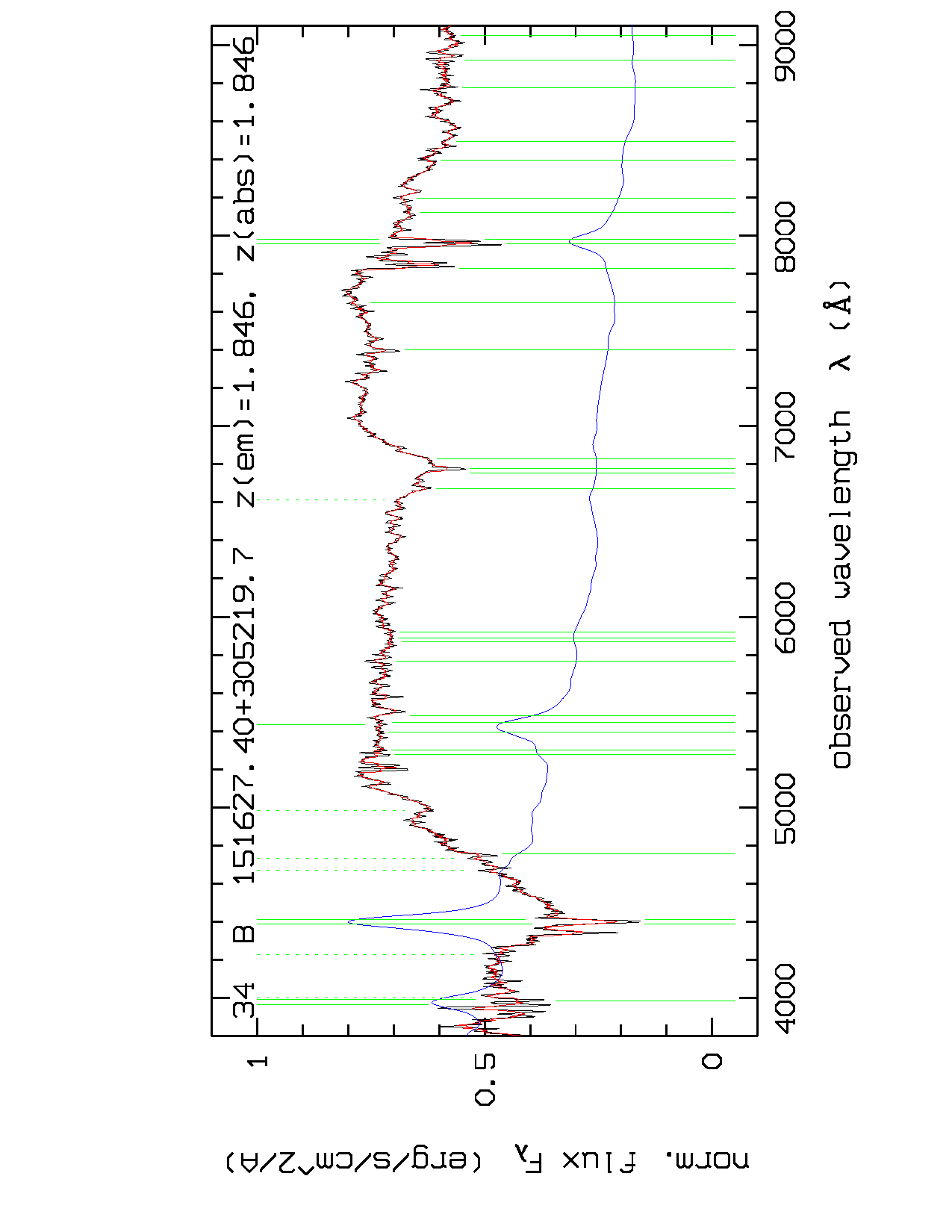}\hfill \=
\includegraphics[viewport=125 0 570 790,angle=270,width=8.0cm,clip]{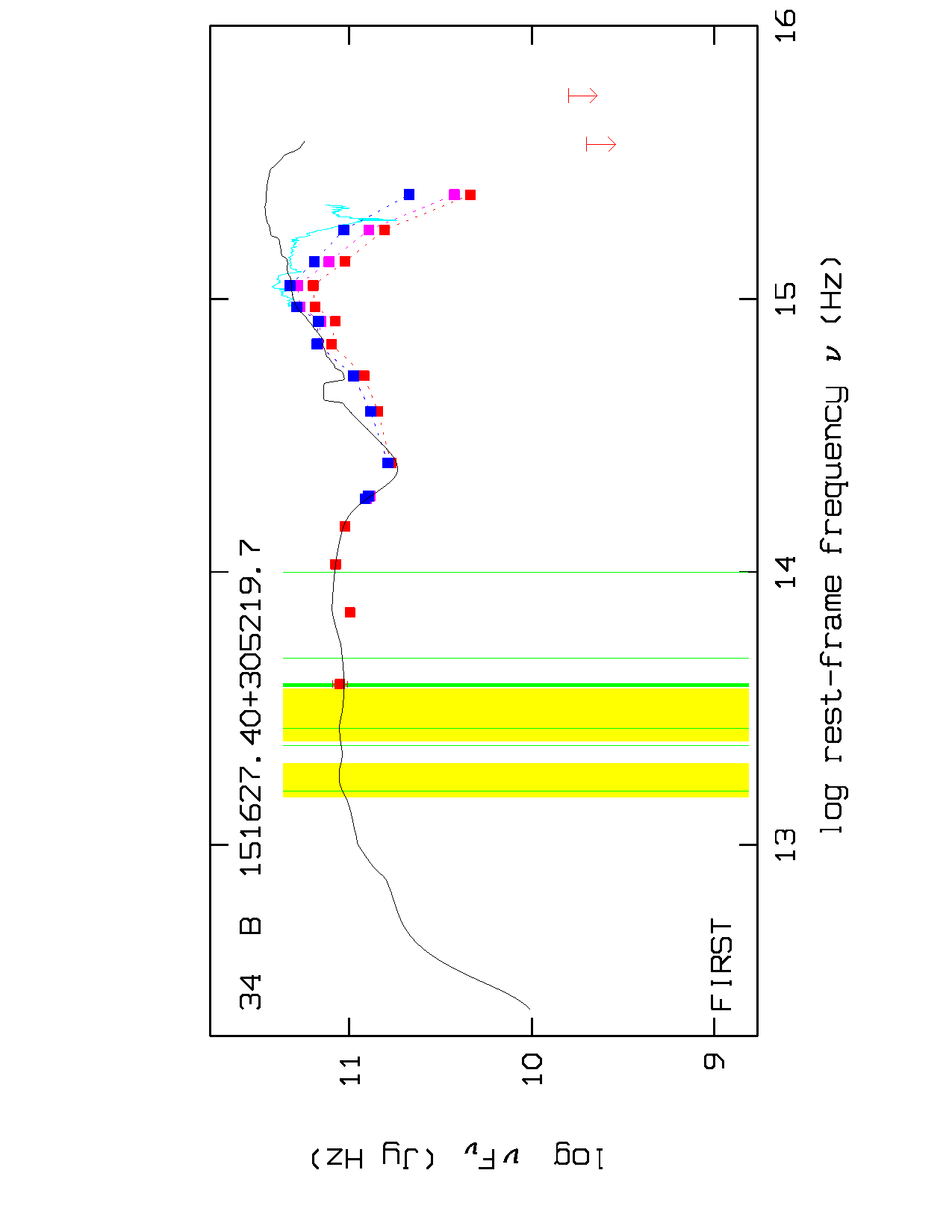}\hfill  \\
\includegraphics[viewport=125 0 570 790,angle=270,width=8.0cm,clip]{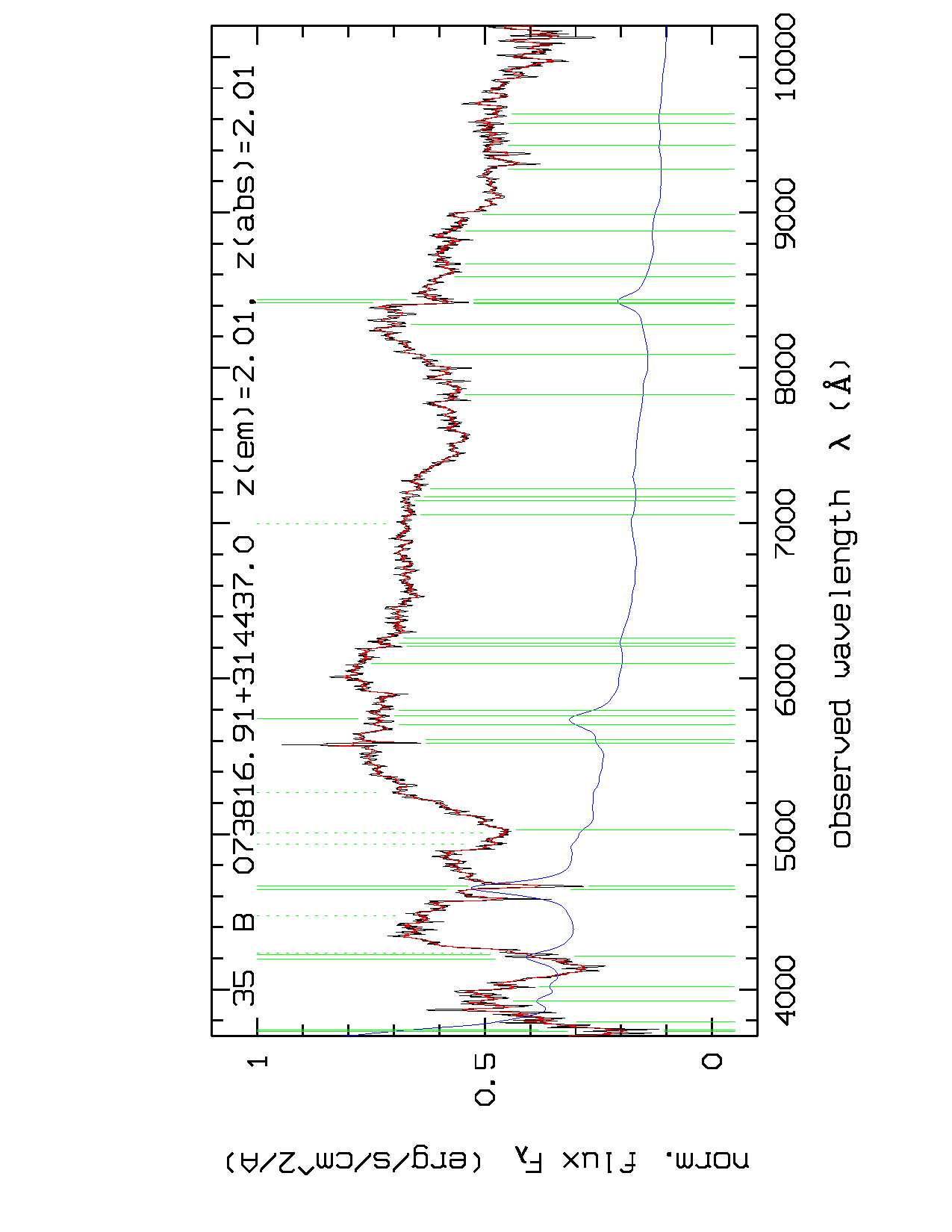}\hfill \=
\includegraphics[viewport=125 0 570 790,angle=270,width=8.0cm,clip]{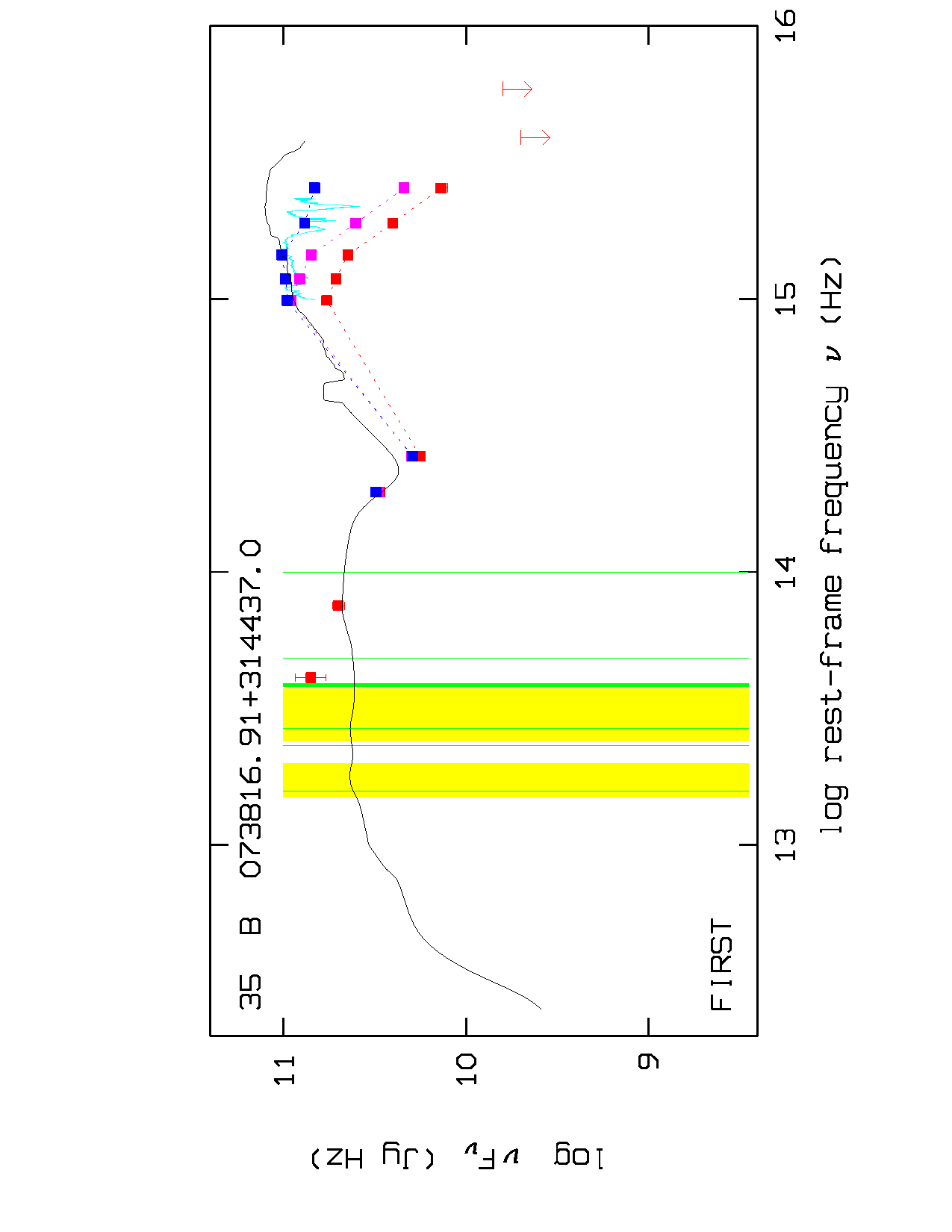}\hfill \\
\includegraphics[viewport=125 0 570 790,angle=270,width=8.0cm,clip]{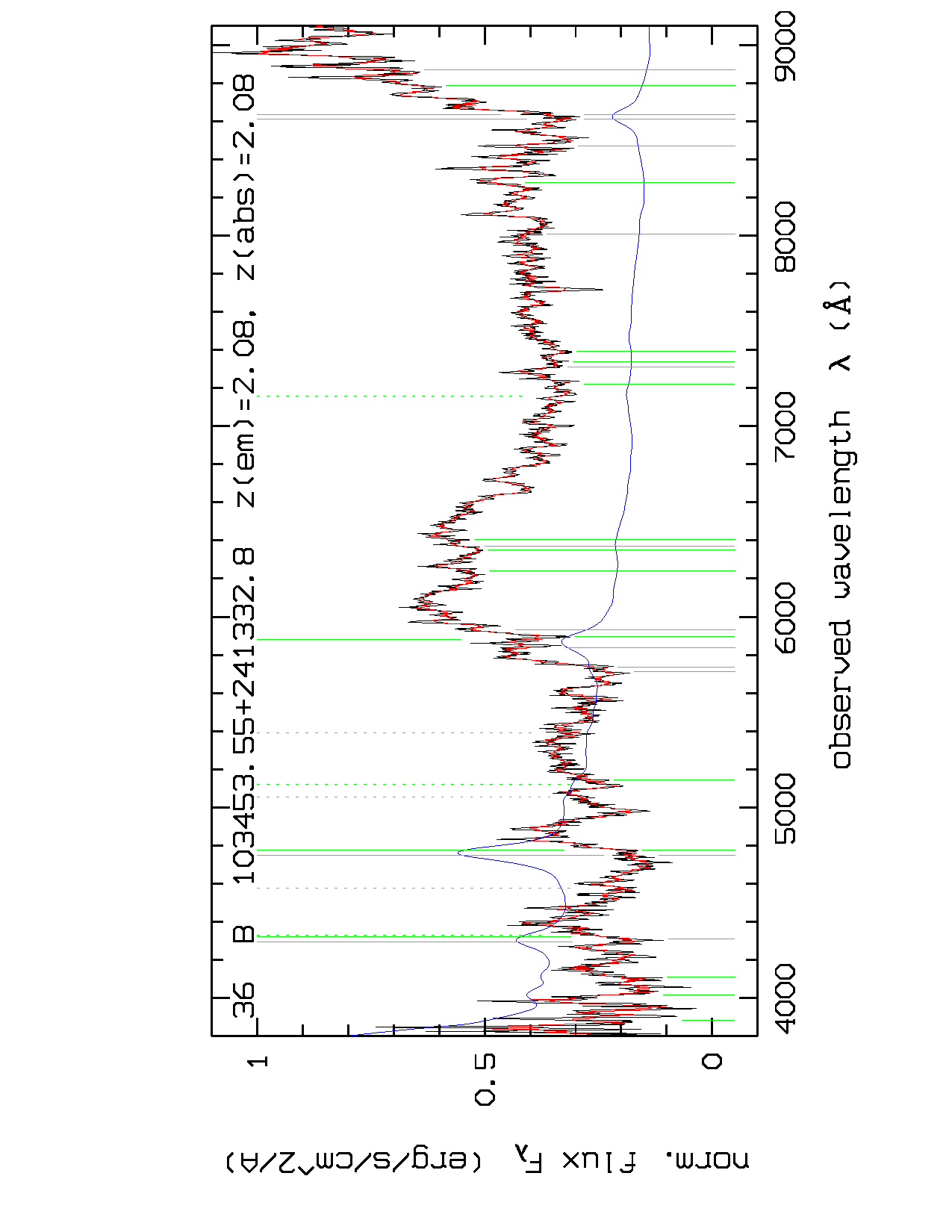}\hfill \=
\includegraphics[viewport=125 0 570 790,angle=270,width=8.0cm,clip]{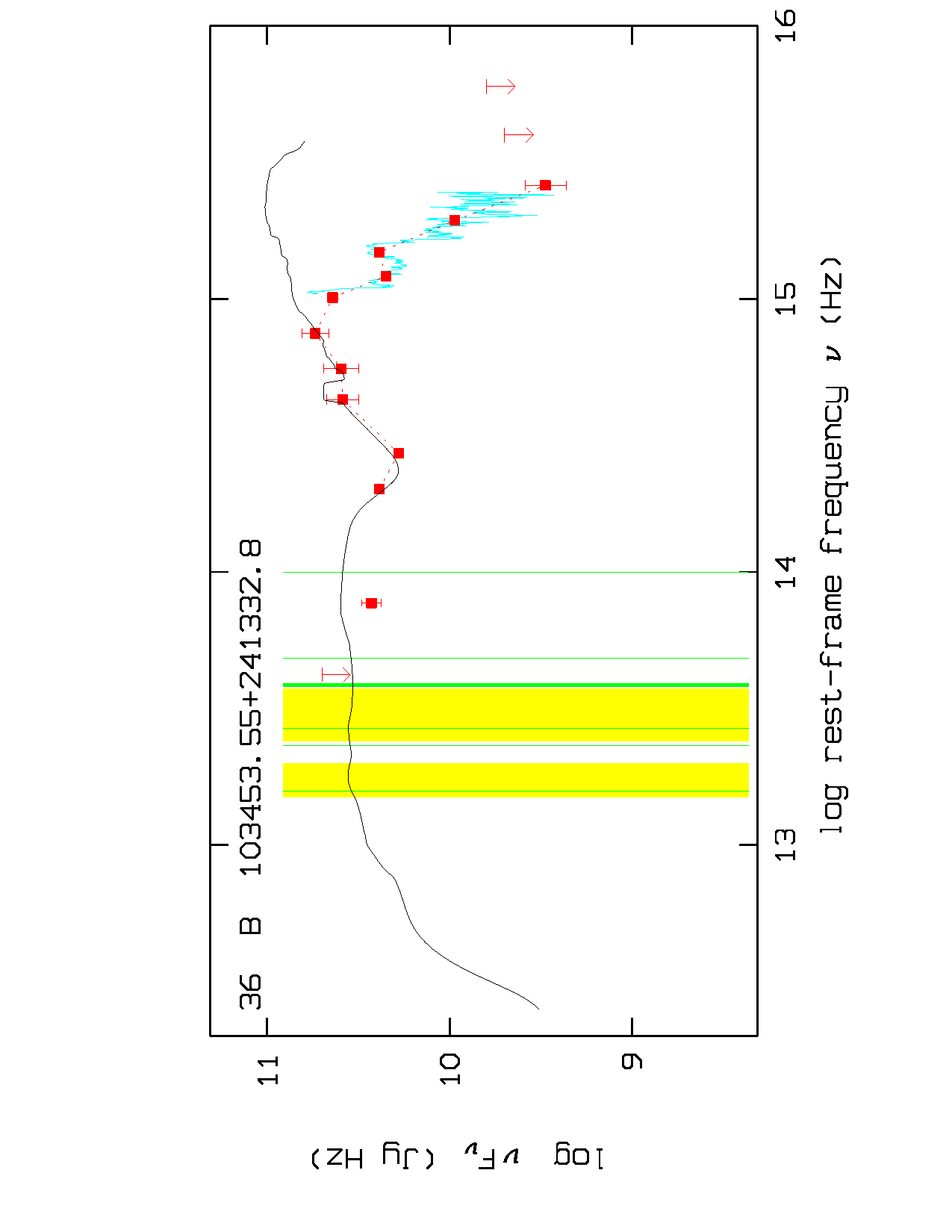}\hfill \\
\includegraphics[viewport=125 0 570 790,angle=270,width=8.0cm,clip]{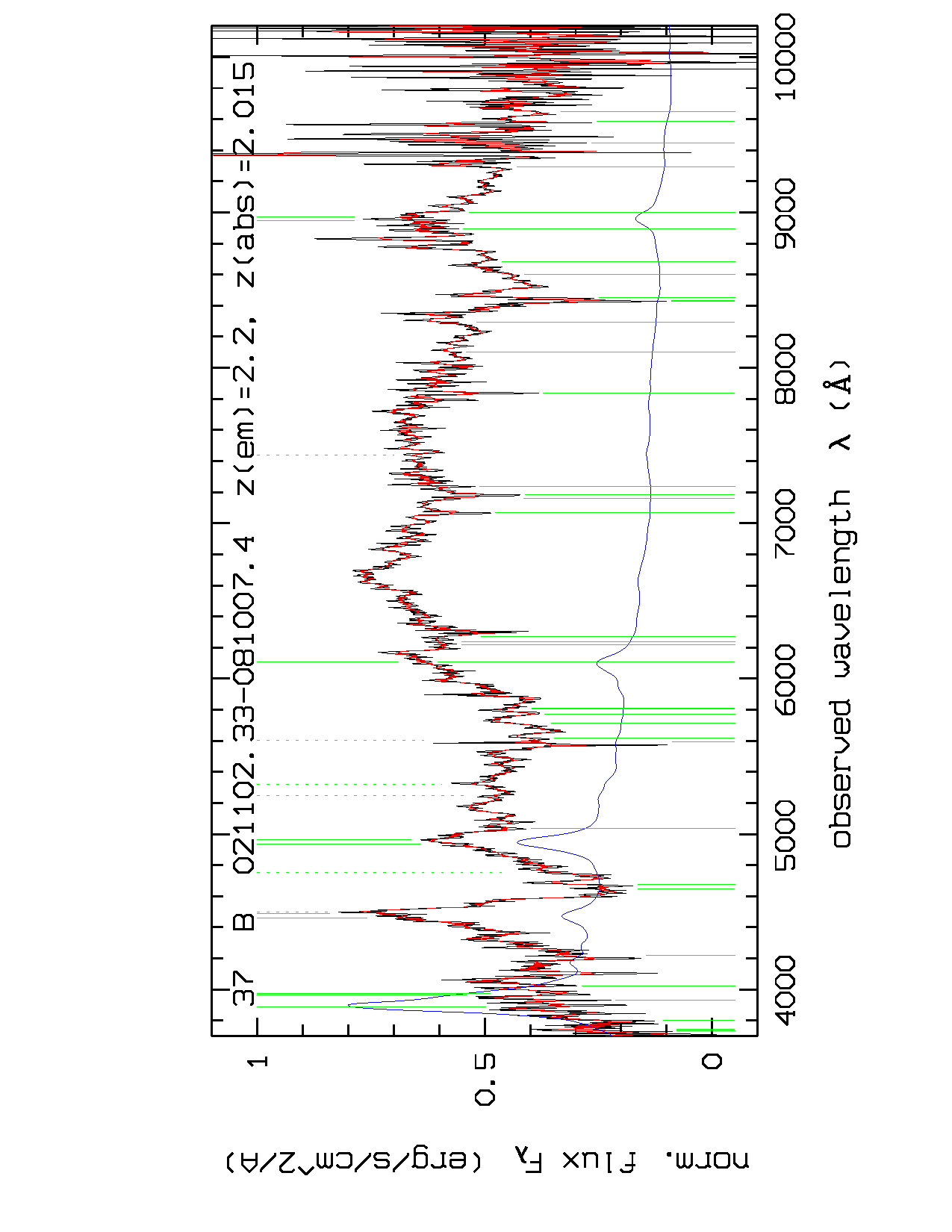}\hfill \=
\includegraphics[viewport=125 0 570 790,angle=270,width=8.0cm,clip]{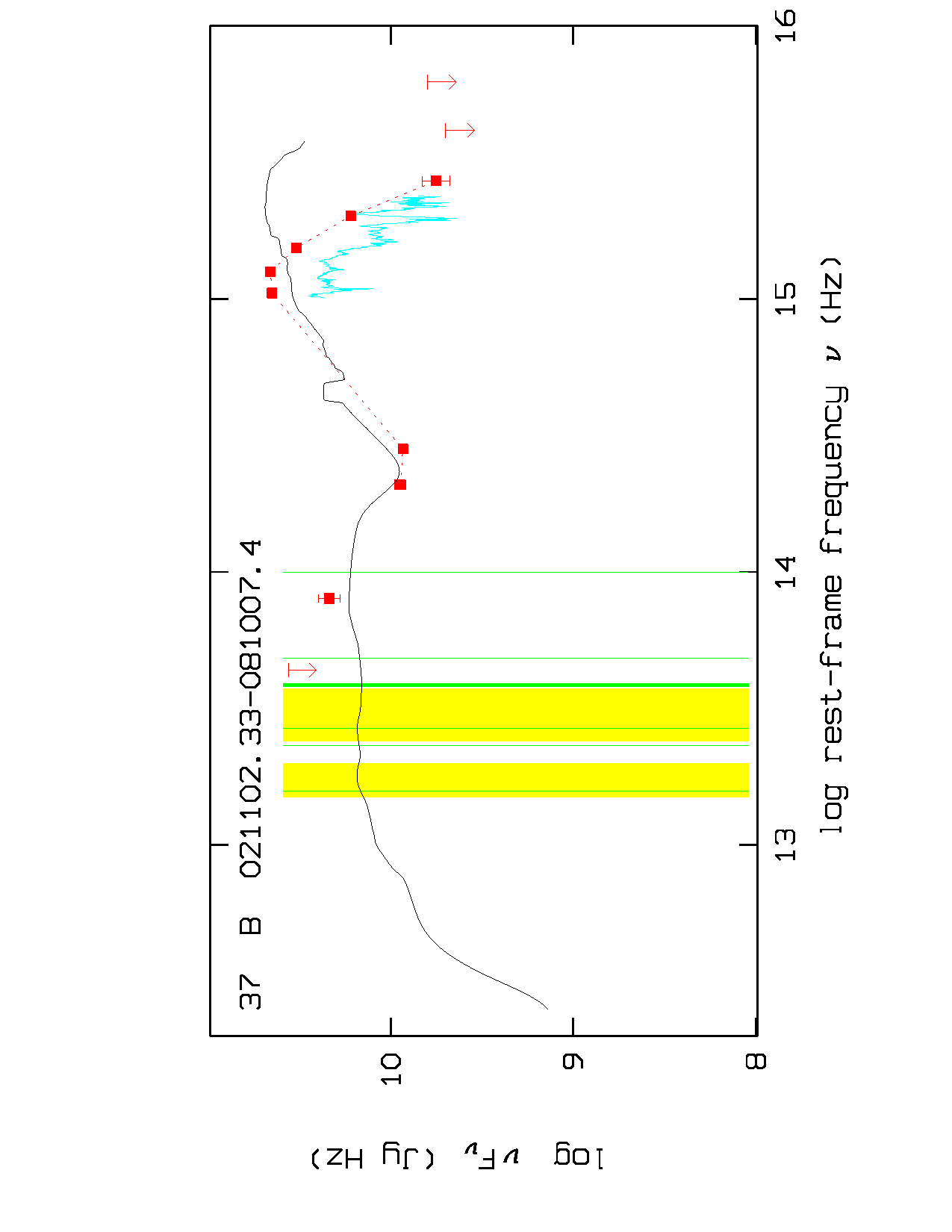}\hfill \\
\includegraphics[viewport=125 0 570 790,angle=270,width=8.0cm,clip]{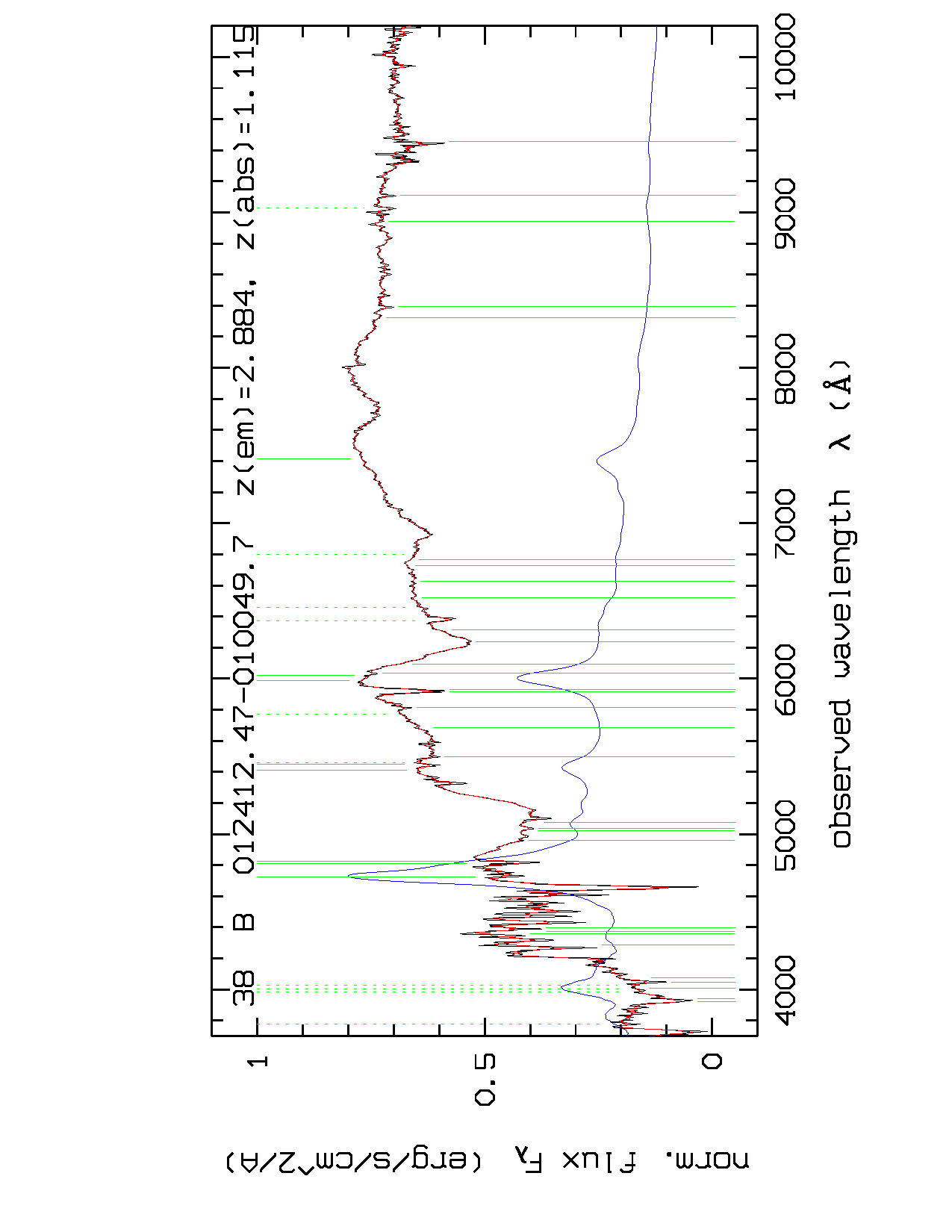}\hfill \=
\includegraphics[viewport=125 0 570 790,angle=270,width=8.0cm,clip]{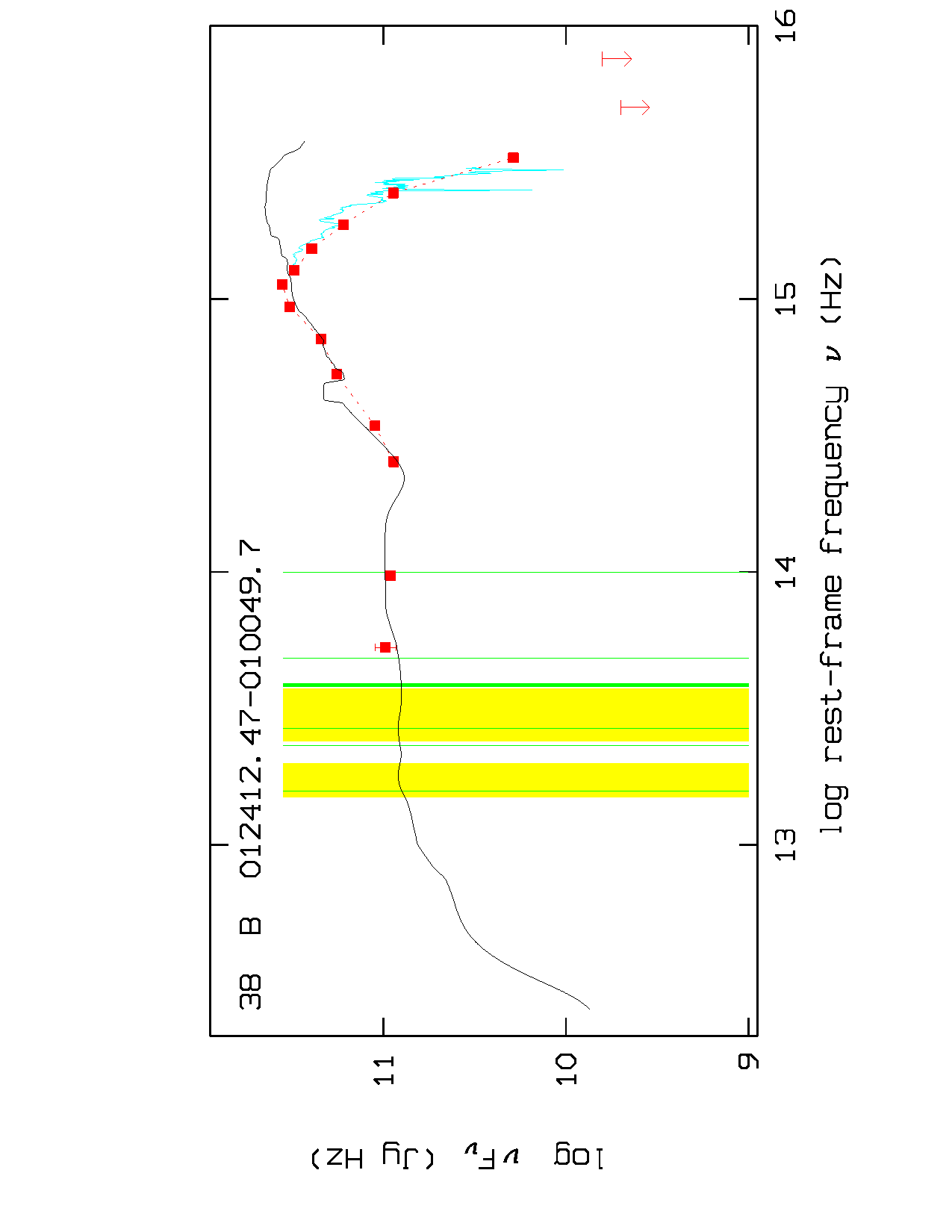}\hfill
\end{tabbing}
\caption{Sample B - continued (2).}
\end{figure*}\clearpage

\begin{figure*}[h]
\begin{tabbing}
\includegraphics[viewport=125 0 570 790,angle=270,width=8.0cm,clip]{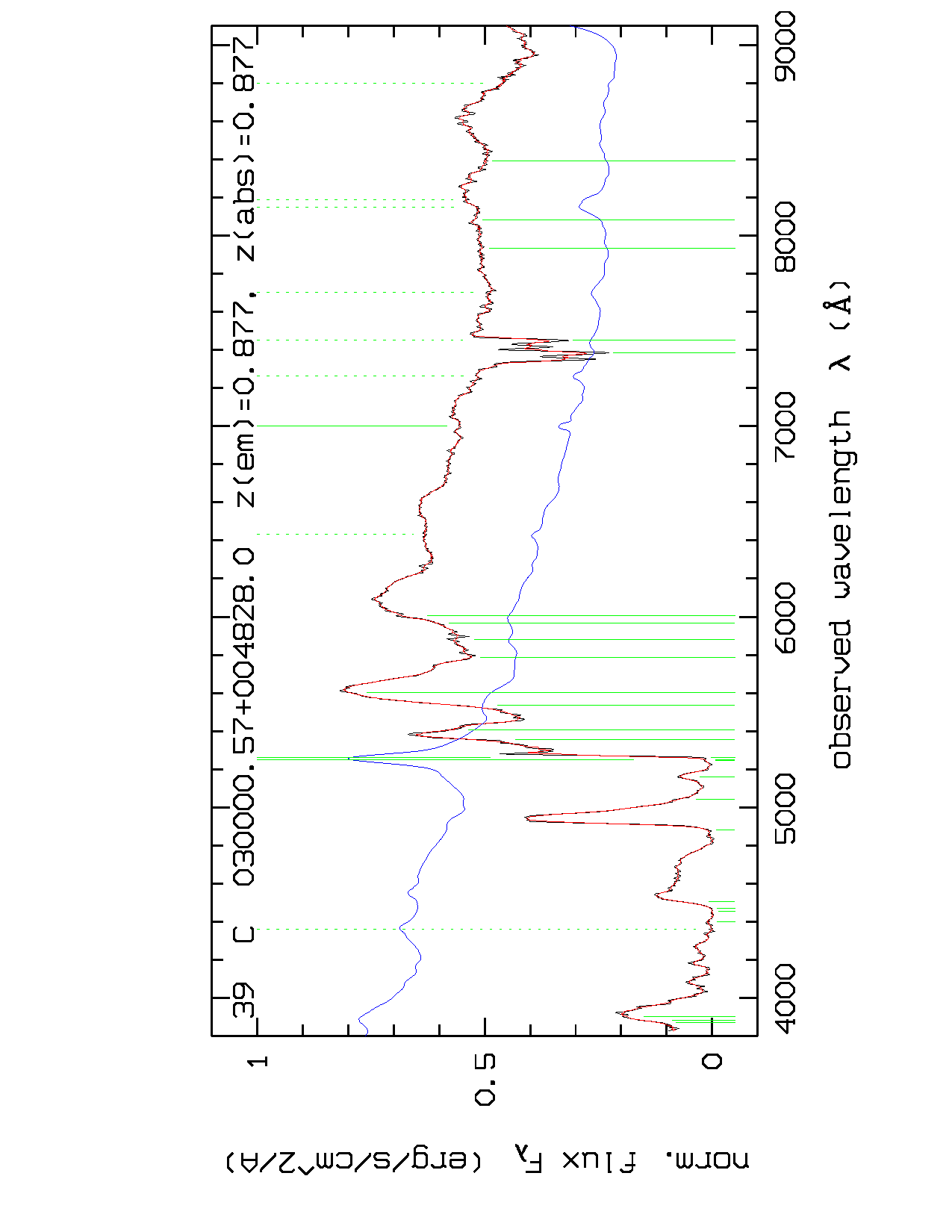}\hfill \=
\includegraphics[viewport=125 0 570 790,angle=270,width=8.0cm,clip]{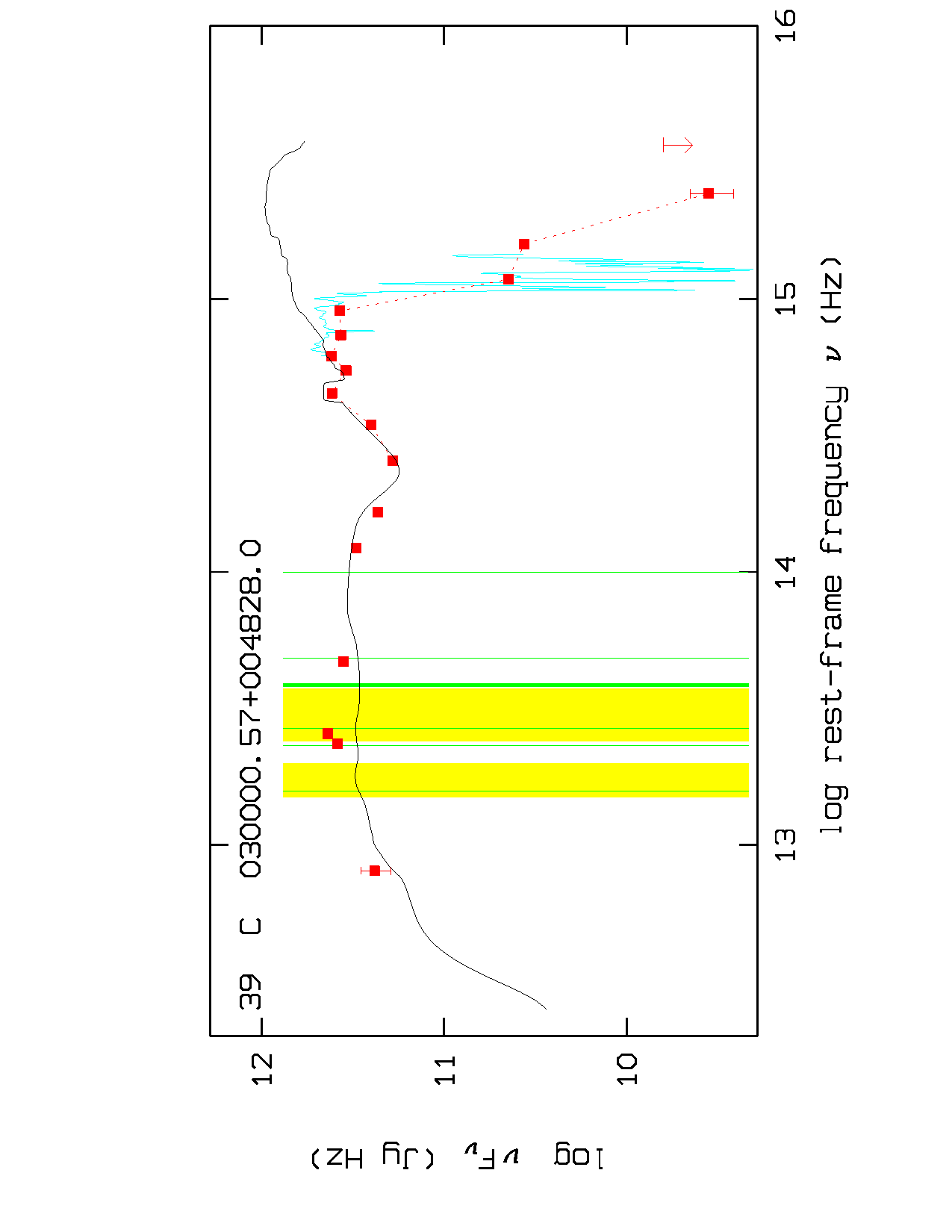}\hfill \\
\includegraphics[viewport=125 0 570 790,angle=270,width=8.0cm,clip]{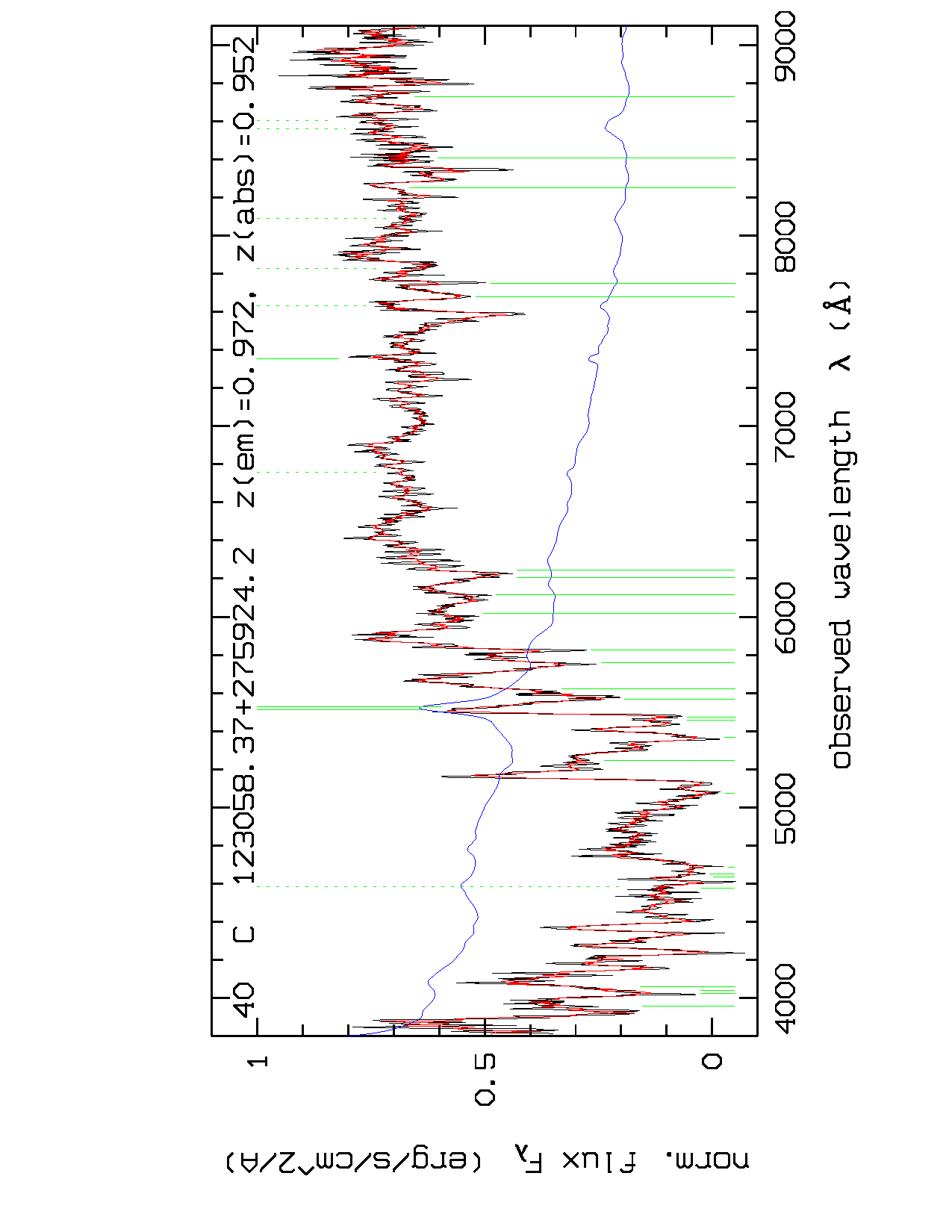}\hfill \=
\includegraphics[viewport=125 0 570 790,angle=270,width=8.0cm,clip]{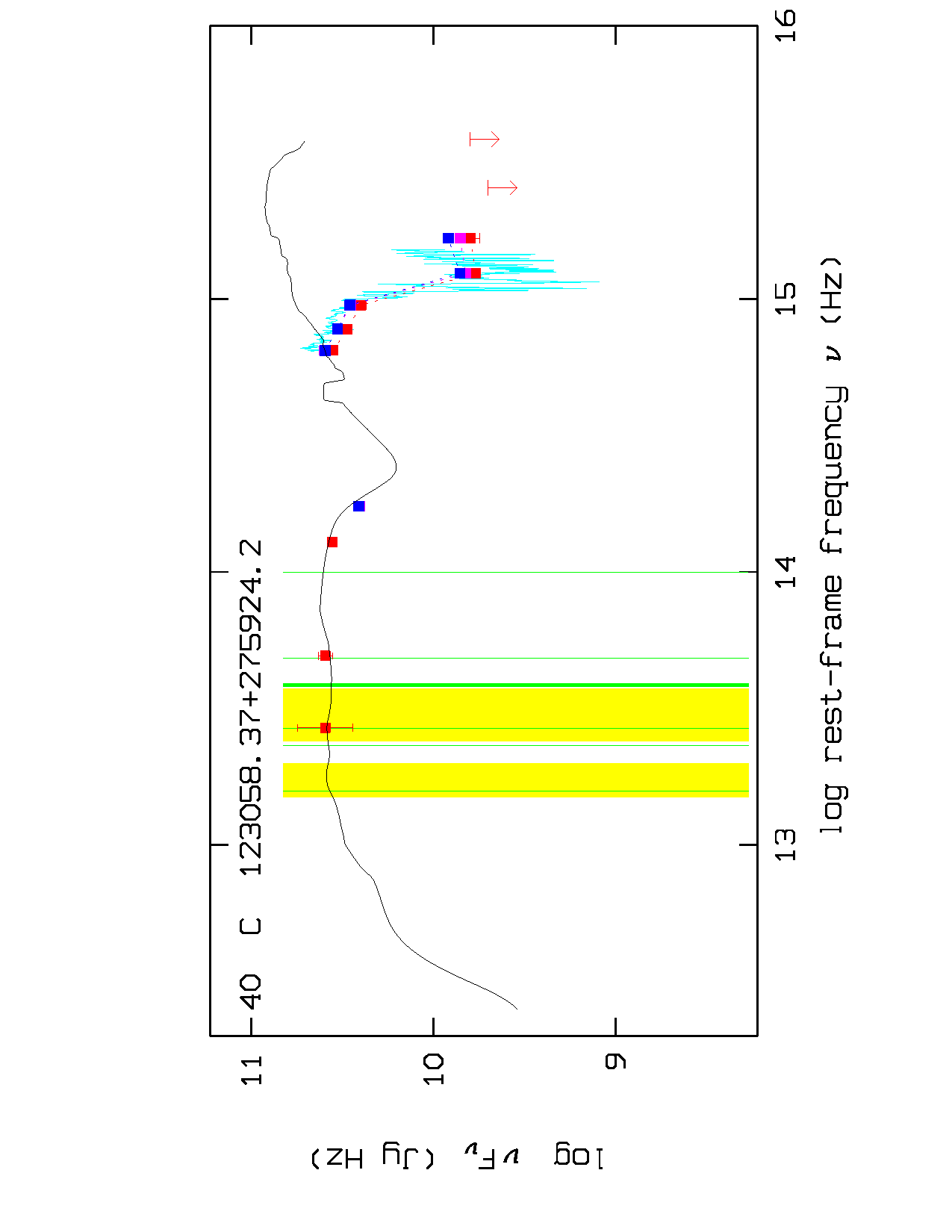}\hfill \\
\includegraphics[viewport=125 0 570 790,angle=270,width=8.0cm,clip]{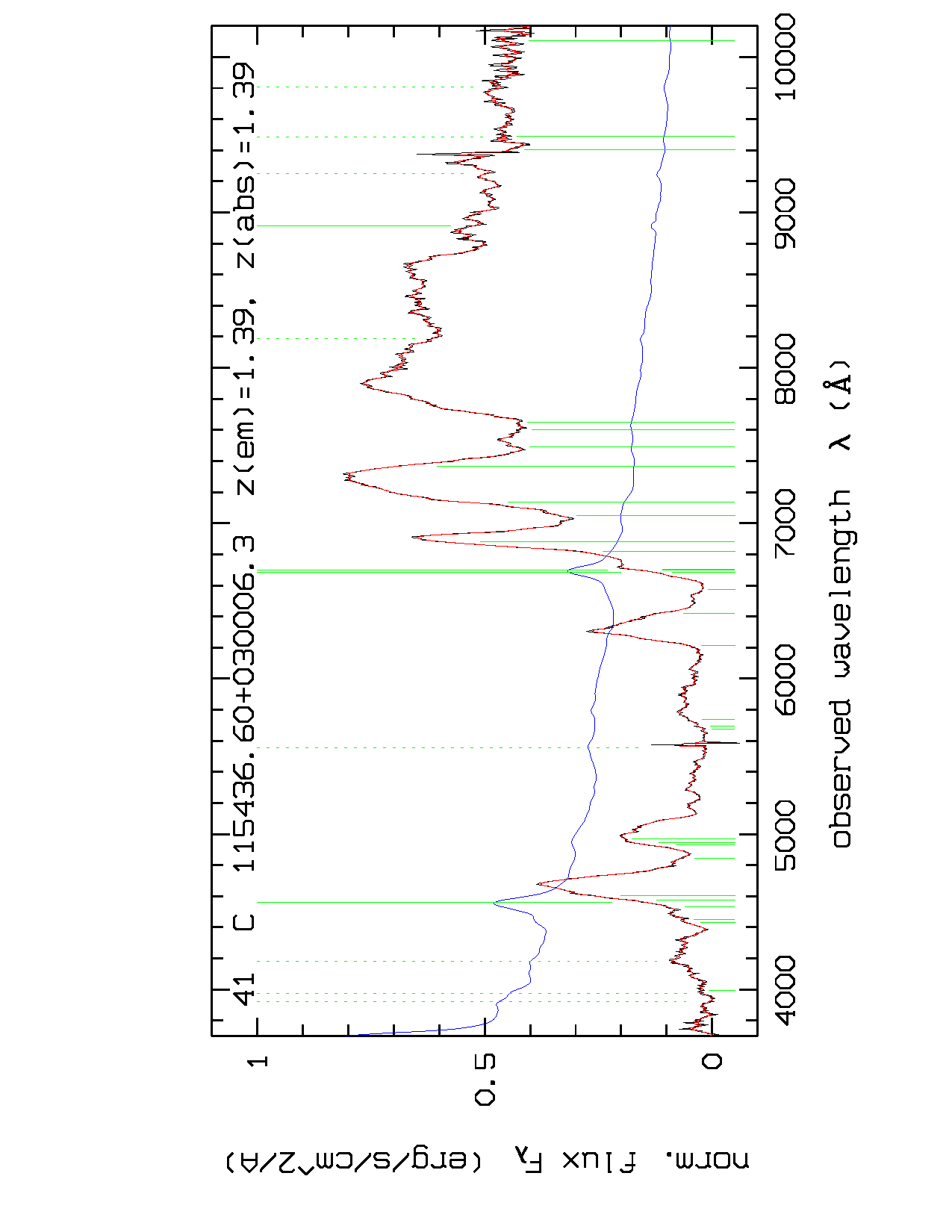}\hfill \=
\includegraphics[viewport=125 0 570 790,angle=270,width=8.0cm,clip]{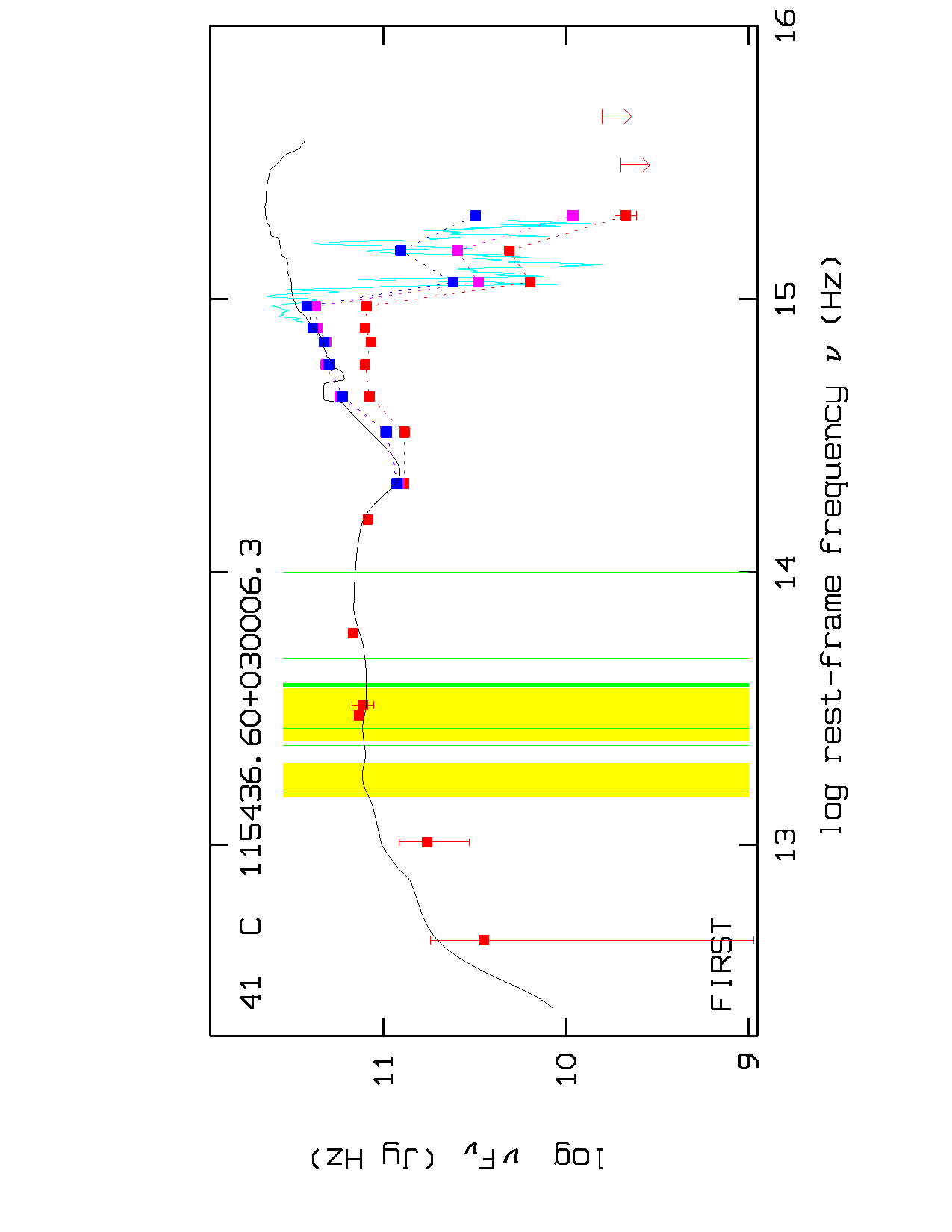}\hfill \\
\includegraphics[viewport=125 0 570 790,angle=270,width=8.0cm,clip]{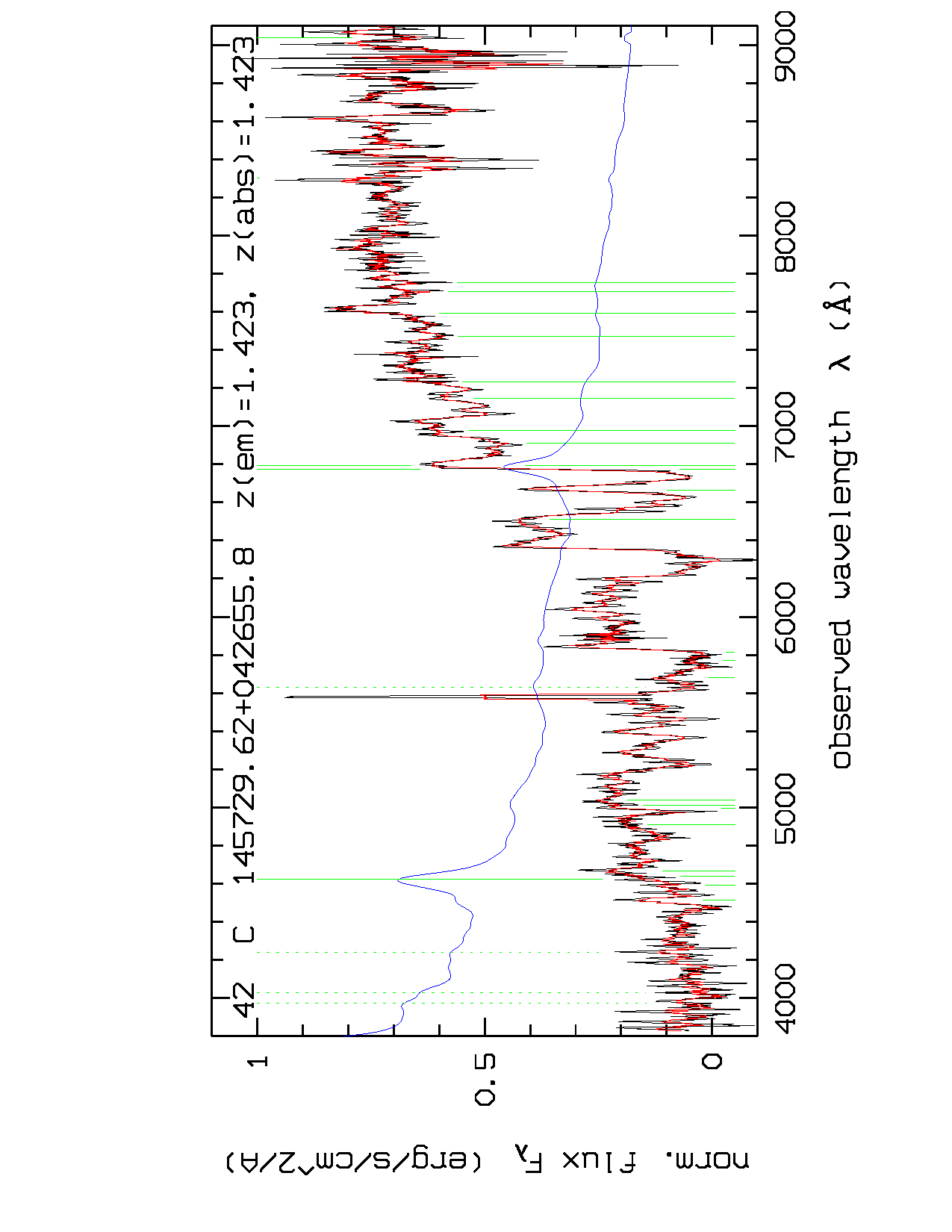}\hfill \=
\includegraphics[viewport=125 0 570 790,angle=270,width=8.0cm,clip]{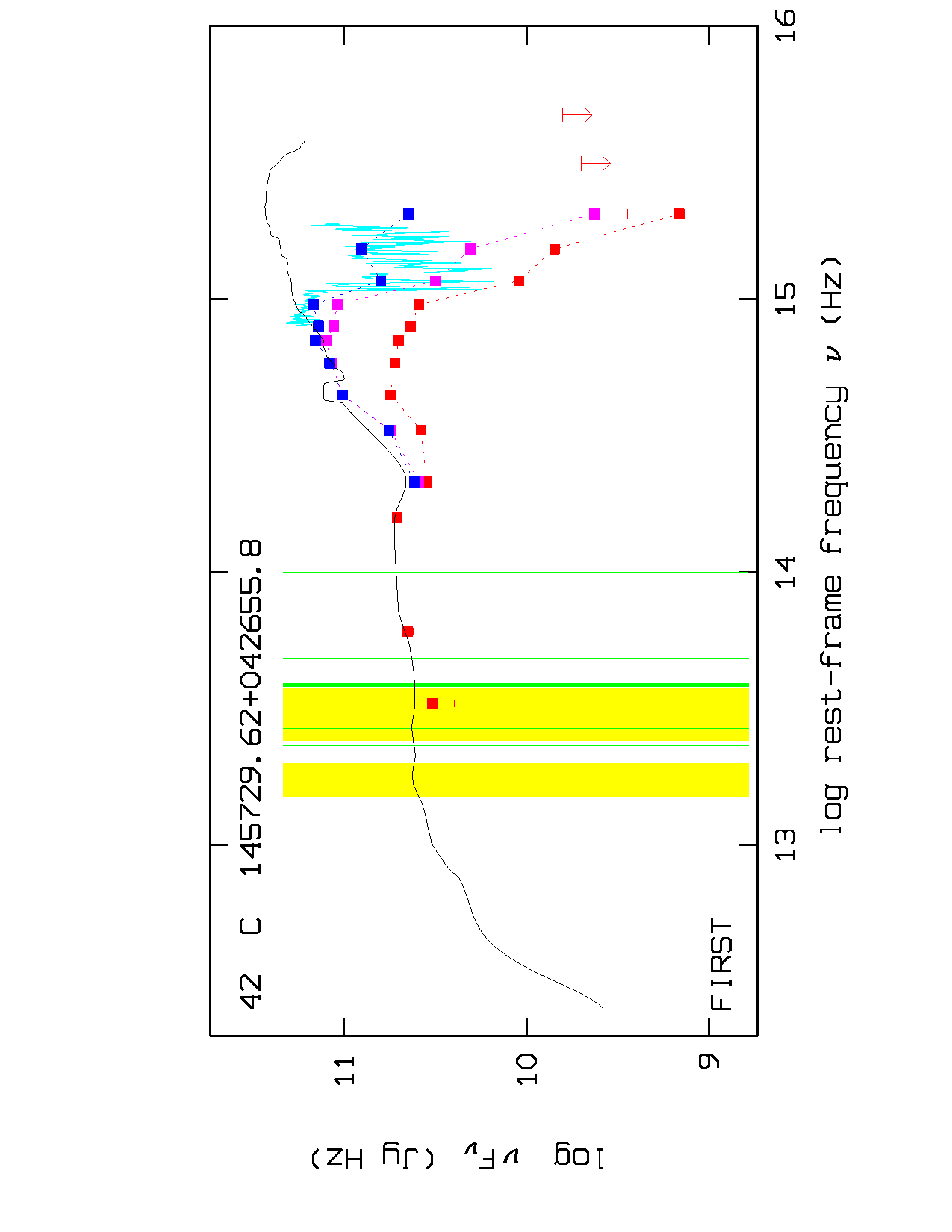}\hfill \\
\includegraphics[viewport=125 0 570 790,angle=270,width=8.0cm,clip]{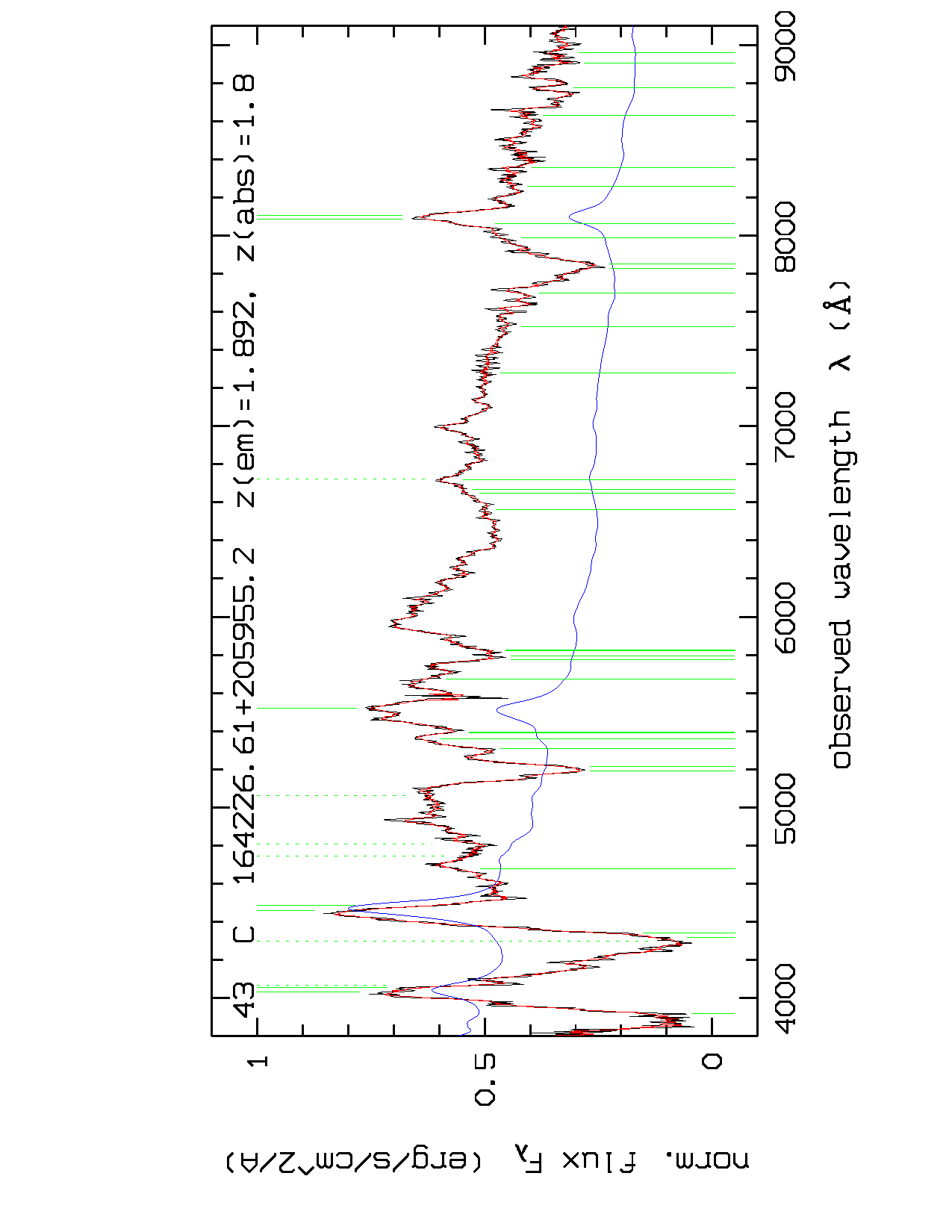}\hfill \=
\includegraphics[viewport=125 0 570 790,angle=270,width=8.0cm,clip]{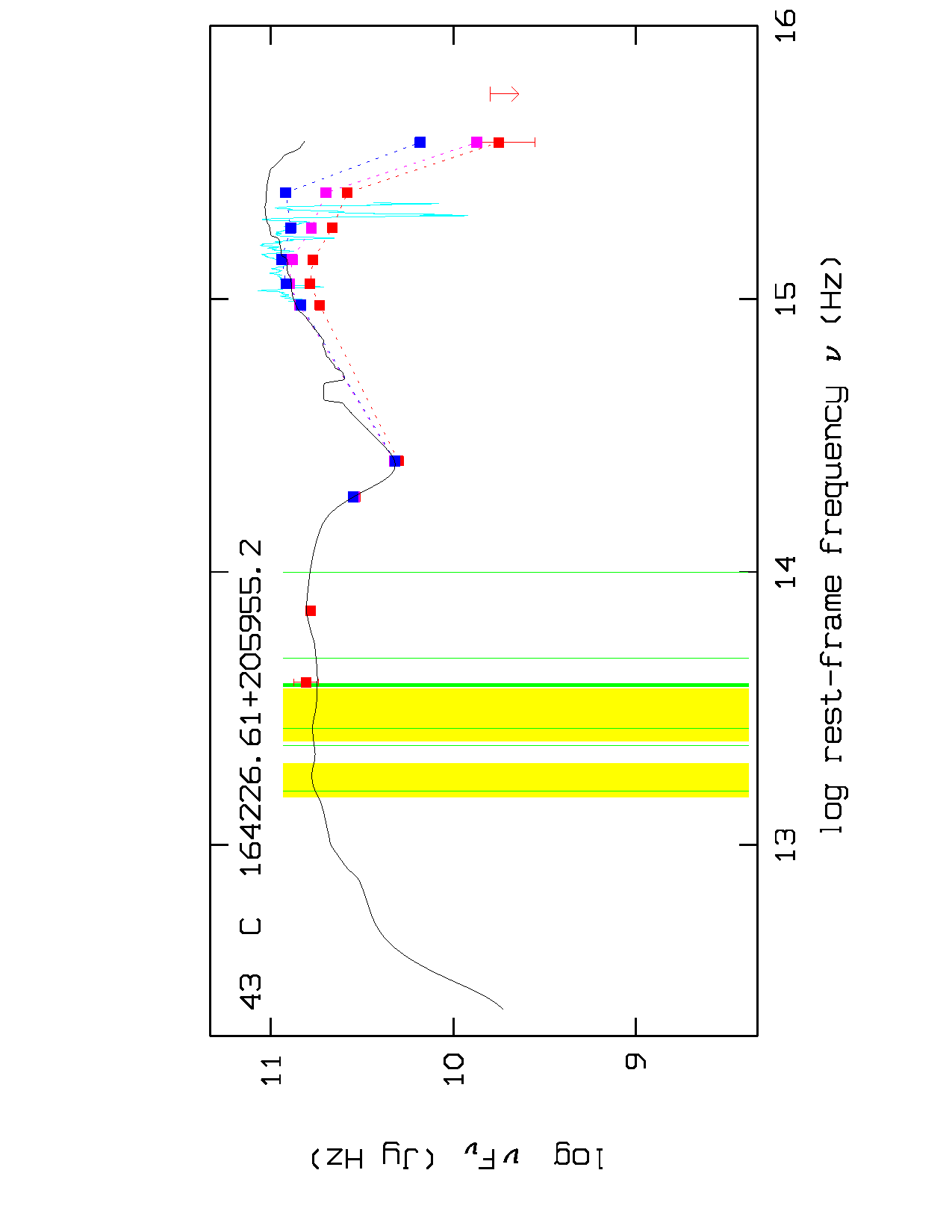}\hfill
\end{tabbing}
\caption{Sample C.}
\end{figure*}\clearpage

\begin{figure*}[h]
\begin{tabbing}
\includegraphics[viewport=125 0 570 790,angle=270,width=8.0cm,clip]{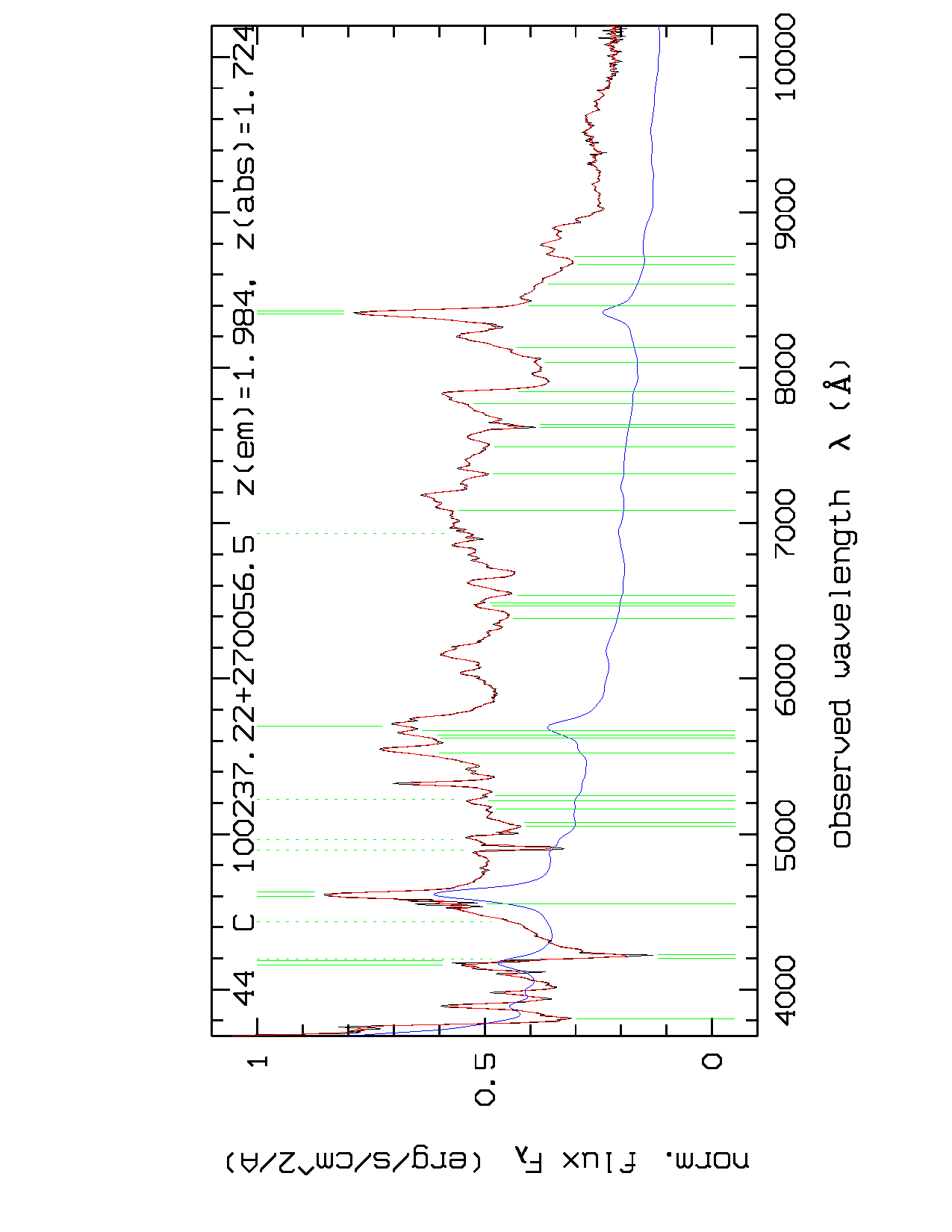}\hfill \=
\includegraphics[viewport=125 0 570 790,angle=270,width=8.0cm,clip]{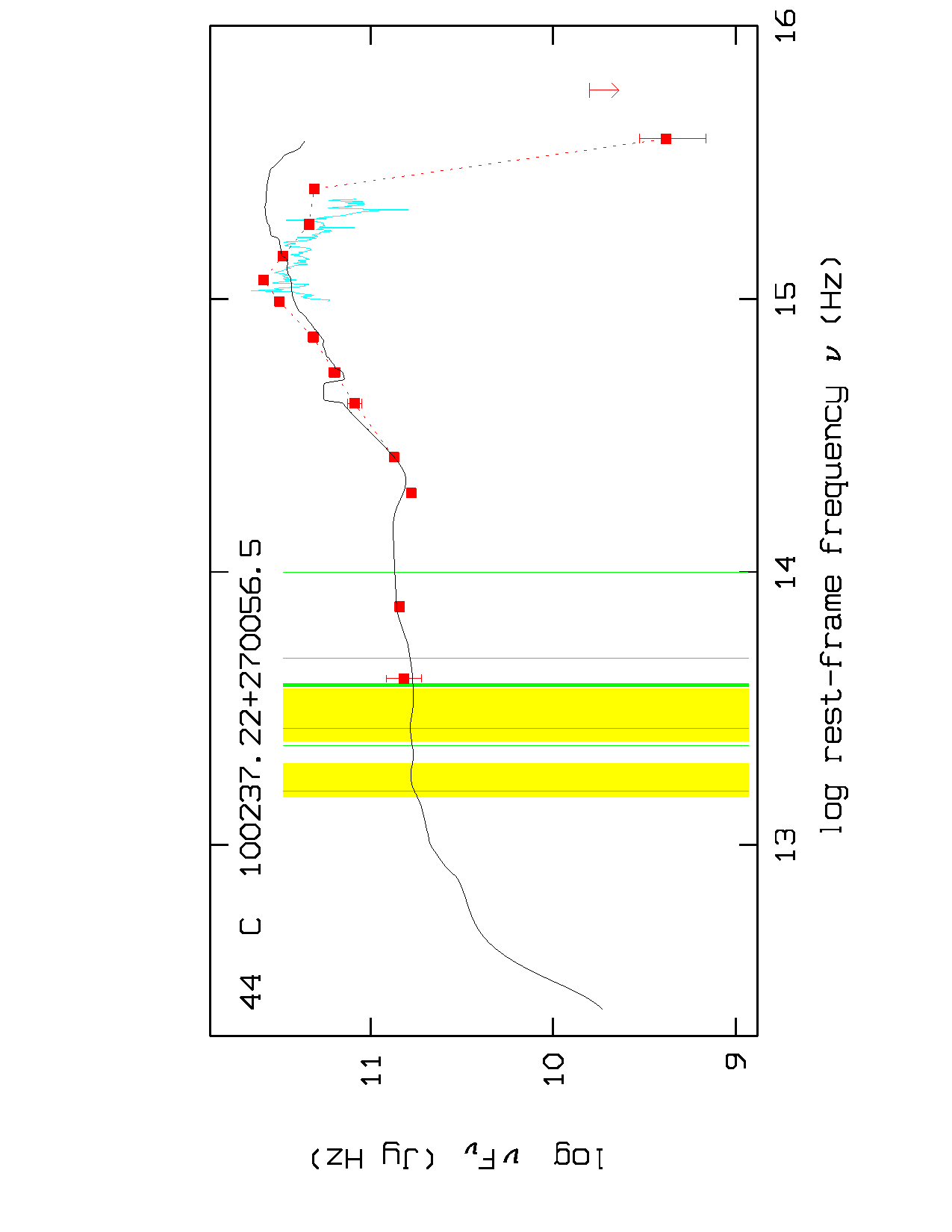}\hfill \\
\includegraphics[viewport=125 0 570 790,angle=270,width=8.0cm,clip]{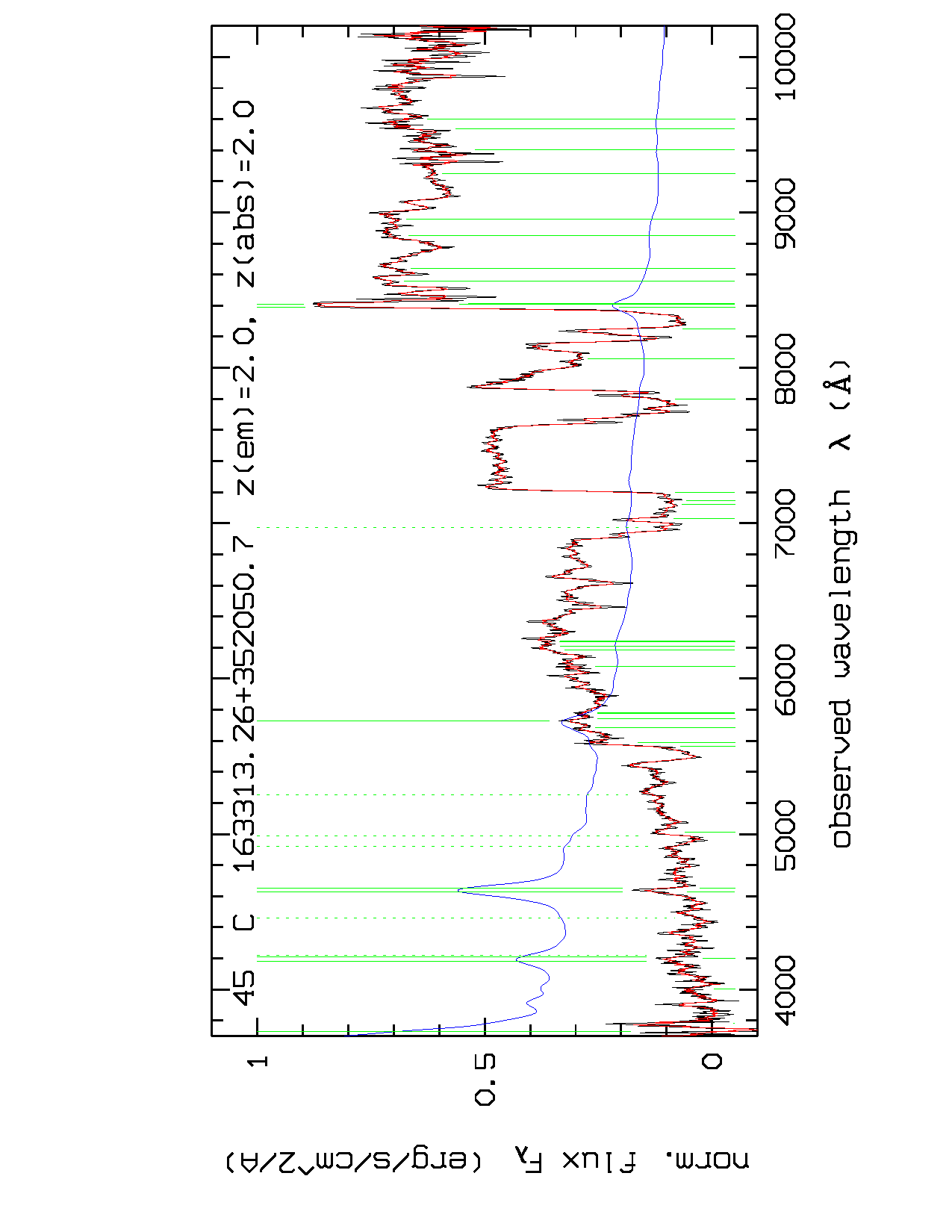}\hfill \=
\includegraphics[viewport=125 0 570 790,angle=270,width=8.0cm,clip]{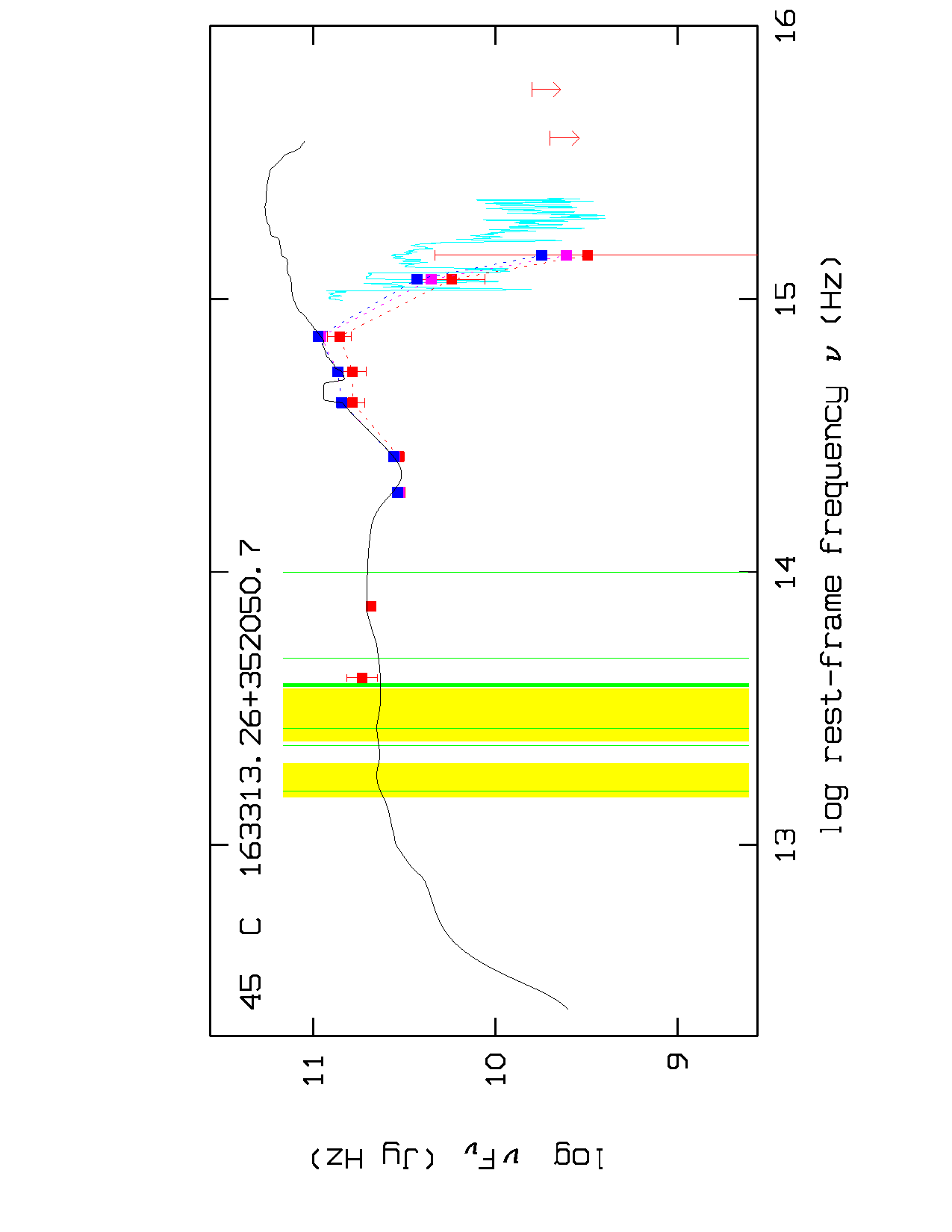}\hfill \\
\includegraphics[viewport=125 0 570 790,angle=270,width=8.0cm,clip]{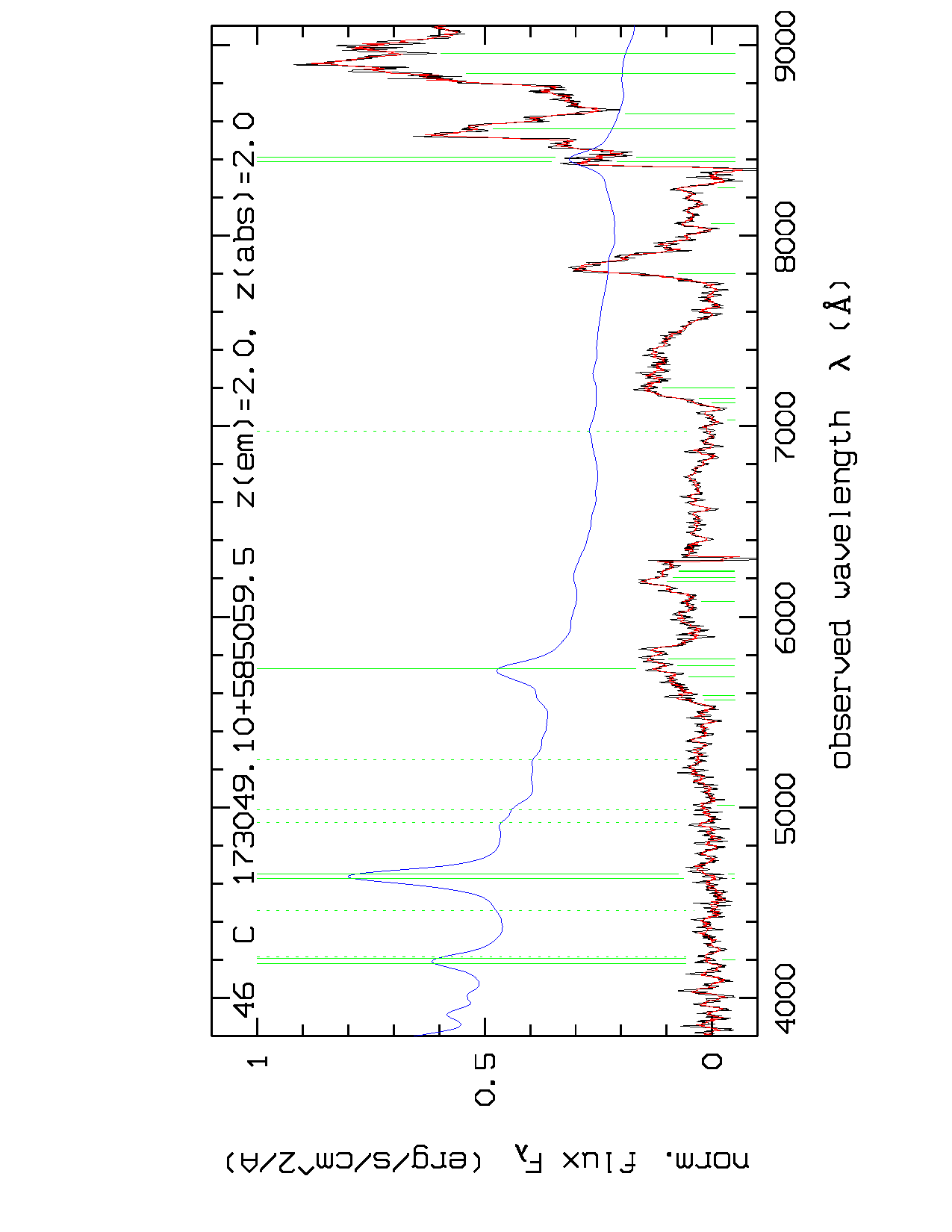}\hfill \=
\includegraphics[viewport=125 0 570 790,angle=270,width=8.0cm,clip]{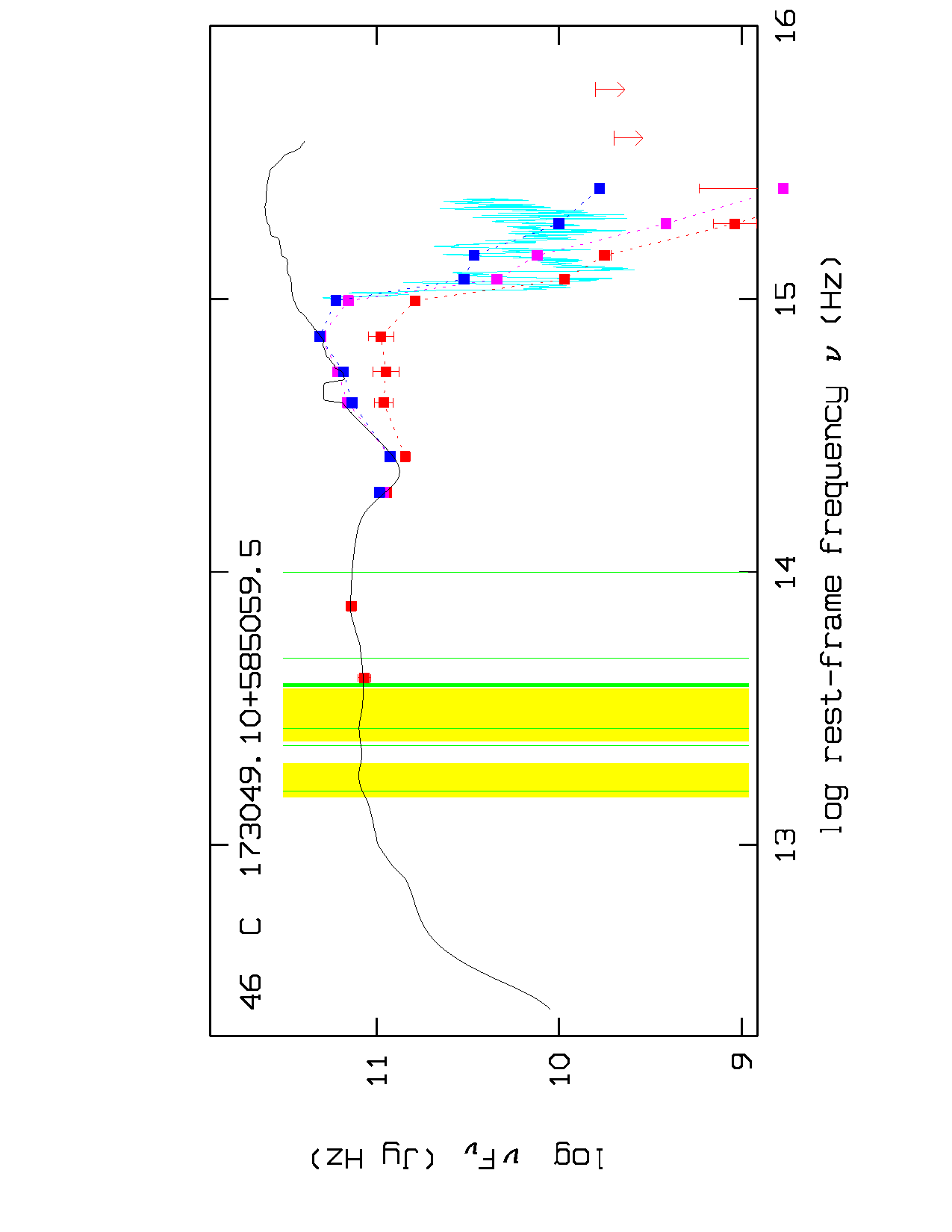}\hfill \\
\includegraphics[viewport=125 0 570 790,angle=270,width=8.0cm,clip]{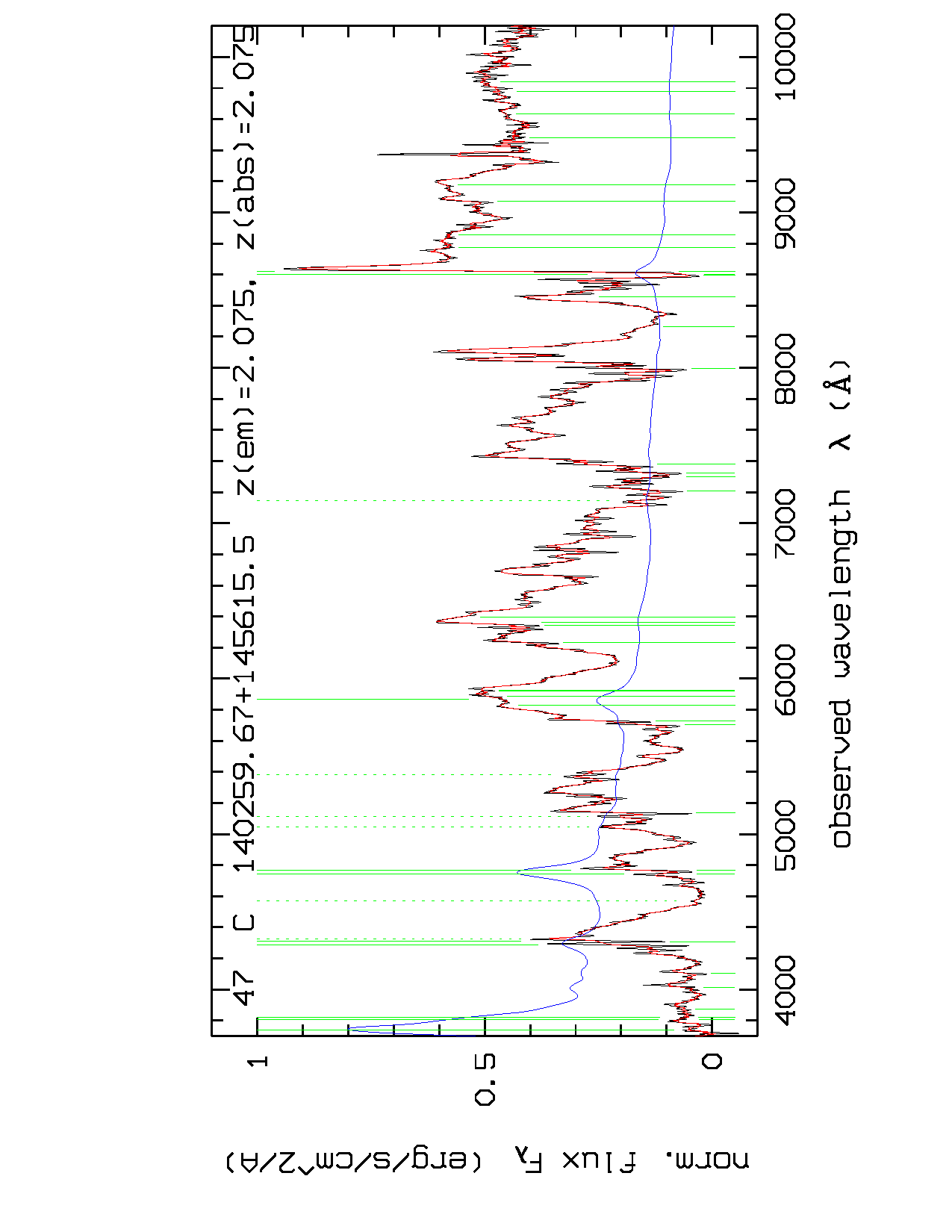}\hfill \=
\includegraphics[viewport=125 0 570 790,angle=270,width=8.0cm,clip]{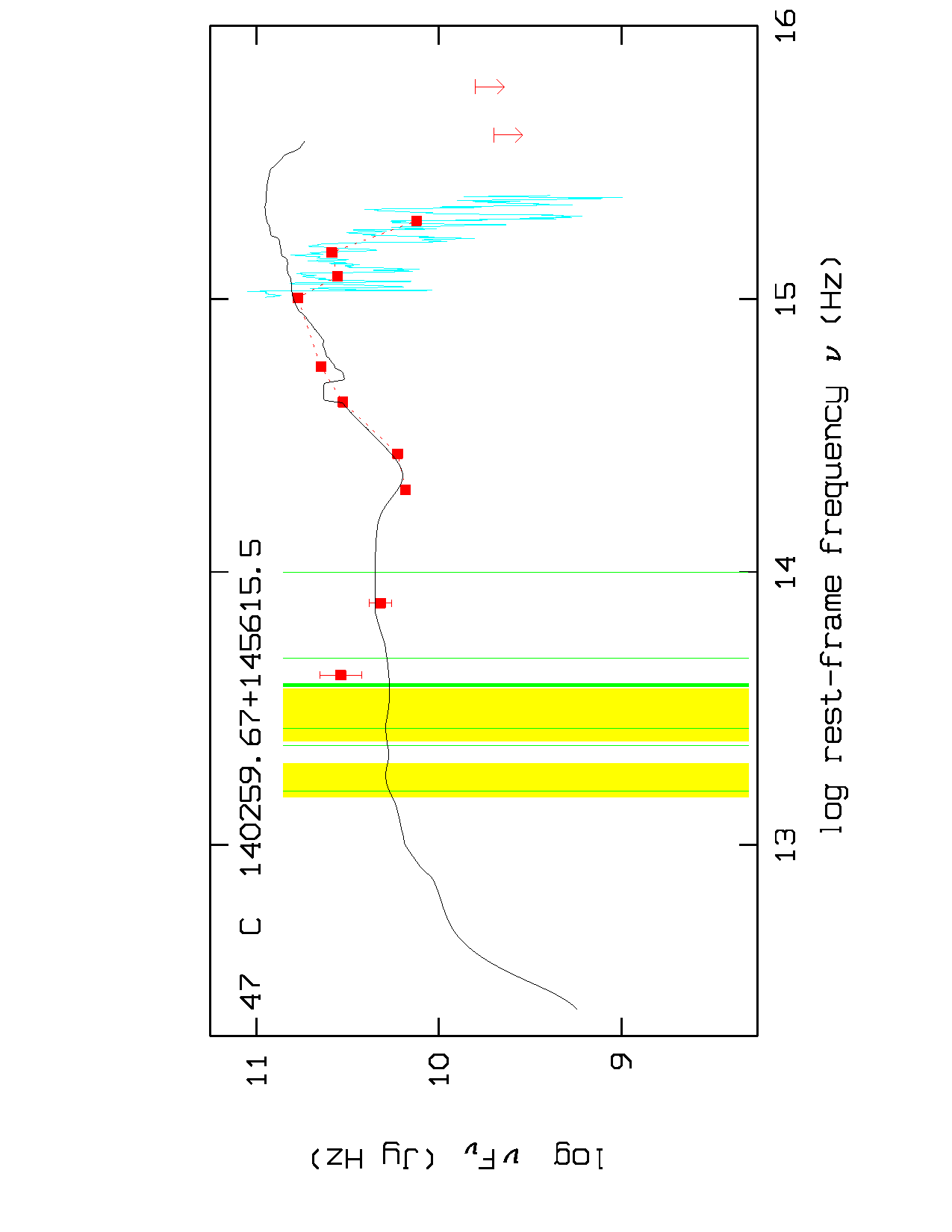}\hfill \\
\includegraphics[viewport=125 0 570 790,angle=270,width=8.0cm,clip]{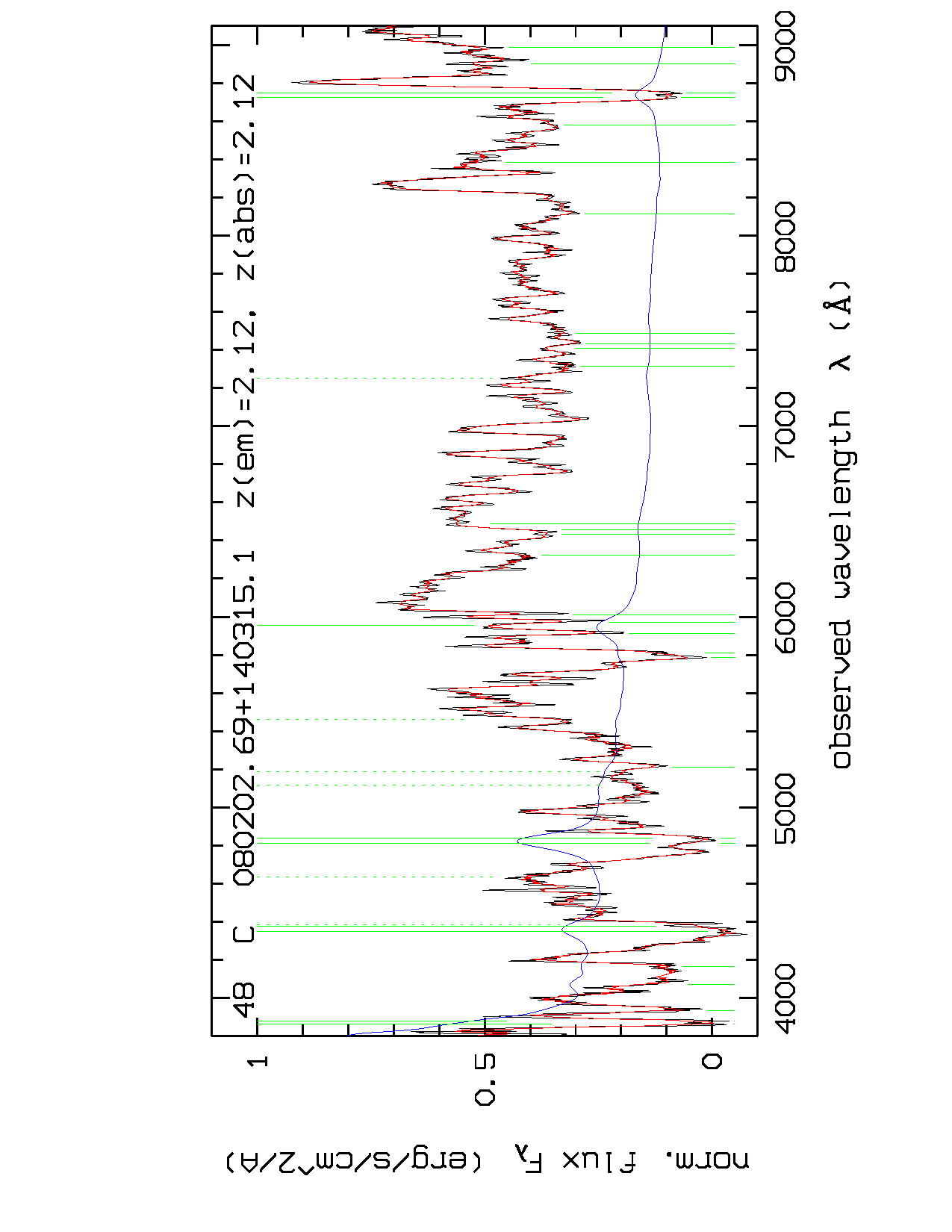}\hfill \=
\includegraphics[viewport=125 0 570 790,angle=270,width=8.0cm,clip]{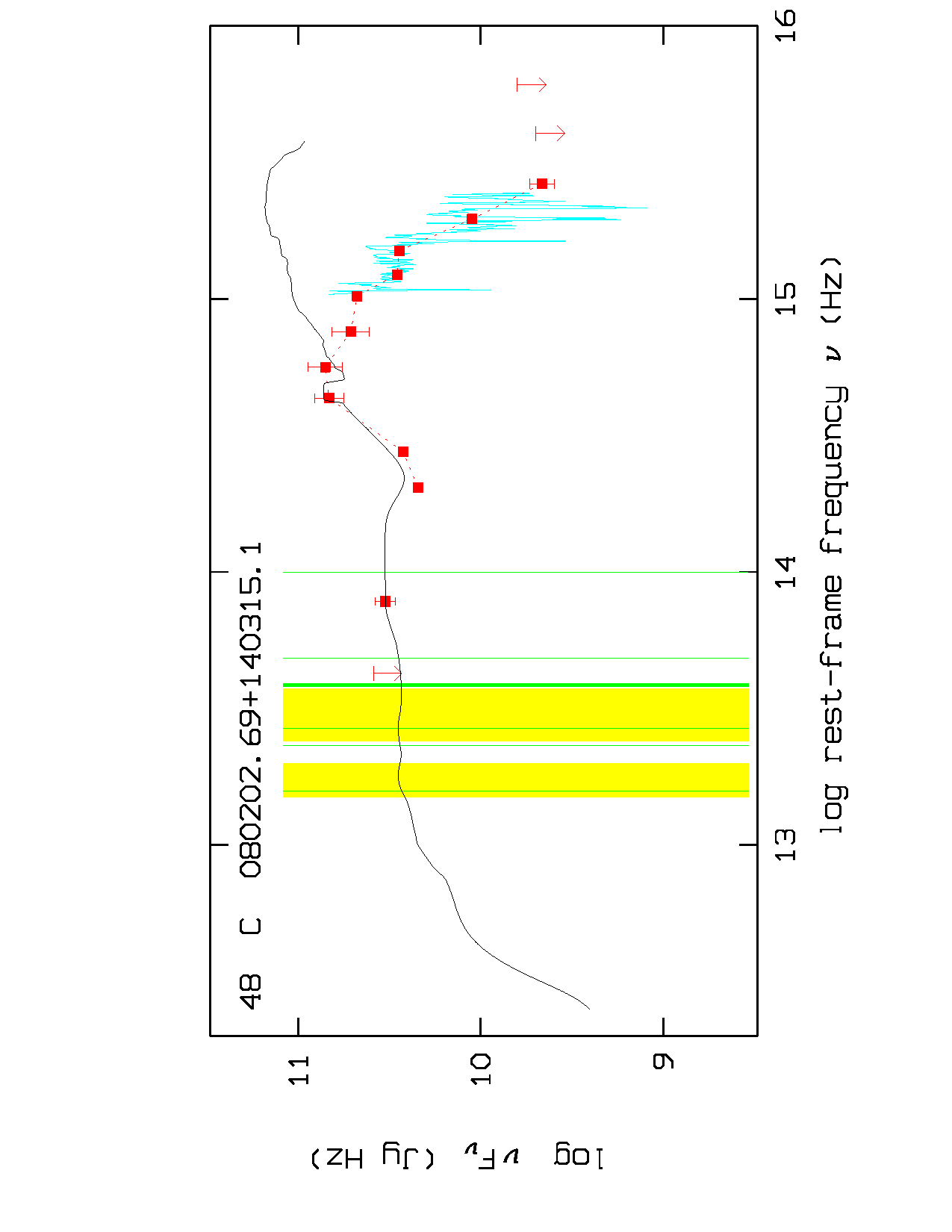}\hfill \\
\end{tabbing}
\caption{Sample C - continued (1).}
\end{figure*}\clearpage

\begin{figure*}[h]
\begin{tabbing}
\includegraphics[viewport=125 0 570 790,angle=270,width=8.0cm,clip]{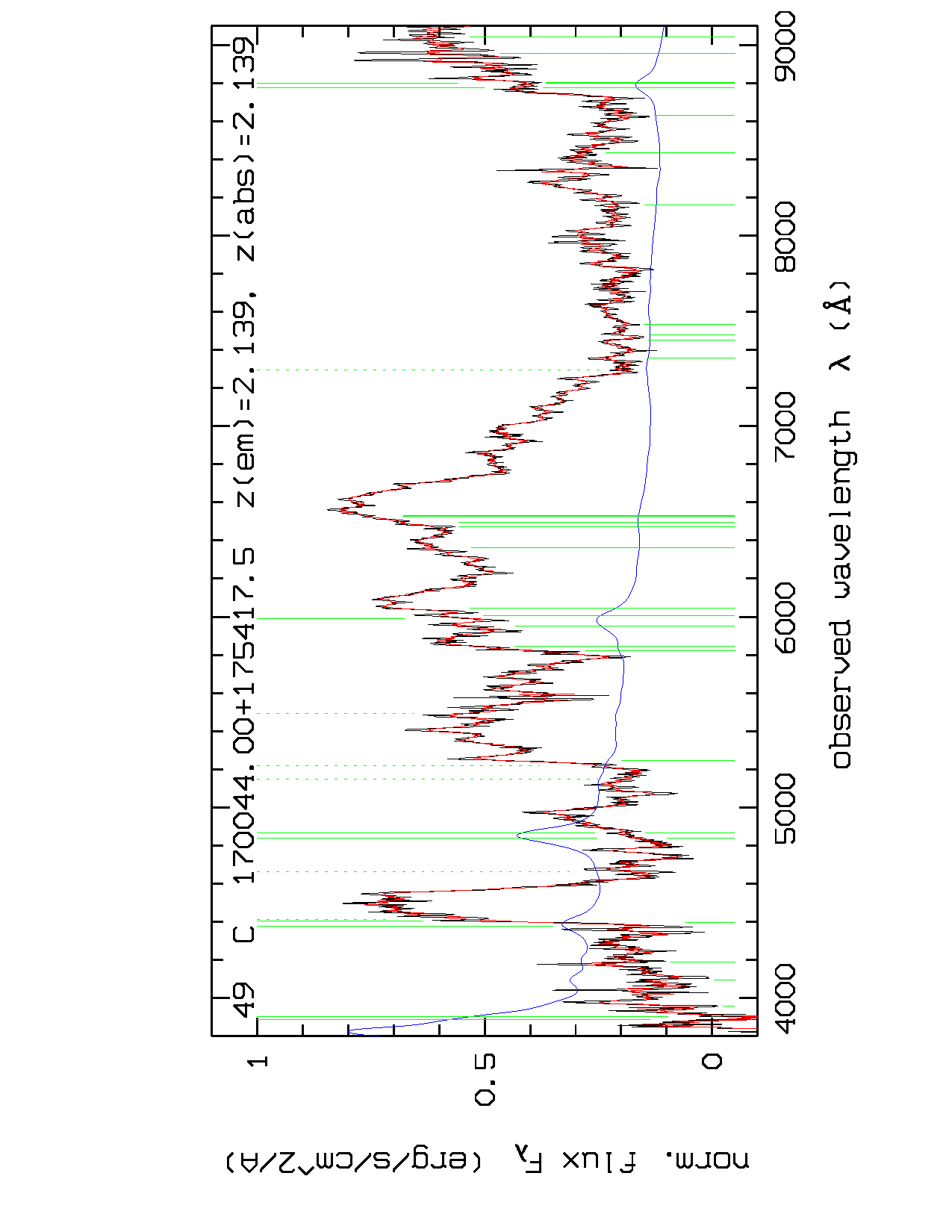}\hfill \=
\includegraphics[viewport=125 0 570 790,angle=270,width=8.0cm,clip]{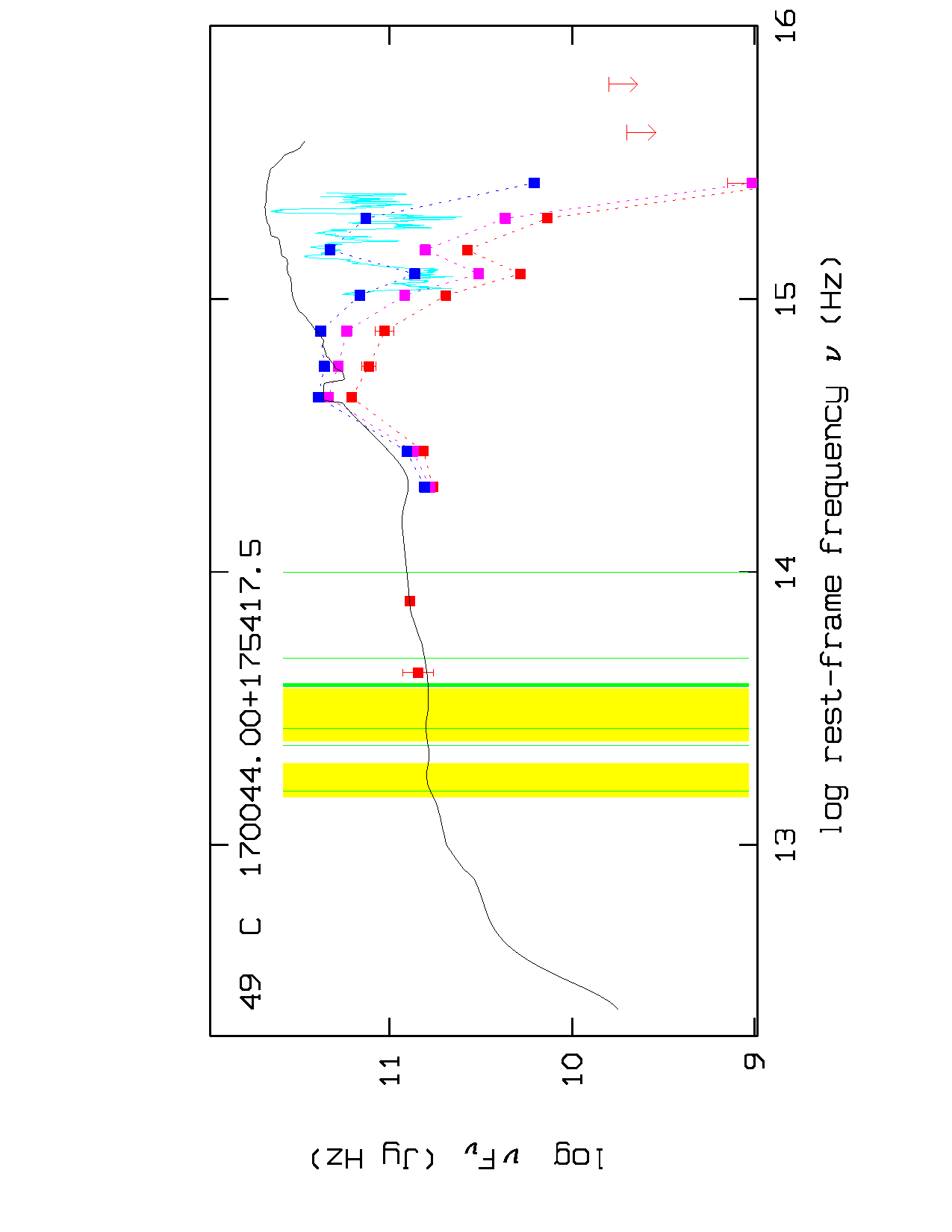}\hfill \\
\includegraphics[viewport=125 0 570 790,angle=270,width=8.0cm,clip]{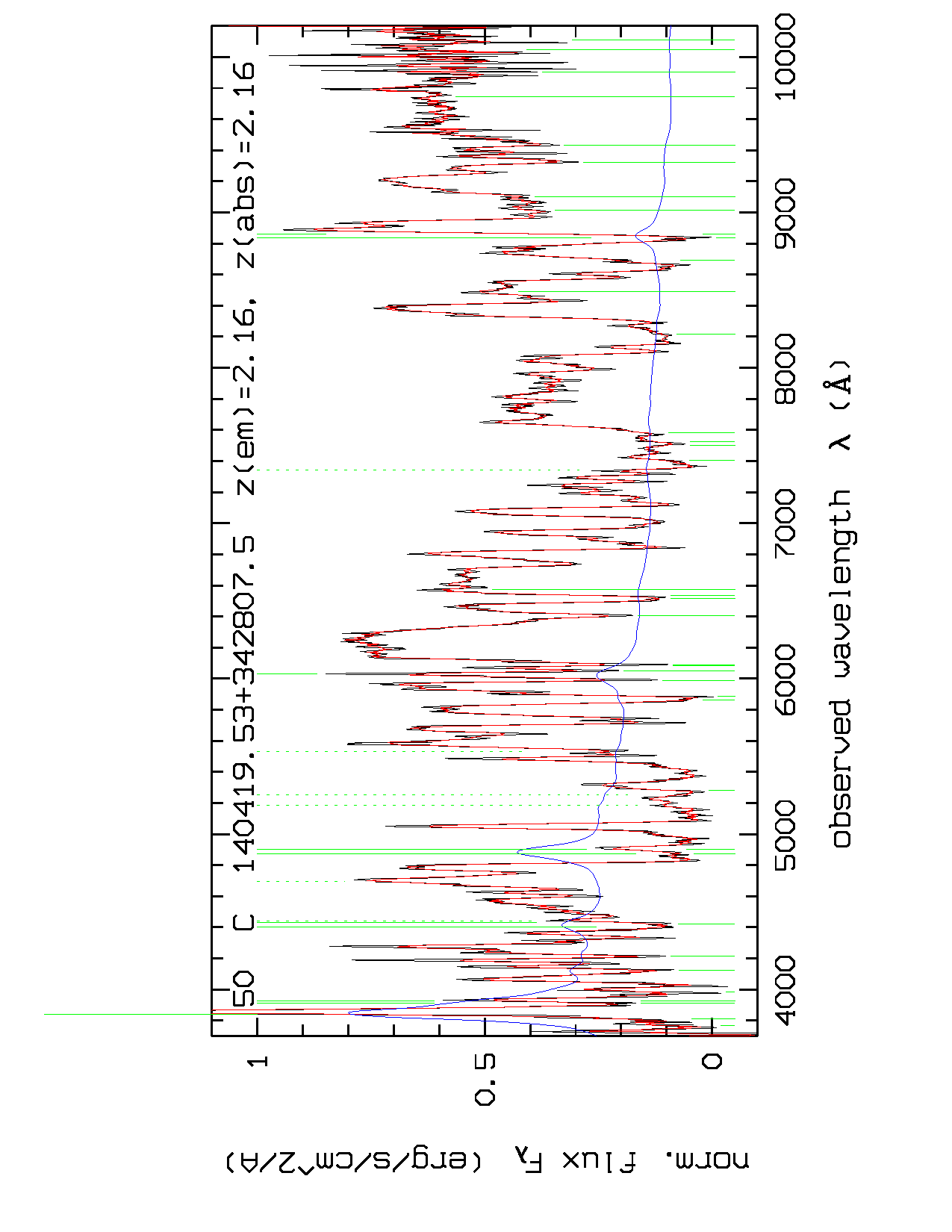}\hfill \=
\includegraphics[viewport=125 0 570 790,angle=270,width=8.0cm,clip]{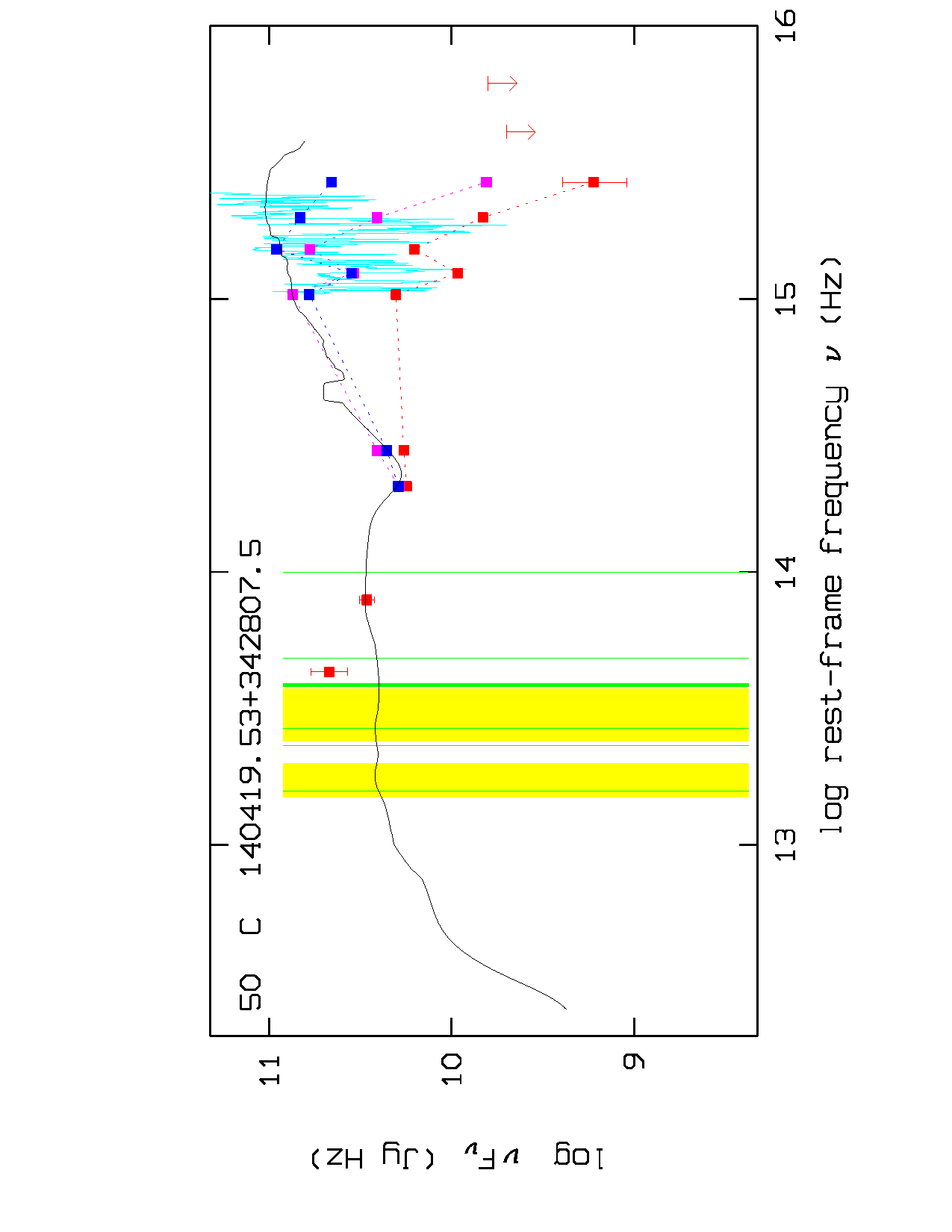}\hfill \\
\includegraphics[viewport=125 0 570 790,angle=270,width=8.0cm,clip]{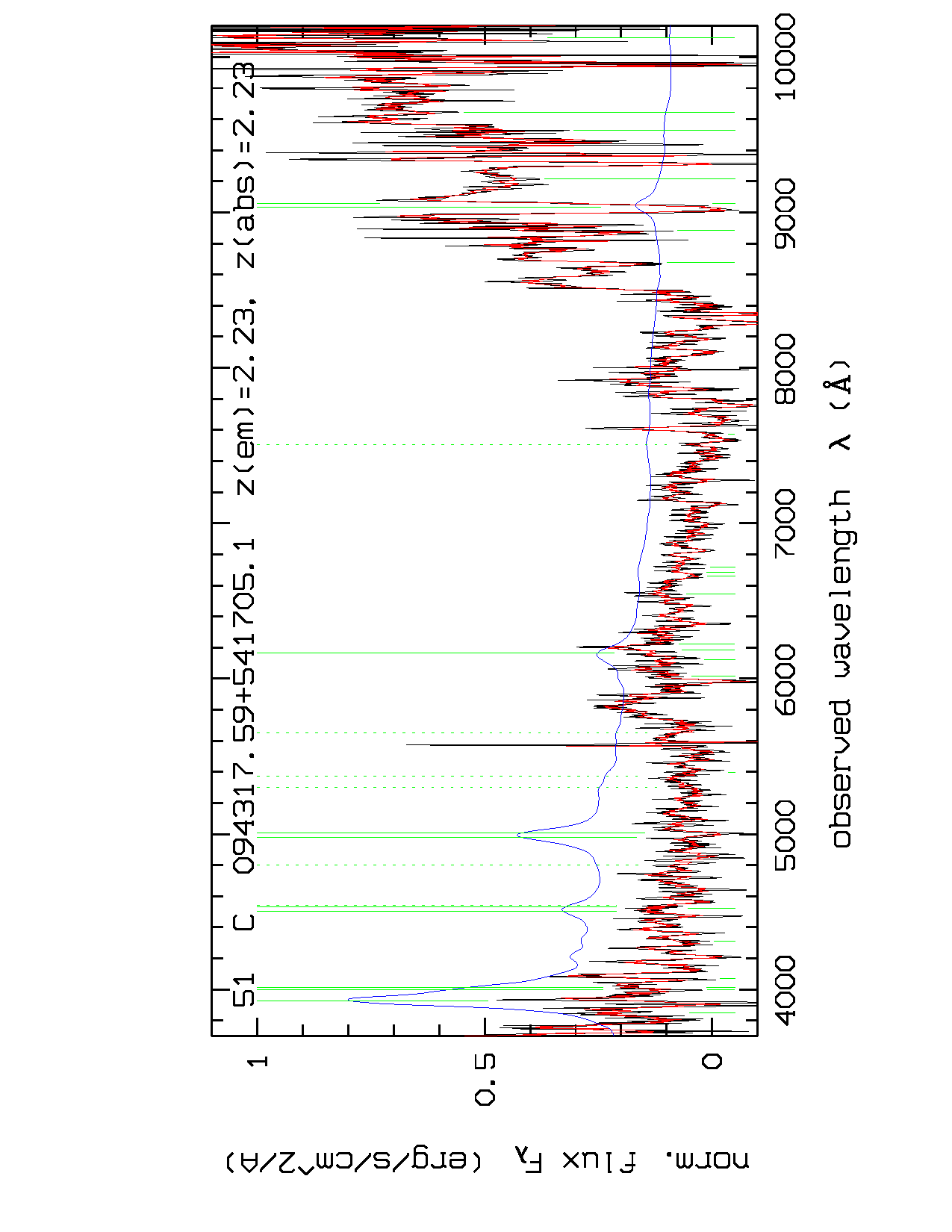}\hfill \=
\includegraphics[viewport=125 0 570 790,angle=270,width=8.0cm,clip]{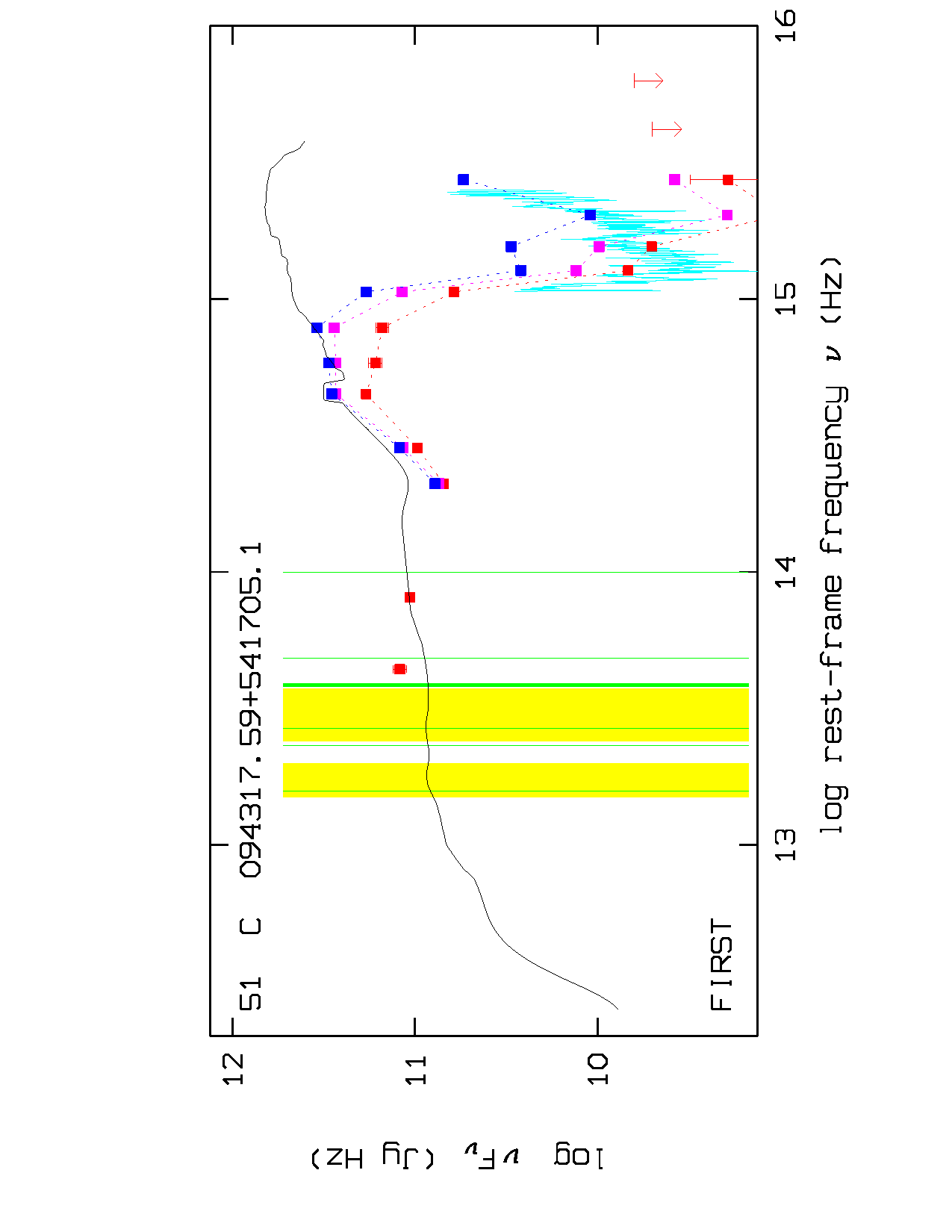}\hfill \\
\includegraphics[viewport=125 0 570 790,angle=270,width=8.0cm,clip]{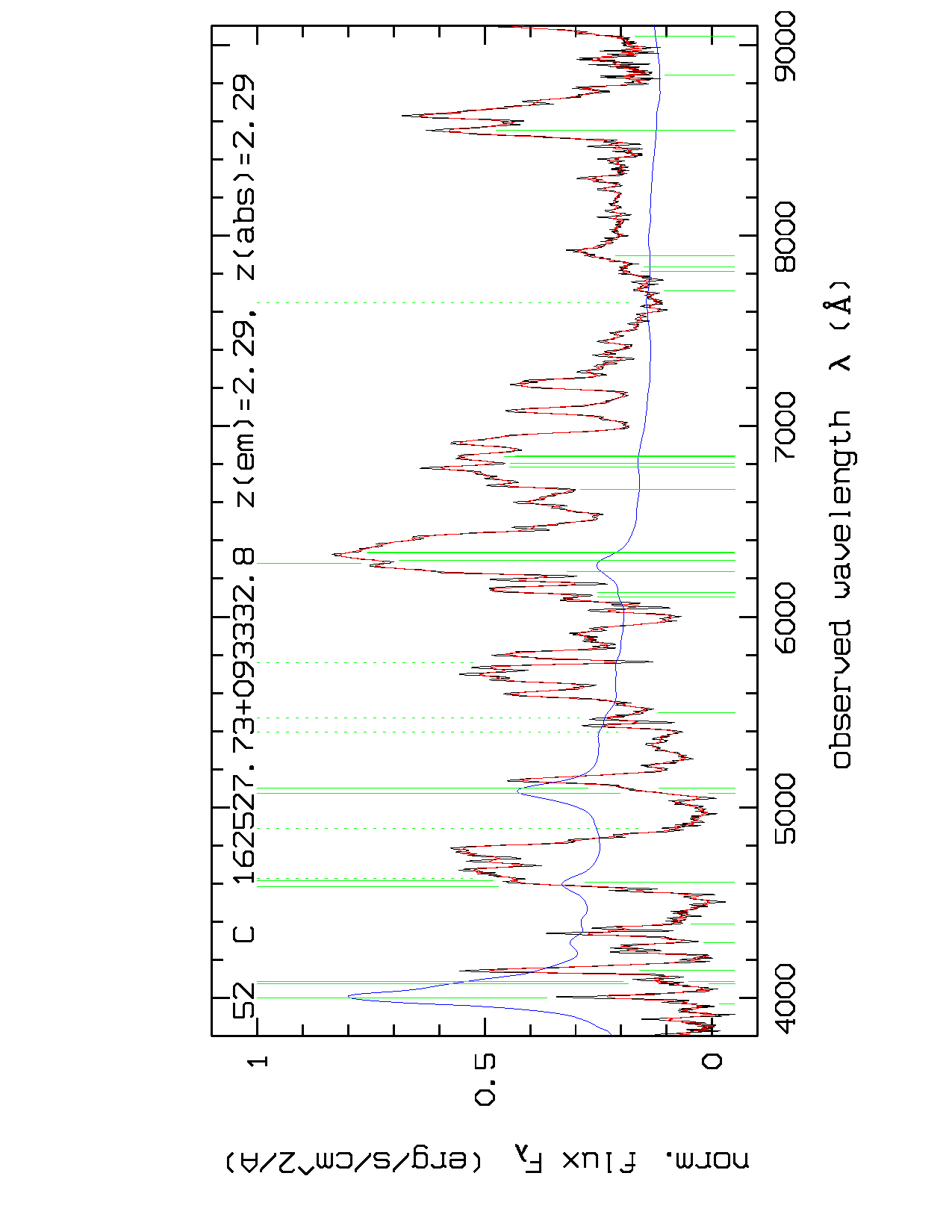}\hfill \=
\includegraphics[viewport=125 0 570 790,angle=270,width=8.0cm,clip]{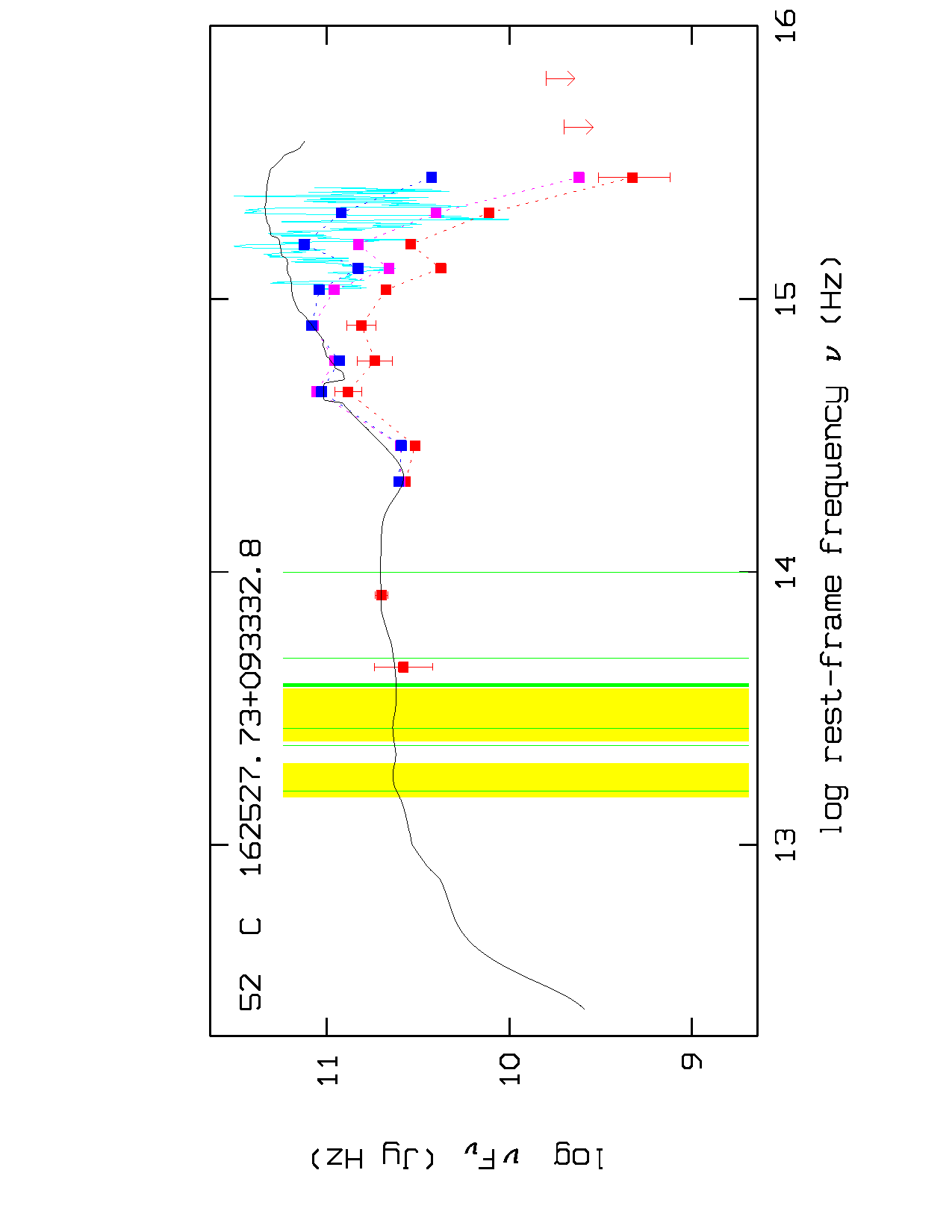}\hfill \\
\includegraphics[viewport=125 0 570 790,angle=270,width=8.0cm,clip]{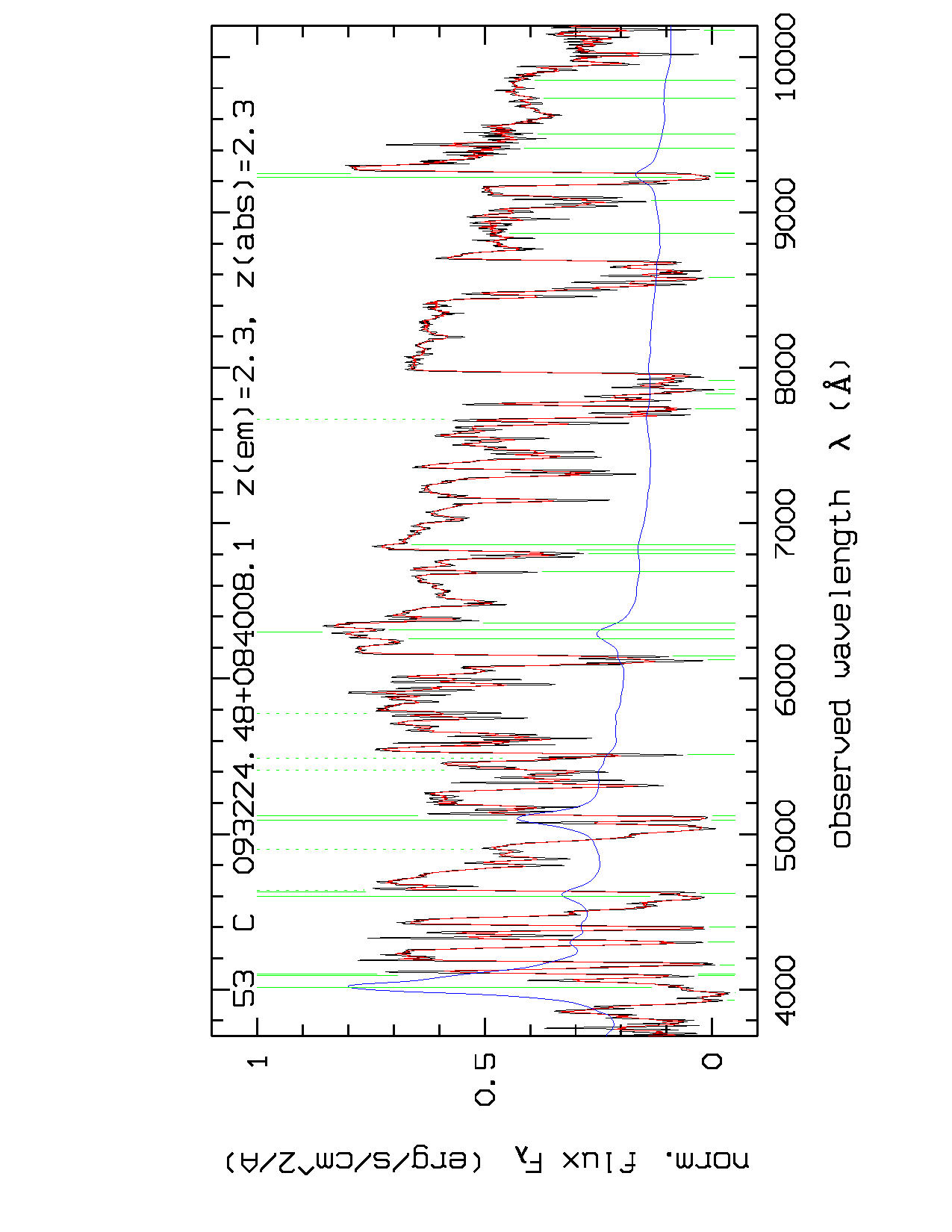}\hfill \=
\includegraphics[viewport=125 0 570 790,angle=270,width=8.0cm,clip]{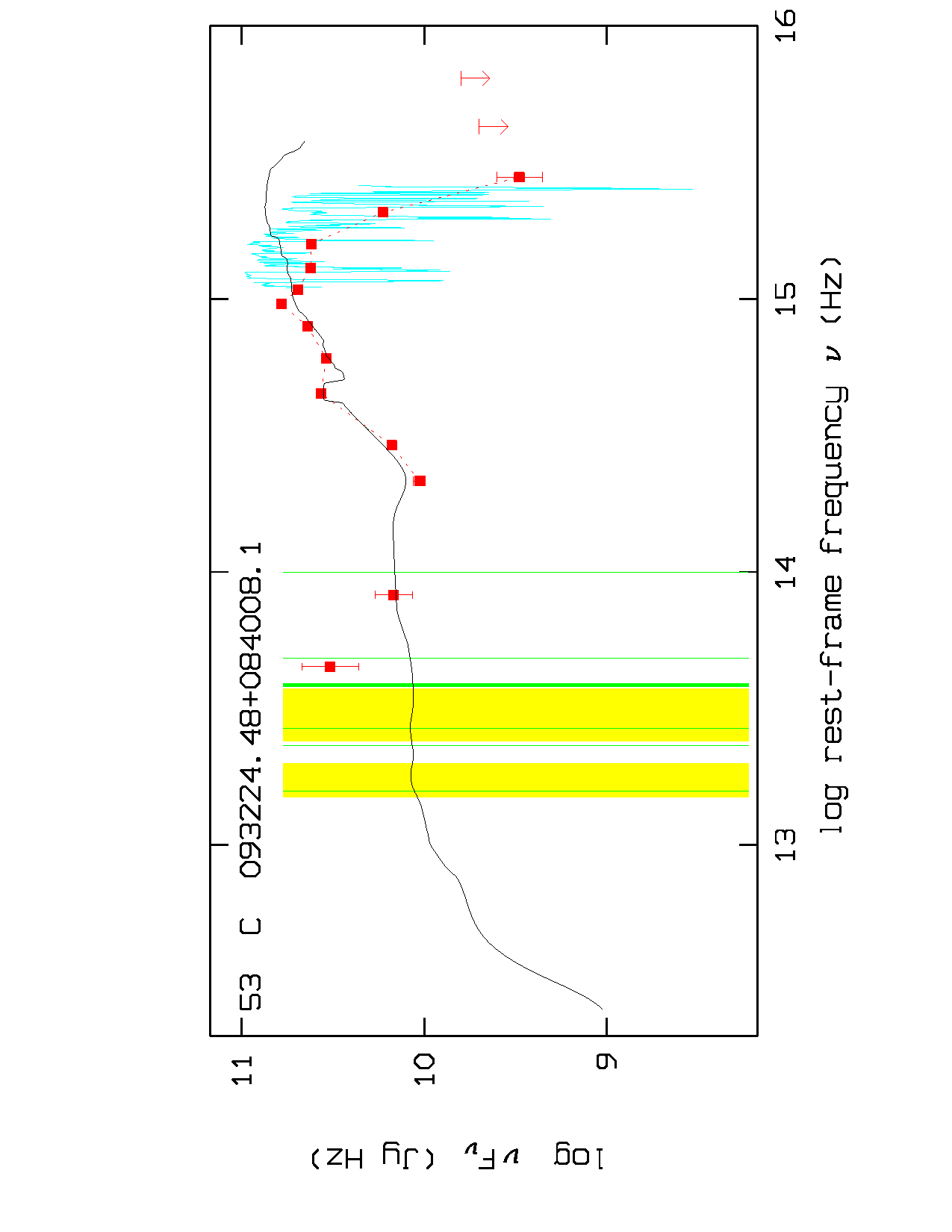}\hfill \\
\end{tabbing}
\caption{Sample C - continued (2).}
\end{figure*}\clearpage

\begin{figure*}[h]
\begin{tabbing}
\includegraphics[viewport=125 0 570 790,angle=270,width=8.0cm,clip]{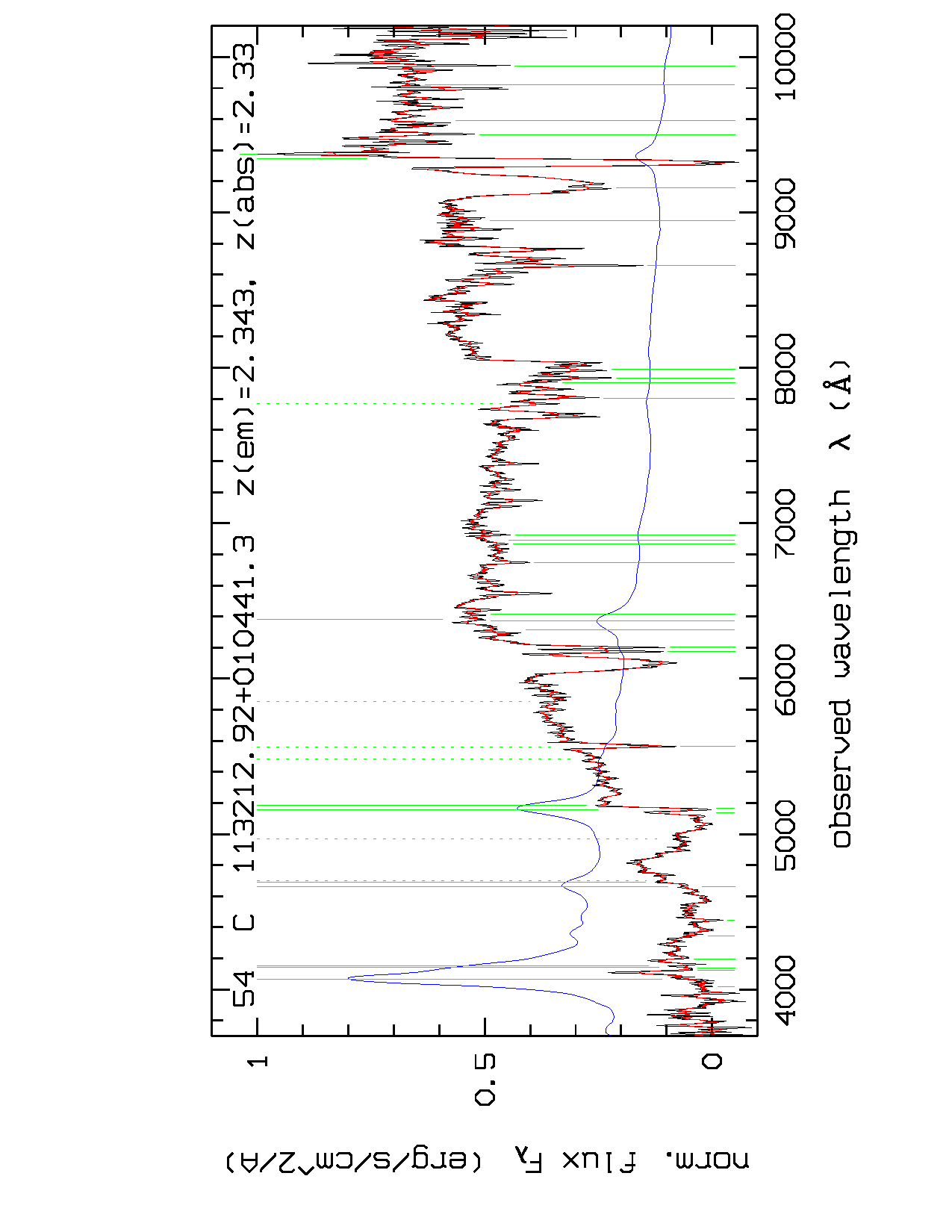}\hfill \=
\includegraphics[viewport=125 0 570 790,angle=270,width=8.0cm,clip]{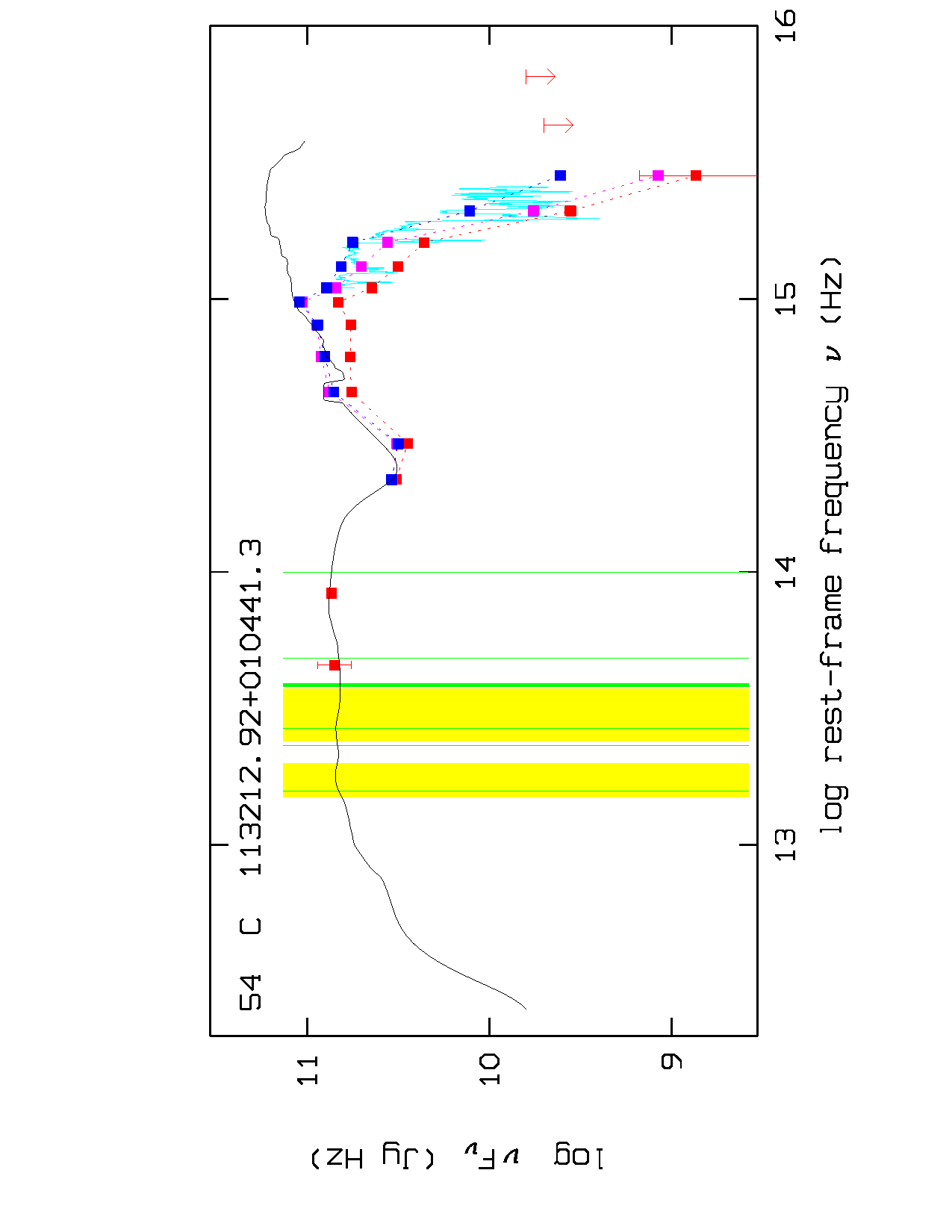}\hfill \\
\includegraphics[viewport=125 0 570 790,angle=270,width=8.0cm,clip]{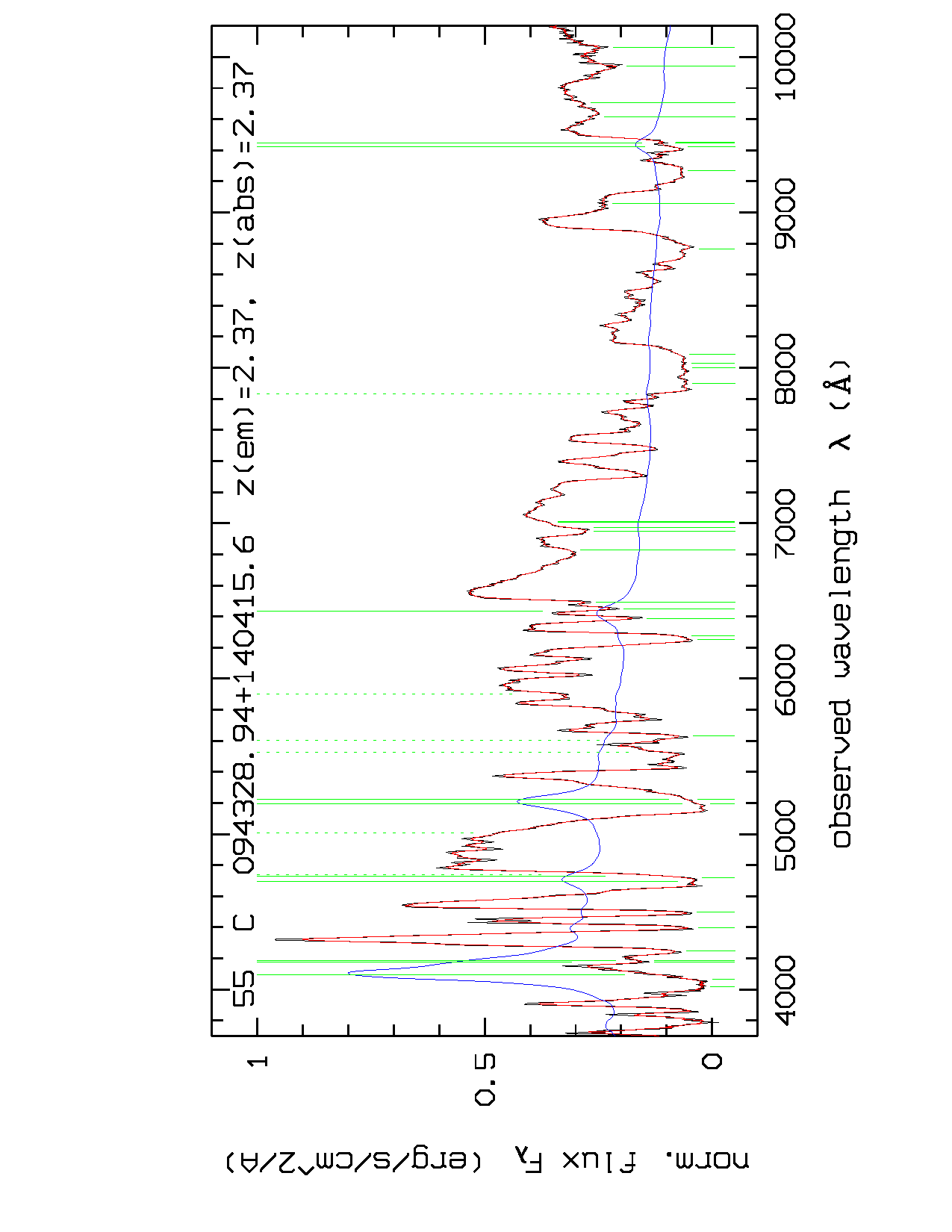}\hfill \=
\includegraphics[viewport=125 0 570 790,angle=270,width=8.0cm,clip]{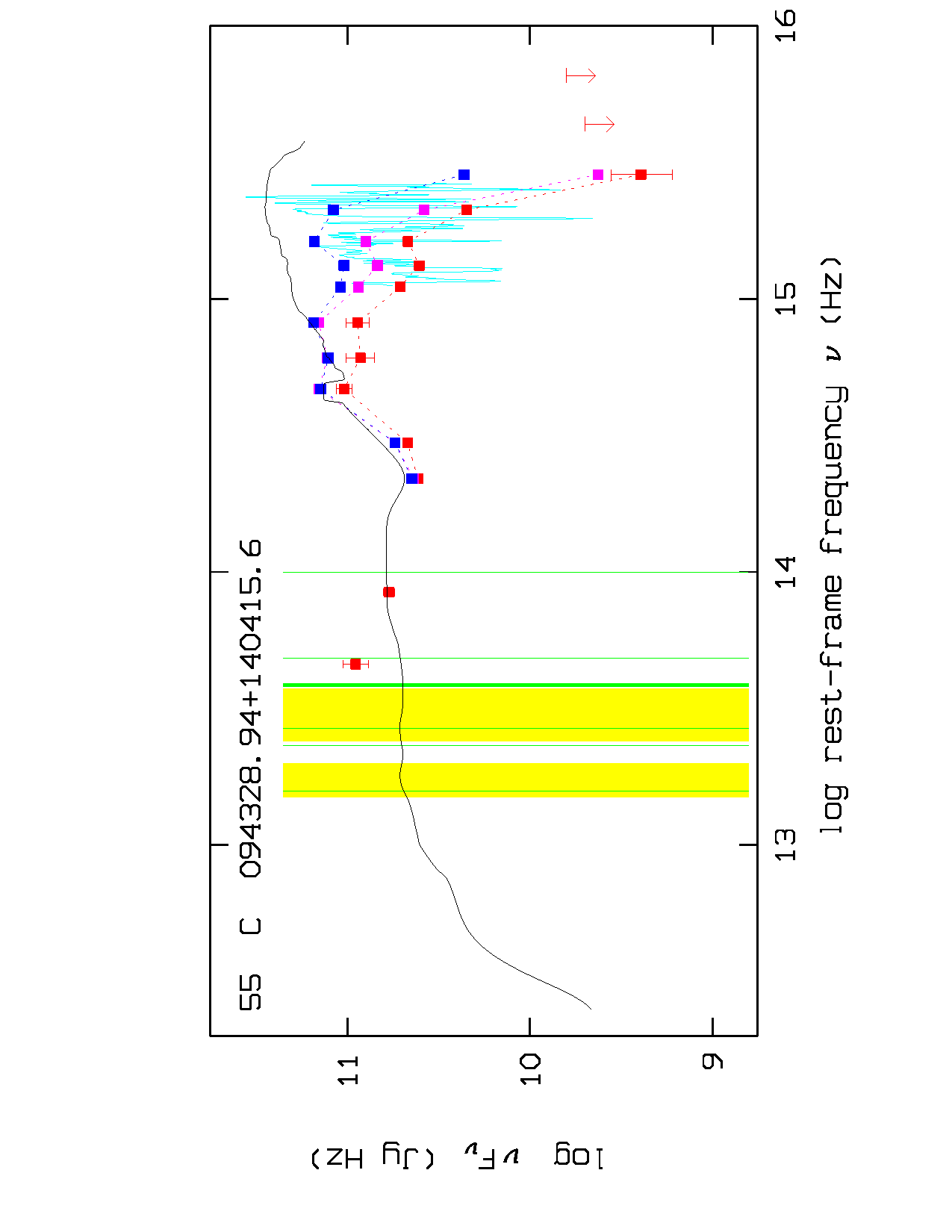}\hfill \\
\includegraphics[viewport=125 0 570 790,angle=270,width=8.0cm,clip]{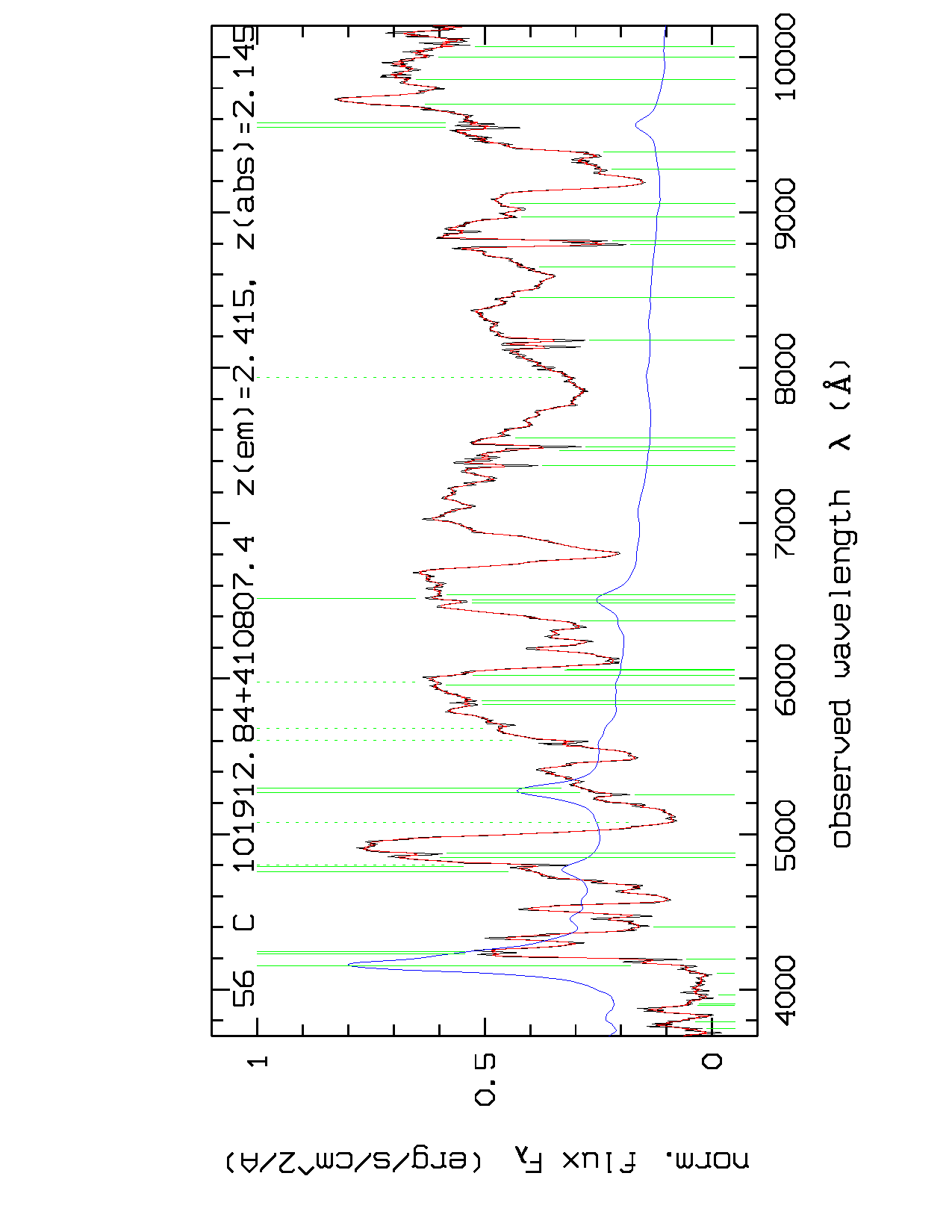}\hfill \=
\includegraphics[viewport=125 0 570 790,angle=270,width=8.0cm,clip]{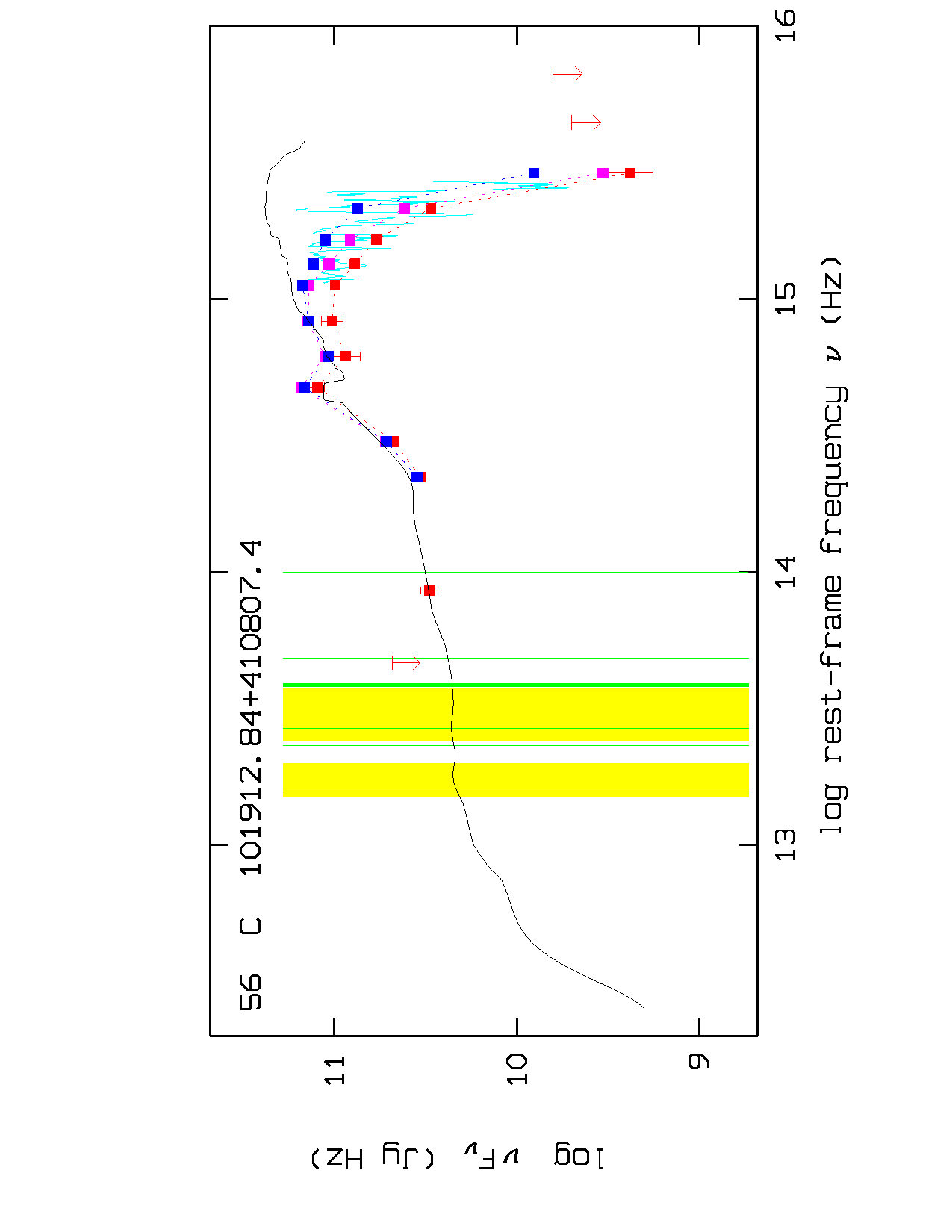}\hfill \\
\includegraphics[viewport=125 0 570 790,angle=270,width=8.0cm,clip]{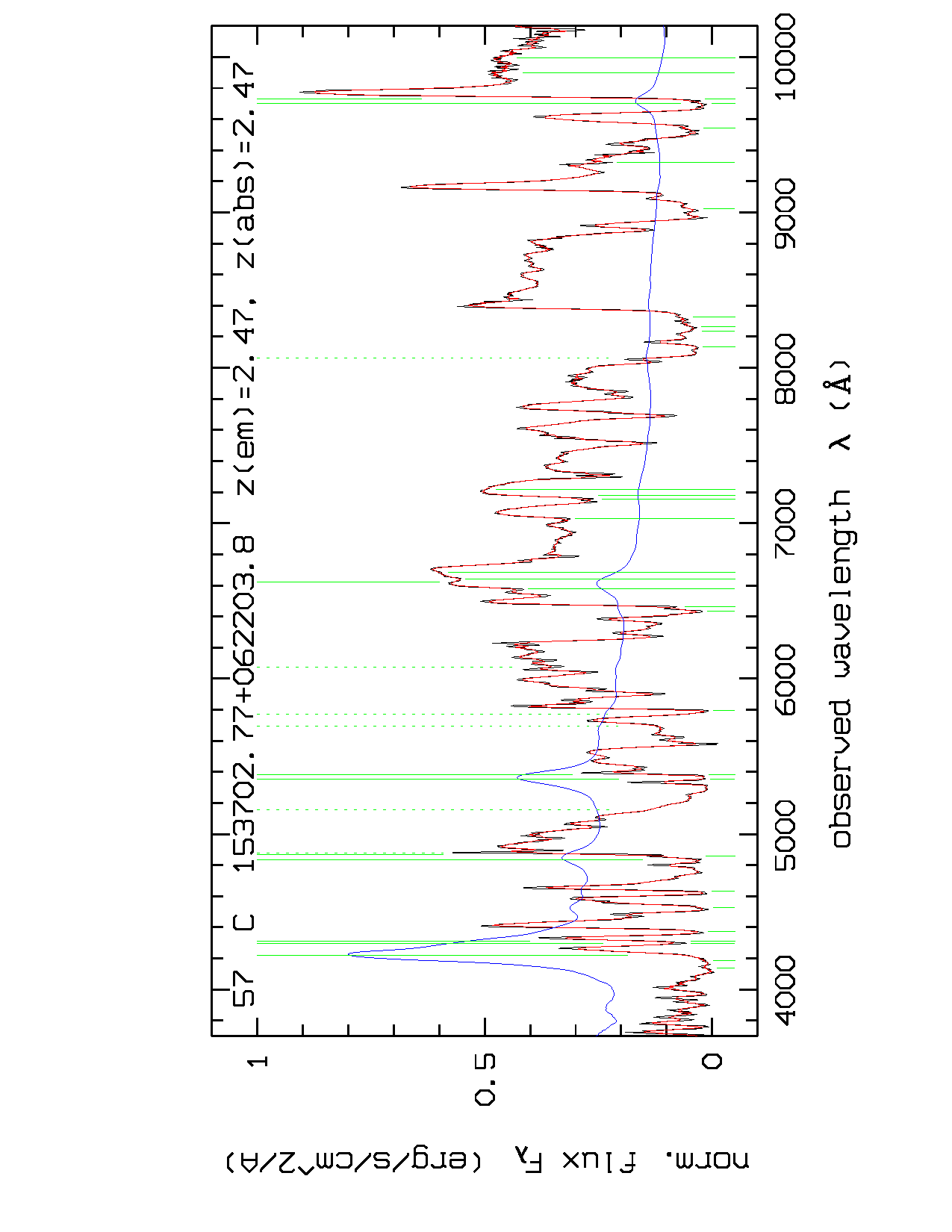}\hfill \=
\includegraphics[viewport=125 0 570 790,angle=270,width=8.0cm,clip]{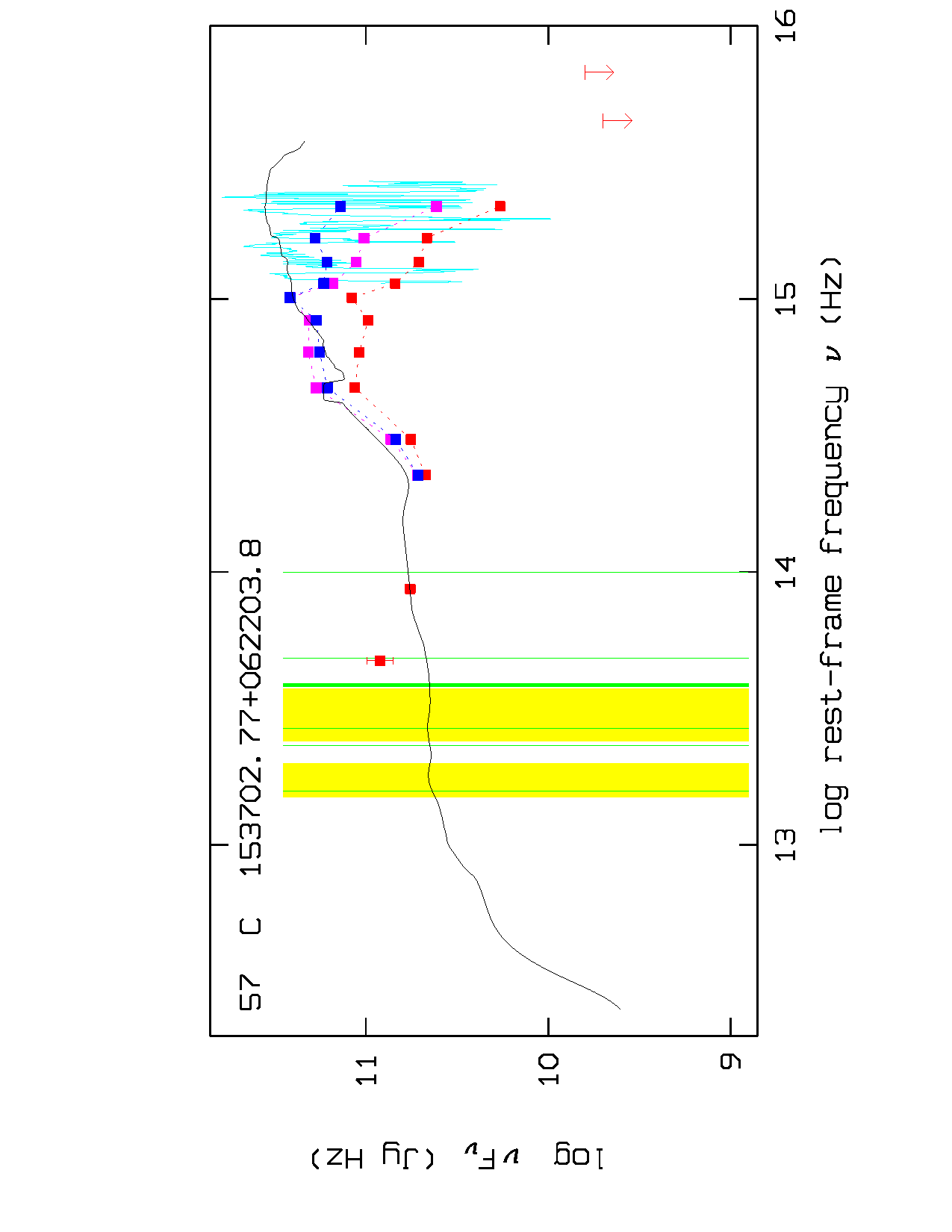}\hfill \\
\includegraphics[viewport=125 0 570 790,angle=270,width=8.0cm,clip]{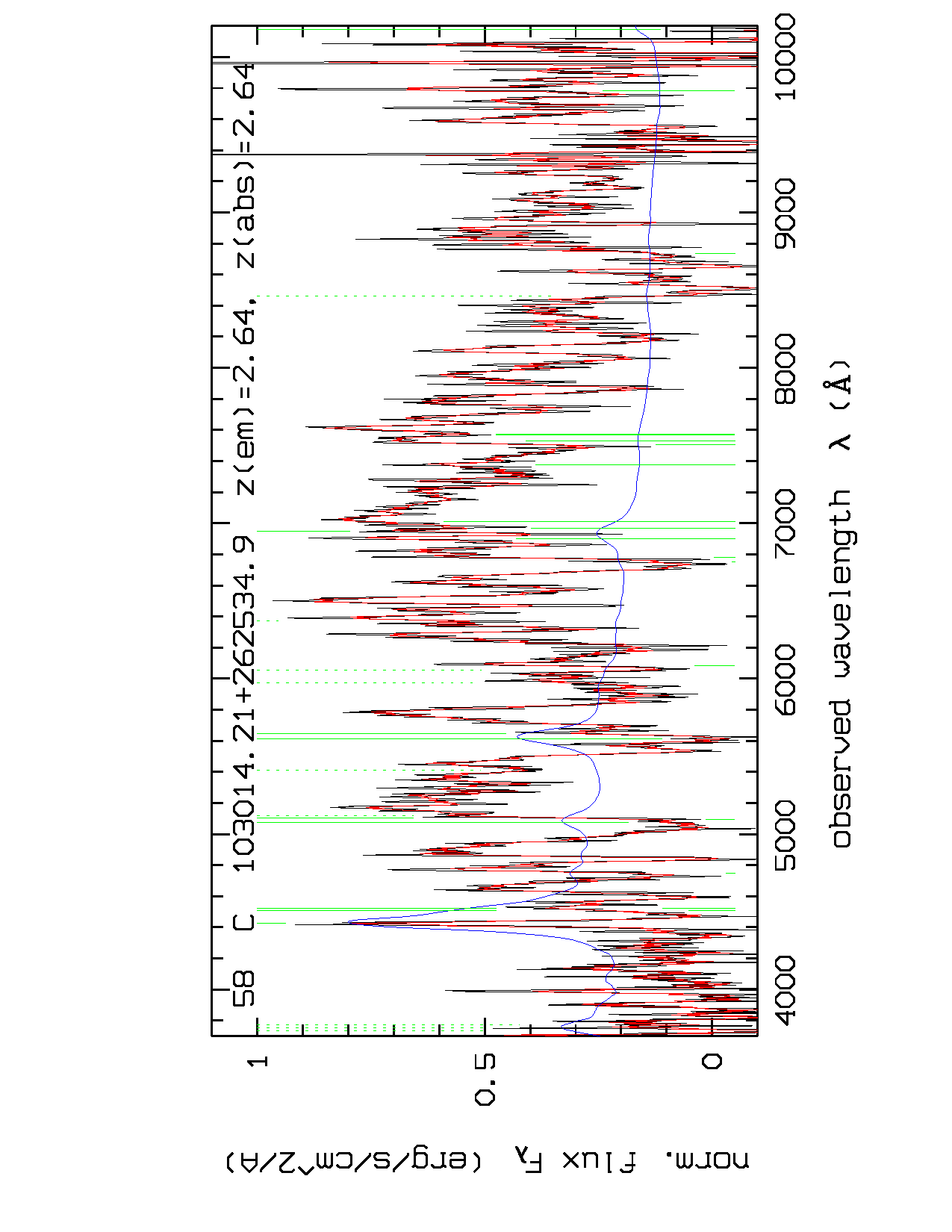}\hfill \=
\includegraphics[viewport=125 0 570 790,angle=270,width=8.0cm,clip]{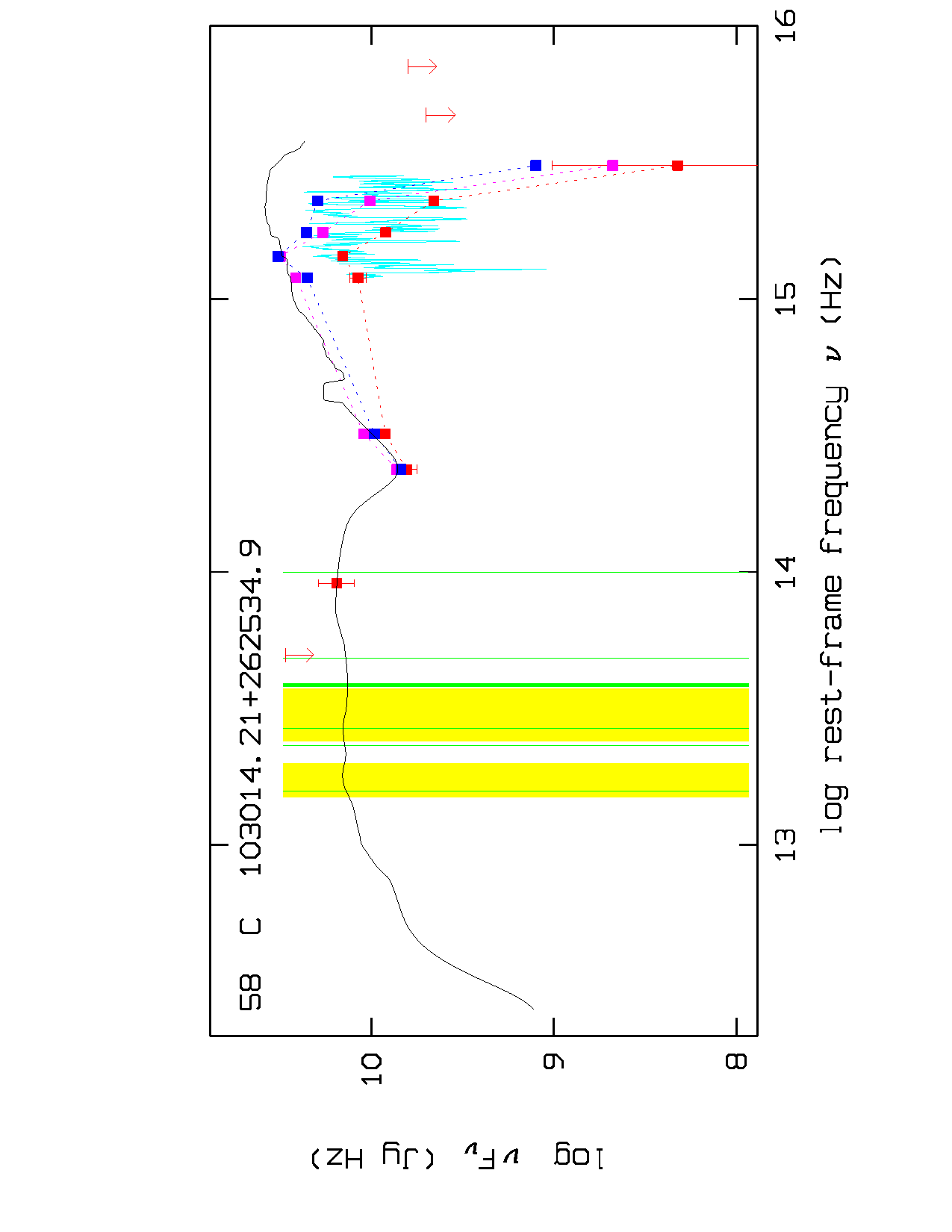}\hfill \\
\end{tabbing}
\caption{Sample C - continued (3).}
\end{figure*}\clearpage

\begin{figure*}[h]
\begin{tabbing}
\includegraphics[viewport=125 0 570 790,angle=270,width=8.0cm,clip]{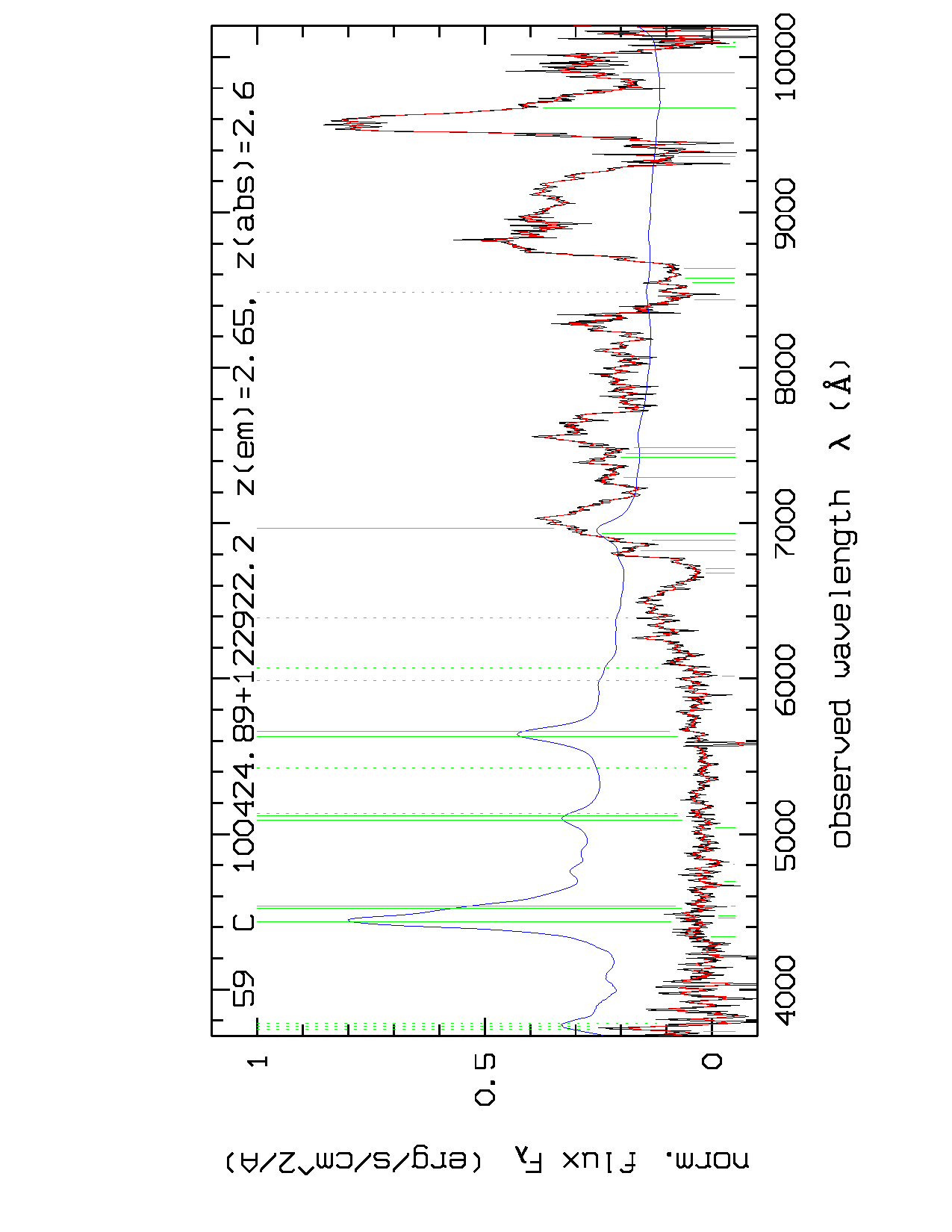}\hfill \=
\includegraphics[viewport=125 0 570 790,angle=270,width=8.0cm,clip]{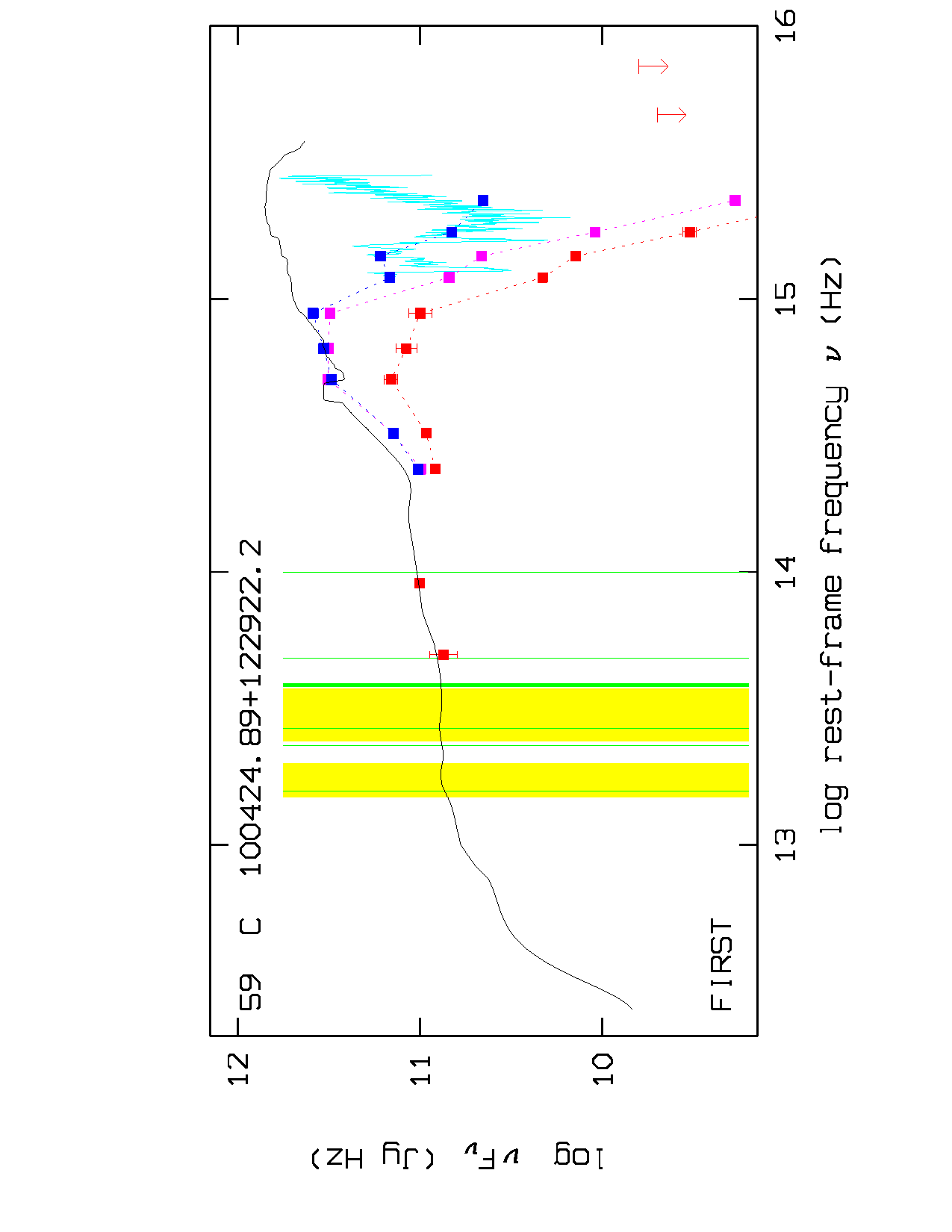}\hfill \\
\includegraphics[viewport=125 0 570 790,angle=270,width=8.0cm,clip]{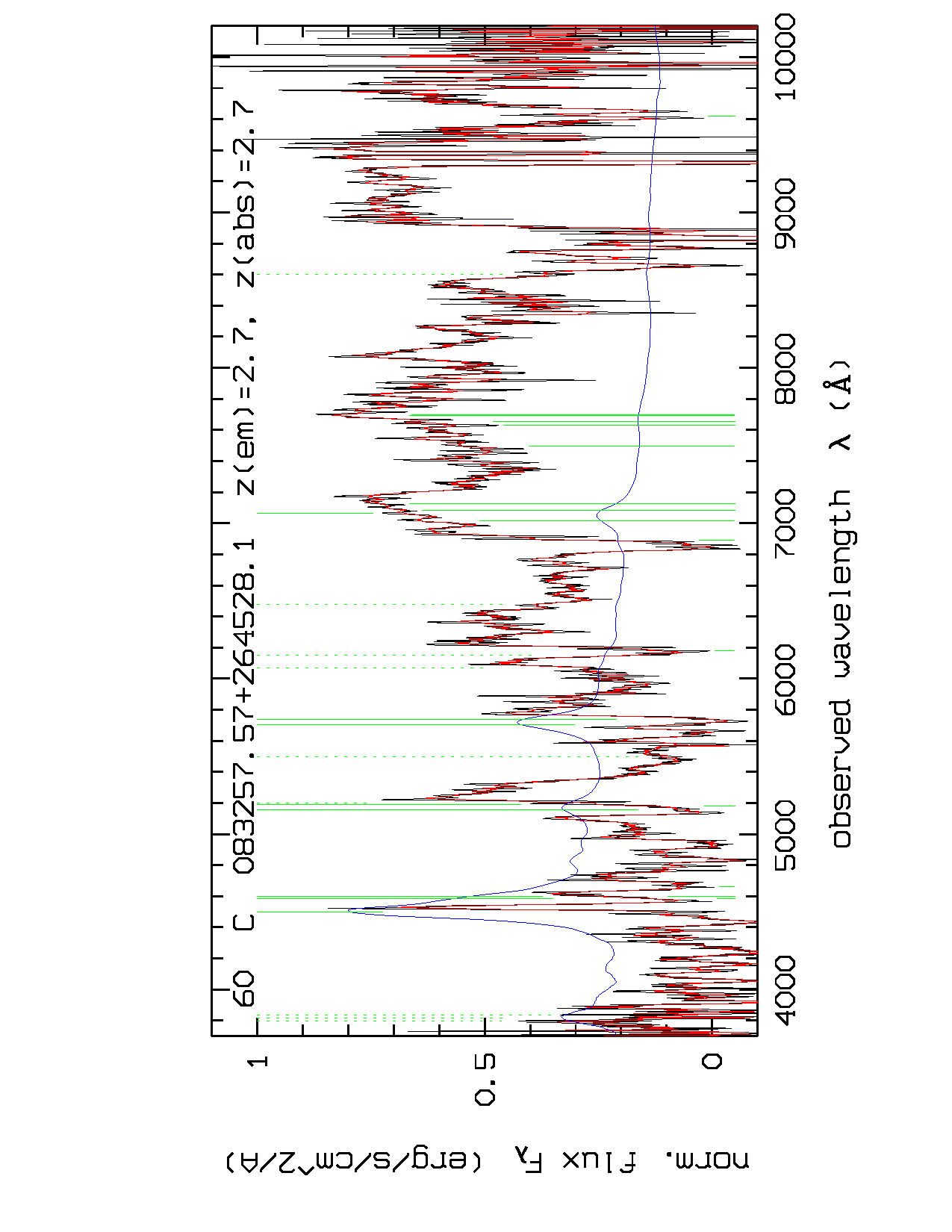}\hfill \=
\includegraphics[viewport=125 0 570 790,angle=270,width=8.0cm,clip]{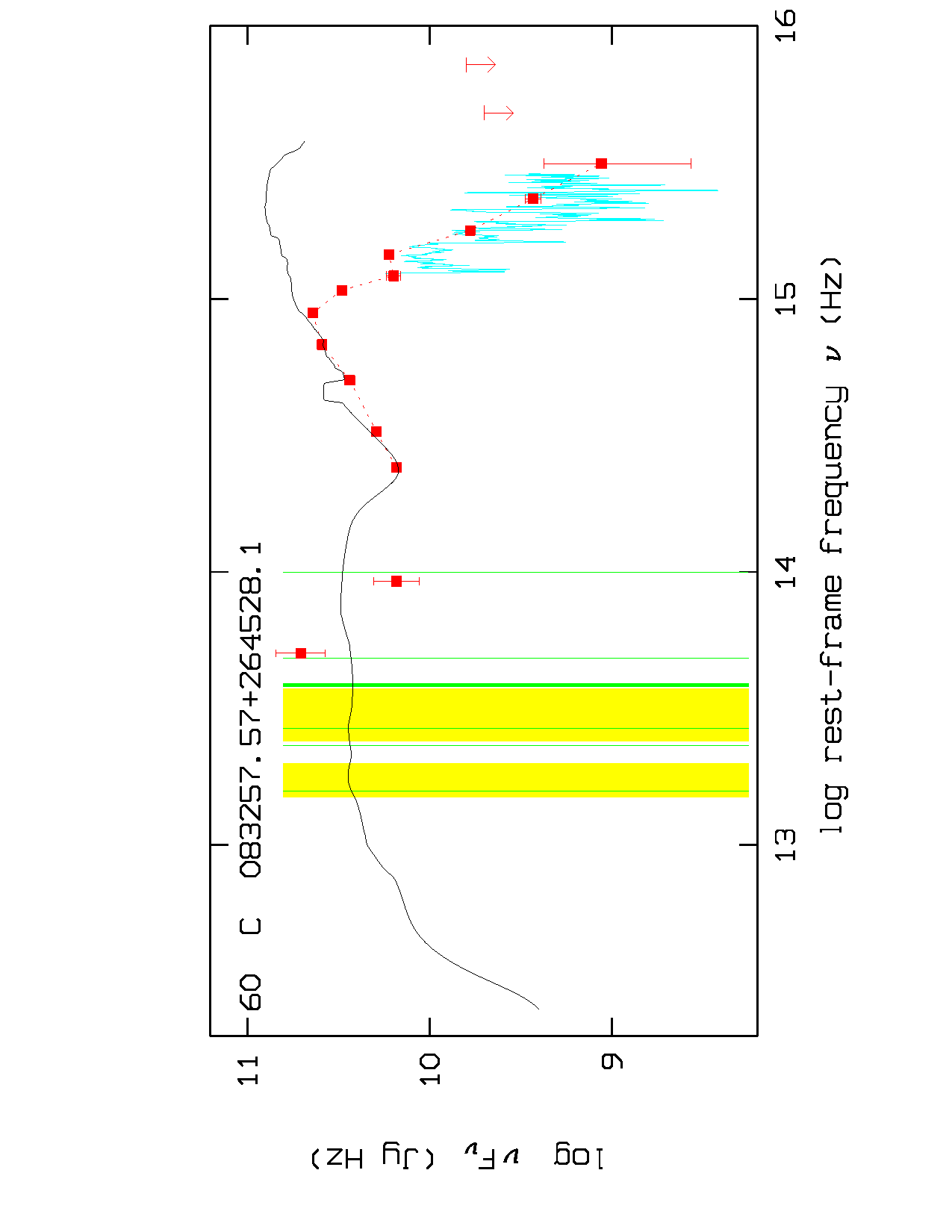}\hfill \\
\includegraphics[viewport=125 0 570 790,angle=270,width=8.0cm,clip]{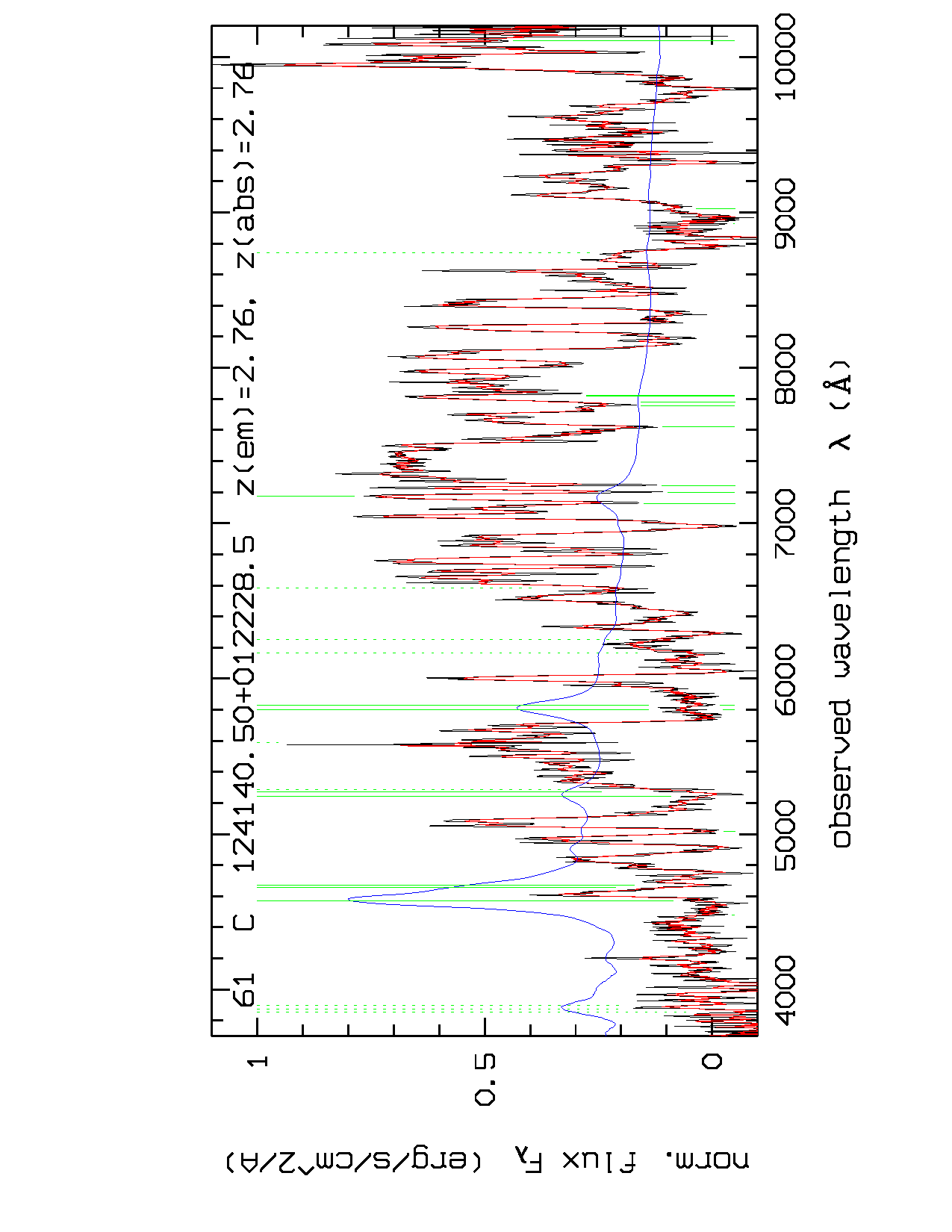}\hfill \=
\includegraphics[viewport=125 0 570 790,angle=270,width=8.0cm,clip]{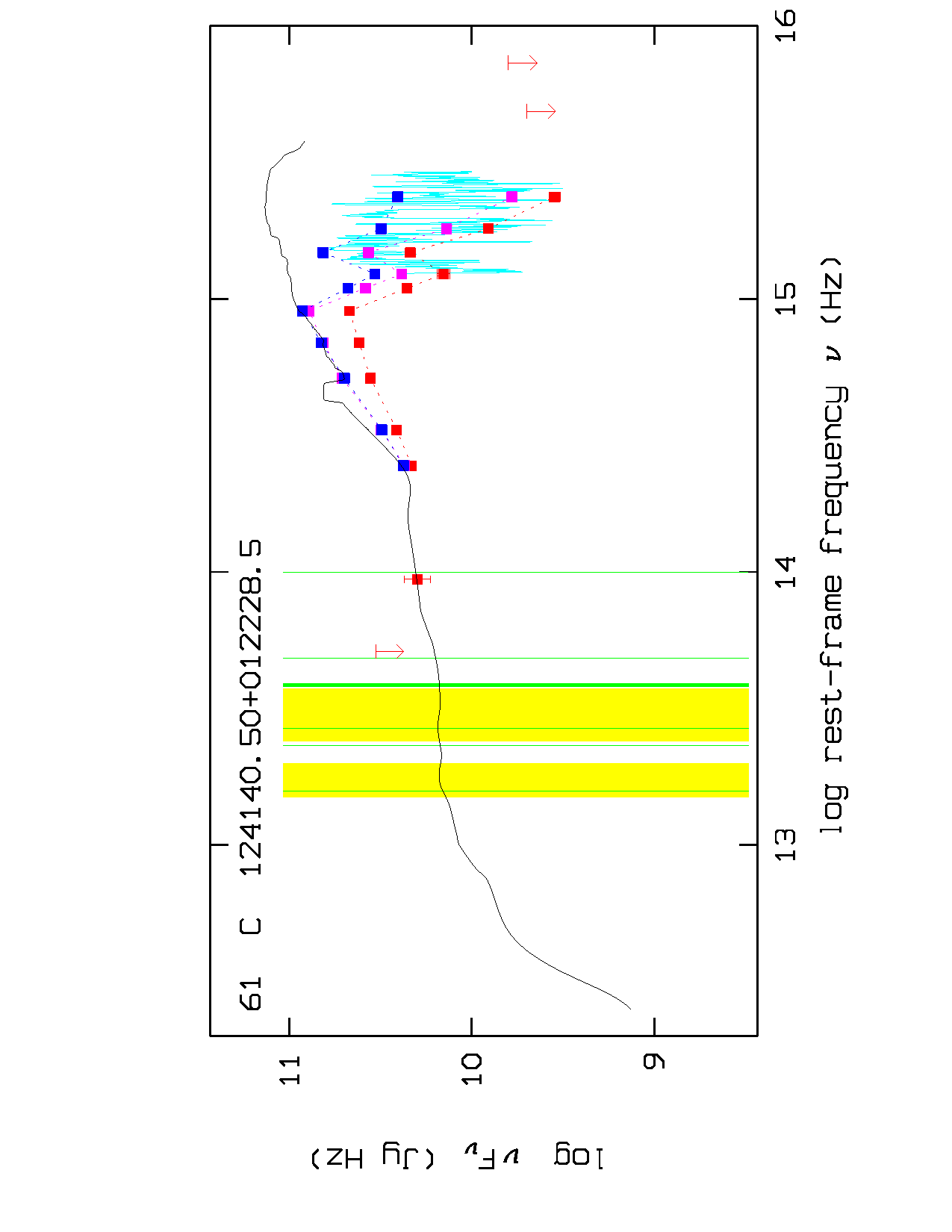}\hfill \\
\includegraphics[viewport=125 0 570 790,angle=270,width=8.0cm,clip]{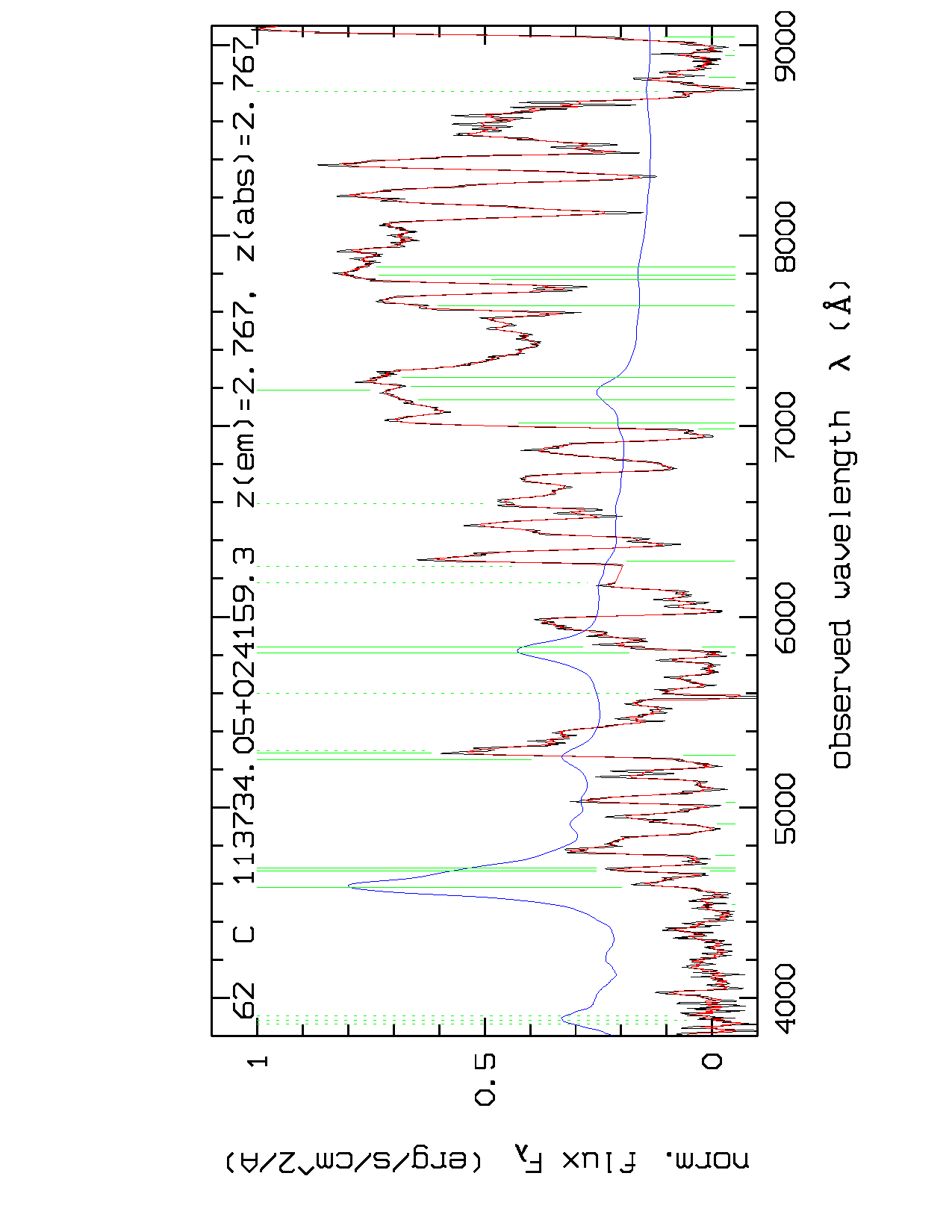}\hfill \=
\includegraphics[viewport=125 0 570 790,angle=270,width=8.0cm,clip]{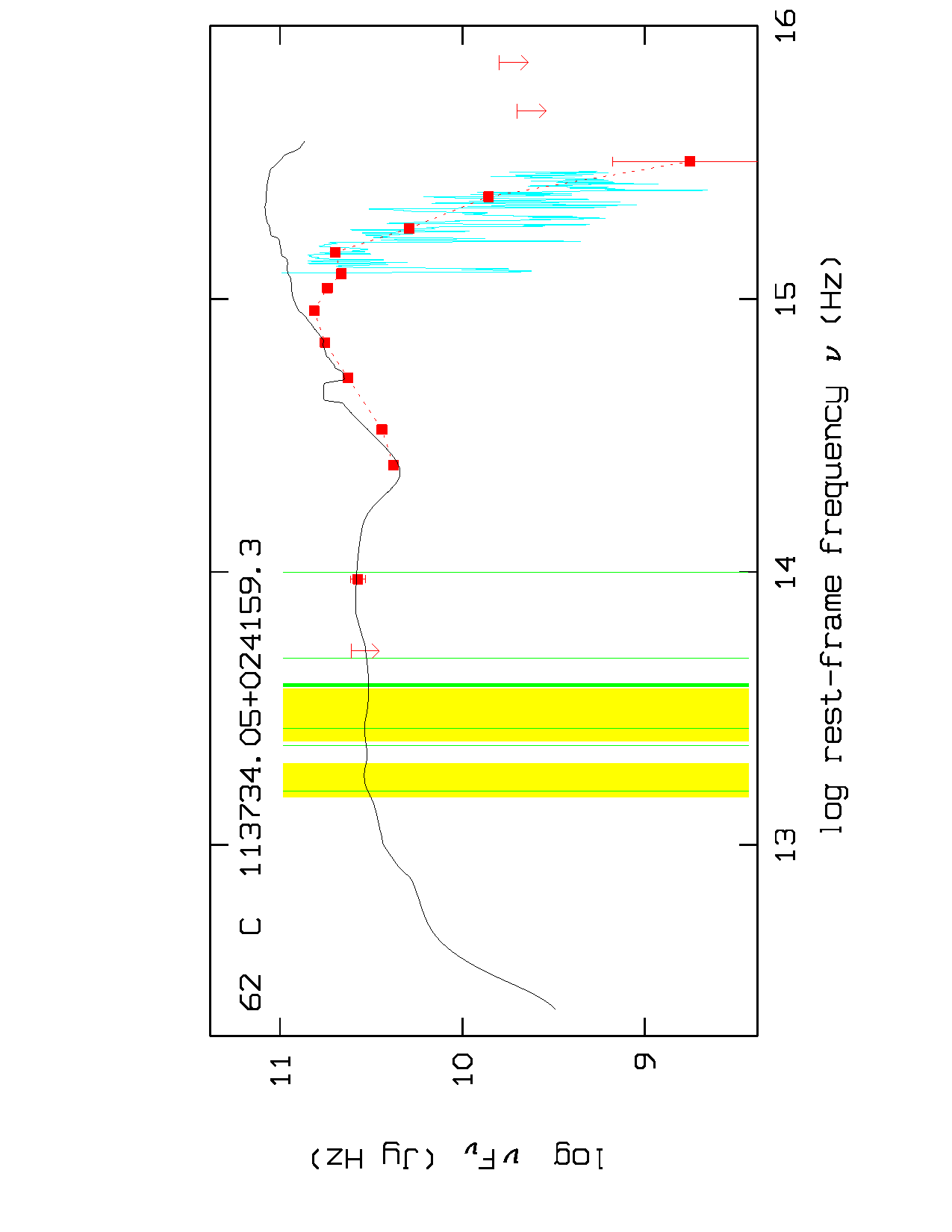}\hfill \\
\includegraphics[viewport=125 0 570 790,angle=270,width=8.0cm,clip]{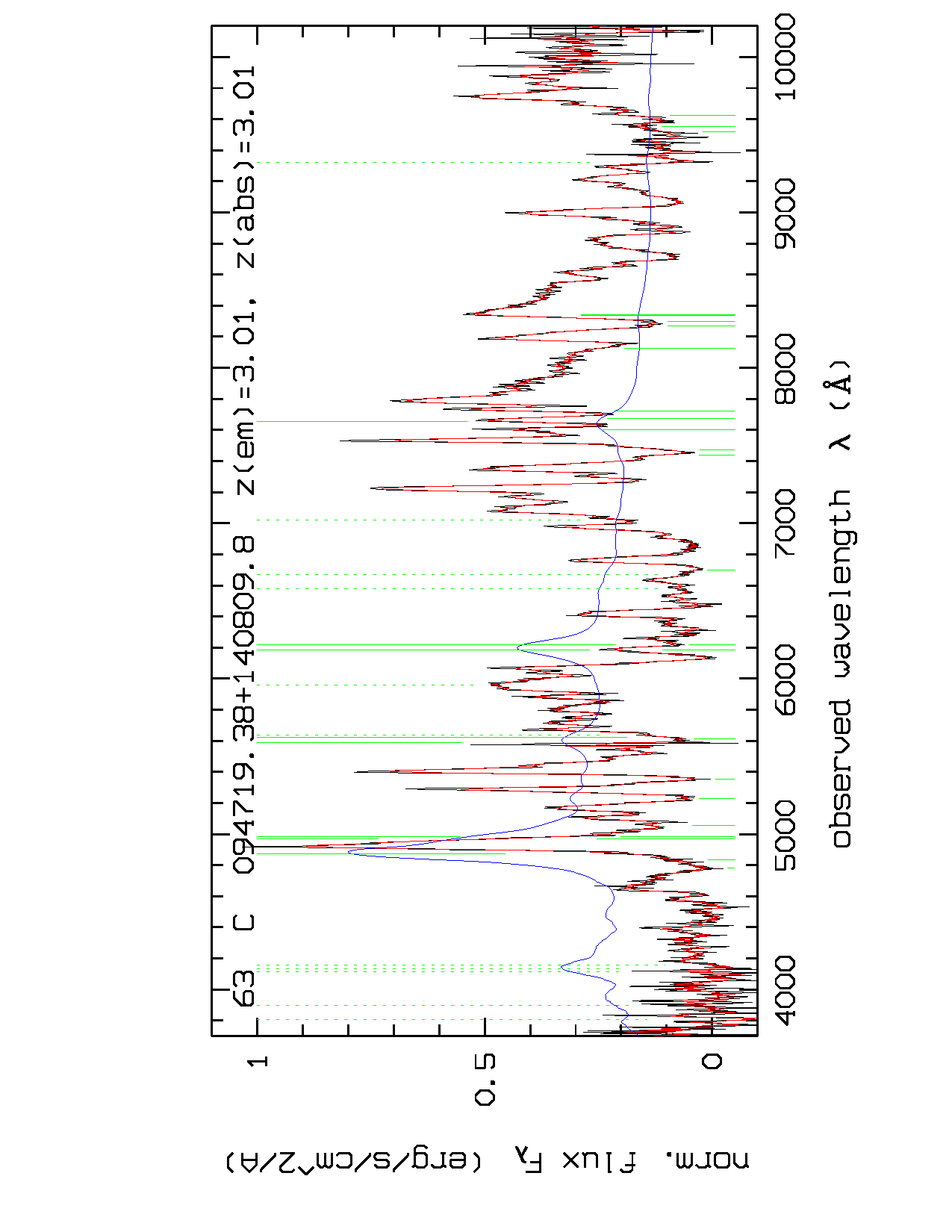}\hfill \=
\includegraphics[viewport=125 0 570 790,angle=270,width=8.0cm,clip]{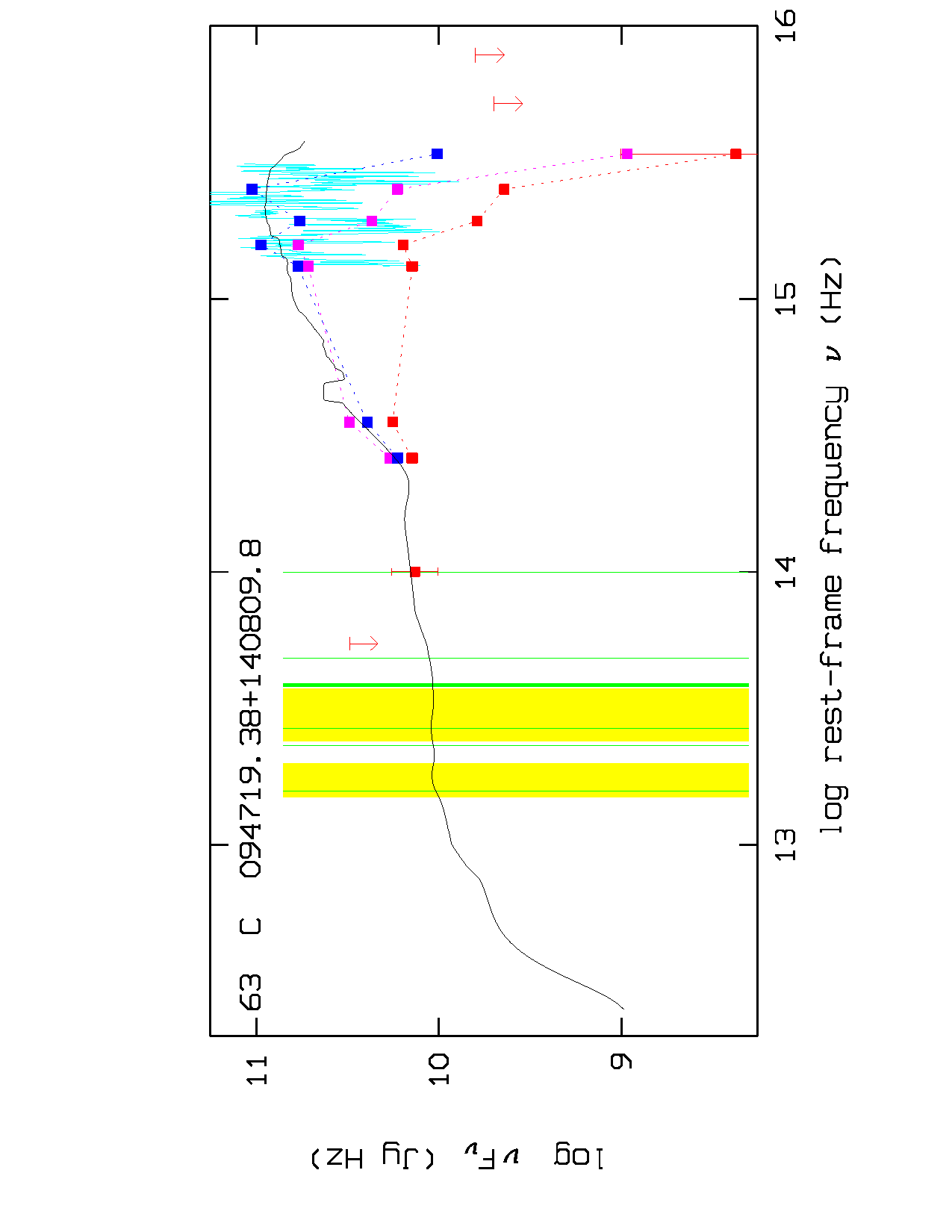}\hfill \\
\end{tabbing}
\caption{Sample C - continued (4).}
\end{figure*}\clearpage

\begin{figure*}[h]
\begin{tabbing}
\includegraphics[viewport=125 0 570 790,angle=270,width=8.0cm,clip]{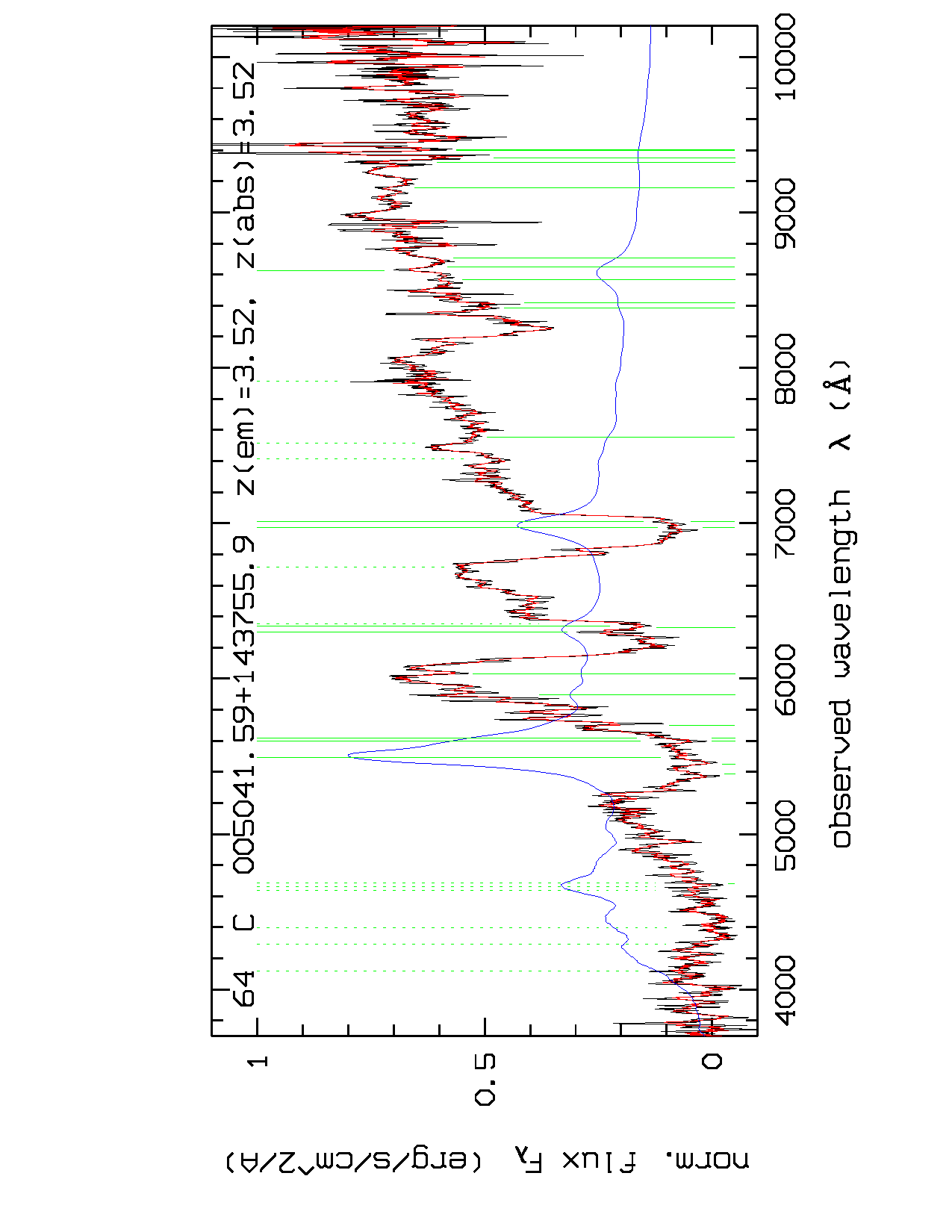}\hfill \=
\includegraphics[viewport=125 0 570 790,angle=270,width=8.0cm,clip]{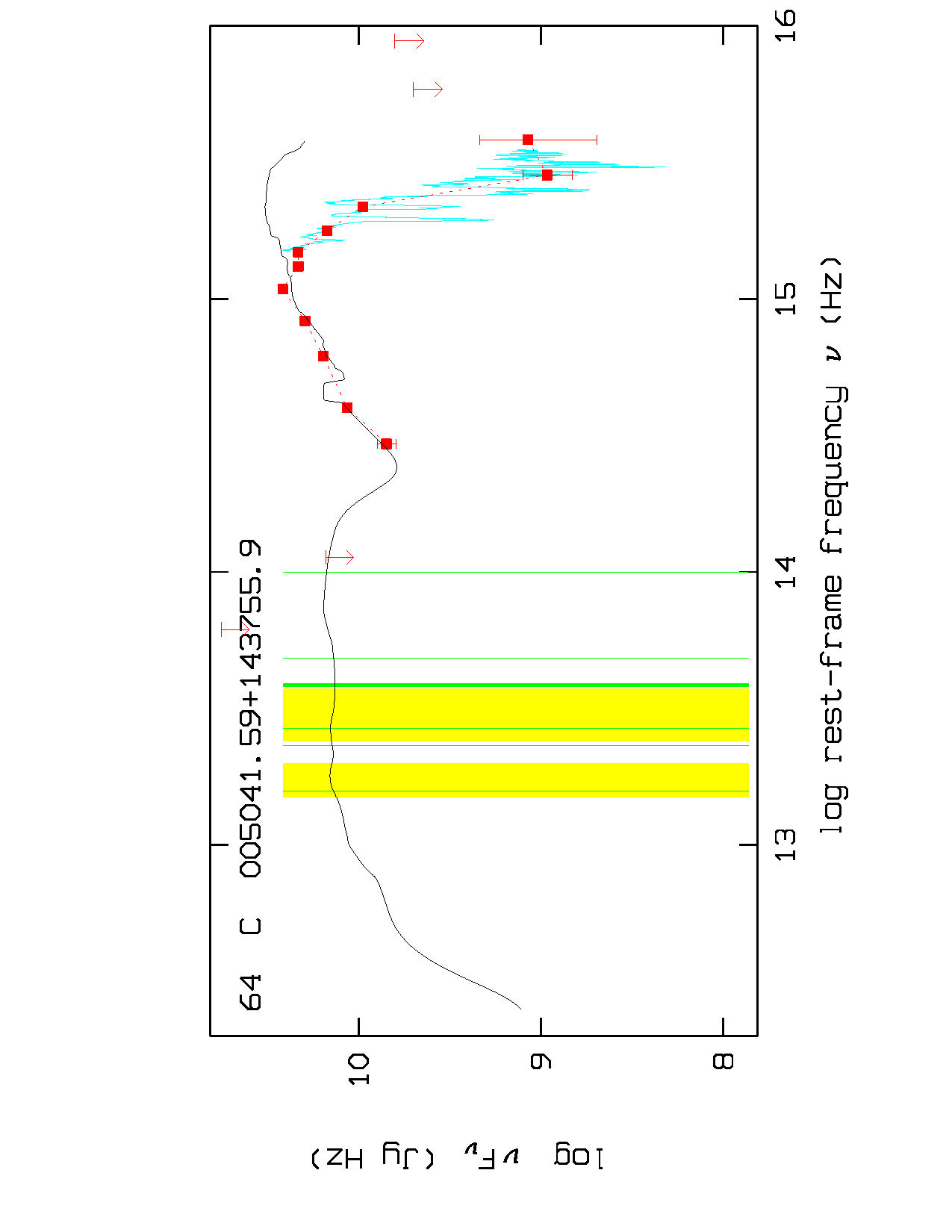}\hfill
\end{tabbing}
\caption{Sample C - continued (5).}
\end{figure*}\clearpage



\end{document}